\documentclass[a4paper,11pt]{article}
\usepackage{jheppub} 
\usepackage{lscape}                     
\usepackage[table]{xcolor}                  
\usepackage{colortbl}
\usepackage{booktabs}                       
\usepackage{multirow,bigdelim}              
\usepackage{etoolbox}
\makeatletter \patchcmd{\maketitle}{\@fpheader}{}{}{} \makeatother
\usepackage{amsmath,amssymb,amsthm,amscd,graphicx,amsfonts}
\usepackage{enumerate}
\usepackage{ytableau}
\usepackage{mathrsfs}
\usepackage{supertabular}
\usepackage{mathtools}
\usepackage{caption}
\usepackage{longtable}
\usepackage{indentfirst}
\usepackage{setspace}

\allowdisplaybreaks
\ytableausetup{mathmode, boxsize=0.4em}

\theoremstyle{definition}

\newtheorem*{conj}{Conjecture}

\newcommand{\und}[1]{{\underline #1}}
\newcommand{\nound}[1]{{ #1}}

\newcommand{\CO}{{\cal O}}

\def\cO{\mathcal{O}}

\def\IZ{{\mathbb Z}}
\def\IR{{\mathbb R}}
\def\IC{{\mathbb C}}
\def\IN{{\mathbb N}}
\def\IP{{\mathbb P}}

\def\IF{{\mathbb F}}

\def\eq{{\epsilon_1}}
\def\et{{\epsilon_2}}

\newcommand{\ep}{\epsilon}

\newcommand{\Qtau}{Q_{\tau}}
\newcommand{\dualCox}{h_G^{\vee}}

\newcommand{\ehgk}{\mathbb{E}_{{h}_{G}^{(k)}}}

\newcommand{\ds}{\displaystyle}

\newcommand{\re}{{\rm e}}
\newcommand{\ri}{\mathsf{i}}

\def\Det{{\rm Det}}
\def\half {{1\over 2}}

\newcommand{\bitem}{\begin{itemize}}
\newcommand{\eitem}{\end{itemize}}
\newcommand{\be}{\begin{equation}}
\newcommand{\ee}{\end{equation}}
\newcommand{\ba}{\begin{aligned}}
\newcommand{\ea}{\end{aligned}}
\newcommand{\ben}{\begin{eqnarray}\displaystyle}
\newcommand{\een}{\end{eqnarray}}

\newdimen\tableauside\tableauside=1.0ex
\newdimen\tableaurule\tableaurule=0.4pt
\newdimen\tableaustep
\def\phantomhrule#1{\hbox{\vbox to0pt{\hrule height\tableaurule width#1\vss}}}
\def\phantomvrule#1{\vbox{\hbox to0pt{\vrule width\tableaurule height#1\hss}}}
\def\sqr{\vbox{%
  \phantomhrule\tableaustep
  \hbox{\phantomvrule\tableaustep\kern\tableaustep\phantomvrule\tableaustep}%
  \hbox{\vbox{\phantomhrule\tableauside}\kern-\tableaurule}}}
\def\squares#1{\hbox{\count0=#1\noindent\loop\sqr
  \advance\count0 by-1 \ifnum\count0>0\repeat}}
\def\tableau#1{\vcenter{\offinterlineskip
  \tableaustep=\tableauside\advance\tableaustep by-\tableaurule
  \kern\normallineskip\hbox
    {\kern\normallineskip\vbox
      {\gettableau#1 0 }%
     \kern\normallineskip\kern\tableaurule}%
  \kern\normallineskip\kern\tableaurule}}
\def\gettableau#1{\ifnum#1=0\let\next=\null\else
\squares{#1}\let\next=\gettableau\fi\next}

\def\IE{\mathbb{E}}

\def\vev#1{\langle #1 \rangle}

\newcommand{\pd}{\partial}

\DeclareMathOperator{\sgn}{sgn}
\def\({\left(}
\def\){\right)}

\numberwithin{equation}{section}

\newcommand{\bs}{\begin{split}}
\newcommand{\es}{\end{split}}
\newcommand{\bdm}{\begin{dmath*}}
\newcommand{\edm}{\end{dmath*}}

\def\mc{\mathcal}
\def\md{\mathbf}
\def\mf{\mathfrak}

\def\({\left(}
\def\){\right)}

\graphicspath{{figs/}}
\def\fq{\mathfrak{q}}

\def\({\left(}
\def\){\right)}
\def\[{\left[}
\def\]{\right]}

\newcommand{\nn}{\nonumber \\}

\def\mc{\mathcal}
\def\md{\mathbf}
\def\mf{\mathfrak}

\def\fg{\mathfrak{g}}
\def\fh{\mathfrak{h}}

\def\fn{\mathfrak{n}}

\newcommand{\bv}{\beta^\vee}
\newcommand{\av}{\alpha^\vee}

\newcommand{\hg}{h^{\vee} _{\fg}}

\newcommand{\nD}{b_4^c}
\newcommand{\nC}{b_2^c}

\newcommand{\mroot}{\vev{{\alpha},{m}}}

\newcommand{\hG}{h^\vee_G}

\title{\boldmath Elliptic Blowup Equations for 6d SCFTs. II:\\ Exceptional Cases}

\author{Jie Gu${}^{a}$, Albrecht Klemm${}^{b,c}$, Kaiwen Sun${^d}$, Xin
  Wang${}^{b,e}$}

\affiliation{
  ${}^a$ D\'epartement de Physique Th\'eorique et Section
  de Math\'ematiques\\
  Universit\'e de Gen\`eve, Gen\`eve, CH-1211 Switzerland\\
  \\
  ${}^b$ Bethe Center for Theoretical Physics and ${}^c$Hausdorff
  Center for Mathematics
  \\Universit\"at Bonn, D-53115 Bonn\\
  \\
  ${}^d$ Scuola Internazionale Superiore di Studi Avanzati (SISSA)\\
  via Bonomea 265, 34136, Trieste, Italy\\
  \\
  ${}^e${Max Planck Institute for Mathematics\\
    Vivatsgasse 7, D-53111 Bonn, Germany} \\}

\emailAdd{jie.gu@unige.ch}
\emailAdd{aklemm@th.physik.uni-bonn.de}
\emailAdd{ksun@sissa.it}
\emailAdd{wxin@mpim-bonn.mpg.de}

\abstract{The building blocks of 6d $(1,0)$ SCFTs include certain rank
  one theories with gauge group $G=SU(3),SO(8),F_4,E_{6,7,8}$. In this
  paper, we propose a universal recursion formula for the elliptic
  genera of all such theories. This formula is solved from the
  elliptic blowup equations introduced in our previous paper. We
  explicitly compute the elliptic genera and refined BPS invariants,
  which recover all previous results from topological string theory,
  modular bootstrap, Hilbert series, 2d quiver gauge theories and 4d
  $\mathcal{N}=2$ superconformal $H_{G}$ theories. We also observe an
  intriguing relation between the $k$-string elliptic genus and the
  Schur indices of rank $k$ $H_{G}$ SCFTs, as a generalization of
  Del Zotto-Lockhart's conjecture at the rank one cases. In a subsequent
  paper, we deal with all other non-Higgsable clusters with matters.}
\begin{document}
\maketitle

\section{Introduction}
\label{sc:intro}

Six is the highest dimension in which representation theory allows for interacting
superconformal quantum theories~\cite{Nahm:1977tg}. Limits of non-perturbative
string theory compactifications~\cite{Ganor:1996mu}  and in particular the decoupling
of gravity in F-theory compactifications  to 6d provided the first
examples~\cite{Klemm:1996hh,Morrison:1996pp} and lead recently
to a complete classification of geometrically engineered 6d superconformal
quantum field theories~\cite{Heckman:2013pva,Heckman:2015bfa,Bhardwaj:2015xxa}. Such a classification in 6d
is highly desirable, as it might lead by further compactifications, to an exhaustive
classification of superconformal theories.

The 6d geometry is the one of an --- in general desingularised ---
elliptic fibration with a contractable configuration of desingularised
elliptic surfaces fibred over a configuration of curves in the
base. In the decoupling limit the volume outside of the configuration
of elliptic surfaces is scaled to infinite size, leaving us with an,
in general reducible, configuration of complex desingularised elliptic
surfaces that can be contracted within a non-compact Calabi-Yau
threefold. Because compact components can be contracted such
geometries are sometimes called local Calabi-Yau spaces. We will call
the above specific ones for short \emph{elliptic non-compact
  Calabi-Yau geometries $X$} and describe them in more detail in
section \ref{sc:geometry}.

The full topological string partition function on these elliptic
non-compact CY geometries has received much attention as it contains
important information about protected states of the 6d superconformal
theories~\cite{Klemm:1996hh,Haghighat:2014vxa}. Solving the
topological string partition function on compact Calabi-Yau manifolds
is currently an open problem. On non-compact Calabi-Yau spaces with an
$U(1)_R$ isometry a refined topological string partition function
$Z({\underline t},\epsilon_1, \epsilon_2)$, which depends on the
K\"ahler parameters ${\underline t}$ and two $\Omega$ background
parameters $\epsilon_1, \epsilon_2$ is defined as generating function
of refined stable pair invariants.\footnote{In this section we
  underline a symbol, if it is a vector. After the introduction
  section, we drop the underline when there is no risk of confusion.}
The refinement of the stable pair
invariants~\cite{MR2545686,MR2552254} and the relation to the refined
BPS invariants $N_{j_l,j_r}^\beta\in \mathbb{N}$ was given
in~\cite{Choi:2012jz,MR3504535}. Here $\beta\in H_2(X,\mathbb{Z})$ is
the degree and the half integer $(j_l,j_r)$ label a spin
representation in the $SU(2)_l\times SU(2)_r$ little group of the 5d
Poincar\'e group, which can be identified with Lefschetz actions on
the moduli space of D2 and D0-branes. On toric non-compact Calabi-Yau
spaces the refined partition function\footnote{The holomorphic all
  genus partition function
  $Z({\underline t}, \lambda)= \exp(\sum_{g=0}^\infty \lambda^{2g-2}
  F_g({\underline t}))$ containing the information of all genus
  Gromov-Witten invariants is obtained as specialisation
  $\epsilon_1=-\epsilon_2$ and $\lambda^2=-\epsilon_1\epsilon_2$.}
$Z({\underline t},\epsilon_1, \epsilon_2)$ can be efficiently
calculated by large $N$
techniques~\cite{Iqbal:2007ii}\footnote{Strictly speaking the refined
  topological vertex applies directly only to geometries which
  engineer $\mathcal{N}=2$ gauge theories, as these have the required
  preferred direction in the torus action. In blow downs and
  transitions of gauge theories geometries with Chern-Simons terms to
  geometries which have no immediate gauge theory interpretation,
  $Z({\underline{ t}},\epsilon_1, \epsilon_2)$ for the latter can
  often be recovered~\cite{Iqbal:2007ii}.}, torus
localisation~\cite{Choi:2012jz}, the integration of the refined
holomorphic anomaly equations~\cite{Huang:2010kf} and a recursive
solution of blowup equations~\cite{Huang:2017mis} generalized from the
G\"{o}ttsche-Nakajima-Yoshioka K-theoretic blowup equations in the
context of 5d $\mathcal{N}=1$ supersymmetric gauge
theories~\cite{Nakajima:2005fg,Gottsche:2006bm,Nakajima:2009qjc}.

The class of elliptic non-compact Calabi-Yau relevant for the $(1,0)$
6d SCFT is non-toric, but has a $U(1)_R$ isometry, and can be viewed
as the borderline case for calculating
$Z({\underline t},\epsilon_1, \epsilon_2)$. Since the techniques based
on toric localisation and large $N$ expansions fail, two related new
methods have been developed.  Similar as in heterotic/Type II duality
one can calculate\footnote{The more supersymmetric case of the
  M-strings has been pioneered
  in~\cite{Haghighat:2013gba}.}~\cite{Haghighat:2014vxa} the
world-sheet elliptic genus of dual 2d quiver $(0,4)$ gauge theories
with supersymmetric localisation
techniques~\cite{Gadde:2013ftv,Benini:2013xpa} leading to
Jeffrey-Kirwan integrals. These elliptic genera
$\IE_d(\tau, {\underline a},{\underline m}, \epsilon_1,\epsilon_2)$
transform as a Jacobi form and are identified with the topological
string partition function $Z_d$ at different winding $d$ of the
base~\cite{Haghighat:2014vxa} up to certain prefactor. The K\"ahler
parameter $\tau$ of the elliptic fibre class becomes the modular
parameter while $(\epsilon_1, \epsilon_2)$ as well as K\"ahler
parameters $\underline a$ of the desingularisations and eventual
further sections $\underline m$ in the elliptic fibration become
elliptic parameters.  The refined holomorphic anomaly equations and
other B-model techniques also apply and lead to a modular bootstrap
approach where different winding contributions $Z_d$ are identified
with meromorphic Jacobi forms with weight zero and an index, which
depends quadratically on the base degree $d$. The $Z_d$ are so
constrained by modularity, the pole -- as well as the refined BPS
structure of the topological string that they can be completely
reconstructed in many
examples~\cite{Huang:2015sta,Gu:2017ccq,DelZotto:2017mee}.

In the 2d approach one needs for higher $d$ to consider ever more
complicated quiver gauge theories, while in the modular approach one
has to deal with more and more complicated rings of weak Jacobi forms.
For this reason we further develop in this paper the recursive
approach based on the elliptic blowup equations~\cite{Gu:2018gmy} for
the calculation of $Z({\underline t},\epsilon_1, \epsilon_2)$ that is
further based on a specialisation of the generalized blowup equation
in \cite{Huang:2017mis} to the elliptic non-compact Calabi-Yau
geometries. The main advantage of this approach is that it needs as
input only\footnote{This holds for all non-compact CY 3-folds studied
  in \cite{Huang:2017mis} and elliptic non-compact CY 3-folds studied
  in \cite{Gu:2018gmy}. For the elliptic non-compact CY 3-folds
  associated to exceptional gauge symmetry studied in this paper, we
  also input $Z_0$ to the blowup equations, which can be easily
  calculated from the intersection and the multi-covering of isolated
  rational curves, see (\ref{eq:Zexp-1}) below.} the classical
topological data of $X$, i.e. the classical triple intersection
numbers as well as the evaluation of the Chern classes on the elements
of the Chow group, and yields with a non ambiguous efficient recursive
procedure the string partition function iteratively in the base degree
$d$ and for each $d$ exact in $(\epsilon_1, \epsilon_2)$ and all other
K\"ahler parameters.

Let us first give a short summary of the structure behind the blowup
equations in four, five and six dimensions. Non-compact Calabi-Yau
spaces with $U(1)_R$ isometry and the (refined) topological string
partition function feature prominently in the geometric engineering
approach~\cite{Katz:1996fh} to 5d and 4d supersymmetric gauge theories
as $Z({\underline t},\epsilon_1, \epsilon_2)$ is related with the
K-theoretic extension of Nekrasov's 4d gauge theory instanton
partition~\cite{Nekrasov:2002qd} on those non-compact Calabi-Yau
spaces~\cite{Nakajima:2005fg,Gottsche:2006bm}, which do engineer
supersymmetric gauge theories. In the geometric engineering approach,
given mirror symmetry, it is physically obvious that world-sheet
instantons and space time instantons are related. Simply because the
former correct the topological string theory or ${\cal N}=2$
supergravity prepotential, while the latter correct the rigid
${\cal N}=2$ or Seiberg-Witten prepotential, which is related to the
former in a well defined limit in the B-model, that decouples gravity
as decribed in~\cite{Katz:1996fh}.  If the geometry engineers five
dimensional $U(N)$ gauge theory, the full correspondence states that
the K-theoretic partition function of the latter is identified with
$Z({\underline t},\epsilon_1, \epsilon_2)$ and provides an alternative
definition~\cite{Nakajima:2005fg,Nakajima:2009qjc}. The K-theoretic
blowup equation for $U(N)$ theories without or with Chern-Simons terms
has been rigorously established in~\cite{Nakajima:2005fg}
and~\cite{Gottsche:2006bm,Nakajima:2009qjc} respectively.

Nakajima and Yoshioka derived the original blowup
equations~\cite{Nakajima:2003pg} in the context of 4d $\mathcal{N}=2$
supersymmetric $SU(N)$ framed gauge instanton calculus, by studying
invariants on moduli space $\widehat M(N,k,n)$ of framed torsion free
sheaves $(E,\Phi)$ on $\widehat { \mathbb{P}}^2$ -- a $\mathbb{P}^1$
blowup of $\mathbb{P}^2$ -- via the Atiyah-Bott localization formalism
w.r.t. an induced toric action
$T=\mathbb{C}^*\times \mathbb{C}^*\times {\rm Gl}_N$ on
$\widehat M(N,k,n)$. Here $\Phi$ is the framing automorphism, $N$ is
the rank of $E$,
$n=\langle c_2(E)-\tfrac{N-1}{2N}c_1^2(E),[\widehat
{\mathbb{P}}^2]\rangle$ and $k=-\langle
c_1(E),[\mathbb{P}^1]\rangle$. A general feature of this calculation
is that the Euler class of the tangent space of $\widehat M(N,k,n)$ at
all relevant fix loci is always a product of contributions from two
fix points at the north and the south poles of the exceptional
$\mathbb{P}^1$, which arise due to the action of the
$\mathbb{C}^*\times \mathbb{C}^*$ on $\widehat{ \mathbb{P}}^2$
parametrized by $\epsilon_1$ and $\epsilon_2$. Upon evaluation of the
Atiyah-Bott localization formula the two contributions yield -- up to
calculable factors -- a sum of products of the original partition
function on $\mathbb{P}^2$ at shifted $\epsilon_{1,2}$ and Coulomb
branch (in type IIA normalisable K\"ahler) parameters. The partition
function on $\widehat { \mathbb{P}}^2$ can be also directly
specialised for $k=0$ to the one on $\mathbb{P}^2$.  The
identification of the two results gives rise to a finite set of
equations for the partition function of the 4d supersymmetric theory
on the Omega background. A similar mechanism applies to the
K-theoretic instanton
calculus~\cite{Nakajima:2005fg,Nakajima:2009qjc}, and leads to blowup
equations for the partition function of the 5d SYM on the Omega
background. The latter setup is directly relevant to the calculation
of $Z({\underline t},\epsilon_1, \epsilon_2)$ on non-compact
Calabi-Yau engineering supersymmetric gauge theories.  The blowup
equations can then be reformulated in terms of the geometric data of
the non-compact Calabi-Yau $X$ and refined topological string
partition function as follows.

Let $C = (C_{ij})$ be the intersection matrix between compact divisor
classes $[\mf D_i]$, $i=1,\ldots,b_4^c$
and compact curve classes $[\Sigma_j]$, $j=1,\ldots, b^c_2$ of
$X$. Then one defines vector
\begin{equation}
  \und{R}_\und{n}=C \cdot \und{n}+\und{r}/2,
\end{equation}
with $\und{n}\in \mathbb{Z}^{b_4^c}$ and
$\und{r}\in \mathbb{Z}^{b_4^c}$ which parametrise the shift of the
K\"ahler parameters. With $|\und{n}|=\sum_{i=1}^{b_4^c} n_i$, the
generalized blowup equations can be cast in the following
form~\cite{Huang:2017mis} (see also section 8 of \cite{Gu:2017ccq} and
\cite{Grassi:2016nnt})
\begin{equation}
  \! \sum_{\und{n}\in \mathbb{Z}^{b_4^c}} \! (-1)^{|\und{n}|} \!
  \widehat{Z}({\und{t}}+\epsilon_1 \und{R}_\und{n},\epsilon_1,
  \epsilon_2-\epsilon_1)  \widehat{Z}(\und{t}+\epsilon_2 \und{R}_\und{n}
  ,\epsilon_1-\epsilon_2, \epsilon_2) =\begin{cases}
    0, \! & \! \und{r} \in {\cal S}_v, \\
    \Lambda(\epsilon_1,\epsilon_2,\und{m}, \und{r}) Z({\underline
      t},\epsilon_1, \epsilon_2)\! & \! \und{r} \in {\cal S}_u .
\end{cases} \ \label{eq:blowupI}
\end{equation}
Here we have separated the K\"ahler parameters $\und{m}$ from the
K\"ahler parameters $\und{t}$ to denote those curve classes that do
not intersect with compact divisors $[\mf D_k]$,
$k=1,\ldots,{b_4^c}$. These $\und m$ correspond to mass parameters in
the gauge theory context, while the other K\"ahler parameters
correspond to Coulomb branch parameters, thus are also called ``true''
parameters. If a local mirror curve exists, e.g. for non-compact toric
Calabi-Yau spaces, the ``true'' K\"ahler parameters are mapped to the
complex structure parameters of the (hyperelliptic) mirror curve (of
genus $g=b_4^c$) and the $\und{m}$ correspond to the residues of the
meromorphic differential $\lambda$. The hat over $Z$ means the
K\"ahler moduli in the instanton partition function have already been
shifted
\begin{equation}
  \widehat{Z}(\und{t},\eq,\et) = Z^{\text{cls}}(\und{t},\eq,\et)
  Z^{\text{inst}}(\und{t}+\pi\ri\und{r},\eq,\et) \ .
\end{equation}
The integral vector $\und{r}$, which we call the
$\und{r}$-field, is consistent with the
checkerboard pattern of refined BPS invariants $N_{j_l,j_r}^{\beta}$,
in other words, they satisfy
\begin{equation}
  2j_l + 2j_r+ 1 \equiv \und{r}\cdot\beta \quad \text{mod}\;\;2
\end{equation}
for non-vanishing $N_{j_l,j_r}^{\beta}$. The set of $\und{r}$-fields
in (\ref{eq:blowupI}) have to be only considered modulo
$2 C\cdot \und{n}$, which leaves two classes of finite sets
${\cal S}_v$ and ${\cal S}_u$. The $\und{r}$-fields in these two sets
are called \emph{vanishing} and \emph{unity} $\und{r}$-fields. It is
important that $\Lambda$, whose form is known, depends beside on
$\epsilon_{1,2}$ only on $\und{r}$-fields in the two classes and the
mass parameters $\und{m}$.

Given the simple form of \eqref{eq:blowupI} and the method of proof in
the gauge theory context ~\cite{Nakajima:2005fg,Nakajima:2009qjc}, it
seems reasonable to
conjecture~\cite{Gu:2017ccq,Huang:2017mis,Gu:2018gmy} that these
equations, called the \emph{generalised blowup equations}, should hold
for the refined partition functions
$Z({\underline t},\epsilon_1, \epsilon_2)$ of all non-compact
Calabi-Yau threefolds with a global $U(1)_R$ symmetry so that the
refined invariants or equivalently the corresponding BPS index for the
space time theory with an $\Omega$ background can be
defined~\cite{Choi:2012jz,MR3504535}. At the technical level the
precise non-trivial claim is that $\mc S_v\cup \mc S_u$ should be
non-empty. In addition it was observed in~\cite{Huang:2017mis} that
the classical topological data of $X$ mentioned above and the genus
zero sector \emph{determine}
$Z({\underline t},\epsilon_1, \epsilon_2)$ recursively, and many
examples were already checked in great detail. In particular in
\cite{Gu:2018gmy} this approach was used to compute the refined BPS
invariants of elliptic non-compact Calabi-Yau geometries associated to
minimal 6d $(1,0)$ SCFTs with gauge group $G=SU(3),SO(8)$. In this
paper we extend this approach to the remaining minimal 6d SCFTs with
exceptional gauge groups $G = F_4, E_{6,7,8}$ and give a universal
description for all minimal building blocks without matter in the
classification of 6d $(1,0)$
SCFTs~\cite{Heckman:2013pva,Heckman:2015bfa}.

The elliptic non-compact Calabi-Yau geometry corresponding to minimal
SCFTs with gauge group $G$ and no matter contains base surface
$\mathcal{O}(-\fn)\to \IP^1$. For $\fn=3,4,5,6,8,12$ which are of
interest in this paper, the Kodaire singularity of the elliptic
fibration gives the gauge group $G=SU(3),SO(8),F_4,E_{6,7,8}$
respectively. We find that for these geometries, the generalized
blowup equations can be uniformly written as the following recursive
relations of the elliptic genera $ \IE_{d}$ of the corresponding 6d
SCFT:
\begin{align}
  \sum_{\substack{\und{\omega}\in \phi_\und{\lambda}(Q^\vee),d_{1,2}\in\IN
  \\
  }}^{\frac{1}{2}||\und{\omega}||^2+d_1+d_2 =d}
  &(-1)^{|\phi_\und{\lambda}^{-1}(\und{\omega})|} \theta_i^{[a]}
    \Big(\fn \tau,(\fn-2)(\epsilon_1+\epsilon_2)
    -\fn\big((\tfrac{1}{2}||\und{\omega}||^2+d_1)\epsilon_1
    +(\tfrac{1}{2}||\und{\omega}||^2+d_2)\epsilon_2- \und{m}\cdot
    \und{\omega}\big)\Big) \nn
  &\times A_{\und{\omega}}(\und{m})
    \IE_{d_1}(\tau, \und{m}-\epsilon_1\und{\omega},
    \epsilon_1,\epsilon_2-\epsilon_1)
    \IE_{d_2}(\tau, \und{m}-\epsilon_2\und{\omega},
    \epsilon_1 -\epsilon_2,\epsilon_2)
    \nn =
  &\begin{cases}
    0, &\quad d\not \in \IN, \\
    \theta_i^{[a]}(\fn\tau,(\fn-2)(\epsilon_1+\epsilon_2))\cdot
    \IE_d(\tau, \und{m},\epsilon_1,\epsilon_2),\quad &\quad d\in \IN.
    \end{cases} \ \label{eq:uv-blowup0}
\end{align}
Here $\und m$ are the Coulomb parameters\footnote{Do not confuse with $\und{m}$ in
  \eqref{eq:blowupI}. The Coulomb parameters are \emph{not} mass
  parameters.} associated to gauge group $G$. The subscript of theta
functions $i$ is $4$ if $\fn$ is odd and $3$ if $\fn$ is even, and the
characteristic $a = k/\fn-1/2$, $k=0,1,\ldots,\fn-1$. Besides,
$\phi_{\und{\lambda}}$ is an embedding of the coroot lattice $Q^\vee$
of $G$ into the weight lattice $P$. Here the $\und{r}$-field is
implicit in $a$ and $\phi_{\und{\lambda}}$. 
Since the
number of different embeddings is $|P:Q^\vee|$, the total number of
non-equivalent blowup equations is $\fn|P:Q^\vee|$. The function
$A_{\und{\omega}}(\und{m})$ is composed of $\theta_1$ and $\eta$
functions, see (\ref{Aomegam}) for the definition. Due to the Jacobi
form nature of every component of the above equations, we call
(\ref{eq:uv-blowup0}) as \emph{elliptic blowup equations}. In fact,
they can be regarded as the natural elliptic lift of the K-theoretic
blowup equations for 5d gauge theories
\cite{Nakajima:2005fg,Keller:2012da}. Moreover, the unity elliptic
blowup equations in \eqref{eq:uv-blowup0} ultimately lead to a
complete solution of the elliptic genera $\IE_d$ in terms of an
universal recursion formula, as will be shown in
\eqref{recursionZd}. The blowup equations are not only effective tools
to calculate the refined partition functions, but also together with
the general constraints from modularity and BPS structure shed some
new light on the structure of $Z(\und{t},\epsilon_1, \epsilon_2)$.  In
particular it is possible to derive from the structure of the blowup
equations the index and the weight of the Jacobi forms that constitute
the building blocks in (\ref{eq:uv-blowup0}). 
The recursive structure also
helps clarify the form of the denominator of elliptic genus in
the modular boostrap approach as discussed in
Appendix~\ref{ap:mod-ansatz}, and predicts many non-trivial relations
among these Jacobi forms, one particular of which is proven in section
\ref{sc:rec}.

In the program of classifying superconformal field theories in various
dimensions the 6d SCFTs play a similar role as 11d $\mathcal{N}=1$
supergravity or more precisely M-theory play for the classification of
supergravity in lower dimensions. For this reason we expect that
various limits as well as suitable expansion of the partition function
$Z(\underline{t},\epsilon_1,\epsilon_2)$ of the 6d $(1,0)$ minimal
SCFTs relate to the protected quantities in lower dimensional
supersymmetric theories.

Since the elliptic blowup equations determine
$Z(\underline{t},\epsilon_1,\epsilon_2)$ in particular the elliptic
genus $\IE_d$ completely, we could make many detailed and indeed
successful checks on our results. We summarise the current status of
the knowledge on the elliptic genera of all 6d $(1,0)$ minimal SCFTs
from various approaches in Table~\ref{tb:summary}. For $\fn=5,6,8,12$
which are of main interest in the current paper, our complete
recursive solution for the elliptic genera from blowup equations
reproduces \emph{all} previous partial results. Since we made many
checks of the elliptic blowup equations based on extensive
calculations, which might yield further insights, we provide the
results of these calculations on a webpage~\cite{kl}.

\begin{table}
  \centering
  {\footnotesize{ \begin{tabular}{|c|c|c|c|c|c|c|c|c|c|}\hline
        \textrm{$n$} & 1& 2& 3 & 4 & 6 & 8& 12& 5 & 7 \\\hline
        features & \textrm{E-strings} & \textrm{M-strings} & $SU(3)$ &
        $SO(8)$ & $E_6$ & $E_7$ & $E_8$ & $F_4$& $E_7+\half{\bf 56}$
        \\\hline 2d quiver & $\mathbb{E}_k\,$\cite{Kim:2014dza,Kim:2015fxa} &
        $\mathbb{E}_k\,$\cite{Haghighat:2013gba} &
        $\mathbb{E}_k\,$\cite{Kim:2016foj} &
        $\mathbb{E}_k\,$\cite{Haghighat:2014vxa} &
        \multicolumn{5}{c|}{?} \\\hline B-model & low genus
        \cite{Huang:2013yta} & ? & \multicolumn{7}{c|}{genus zero
          \cite{Haghighat:2014vxa}} \\ \hline modular bootstrap &
        \multicolumn{2}{c|}{$\IE_k$ \cite{Gu:2017ccq,Duan:2018sqe}} &
        \multicolumn{2}{c|}{$\mathbb{E}_1\,$\cite{DelZotto:2016pvm,DelZotto:2017mee}}
        & \multicolumn{5}{c|}{$\mathbb{E}_1$ with fugacities off
          \cite{DelZotto:2016pvm,DelZotto:2018tcj}} \\\hline
        topological vertex & $\mathbb{E}_k\,$\cite{Kim:2015jba} &
        $\mathbb{E}_k\,$\cite{Haghighat:2013gba} &
        \multicolumn{5}{c|}{$\mathbb{E}_k^{q\to
            0}\,$\cite{Hayashi:2017jze}} & \multicolumn{2}{c|}{?} \\
        \hline Hilbert series & - & - &
        \multicolumn{6}{c|}{$\mathbb{E}_k^{q\to 0}$
          \cite{Benvenuti:2010pq,Hanany:2012dm,Cremonesi:2014xha}} & -
        \\\hline HL index & - & - & ? &
        \multicolumn{4}{c|}{$\mathbb{E}_k^{q\to 0}$
          \cite{Gadde:2011uv,Gaiotto:2012uq}} & \multicolumn{2}{c|}{-}
        \\\hline twisted $H_G$ theories & - & - & ? &
        \multicolumn{2}{c|}{$\mathbb{E}_1\,$\cite{Putrov:2015jpa,DelZotto:2016pvm,Gadde:2015xta}}
        & $\mathbb{E}_1\,$\cite{Agarwal:2018ejn} & ? &
        \multicolumn{2}{c|}{-} \\\hline domain walls &
        $\mathbb{E}_{1,2}\,$\cite{Haghighat:2014pva}
        $\mathbb{E}_3\,$\cite{Cai:2014vka} & $\mathbb{E}_k$
        \cite{Haghighat:2014pva} & \multicolumn{7}{c|}{-} \\\hline 5d
        blowup equations & \multicolumn{2}{c|}{trivial} &
        \multicolumn{6}{c|}{$\mathbb{E}_k^{q\to 0}$
          \cite{Keller:2012da}} & $\mathbb{E}_k^{q\to 0}\,$\cite{GKSW3} \\\hline 6d blowup equations &
        $\mathbb{E}_k$ \cite{Gu:2017ccq,Huang:2017mis,GHSWunpublished}
        & $\mathbb{E}_k\,$\cite{GHSWunpublished}  & \multicolumn{2}{c|}{$\mathbb{E}_k$ \cite{Gu:2018gmy}} &
        \multicolumn{4}{c|}{$\mathbb{E}_k$, \textrm{current paper}} &
        $\mathbb{E}_k\,$\cite{GKSW3} \\\hline
                 \end{tabular}}}
             \caption{Known results on the elliptic
               genera of 6d minimal (1,0) SCFTs from various approaches. Here
               $q=\Qtau = \re^{2\pi\ri\tau}$ is the modular
               parameter. - means the method does not apply, and ?
               means possible applicable but results not yet
               attained. HL means Hall-Littlewood.} \label{tb:summary}
\end{table}

Part of these checks indicated in Table~\ref{tb:summary} are quite
obvious as for example the 5d limit gives a good confirmation of
our results. Others are highly non-trivial and indicate new exciting
connections to the protected quantities in lower dimensional theories.
For example one of the most important tools for the analysis of the
spectra and phenomena like Seiberg duality in four dimensional SCFTs
are the superconformal indices for $\mathcal{N}=1,2,4$ SCFT, which
count operators in the chiral rings of these theories. These indices have in turn various
limits such as Macdonald indices, Hall-Littlewood indices and Schur
indices which are relatively easy to compute. As explained in section
\ref{sec:HGk} the latter two occur in a quite non-trivial manner in
the expansion of the elliptic genera that we can efficiently
calculate. This surprising relation between elliptic genera and
superconformal indices was found for the rank one $H_G$ theories in
\cite{DelZotto:2016pvm}. We will push the study on such relation for
all rank two and even some rank three cases. This not only sheds light
on the structure of these objects, but also allows to calculate them
efficiently for example in theories with no Lagrangian description in
which other methods are quite difficult to carry through.

This paper is organized as follows: in Section \ref{sc:gbp}, we review
the geometric construction of elliptic non-compact Calabi-Yau
threefolds that engineer 6d $(1,0)$ minimal SCFTs, the basic
properties of the generalized blowup equations in
\cite{Huang:2017mis}, and the de-affinisation procedure which was
essentially already used in \cite{Gu:2018gmy} to obtain elliptic
blowup equations for $G=SU(3)$ and $SO(8)$. In Section \ref{sc:ebp},
we discuss both unity and vanishing elliptic blowup equations in
detail, and derive a universal recursion formula for the elliptic
genera of all minimal SCFTs with $G=SU(3),SO(8),F_4,E_{6,7,8}$. We
also prove two important properties of the elliptic blowup equations,
i.e. modularity and universality. In Section \ref{sc:eg}, we
explicitly show for each $G$ the one- and two-string elliptic genera
computed from our universal recursion formula and also some relevant
information for the blowup equations, such as triple intersection
numbers and the $\und r$-fields. In Section \ref{sec:HGk}, we discuss
a surprising relation between the elliptic genera of 6d minimal SCFTs
and the Hall-Littlewood indices and Schur indices of 4d
$\mathcal{N}=2$ superconformal $H_G$ theories, as was revealed for
rank one in \cite{DelZotto:2016pvm}. We find analogous relation indeed
exist for rank two and higher. Finally, in Section \ref{se:outlook},
we discuss various possible application and future
directions. 
In a series of appendices we explain our
convention, some technical details, and collect more results on
elliptic genera and refined BPS invariants too lengthy to be put in
the main text.

\section{Elliptic non-compact CY 3-folds and generalised blowup equations}
\label{sc:gbp}

The \emph{generalised blowup equations} proposed
in~\cite{Huang:2017mis} (see also section 8 of~\cite{Gu:2017ccq} and
\cite{Grassi:2016nnt}) generalise the K-theoretic blowup equations of
Nakajima and Yoshioka~\cite{Nakajima:2005fg,Nakajima:2009qjc} for 5d
SYM theories to all non-compact Calabi-Yau geometries that have an
$U(1)_R$ isometry which may or may not engineer 5d supersymmetric
field theories. In subsection \ref{sc:geometry} we describe the
geometric data of the non-compact elliptic Calabi-Yau threefolds
associated to the minimal 6d SCFTs with pure gauge bulk theory.  With
this input and $b_i^{ns}$ calculated in subsection \ref{sc:Zpert}, the
$r$ fields can be determined in subsection~\ref{sc:adm} and the
generalised blowup equations can be expanded to extract BPS
constraints that allow for solution of refined BPS invariants as in
\cite{Huang:2017mis}.

We then consider the expansion of the partition partition function in the
base degrees and describe how to recast the generalised blowup equations,
with some additional input, as functional equations of elliptic
genera of the 6d SCFTs. The latter, which we call the \emph{elliptic
blowup equations}, will be discussed in full detail in the next section.

\subsection{Geometry of elliptic fibrations}
\label{sc:geometry}

In this paper we are specifically interested in the non compact
elliptic Calabi-Yau threefolds on which F-theory compactification
yields minimal 6d $(1,0)$ SCFTs with $\fn=3,4,5,6,8,12$ so that the
bulk theory has a pure gauge group $G$.

We discuss some generic features of these Calabi-Yau threefolds, in
particular the compact curves and the compact divisors in the
Calabi-Yau. The compact curves and compact divisors in these
geometries are best illustrated in \cite{DelZotto:2017pti} and
summarised in \cite{Gu:2018gmy}. Let us go over them here
quickly. Suppose the gauge group $G$ has rank $\mathrm{rk}(G)$ and its
associated Lie algebra is $\fg$, then there are
$\nC = \mathrm{rk}(G)+2$ linearly independent compact curves in the
Calabi-Yau threefold.\footnote{From now on, we simply denote
  $\mathrm{rk}(G)$ as $\mathrm{rk}$ to lighten the notation.}  One of
them is the $-\fn$ curve $\Sigma_B$ in the base, while the remaining
$\mathrm{rk}+1$ curves $\Sigma_I$ $(I=0,1,\dots,\mathrm{rk})$ are
$\IP^1$s resulting from resolution of the singular elliptic fiber
fibered over $\Sigma_B$. These $r+1$ curves intersect with each other
according to the affine Dynkin diagram of $\fg$, where each node
corresponds to a rational curve of self-intersection $-2$ and two
curves intersect with intersection number $1$ if the corresponding
nodes are linked. We denote the curve corresponding to the affine node
by $\Sigma_0$, and it is the only $\IP^1$ which intersects with
$\Sigma_B$. The linear combination
\begin{equation}\label{eq:Sigma-sum}
  \sum_{I=0}^{\mathrm{rk}} a_I[\Sigma_I] = [\delta]
\end{equation}
with $a_I$ the marks of $\hat{\fg}$ is homologous to the generic
elliptic fiber. We denote the complexified K\"ahler parameters of
$\Sigma_B$ and $\Sigma_I$ by $t_B$ and $t_I$ respectively.

\begin{figure}
  \centering
  \includegraphics[width=0.5\linewidth]{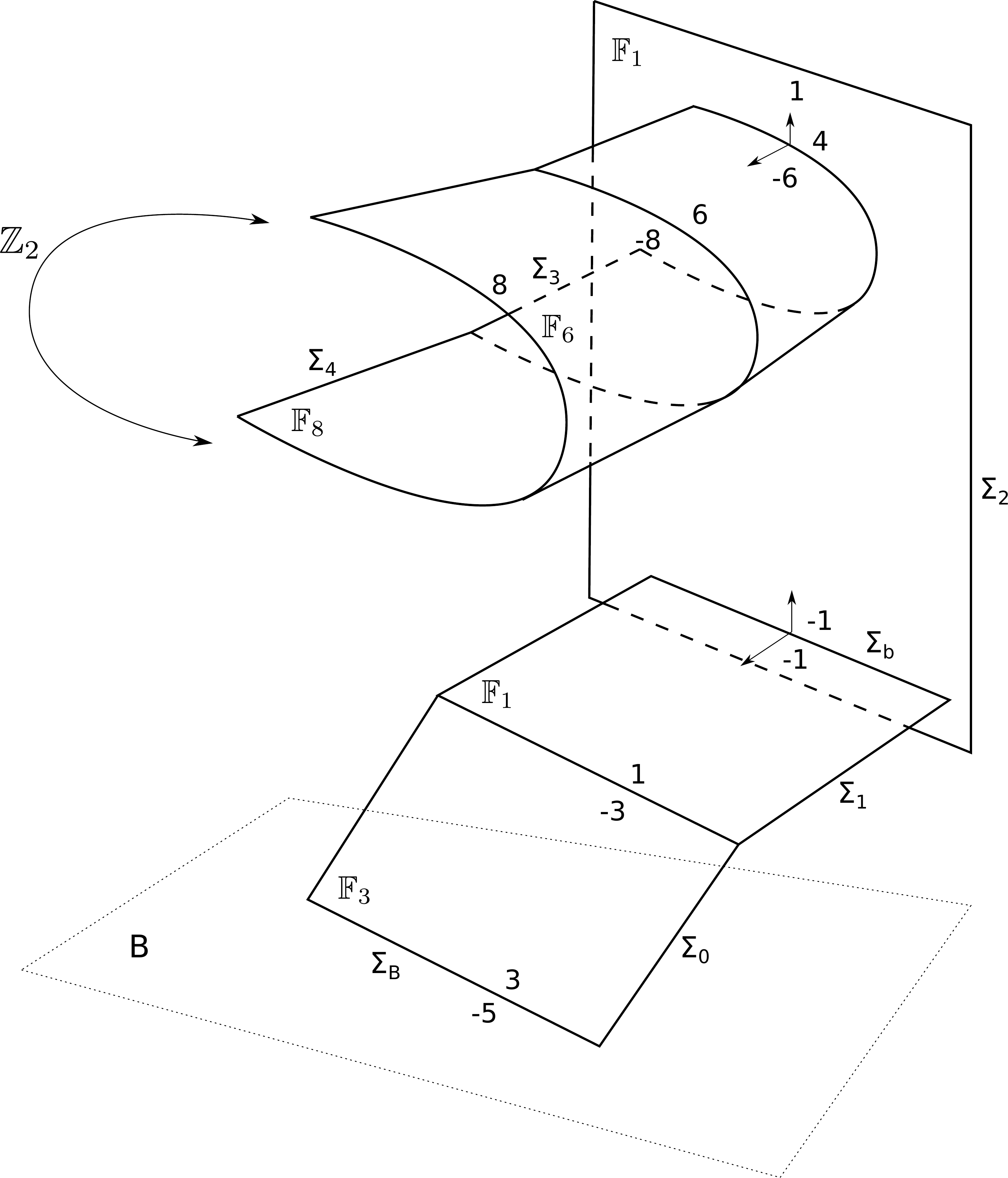}
  \caption{Compact curves and compact divisors in the $\fn=5$ local
    elliptic Calabi-Yau from~\cite{DelZotto:2017pti}. The indices $(-n,n-2)$ on each rational section $S$ or $S'=S+k F$ of $\mathbb{F}_k$ are the degrees
    of
    its normal bundles ${\cal O}(-n)\oplus {\cal  O}(n-2)$ in the corresponding direction. $\mathbb{F}_6$ meets $\mathbb{F}_1$
    in a double section. Globally the fibration has a $\IZ_2$ monodromy encirceling
    $\Sigma_B$ that corresponds to an outer automorphism of  the $E_6$ Dynkin diagram and folds its sphere tree to an
    $F_4$ type sphere tree over $\Sigma_B$.Note that the curve $\Sigma_b$ plays the role of a Mori
    cone generator and is related to the base curve $\Sigma_B$ by
    \eqref{eq:SBSb-n5}. }
  \label{fg:F4}
\end{figure}

The $\nD = \mathrm{rk}+1$ vertical compact divisors $\mf D_I$ for
$I=0,1,\ldots$ are fibrations of $\Sigma_I$ over $\Sigma_B$. They are
argued in \cite{DelZotto:2017pti} to be Hirzebruch surfaces of various
degrees, and the $\Sigma_I$ are the $\IP^1$ fibers of these Hirzebruch
surfaces. It is then easy to deduce that the
$(\mathrm{rk}+2)\times (\mathrm{rk}+1)$ matrix $C$ encoding the
intersections between $\Sigma_I$, $\Sigma_B$ and $\mf D_I$ is given by
\begin{equation}\label{eq:matC}
   C = \begin{pmatrix}
  - \widehat{ A}\\
    -\fn,0\ldots,0
    \end{pmatrix}
\end{equation}
where $\widehat{ A}$ is the affine Cartan matrix of $G$. We
illustrate compact curves and compact divisors in the example of the
$\fn=5$ model with $G=F_4$ in Figure~\ref{fg:F4}.

The topological string partition function on elliptic fibrations can
be expanded in terms of the base degree $d$ w.r.t. the
base curve $\Sigma_B$ labelled by  $Q_B = \re^{t_B}$
\begin{equation}\label{eq:Zexp-1}
 Z = Z^{\text{cls}} Z_0\bigg(1+\sum_{d=1}^\infty Z_d Q_B^d \bigg) \ .
\end{equation}
$Z^{\text{cls}}$ comes from the degree zero maps and depends hence on
the classical topological data of $X$.  We discuss them in
section~\ref{sc:Zpert}.  $Z_0(t_I,\epsilon_-,\epsilon_+)$ gets
contributions from the rational curves $\Sigma_I$ in the elliptic fibre
that form the affine Dynkin diagram.  These can be
directly calculated from the geometry reflecting the affine group
structure, i.e. the intersection matrix $\hat A$ in (\ref{eq:matC})
using $N^{[\Sigma_I]}_{0,\frac{1}{2}}=1$, $I=0,\ldots r$ for
rational curves~\cite{Choi:2012jz} as well as the general multi cover formula (\ref{BPSII}),
which leads to (\ref{eq:1-loop}). These contribution of isolated rational
curves can be also calculated as one loop correction to the
gauge coupling~\cite{Katz:1996fh}, which is the
reason that $Z_0$ is sometimes identified as
$Z_0=Z^{\text{1-loop}}$. The coefficients $Z_{d>0}$ in the expansion,
on the other hand, encode the BPS invariants that do wrap the base
curve $\Sigma_B$, and are rather difficult to compute.

At this point a clarification of subtlety is in order. The curve
classes $\Sigma_B,\Sigma_I$ actually do not give a good basis for
computing the BPS invariants, as they are not all Mori cone
generators. To remedy this, one should keep $\Sigma_I$ and replace
$\Sigma_B$ by the $\IP^1$ base of the Hirzebruch surface with the
lowest degree in the chain, so that it cannot be expressed as linear
combinations of other curves with non-negative coefficients. We will
illustrate this point in example section~\ref{sc:eg}.

The topological string partition function is identical with the BPS
partition function of the corresponding 6d SCFT in the tensor branch,
put on the Omega background $\IR^4\times_{\epsilon_1,\epsilon_2}
T^2$. In the latter point of view, it is more natural to use another
set of K\"ahler parameters $t_{\text{ell}}, \tau, m_i$
$(i=1,\ldots,\mathrm{rk})$, which are related to $t_B,t_I$ by
\begin{equation}\label{eq:telltB}
  t_{\text{ell}} = t_B - \frac{\fn-2}{2}\tau\ , \quad \tau = \sum_{I=0}^{\mathrm{rk}} a_I
  t_I \ ,\quad m_i = t_i,\quad i=1,\ldots,\mathrm{rk} \ .
\end{equation}
$t_{\text{ell}}$ is defined such that the coefficients $\IE_d$
in the expansion of the BPS partition function in terms of
$Q_{\text{ell}} = \re^{t_{\text{ell}}}$
\begin{equation}\label{eq:Zexp-2}
  Z = Z^{\text{cls}} Z^{\text{1-loop}} \bigg(1+\sum_{d=1}^\infty
  \IE_d Q_{\text{ell}}^d \bigg)
\end{equation}
are elliptic genera of the self-dual strings present in the 6d
SCFT. $\tau$ measures the volume of the generic elliptic fiber
$[\delta]$, and since $\delta$ intersect with no compact divisor, it
is a mass parameter of the theory. In addition, it is also identified
with the complex structure modulus of the torus $T^2$. $m_i$ are now
interpreted as the Wilson loops of the vector multiplets in $T^2$.

Note that the 6d SCFT can be reduced to a 5d pure SYM with the same
gauge group $G$ if we decompactify $\Sigma_0$ and send its volume
$-t_0$ to infinity. In the resulting 5d theory, the only mass
parameter is the instanton counting parameter $t_\fq$, and we find, by
looking for curve class not intersecting with divisors, that
\begin{equation}\label{eq:tq}
  t_\fq = t_{\text{ell}} - \frac{\fn-2}{2}\tau \ .
\end{equation}

We finally comment that in light of the correspondence between $m_i$
and nodes in Dynkin diagram of $\fg$, we can collect the K\"ahler
moduli $m_i$ into a single vector $\nound{m}$ taking value in the
complexified Cartan subalgebra $\fh_{\IC} = \IC^{\mathrm{rk}}$ with
\begin{equation}\label{eq:dm}
  \nound{m} = \sum_{i=1}^{\mathrm{rk}} m_i \nound{\omega}_i \ ,\quad
  m_i = \vev{\nound{\alpha}_i, \nound{m}} \ ,\;\;  i=1,\ldots,\mathrm{rk} \ .
\end{equation}
where $\nound{\omega}_i$ are the fundamental weights,
$\nound{\alpha}_i$ the simple roots of $\fg$, and $\vev{,}$
the natural pairing between $\fh_{\IC}$ and $\fh_{\IC}^*$. This allows
a reformulation of the generalised blowup equations in terms of Lie
algebraic data, which we use heavily in the uniform formula
\eqref{eq:uv-blowup0}. The convention of Lie algebra we use is given
in appendix~\ref{ap:Lie}.

\subsection{Semiclassical partition functions}
\label{sc:Zpert}

We summarise the computation of the semiclassical partition functions
here. These are the minimal initial data one needs in
order to extract refined BPS invariants from the generalised blowup
equations. First of all the semiclassical contribution
$Z^{\text{cls}}(\nound{t},\eq,\et) = \exp(F^{\text{cls}}(\nound{t},\eq,\et))$
can be written as
\begin{equation}\label{eq:Fcls}
  \begin{aligned}
    F^{\text{cls}}(\nound{t},\eq,\et) =
    &\frac{1}{\eq\et}F^{\text{cls}}_{(0,0)} (\nound{t}) +
    F^{\text{cls}}_{(1,0)}(\nound{t}) -
    \frac{(\eq+\et)^2}{\eq\et}F^{\text{cls}}_{(0,1)}(\nound{t}) \\
    =
    &\frac{1}{\eq\et}\bigg(\frac{1}{6}\sum_{i,j,k=1}^{\nC}\kappa_{ijk}t_it_jt_k\bigg)
    +\sum_{i=1}^{\nC} b_i^{\text{GV}}t_i
    -\frac{(\eq+\et)^2}{(\eq\et)}\sum_{i=1}^{\nC} b_i^{\text{NS}}t_i \
    .
  \end{aligned}
\end{equation}
The coefficients $\kappa_{ijk}$ are the triple intersection numbers of
divisors $J_i$ Poincar\'e dual to the curve classes $\Sigma_i$ with
volumes $t_i$. $b_i^{\text{GV}}$ are intersections of the divisors
$J_i$ with the second Chern class of the Calabi-Yau threefold. The
coefficients $b_i^{\text{NS}}$, on the other hand, do not have a
geometric meaning, they are usually computed by the refined
holomorphic anomaly equations \cite{Huang:2010kf,Huang:2011qx}, which
are difficult to apply here. The Nekrasov partition function of a 5d
pure SYM also has a semiclassical contribution which takes the same
form as \eqref{eq:Fcls}, and the linear coefficients are subject to
the relation
\begin{equation}\label{eq:b-rel}
  b^{\text{GV}}_i + b^{\text{NS}}_i = 0 \ .
\end{equation}
In our previous paper \cite{Gu:2018gmy} we argued that
$b_i^{\text{GV}}, b_i^{\text{NS}}$ of the minimal 6d SCFTs with
$G = SU(3), SO(8)$ can be computed by uplifting the semiclassical
Nekrasov partition function of the 5d pure SYM with the same gauge $G$
aided by the nontrivial automorphism of the affine Dynkin diagrams,
and we found these coefficients also satisfy the relation
\eqref{eq:b-rel}. In the remaining minimal SCFTs with
$G=F_4,E_6,E_7,E_8$, not all the affine Dynkin diagams have a
non-trivial automorphism, and the method of uplifting does not always
work. Instead we assume \eqref{eq:b-rel} to be true and only compute
$b^{\text{GV}}_i$ by geometric means.\footnote{Up to a irrelevant
  $\tau$ term, the numbers $b^{\text{GV}}_i,b^{\text{NS}}_i$ can be
  also predicted from blowup equations by requiring the consistency of
  BPS invariants. For all the minimal SCFTs, these predictions agree
  with the values computed from the method we will describe later. }

\begin{table}
$$
 \begin{array}{crrrr|rrrrrrr|rrrrrrrr|}  
    D &\multicolumn{4}{c}{\nu_i^*}     &l^{(1)}& l^{(2)}& l^{(3)} & l^{(4)} & l^{(5)} & l^{(6)} & l^{(7)}
    &l_{ F_4}^{(0)}& l_{ F_4}^{(1)}&  l_{ F_4}^{(2)}&l_{ F_4}^{(3)}&l_{ F_4}^{(4)}&l_{ F_4}^{(b)} \\
    D_0    &          0&  0&  0&  0\phantom{\ }&    -2& 0   & 0 &0 &  0     &   0  & 0 \phantom{\ }  &  0 & 0  & 0 & 0& -2& 0 \\
    D_1    &        -1&   0&   0&   0\phantom{\ }&  0& -2   & 0 &0 &  0     &   0  & 1 \phantom{\ }  &  1 & 0  & 0 & 0&  0&  0 \\
    D_2    &         0&  -1&   0&   0\phantom{\ }&  1&  0   & 0 &0 &  0     &   0  & 0  \phantom{\ } &  0 & 0  & 0 & 0&  1&  0\\
    D_3    &         0&   1&   0&  -1\phantom{\ }&  0& 3   & 0 &0 &   0     &    1 & -2 \phantom{\ } &  -2& 1  & 0 & 0&  0&  0\\
    D_4    &         1&   2&   0&  -2\phantom{\ }&  2& 0   & 0 &0 &   0     &    -2 & 1 \phantom{\ } &  1 &-2  & 0 & 0&  2&  0\\
    S'     &         2&   3&   0&  -1\phantom{\ }&  0& 1   & -2&0 &   1     &    0 & 0  \phantom{\ } &  0 & 0  &-2 & 1&  0&  0\\
    S''    &         2&   3&   0&  -2\phantom{\ }&  1& -2   &  1&-1&   0    &    0 & 0  \phantom{\ } &  0 & 0  & 1 &-2&  1&  -1\\
    S'''   &         2&   3&   0&  -3\phantom{\ }& -2& 0   & 0 &-1&  0      &   1  & 0  \phantom{\ } &  0 & 1  & 0 & 1& -2&  -1\\
    K      &         2&   3&   0&   0\phantom{\ }&  0& 0   & 1 &0 &  -2     &    0 &  0 \phantom{\ } &  0 & 0  & 1 & 0&  0& 0  \\
    F      &         2&   3&  -1&  -5\phantom{\ }&  0& 0   & 0 &1 &  0      &   0  & 0  \phantom{\ } &  0 & 0  & 0 & 0&  0&  1 \\
    S      &         2&   3&   0&   1\phantom{\ }&  0& 0   & 0 &0 &  1      &    0 & 0  \phantom{\ } &  0 & 0  & 0 & 0&  0&  0 \\
    F      &         2&   3&   1&   0\phantom{\ }&  0& 0   & 0 &1 &  0      &   0  & 0  \phantom{\ } &  0 & 0  & 0 & 0&  0& 1  \\
   \end{array} \ .
$$
  \caption{The toric data of $\IP_\Delta$ for
    $G=F_4$.}\label{tb:PD4}
\end{table}
To compute $\kappa_{ijk}$ and $b^{\text{GV}}_i$, we need to embed the
Calabi-Yau threefold $X \to \cO_{\IP^1}(-\fn)$ in a compact
Calabi-Yau, for instance the elliptic fibration $\hat{X}$ over
$\IF_{\fn}$ with a single section, and first compute these
intersection numbers in the compact geomtry. The compact Calabi-Yau
$\hat{X}$ can be realised as a hypersurface in a toric variety
\cite{Haghighat:2014vxa}. Let us look at an example in detail. The
threefold $\pi: X\to \cO_{\IP^1}(-5)$ is the zero loci of the section
of the anti-canonical bundle of the toric variety $\IP_\Delta$, whose
toric data is given in Table~\ref{tb:PD4}
\cite{Haghighat:2014vxa}.\footnote{See also the geometric description
  in \cite{Esole:2017rgz}.}  It has 7 Mori cone generators with
charges $l^{(i)}= (l^{(i)}_n)$ for $i=1,\ldots,7$ (We will use the
same notation for both the curves and their toric charges). Note the
number of Mori cone generators is $(\mathrm{rk}+3)$. Among these
curves $l^{(4)},l^{(5)}$ are identified as the $(-5)$ curve (i.e.\
$\Sigma_B$) and the $(0)$ curve in the base $\IF_5$\footnote{Strictly
  speaking, the $(0)$ curve should be corrected by certain linear
  combination of other curves. But after decompactifying this curve to
  arrive at the non-compact threefold $X$, this difference
  disappears.}, while the other five toric curves combine linearly
into $\Sigma_I$ for $I=0,1,\ldots,4$ \cite{Haghighat:2014vxa}
\begin{equation}
  \begin{gathered}
    \Sigma_0 = l^{(3)} \ ,\;
    \Sigma_1 = l^{(2)}+l^{(6)} + 2l^{(7)} \ ,\;
    \Sigma_2 = l^{(1)} \ , \;
    \Sigma_3 =  l^{(6)}\ ,\;
    \Sigma_4 = l^{(7)}\ .
  \end{gathered}
\end{equation}
The toric divisors listed in the first column of Table~\ref{tb:PD4} can
also be identified. $D_0$ is associated to the canonical bundle of
$\IP_\Delta$. When $\hat{X}$ is written in the Weierstrass form,
$D_1,D_2$ correspond to the divisors $x=0$ and $y=0$ respectively,
while $K$ is the zero section at $x\to\infty$. $F,S',S$ are
respectively the vertical divisors pulled back from the $(0)$, $(-5)$,
and the $(5)$ curves in $\IF_5$ in the base. Since the elliptic
fibration over $\pi(S') = \Sigma_B$ factorises to an intersecting tree
of Hirzebruch surfaces $\mf D_I$ for $I=0,1,\ldots,4$, $S'$ should
actually be identified with $\mf D_0$. The remaining four divisors
$D_3,D_4,S'',S'''$ are identified (up to linear combination) as the
exceptional divisors $\mf D_i$ for $i=1,\ldots,4$. Only $\mathrm{rk}(F_4)+3=7$ of
these divisors are linearly independent.

Once the toric data of $\IP_\Delta$ are specified, there are standard
techniques in toric geometry to compute the triple intersection
numbers $\hat{\kappa}_{ijk}$ of the divisors $\hat{J}_i$, and the
intersection numbers with $c_2(\hat{X})$ are given by
\cite{Hosono:1994av} \footnote{We use the same notation for the
  divisor $\hat{J}_i$ and its dual 2-form.}
\begin{equation}\label{eq:bGV-comp}
  -24\hat{b}^{\text{GV}}_i = \int_{\hat{X}}c_2(\hat{X})\wedge \hat{J}_i =
  \frac{1}{2}\sum_{jk}\hat{\kappa}_{ijk}
  (l^{(j)}_0l^{(k)}_0-\sum_{n>0}l^{(j)}_nl^{(k)}_n) \ .
\end{equation}
Alternatively, $\hat{\kappa}_{ijk}$ can be computed by the special
geometry relation. For a compact Calabi-Yau threefold realised as a
hypersurface in a toric variety one can define the deformed
fundamental period $\hat{\omega}_0(z;\rho)$ as a holomorphic function
of the Batyrev coordinates $z_i$ and compute the homogeneous A- and
B-periods (see for instance \cite{Hosono:1993qy})
\begin{equation}
  \begin{aligned}
    \hat{\Pi}^{(1)}_i(z) &= \pd_{\rho_i}\hat{\omega}_0(z;\rho) |_{\rho_i = 0} \ ,\\
    \hat{\Pi}^{(2)}_i(z) &= \frac{1}{2}
    \hat{\kappa}_{ijk}\pd_{\rho_j}\pd_{\rho_k}\hat{\omega}_0(z;\rho)|_{\rho_i=0}
    \ .
  \end{aligned}
\end{equation}
They are interpreted as the masses of the D2-, D4-branes supported on
the curves $l^{(i)}$ and the dual divisors $\hat{J}_i$. The affine A- and
B-periods defined by
\begin{equation}
  \hat{t}_i = \hat{\Pi}_i^{(1)}/\hat{\omega}_0(z;0) \ ,\;
  \hat{F}_i = \hat{\Pi}_i^{(2)}/\hat{\omega}_0(z;0)
\end{equation}
satisfy the relation
\begin{equation}
  \hat{F}_i = \frac{\pd \hat{F}_{(0,0)}}{\pd \hat{t}_i} \ .
\end{equation}
The existence of the prepotential $\hat{F}_{(0,0)}$ uniquely fixes the
coefficients $\hat{\kappa}_{ijk}$.

The non-compact Calabi-Yau $X$ is obtained by decompactifying
$\hat{X}$ in the direction of the $(0)$-curve in the base. In
practise, this corresponds to taking the K\"ahler parameter in the
decompactified direction $t_{\text{dc}}$ to infinity in A-model or
taking the corresponding complex structure parameter $z_{\text{dc}}$
to zero in B-model. In A-model, this limit can be understood as taking
some of the compact $(1,1)$ cycles to infinite size, keeping the other
$(1,1)$ cycles finite. The periods of the geometry will be rearranged
so that only one A-period and some B-periods go to infinity. In our
current case, the B-period $\frac{\partial}{\partial \tau} F_{(0,0)}$
goes to infinity, while the corresponding A-period remains finite, and
becomes the elliptic fiber parameter $\tau$. We can then integrate
over the new periods to get the triple intersection numbers of the
non-compact geometry. However, this method will always have a
integration constant term $\tau^3$ unfixed, which is very important
for the refined BPS invariants in the $\tau$ direction. To determine
the $\tau^3$ term, we study the normalization scheme of $\tau$
derivative of the genus zero free energy
$\frac{\partial}{\partial \tau} F_{(0,0)}$.

In the example of the $G=F_4$ model visited above, this is $z_5$
associated to the curve $l^{(5)}$.  In the limit $z_{\text{dc}}\to 0$
the affine A-period associated to the $(0)$-curve diverges, while the
other $\mathrm{rk}+2$ affine A-periods remain finite. We can choose a
basis of the latter to be\footnote{They are the limit
  $z_{\text{dc}}\to 0$ of proper linear combinations of $\hat{t}_i$ of
  the compact Calabi-Yau. Similarly $F_J$ defined below are the limit
  $z_{\text{dc}}\to 0$ of linear combinations of $\hat{F}_i$.}
\begin{equation}
  (t_I) \ ,\quad I=0,1,\ldots,\mathrm{rk},B \ ,
\end{equation}
which correspond to the curve classes
$\Sigma_0,\ldots,\Sigma_r, \Sigma_B$ discussed in the previous
section. At the same time, both the zero section of the elliptic
fibration and the vertical disivor of the $(0)$-curve become infinite
in volume. Therefore in the limt $z_{\text{dc}} \to 0$ two B-periods
diverge, and only $\mathrm{rk}+1$ affine B-periods remain finite. We choose a
basis
\begin{equation}
  (F_J) \ ,\quad J=0,1,\ldots,\mathrm{rk} \ ,
\end{equation}
which correspond to the divisor classes $\mf D_I$ with
$I=0,1,\ldots,\mathrm{rk}$. The special geometry relation of the non-compact
Calabi-Yau $X$ dictates
\begin{equation}\label{eq:sg-local}
  F_J = \sum_{I=0,\ldots,\mathrm{rk},B} C_{JI} \frac{\pd F_{(0,0)}}{\pd t_I} \ ,
  \quad J=0,1,\ldots,\mathrm{rk} \ ,
\end{equation}
where $C_{JI}$ are the components of the divisor-curve intersection
matrix \eqref{eq:matC}. The identity \eqref{eq:sg-local} allows the
computation of the semi-classical components of $F_{(0,0)}(t)$ up
to a term proportional to $\tau^3$; in other words, the intersection
number $\kappa_{\tau\tau\tau}$ can not be fixed by
\eqref{eq:sg-local}. Since $\tau$ is a mass parameter, the term
$\tau^3$ can always be factored out of the blowup equations and it is
not of importance to us. Nevertheless, in Appendix~\ref{ap:norm} we
will introduce a normalisation scheme which fixes such a term in a
reasonable way, and we adopt this normalisation scheme in the example
section~\ref{sc:eg}. In any case, for all the minimal 6d SCFTs with a
pure gauge theory in the bulk which can be reduced to a 5d pure SYM,
we find that up to $\tau^3$
\begin{equation}\label{eq:pertF0-g}
  F^{{\text{cls}}}_{(0,0)}(\nound{t}) = -\sum_{\alpha \in
    \Delta_+}\bigg(\frac{\mroot^3}{6} + \frac{t_\fq}{2\hG}\mroot^2
  \bigg) - \frac{((\fn-2)\tau+t_\fq)^3-t_\fq^3}{6\fn(\fn-2)}  + \ldots
\end{equation}
where $\nound{m}, t_\fq$ are defined in \eqref{eq:dm} and \eqref{eq:tq}
respectively. We recognise the sum over positive roots is from the
Nekrasov prepotenital of the 5d pure SYM. We can also massage
\eqref{eq:pertF0-g} into a more suggestive form
\begin{equation}\label{eq:pertF0-uni}
  F^{{\text{cls}}}_{(0,0)} = \frac{1}{2\fn}t^2_{\rm ell}\tau +
  \frac{1}{2}t_{\rm ell}(m,m) - \frac{\fn-2}{4}\tau(m,m) + \ldots
\end{equation}
up to $\tau^3$ and cubic terms in $m_i$. Here $(,)$ is the
invariant bilinear form on $\fh_{\IC}$. See appendix~\ref{ap:Lie} for
our convention.

As for $F^{\text{cls}}_{(1,0)}(\nound{t})$ and the intersection
numbers with $c_2(X)$, we use the same formula \eqref{eq:bGV-comp}
with $\hat{\kappa}_{ijk}$ replaced by the triple intersection numbers
of the non-compact Calabi-Yau and $l^{(i)}$ replaced by the toric
charges of $\Sigma_0,\ldots,\Sigma_r,\Sigma_B$. In the example of
$G=F_4$ discussed above, one has the toric charges
$l_{ F_4}^{(0)}, l_{ F_4}^{(1)}, l_{ F_4}^{(2)},l_{ F_4}^{(3)},l_{
  F_4}^{(4)},l_{ F_4}^{(b)}$ as in Table~\ref{tb:PD4}.

As in the case of the prepotential, one cannot determine the pure mass
term proportional to $\tau$ which is irrelevant, although it can be
fixed by the same normalisation scheme if one wishes. We have checked
that $F^{\text{cls}}_{(1,0)}(\nound{t})$ computed in this way reduces
correctly to the semiclassical Nekrasov partition function when the 6d
SCFT is reduced to the 5d pure SYM.

\subsection{Determination of $r$ fields}
\label{sc:adm}

In general, the $\nound{r}$ fields associated to a
non-compact Calabi-Yau can be determined by the method in
\cite{Huang:2017mis}. Here we give a brief description of it.  As
proved in \cite{Huang:2017mis}, even without any assumption or
constraint put on $\Lambda$ (for instance it can depend on all
K\"ahler moduli), the $\Lambda$ as \emph{defined} by
(\ref{eq:blowupI}) must be quasi-modular of weight zero under the
Siegel modular transformations of
$\tau_{ij}=\frac{\partial^2}{\partial t_i\partial t_j}F_{(0,0)}$,
where $t_i$ are the true K\"ahler parameters.\footnote{Note that this
  $\tau_{ij}$ parameter is different from the $\tau$ parameter of the
  elliptic fiber.} Let us expand this
$\Lambda$ 
in terms of all the exponentiated K\"ahler moduli
$\re^{\nound{t}_i}$. The leading terms, which come from
$Z^{\text{cls}}$, read
\be \ba \log\Lambda\sim
&\log
    \Big(Z^{\rm cls}(\epsilon_1,\epsilon_2-\epsilon_1)
    Z^{\rm cls}(\epsilon_1-\epsilon_2,\epsilon_2)/Z^{\rm
    cls}(\epsilon_1,\epsilon_2)\Big)  \\
 = & \bigg(-\frac{1}{6}
    \sum_{i,j,k=1}^{b_2^c}\kappa_{ijk}R_iR_jR_k
    +\sum_{i=1}^{b_2^c}(b_i^{\rm GV} - b_i^{\rm NS})R_i\bigg) (\epsilon_1+\epsilon_2)
    + \sum_{k=1}^{b_2^c}\bigg(-\frac{1}{2}\sum_{i,j=1}^{b_2^c}\kappa_{ijk}R_iR_j\bigg) t_k
    \\[-2mm]
 = &:f_0(\und{n})(\epsilon_1+\epsilon_2) + \sum_{k=1}^{b_2^c}
    f_k(\und{n}) t_k \ ,
    \label{eq:fk}
    \ea
    \ee
which are linear in $t_i$.
It implies that $\Lambda$ can be expanded as a well-defined power
series in $e^{\nound{t}_i}$.
Now let us assume that for appropriate choice of the $\nound{r}$-field, this
power series with all the instanton contributions taken into account
truncates at finite orders
for all true $t_i$\footnote{This assumption is the most
  natural consideration for the generalized blowup equations compared
  with the initial Nakajima-Yoshioka blowup equations.},
which actually implies that the $\Lambda$ does not depend on any of
the true K\"ahler parameters as stated in the introduction, since
otherwise it can not be of modular weight zero.
Needless to say, this assumption puts very strong constraint on the
choice of the $\nound{r}$-field, and these are the $\nound{r}$-fields
we are interested in.

In our current cases of elliptic non-compact Calabi-Yau threefolds,
there is only one mass parameter which is $\tau$. The strong
constraint then means that for the $\nound{r}$ fields we are
interested in
only the lowest order of K\"ahler
parameters contributes. Then we can simply define $\Lambda$
\begin{equation}\label{rcond}
  \Lambda(\tau,\epsilon_1,\epsilon_2)= \sum_{\und{n} \in \mathcal{I}}(-1)^{|\und{n}|}
  \re^{f_0(\und{n})} \re^{f_k(\und{n})t_k}  \ ,
\end{equation}
where $\mathcal{I}$ is the set of integral vectors $\und{n}$ that
minimize all the $f_k(\und{n})$ for true K\"ahler parameters
\emph{simultaneously} after subtracting mass parameters. If the
minimal values for one $\nound{r}$ are not zero simultaneously, then
it must be a vanishing $\nound{r}$ or an incorrect $\nound{r}$.

In the case of elliptic non-compact Calabi-Yau threefolds associated to
minimal 6d SCFTs, the $\tau$ parameter is always some combinations of
K\"ahler parameters in the fiber direction and it will be a little bit
subtle to subtract the mass parameter $\tau$. We can first consider
the $\tau$ irrelevant K\"ahler parameter $t_{\text{ell}}$, and the
minimum of the associated $f_{\text{ell}}(\und{n})$, and then check
the solved $\nound{r}$ fields with the condition (\ref{rcond}). As
also shown in \cite{Gu:2018gmy}, the existence of the minimum of
$f_{\text{ell}}(\und{n})$, already suffices to fix all the $\nound{r}$
fields. In particular, similar to $\nound{t}$, we decompose
$\nound{r}$ into components $r_0,r_1,\ldots,r_{\mathrm{rk}},r_B$. Then
the weak consistency condition implies the \emph{admissibility}
condition, which we will prove shortly
\begin{equation}\label{eq:rtau0}
  r_\tau = \sum_{I=0}^\mathrm{rk} a_I r_I = 0 \ .
\end{equation}
It means the component of $\nound{r}$ in the direction of the elliptic
fiber must vanish. Recall that we only consider $\nound{r}$ modulo
$2C\cdot\und{n}$ for
$\forall\und{n}\in \IZ^{b_4^c} = \IZ^{\mathrm{rk}+1}$. The
intersection matrix $C$ defines the injection
$\IZ^{\mathrm{rk}+1} \hookrightarrow\IZ^{\mathrm{rk}+2}$. The
$\nound{r}$ fields that satisfy \eqref{eq:rtau0} can only take value
in the dimension zero quotient lattice $\Gamma/\IZ^{\mathrm{rk}+1}$
with $\Gamma = \IZ^{\mathrm{rk}+2}|_{r_\tau = 0}$, and they are thus
finite in number. Finally we impose the BPS checkerboard pattern
condition
\begin{equation}\label{eq:checkerboard}
  2j_l + 2j_r+ 1 \equiv {r}\cdot \beta \quad \text{mod}\;2
\end{equation}
to remove half of them. We comment that the checkerboard condition can
be written down without computing any BPS invariants. As argued in
\cite{Gu:2018gmy}, any rational curve $\beta$ in the Calabi-Yau with
normal bundle $\mc O(-n)\oplus\mc O(n-2)$ must have
\begin{equation}
  2j_l+2j_r + 1 \equiv n \quad \text{mod}\;2 \ .
\end{equation}
This implies the $\nound{r}$-field always satisfies
\begin{equation}\label{eq:rn}
  \nound{r} \equiv (0,\ldots,0,\fn) \quad \text{mod}\;2\ ,
\end{equation}
for minimal 6d (1,0) SCFTs.
The argument above not only establishes the finiteness of admissible
$\nound{r}$, but also provides a guideline on how to determine them. We
find all of them for the $G=SU(3),SO(8)$ models in \cite{Gu:2018gmy}
and for the remaining models with $G=F_4,E_{6,7,8}$ in the example
section \ref{sc:eg} of this paper. In all these examples we checked
that they satisfy the stronger consistency condition. Here we
summarise their numbers in Table~\ref{tb:r-groups}.

\begin{table}
  \centering
  \begin{tabular}{*{7}{>{$}c<{$}}}\toprule
    \fn & 3 & 4 & 5 & 6 & 8 & 12 \\\midrule
    \#(\nound{r}) & 9 & 16 & 5 & 18 & 16 & 12 \\
    |P:Q^\vee| & 3 & 4 & 1 & 3 & 2 & 1 \\\bottomrule
  \end{tabular}
  \caption{Numbers and sizes of groups of $\nound{r}$-fields giving rise
    to the same embedding $\phi_\lambda$.}
  \label{tb:r-groups}
\end{table}

Let us prove the admissibility condition \eqref{eq:rtau0}. An
important ingredient of the blowup equations \eqref{eq:blowupI} are
the shifts of K\"ahler moduli $m_{i}$ $(i=1,\ldots,\mathrm{rk})$ by
\begin{equation}
  R_{i} = \sum_{J=0}^\mathrm{rk} C_{i,J} n_J + \frac{1}{2} r_{i} \ ,
\end{equation}
where $n_J$ take value in $\IZ$. We collect them into a single
vector just like $\nound{m}$
\begin{equation}\label{eq:dR}
  {R} = \sum_{i=1}^\mathrm{rk} \Big(-\sum_{J=0}^\mathrm{rk}\hat{A}_{i,J} n_J +
  \frac{1}{2}r_i\Big)\,\omega_i
  = - \alpha^\vee + \lambda + n_0\theta \ ,
\end{equation}
where we have used the generic form \eqref{eq:matC} of the
intersection matrix $C$ and the following Lie algebraic notation
\begin{equation}\label{eq:lambda-alpha}
  \lambda = \frac{1}{2}\sum_{i=1}^\mathrm{rk} r_i \omega_i \ ,\quad
  \alpha^\vee = \sum_{i=1}^\mathrm{rk} n_i \alpha_i^\vee \ .
\end{equation}
The function $f_{\text{ell}}(\und{n})$, which we also denote by
$P_G(\und{n})$ due to its particular importance, then has the form
\begin{equation}\label{eq:PG-1}
  f_{\text{ell}}(\und{n}) =
  \frac{1}{\fn}R_{\text{ell}}R_{\tau} + \frac{1}{2}({R},{R})
  =: P_{G}(\und{n})\ .
\end{equation}
where $R_{\text{ell}}, R_{\tau}$ are the shifts of
$t_{\text{eff}}, \tau$ given respectively by
\begin{equation}
  R_{\text{ell}} = -\fn n_0 + \frac{1}{2}r_{\text{ell}}\ , \quad
  R_{\tau} = \frac{1}{2}r_\tau =
  \frac{1}{2}\sum_{I=0}^\mathrm{rk} a_I r_I \ .
\end{equation}
A little algebraic manipulation leads to
\begin{equation}\label{eq:PG-2}
  P_{G}(\und{n})  =
  \frac{1}{2}\sum_{I,J=0}^\mathrm{rk} n_In_J(\alpha_I^\vee,\alpha_J^\vee)
  -\frac{1}{2}\sum_{I,J=0}^\mathrm{rk} r_In_J(\widehat{\omega}_I,\alpha_J^\vee)+ \ldots
\end{equation}
where $\ldots$ denote the terms that are independent from $\und{n}$.  If we demand that this
function have a minimum, all the derivatives $\pd_{n_I}P_G(\und{n})$
must have a common zero. Multiplying each of them with comark $a_I$
and adding them up, we immediately arrive at the admissibility
condition \eqref{eq:rtau0}\footnote{This condition was found in
  \cite{Gu:2018gmy} by an intuitive geometric argument for minimal
  theories with a gauge group $G$ of the ADE type. Here we prove it
  for all gauge groups including the non-ADE types.}.
Let us make some remarks here. As we will see in section \ref{sc:eg},
all the admissible $\nound{r}$-fields are such that the components
$r_i$ ($i=1,\ldots,\mathrm{rk}$) are even integers. It is clear then that
$\lambda$ associated to $\nound{r}$ as well as the shift $\nound{R}$ have
nice interpretation as weight vectors of $\fg$. And $P_G(\und{n})$ is
nothing else but half the norm square of $\nound{R}$ due to \eqref{eq:rtau0}
\begin{equation}
  P_G(\und{n}) =\frac{1}{2}(\nound{R},\nound{R}) \ .
\end{equation}
Furthermore $\lambda$ defines an embedding of $Q^\vee$ into $P$
\begin{equation}\label{eq:QvP}
  \begin{aligned}
    \phi_\lambda: Q^\vee &\hookrightarrow P \\
    \alpha^\vee &\mapsto \phi_\lambda(\alpha^\vee) = -\alpha^\vee +
    \lambda
  \end{aligned}
\end{equation}
and $\nound{R}$ takes value in the image $\phi_\lambda(Q^\vee)$ if
$n_0=0$.  The number of inequivalent embeddings is the index
$|P:Q^\vee|$, which also happens to be the order of the automorphism
group of the associated Dynkin diagram. We also list these numbers in
Table~\ref{tb:r-groups}. The reader may notice the curious relation
\begin{equation}\label{eq:nPQr}
  \fn \cdot |P:Q^\vee| = \#(\nound{r}) \ ,
\end{equation}
whose meaning will be clear in section~\ref{sc:ebp}.

\subsection{De-affinisation}\label{deaff}

Once the semiclassical piece $Z^{\text{cls}}$ and the
$\nound{r}$-fields are known, we can start solving refined BPS
invariants by expanding the blowup equations \eqref{eq:blowupI} in
terms of K\"ahler moduli and extracting constraint equations of BPS
invariants at each order. Alternatively since the $Z^{\text{1-loop}}$
piece is rather easy to compute for the minimal 6d SCFTs, which reads
\cite{Gu:2018gmy}\footnote{Here $\mathrm{PE}$ is the plethystic exponent operator defined as
\be
\mathrm{PE}[f(x)]=\exp\bigg[\,\sum_{n=1}^\infty\frac{1}{n}f(x^n)\,\bigg].
\ee
Note we only consider the contribution from vector multiplets to the one-loop partition function here. The tensor multiplets
actually also contribute to one-loop partition function, but their contribution does not depend on the gauge parameter $m$, i.e. it is pure $\tau$ terms
which will decouple from the blowup equations. Thus we do not consider them here. See more discussion in section 2.4 in \cite{Gu:2018gmy}.}
\begin{equation}\label{eq:1-loop}
  Z^{\text{1-loop}}(\nound{m},\epsilon_1,\epsilon_2)
  =
  \,{\rm PE}\Bigg[-\frac{q_R+q_R^{-1}}
  {\big(q_1^{1/2}-q_1^{-1/2}\big)\big(q_2^{1/2}-q_2^{-1/2}\big)}
  \sum_{\alpha \in \Delta_+}  \(\re^{\mroot}
  +Q_\tau\re^{-\mroot}\) \frac{1}{1-Q_\tau}\Bigg] \ ,
\end{equation}
we can also plug in this piece of information, expand the partition
functions in the blowup equation \eqref{eq:blowupI} only in terms of
$Q_{\text{ell}}$, and obtain recursion relations of elliptic genera
\begin{equation}\label{eq:DDE}
\ba
  \sum_{P_G(\und{n})+d_1+d_2 = d} D^{',\text{cls}}\, D^{\text{1-loop}}&\,
  \IE_{d_1}(\tau,m+\eq R_{m}(\und n),\epsilon_1,\epsilon_2-\epsilon_1)
  \IE_{d_2}(\tau,m+\et R_{m}(\und n),\epsilon_1-\epsilon_2,\epsilon_2)\\
  &=\Lambda \,\IE_{d}(\tau,m,\epsilon_1,\epsilon_2) \ ,
  \ea
\end{equation}
which allows the solution of the elliptic genera in compact
formulas. Here
\begin{equation}
  D^{',\text{cls}} =(-1)^{|\und{n}| + (d_1+d_2-d)r_{\text{ell}}}
  \exp\bigg(f_0(\und{n})(\epsilon_1+\epsilon_2)
  +\sum_{I=0}^\mathrm{rk} f_I(\und{n})t_I +
  (d_1\epsilon_1+d_2\epsilon_2)R_{\text{ell}}(\und{n})\bigg)
\end{equation}
collects contributions from the semiclassical partition function (as
well as the shift of $t_{\text{ell}}$), and
\begin{equation}
  D^{\text{1-loop}} =
  Z^{\text{1-loop}}(\tau,m+\eq R_{m}(\und n),\epsilon_1,\epsilon_2-\epsilon_1)
  Z^{\text{1-loop}}(\tau,m+\et R_{m}(\und n),\epsilon_1-\epsilon_2,\epsilon_2) /
  Z^{\text{1-loop}}(\tau,m,\epsilon_1,\epsilon_2)
\end{equation}
is the contribution from the one-loop partition function. The elliptic
blowup equations can be put in an elegant form by partially resumming
the left hand side of \eqref{eq:DDE}. With \eqref{eq:PG-2} and the
admissibility condition \eqref{eq:rtau0}, one can
show that the polynomial $P_G(\und{n})$ that characterises the
summation index $\und{n}$ is invariant under the translation
\begin{equation}\label{eq:aI-shift}
  n_I \to n_I + a_I^\vee k \ ,\quad k\in\IZ \ .
\end{equation}
Besides, in the components $Z^{\text{1-loop}}$ and $\IE_d$ the
dependence on $\und{n}$ only appears in the shifts
\begin{equation}
  R_I = -\sum_{J=0}^\mathrm{rk} \hat{A}_{I,J}n_J + \frac{1}{2}r_I \ ,
  \quad I=0,1,\ldots,\mathrm{rk} \ ,
\end{equation}
which are also invariant under \eqref{eq:aI-shift}. As a result, we
can decompose the summation index
\begin{equation}
  \und{n} = \hat{\und{n}} + \nound{a}^\vee k
\end{equation}
with the zeroth component of $\hat{\und{n}}$ fixed to zero, i.e.\
$\hat{\und{n}}=(0,n_1,n_2,\ldots,n_\mathrm{rk})$, a step we call
``de-affinisation'', and perform the infinite sum on the left hand
side of \eqref{eq:DDE} in two steps. In the first step, we factor
out $D^{\text{1-loop}}$, $\IE_{d_1,d_2}$ and only sum
$D^{',\text{cls}}$ over $k\in \IZ$. Due to the quadratic nature of the
polynomials $f_0(\und{n}), f_I(\und{n})$, and the relation of $t_I$
with $\tau$ \eqref{eq:telltB}, this first summation in fact produces a
theta function with characteristics \cite{Gu:2018gmy}.
In the second step, we sum over $d_1,d_2$ and $\hat{\und{n}}$ that satisfy
\begin{equation}
  P_G(\hat{\und{n}}) + d_1 + d_2 = d \ .
\end{equation}
Instead of $\hat{\und{n}}$ we can treat $\nound{R}(\hat{\und{n}})$ as
the summation index, which as we argued before is now interpreted as a
weight vector in $\phi_{\lambda}(Q^\vee)$ determined by $\nound{r}$. In
addition, the one-loop contribution $D^{\text{1-loop}}$ also turns out
to be a quotient of theta functions \cite{Gu:2018gmy}, while the
elliptic genera themselves are meromorphic Jacobi forms. Therefore in
the end, the elliptic blowup equations can be presented as beautiful
equations of Jacobi forms with a sum over the shifted coroot lattice
$\phi_\lambda(Q^\vee)$. These equations are the highlights of the next
section. We will present these equations in the beginning of the next
section, and then discuss their properties and how to solve elliptic
genera from them.

\section{Elliptic blowup equations}
\label{sc:ebp}

In this section, we first present the elliptic blowup equations for 6d
minimal $\mathcal{N}=(1,0)$ SCFTs with
$G=SU(3), SO(8), F_4, E_6, E_7, E_8$, and discuss their two
interesting properties, the modularity and the universality. The first
property in particular serves as a strong support for the validity of
the elliptic blowup equations to arbitrary degrees. Then we
distinguish two cases with $\Lambda(\eq,\et,m,r)$ non-vanishing or
identically vanishing, and discuss these two cases in detail. In
particular, the blowup equations in the first case with non-vanishing
$\Lambda$ allow us to write down an exact and universal recursion
formula for elliptic genera, thus offering a complete solution to the
elliptic genera.

Let us first fix some conventions. In the following whenever there is no risk of confusion
we will use the dot to denote both the invariant bilinear form on
$\fh$ or $\fh^*$ and the natural inner product between $\fh$ and
$\fh^*$
\begin{equation}
  \alpha\cdot \beta = (\alpha,\beta) \ ,\;\text{or}\;
  \alpha\cdot \beta = \vev{\alpha,\beta}\ .
\end{equation}
We define the norm square
\begin{equation}
  ||\alpha||^2 = \alpha\cdot\alpha \ .
\end{equation}
For a coroot $\beta^\vee \in Q^\vee$, we also define
\begin{equation}
  |\bv| = \sum_{i=1}^\mathrm{rk} \bv_i \ ,\quad \text{with}\ \;
  \bv = \sum_{i=1}^\mathrm{rk} \bv_i \av_i \ .
\end{equation}
Besides, for a vector $m$ representing the K\"ahler paramters associated to a Lie algebra, we denote $m_{\alpha}=m\cdot\alpha$ for short.

Following the de-affinisation procedure described in Section \ref{deaff}, we derive the \emph{elliptic blowup equations} as
\begin{equation}\label{eq:uv-blowup}
  \begin{aligned}
    &\sum_{\omega\in\phi_\lambda(Q^\vee),d_{1,2}\in\IN}
    ^{\frac{1}{2}||\omega||^2+d_1+d_2 =d}
    (-1)^{|\phi_\lambda^{-1}(\omega)|} \cdot \theta_i^{[a]}
    \Big(\fn
    \tau,(\fn-2)(\eq+\et)-\fn\big((\tfrac{1}{2}||\omega||^2+d_1)\eq
    +(\tfrac{1}{2}||\omega||^2+d_2)\et
    - m_\omega\big)\Big)    \\[-1ex]
    &\phantom{===-----}\times A_{\omega}(m) \cdot \IE_{d_1}(\tau,
    m-\epsilon_1\omega, \epsilon_1,\epsilon_2-\epsilon_1) \cdot
    \IE_{d_2}(\tau, m-\epsilon_2\omega, \epsilon_1
    -\epsilon_2,\epsilon_2) \\[1ex]
    &\phantom{=========}=\begin{cases}
      \theta_i^{[a]}(\fn\tau,(\fn-2)(\epsilon_1+\epsilon_2))\cdot
      \IE_d(\tau, m,\epsilon_1,\epsilon_2),
      & \quad\text{fixed}\ d\in \IN, \\
      0, &\quad \text{fixed}\ d\not \in \IN.
          \end{cases}
  \end{aligned}
\end{equation}
Here the subscript of theta functions $i$ is $4$ if $\fn$ is odd and
$3$ if $\fn$ is even, and the characteristic $a=k/\fn-1/2$, $k=0,1,\ldots, \fn-1$. The factor $A_{\omega}(\nound{m})$ is given by
\begin{equation}\label{Aomegam}
  A_{\omega}( m) = \prod_{\alpha \in \Delta_+}
  \breve{\theta}(m_\alpha,-\alpha\cdot \omega)
  \ ,
\end{equation}
where we denote for $R\ge 0$,
\be
\breve{\theta}(z,R)=\prod_{\displaystyle \substack{m,n\geq 0\\m+n\leq R-1}}\frac{\eta}{\theta_1(z+m\eq+n\et)}\prod_{\displaystyle \substack{m,n\geq
0\\m+n\leq R-2}}\frac{\eta}{\theta_1(z+(m+1)\eq+(n+1)\et)},
\ee
and for $R\le 0$, $\breve{\theta}(z,R)=\breve{\theta}(z,-R)|_{\epsilon_{1,2}\to-\epsilon_{1,2}}$. Note the $\alpha\cdot \omega$ in (\ref{Aomegam}) is
guaranteed to be an integer, as requested for the definition of $\breve{\theta}(z,R)$. See more about the origin of $\breve{\theta}(z,R)$ function in
appendix~\ref{ap:ids}. Let us show some examples here:
\begin{equation}
  \begin{aligned}
  \breve{\theta}&(z,0)
    = 1 \ ,\\
    \breve{\theta}(z,&\pm 1)
    = \frac{\eta}{\theta_1(z)} \ ,\\[-1mm]
    \breve{\theta}(z,\pm 2) =&\frac{\eta^4}{\theta_1(z)\theta_1(z \pm \eq)
      \theta_1(z \pm \et) \theta_1(z
      \pm\eq\pm\et)
      } \ . \\
  \end{aligned}
\end{equation}

The dependence on the $\nound{r}$-field in (\ref{eq:uv-blowup}) is related to the choice of $a$. If we choose basis
of K\"ahler moduli $(t_{\text{ell}},\tau,m_i)$, $i=1,\ldots,\mathrm{rk}$ as in
\eqref{eq:telltB}, the corresponding components of the $\nound{r}$-field
are $(r_{\text{ell}},r_\tau,r_i)$, $i=1,\ldots,\mathrm{rk}$. The first
component $r_{\text{ell}}$ controls the characteristic $a$ through
$a = r_{\text{ell}}/2\fn$, thus the latter can take any of the
following values
\begin{equation}
  a = -\frac{\fn-2k}{2\fn} \ ,\quad k=0,1,\ldots, \fn-1 \ .
\end{equation}
The component $r_\tau$ vanishes due to the admissibility condition
\eqref{eq:rtau0}. The remaining components $r_i$, $i=1,\ldots,\mathrm{rk} $
always correspond to a weight vector $\lambda \in P$ through
\eqref{eq:lambda-alpha}, which in turn induces the embedding
$\phi_{\lambda}: Q^\vee \hookrightarrow P$, and the summation index
vector $\omega\in \phi_\lambda(Q^\vee)$ plays the role of the shift
vector $-\nound{R}$. The number of different embeddings is
$|P:Q^\vee|$. The total number of different blowup equations is then
$\fn|P:Q^\vee|$, which explains the numerology found in
\eqref{eq:nPQr}.

\subsection{Modularity of elliptic blowup equations}
\label{sc:mod}

In this section, we provide evidence for the elliptic blowup equations
\eqref{eq:uv-blowup} by showing that the components of the elliptic
blowup equations transforms correctly as weak Jacobi forms. This is
established by showing that the weight and the index, in general a
quadratic polynomial, of the corresponding components in
(\ref{eq:uv-blowup}) match the predictions for the index and weight
made from the 2d and the 6d anomaly polynomial or from the
transformation properties of the refined topological string partition
function under the $S$ and $T$ monodromies of the Calabi-Yau space
$X$, see \cite{Huang:2015sta} and more generally
\cite{Schimannek:2019ijf}. In general the blowup equations give
interesting identities for Jacobi forms, one example is proven in
section \ref{sc:rec}, see also (\ref{vanish0}). In the fortuitous
cases where the expressions of $\IE_{d}$ are already known, for
instance the $G=SU(3),SO(8)$ models
\cite{Kim:2016foj,Haghighat:2014vxa}, we can plug in their
expressions, and verify these identities by small $Q_\tau$ expansion.

It is easy to see that each term in the summation of \eqref{eq:uv-blowup} has weight
$1/2$ as both $A_\omega$ and $\IE_d$ are of weight zero and $\theta_i$
has weight $1/2$. The identification of the modular indices requires a bit of computation, which is independent from the
characteristic $a$. The basic idea is to repeatedly use the fact that
$\theta(N\tau,Nz)$ is of index $N/2$. Let us denote
\begin{equation}
  d_0 = \tfrac{1}{2}||\omega||^2 \ .
\end{equation}
It is easy to see the theta function $\theta_i^{[a]}$ on the left hand side has modular
index
\begin{equation}
  \mathrm{Ind}_{d_0}^{G}=
  \frac{\fn}{2}\(\Big(d_0+d_1-\frac{\fn-2}{\fn}\Big)\eq+
  \Big(d_0+d_2-\frac{\fn-2}{\fn}\Big)\et - {m}_\omega\)^2.
\end{equation}
Using (\ref{indbreve}), the modular index polynomial of $A_\omega({m})$ can be computed as
\begin{equation}
  \begin{aligned}
    \mathrm{Ind}_{A}^{G}=
    &-\frac{\fn}{2}
      m_\omega^2-\frac{\fn}{2}
      d_0 m\cdot m +\(\fn d_0+2-\fn\)(\eq+\et) {m}_\omega\\
    &-\frac{d_0}{2}\(\fn d_0+2- \fn\)(\ep_1^2+\eq\et+\ep_2^2).
  \end{aligned}
\end{equation}
The elliptic genus $\IE_d(\tau,m,\epsilon_1,\epsilon_2)$ is known to have the modular
index \cite{DelZotto:2016pvm,DelZotto:2017mee}
\begin{equation}\label{eq:ind-Ed}
  \mathrm{Ind}_d^G = -\frac{1}{2}d(\fn-2)(\eq+\et)^2
  +\frac{1}{2}d (\fn\, d-\fn+2)\eq\et -\frac{1}{2}\fn\, d {m}\cdot
  \md\, m\ .
\end{equation}
Thus the modular index polynomials of $\IE_{d_1}(\tau,
    m-\epsilon_1\omega, \epsilon_1,\epsilon_2-\epsilon_1)$ and $\IE_{d_2}(\tau, m-\epsilon_2\omega, \epsilon_1
    -\epsilon_2,\epsilon_2)$ can be computed respectively as
\begin{align}
  \mathrm{Ind}_{d_1}^{G}=
  &\frac{1}{2}\(-(\fn-2)d_1\ep_2^2+(\fn\, d_1-\fn+2)d_1\eq(\et-\eq)
  -\fn d_1\({m} -\eq\omega\)^2\) \ ,\\
  \mathrm{Ind}_{d_2}^{G}=
  &\frac{1}{2}\(-(\fn-2)d_2\ep_1^2+(\fn\, d_2-(\fn-2))d_2(\eq-\et)\et
  -\fn d_2 \({m} - \et\omega\)^2\) \ .
\end{align}
Using $d=||\omega||^2/2+d_1+d_2$, we find that the four components on the left hand side of \eqref{eq:uv-blowup} has
total modular index polynomial as
\begin{equation}\label{eq:ind}
  \mathrm{Ind}_{d_0}^{G}+\mathrm{Ind}_{A}^{G}
  +\mathrm{Ind}_{d_1}^{G}+\mathrm{Ind}_{d_2}^{G}
  =
  -\frac{(\fn-2)(\fn-2+d \fn)}{2\fn}(\eq+\et)^2
  +\frac{d(d\fn-\fn+2)}{2}\eq\et
  -\frac{d \fn}{2} {m}\cdot {m} \ ,
\end{equation}
which is independent from $\omega,d_1,d_2$ individually but only depends on their combination $d\;$! This highly nontrivial fact guarantees the modularity
of elliptic blowup equation, which means in summation of the left hand side of \eqref{eq:uv-blowup} all terms share the same modular index, thus transform
as whole Jacobi form together!
In the case that $d\not\in\IN$ where the right hand side of
\eqref{eq:uv-blowup} vanishes, this is the end of the story. If
$d\in\IN$ so that the right hand side of \eqref{eq:uv-blowup} is
non-vanishing, we still need to show the right hand side also shares
the same index polynomial. The index polynomial of $\Lambda$ is simply,
\begin{equation}
  \mathrm{Ind}_{\Lambda}^{G}=\frac{(\fn-2)^2}{2\fn}(\eq+\et)^2.
\end{equation}
which together with \eqref{eq:ind-Ed} indeed sum up to \eqref{eq:ind}.

\subsection{Universality of elliptic blowup equations}
\label{sc:unv}

We demonstrate here an interesting property of the elliptic blowup
equations. The $\fn$ blowup equations \eqref{eq:uv-blowup} with a
fixed embedding $\phi_\lambda$ can be ordered by the characteristics
$a$ of the theta functions, where for two consecutive equations $a$
differ by $1/\fn$. We claim that if two consecutive unity blowup
equations are valid, the other equations must hold automatically. We
call this the universality of the elliptic blowup equations.\footnote{The following argument assumes the form of refined BPS expansion. Thus the
universality here does not contradict with our statement that choosing arbitrary three unity $r$ fields, i.e. three different characteristics $a$, one is able to use the
blowup equations to solve out the elliptic genus.}

Theta functions with characteristics have the following properties
\begin{align}
  &\theta_3^{[\frac{m}{2\fn}]}(\fn\tau,z) =
    \re^{\pi\ri\tau \frac{m^2}{4\fn} + \pi\ri z\frac{m}{\fn}}
    \theta_3(\fn\tau,z+\tfrac{1}{2}m\tau) \ , \quad m\in \IZ\ ,\\
  &\theta_4^{[\frac{m}{2\fn}]}(\fn\tau,z) =
    \re^{\pi\ri\tfrac{m}{2\fn}+\pi\ri\tau\tfrac{m^2}{4\fn}+\pi\ri
    z\tfrac{m}{\fn}}
    \theta_4(\fn\tau,z+\tfrac{1}{2}m\tau) \ , \quad m\in \IZ \ .
\end{align}
Therefore, shifting $z$ by $\tau$ is equivalent to shifting the
characteristic of these theta function by $1/\fn$. Then starting from
one unity blowup equation, let us shift $\eq$ by $\tau$ and
check how various Jacobi forms in \eqref{eq:uv-blowup} change.
\begin{itemize}
\item $\theta_i^{[a]}(\fn \tau, \ldots)$: the elliptic parameter
  changes by $-2\tau+\text{integer}\cdot \fn \tau$. The shift
  $\text{integer}\cdot \fn \tau$ can be removed at the expense of an
  additional exponential factor due to quasi-periodicity of the theta
  function, while the shift $-2\tau$, as we have argued, is equivalent
  to shifting $a$ by $-2/\fn$.
\item $A_{\omega}(m)$: it is a product of factors like
  $\theta_1(\tau, m_\alpha+ m\eq+n\et)$, $m,n\in\IZ$, therefore is invariant
  under this shift up to an exponential factor.
\item $\IE_k$: we first argue that under the shift $\eq\to \eq+1$,
  $\IE_k(\tau,\nound{m},\eq,\et)$ is invariant up to an exponential
  factor. The refined BPS are defined from the topological string free
  energy as\footnote{Here $q_j\equiv\exp(2 \pi i \epsilon_j)$ $j=1,2$
    and $q_{l}\equiv\sqrt{q_1/q_2}$, $q_{r}\equiv\sqrt{q_1 q_2}$. We
    will also use $v\equiv q_r$, $x\equiv q_l$ in Section \ref{sc:eg}
    and \ref{sec:HGk} to make contact with the literature such as
    \cite{DelZotto:2016pvm,DelZotto:2018tcj}.}
  \begin{equation}
    F^{\text{inst}}(\nound Q,\eq,\et) =
    \sum_{j_l,j_r\geq 0}\sum_{w\geq 1,\nound \beta}
    (-1)^{2(j_l+j_r)} N_{j_l,j_r}^{\nound \beta}
    \frac{\chi_{j_l}(q_l^w)\chi_{j_r}(q_r^w)}
    {w(q_1^{w/2}-q_1^{-w/2})(q_2^{w/2}-q_2^{-w/2})} \nound Q^{w \nound \beta},
    \label{BPSI}
  \end{equation}
 where
 \be
 \chi_j(q)=\frac{q^{2j+1}-q^{-2j-1}}{q-q^{-1}}.
 \ee
Exponentiating (\ref{BPSI})  the instanton partition function reads
\begin{equation}
Z^{\text{inst}}(\nound Q,\eq,\et) =\prod_\beta \prod_{j_{l/r}=0}^\infty \prod_{m_{l/r}=-j_{l/r}}^{j_{l/r}}\prod_{m_1,m_2=1}^\infty
\left(1-q_l^{m_l} q_r^{m_r} q_1^{m_1-\frac{1}{2}} q_2^{m_2-\frac{1}{2}} Q^{\beta}\right)^{ (-1)^{2(j_l+j_r)} N^\beta_{j_lj_r}}\ .
\label{BPSII}
\end{equation}
Using (\ref{BPSI}) and the checkerboard pattern identity
  \begin{equation}
    2j_l + 2j_r + 1 \equiv \nound r\cdot \nound \beta\quad \,(\text{mod}\, 2) \
  \end{equation}
  for non-vanishing BPS invariants one can show that the refined BPS partition
  function is invariant under the combined transformation
  \begin{equation}
    (\eq,\et,\nound t) \to (\eq+1,\et,\nound
    t+\nound r) \ .
  \end{equation}
  Since the $\nound r$-vector for the minimal 6d SCFTs in terms of the
  K\"ahler moduli $t_I,t_B$ have components
  \begin{equation}
    \nound r \equiv ( 0,\ldots, 0,\fn )\quad\mathrm{mod}\ (2\IZ)^{\mathrm{rk}+2},
  \end{equation}
  the elliptic genus $\IE_k(\tau,\nound{m},\eq,\et)$ is invariant under
  the shift $\eq \to \eq+1$, at most up to a sign if $\fn$ is odd. In
  fact
  \begin{equation}\label{eq:e1shift}
    \IE_k(\tau,\nound{m},\eq+1,\et) = (-1)^{\fn k} \IE(\tau,\nound{m},\eq,\et)
    \ .
  \end{equation}
  Together with the modular property of $\IE_k$, this implies
  $\IE_k(\tau,\nound{m},\eq,\et)$ is quasi-periodic for
  $\eq \to \eq+ \tau$. Similarly, one can show that
  $\IE_k(\tau,\nound{m},\eq,\et)$ is quasi-periodic for
  $\et \to \et+\tau$ as well. As examples one could inspect the
  expressions of $\IE_k$ for the $\fn=3,4$ models
  \cite{Kim:2016foj,Haghighat:2014vxa}, which are composed of
  $\theta_1(\tau,\text{integer}\cdot
  \eq+\text{integer}\cdot\et+\ldots)$ and therefore are indeed
  quasi-periodic for $\epsilon_{1,2} \to \epsilon_{1,2}+\tau$.  Now if
  we forget for the moment the shift on the mass parameters, the three
  instances of elliptic genera
  $\IE_k(\tau,\nound{m} - \eq \omega, \eq,\et-\eq)$,
  $\IE_k(\tau,\nound{m} - \et \omega, \eq-\et,\et)$,
  $\IE_k(\tau,\nound{m},\eq,\et)$ in \eqref{eq:uv-blowup} should
  already be invariant under $\eq\to \eq+\tau$ up to an exponential
  factor. The shift on the mass parameters $\nound{m} - \eq \omega$ in
  the first instance of $\IE_k$ means that upon $\eq\to\eq+\tau$ its
  elliptic parameter is in addition shifted by $\tau$ times a weight
  vector, which can also be removed at the expense of an additional
  exponential factor \cite{DelZotto:2017mee}\footnote{It is
    established in section~4 of \cite{DelZotto:2017mee} that the
    elliptic genera of 6d SCFTs with a pure gauge bulk theory in fact
    consist of special Weyl invariant Jacobi forms, which, among other
    things, are quasi-periodic if the elliptic parameter is shifted by
    $\tau$ times a weight vector.}.
\end{itemize}
In summary, the shift $\eq \to \eq+\tau$ is equivalent to shifting the
characteristic $a$ of the theta functions
$\theta_i^{[a]}(\fn\tau,\ldots)$ by $-2/\fn$ and in addition
multiplying each term in \eqref{eq:uv-blowup} by an exponential
factor. These exponential factors are determined by the index
polynomial of each term, which as a consequence of section~\ref{sc:mod}, should
be identical. Thus all the exponential factors can be factored out and
removed, and we are left again with a unity blowup equation where the
characteristic is shifted by $-2/\fn$. This immediately indicates that
starting from two consecutive unity blowup equations, we can obtain
all the other unity blowup equations, hence the universality property.

\subsection{Unity blowup equations}
\label{sc:unity}

The elliptic blowup equations depend on the choice of the
weight vector $\lambda$ and they take different forms depending on if
$\lambda\in Q^\vee$ or not. We first consider the former case where
$\phi_\lambda(Q^\vee)$ coincides with $Q^\vee$. We can denote $\omega$ as vector $\bv$ in the coroot lattice. Then $||\bv||^2/2$ and thus $d$ are always nonnegative integers, and as a result the right
hand side of \eqref{eq:uv-blowup} does not vanish
\begin{equation}\label{eq:u-blowup}
\begin{aligned}
  \sum_{\bv\in Q^\vee,d_{1,2}\in \IN}^{\frac{1}{2}||\bv||^2+d_1+d_2 =
    d} &(-1)^{|\bv|} \theta_i^{[a]} \Big(\fn
  \tau,(\fn-2)(\eq+\et)-\fn\big((\tfrac{1}{2}||\bv||^2+d_1)\eq+(\tfrac{1}{2}||\bv||^2+d_2)\et
  -m_{\bv}\big)\Big)\\[-1ex]
  &\times A_{\bv}(\nound{m}) \IE_{d_1}(\tau,\nound{m}-\eq\bv,
  \eq,\et-\eq) \IE_{d_2}(\tau,\nound{m}-\et\bv, \eq -\et,\et)
  \\[+2mm]
  \phantom{}=&\,\theta_i^{[a]}(\fn\tau,(\fn-2)(\eq+\et))\cdot
  \IE_d(\tau,\nound{m},\eq,\et) \ ,\quad\quad d\in\IN \ .
  \end{aligned}
\end{equation}
We say these elliptic blowup equations are of the \emph{unity} type
following the nomenclature in \cite{Huang:2017mis}. Since the number of embedding
$Q^\vee \hookrightarrow \phi_\lambda(Q^\vee) = Q^\vee$ is unique, the
number of unity blowup equations is the same as the range of
characteristics $a$, which is $\fn$. We also point out that, using the property that the leading $\Qtau$ order of
$\IE_d$ is $-d h^\vee_G/6$,\footnote{See more discussion in Section \ref{sc:ind-gen}.} these
equations in the leading order of $Q_{\tau}$ boil down to the identity
\begin{equation}\label{eq:conj1}
  \sum_{\alpha\in\Delta^+(G)} (\beta\cdot\alpha)^2 = h_G^\vee
  \,||\beta||^2 \ ,
\end{equation}
which is guaranteed by the Lie algebraic identity \eqref{eq:kk-id}.

The unity blowup equations are particular interesting as they allow us
to write down recursion formulas for the elliptic genera.

\subsubsection{Recursion formulas for elliptic genera}
\label{sc:rec}

The unity blowup equations \eqref{eq:u-blowup} can be put in the
following more suggestive form
\begin{equation}\label{eq:ubs}
\begin{aligned}
  \theta_i^{[a]}(\fn\tau,(\fn-2-d\fn)\eq +(\fn-2)\et)
    \IE_{d}(\tau,\nound{m},\eq,\et-\eq)
  &\\
  +\,\theta_i^{[a]}(\fn\tau,(\fn-2)\eq +(\fn-2-d\fn)\et)
  \IE_{d} (\tau,\nound{m},\eq-\et,\et)
  &\\
  -\,\theta_i^{[a]}(\fn\tau,(\fn-2)(\eq+\et))
  \IE_{d}(\tau,\nound{m}, \eq,\et)
  &= I_{d}(\IE_{<d}) \ .
\end{aligned}
\end{equation}
where $I_{d}(\IE_{<d})$ only contains the elliptic genera of degrees lower
than $d$. Since three copies of $\IE_d$ on the left hand side do not
depend on the characteristic $a$, if we have three such equations with
different $a$, which is indeed the case for all the minimal 6d
$\mathcal{N}=(1,0)$ SCFTs with pure gauge bulk theory, we can solve
$\IE_d(\tau,\md,\eq,\et)$ in terms of elliptic genera with lower
number of strings; in other words, we obtain recursion formulas for
elliptic genera.

Let us use the short hand notation\footnote{Note
  $d_0=||\alpha^{\vee}||^2/2$ is always implied. When $d_0=0$,
  $\theta_{i,\{d_0,d_1,d_2\}}^{[a]}$ does not depend on
  $\alpha^{\vee}$ since $\alpha^{\vee}=0$.}
\begin{equation}
  \theta_{i,\{d_0,d_1,d_2\}}^{[a]}=
  \theta^{[a]}_i\(\fn\tau, \fn m_{\alpha^{\vee}}
  +(\fn-2)(\eq+\et)-\fn((d_0+d_1)\eq+(d_0+d_2)\et)\) \ ,
\end{equation}
where $m_{\alpha^\vee} = {m}\cdot \alpha^\vee$, and furthermore define
\begin{equation}
  D_d=\Det\left(
    \begin{array}{ccc}
      \theta_{i,\{0,d,0\}}^{[a_1]}
      &\ \theta_{i,\{0,0,d\}}^{[a_1]} &\ \theta_{i,\{0,0,0\}}^{[a_1]}\\
      \theta_{i,\{0,d,0\}}^{[a_2]}
      &\ \theta_{i,\{0,0,d\}}^{[a_2]} &\ \theta_{i,\{0,0,0\}}^{[a_2]}\\
      \theta_{i,\{0,d,0\}}^{[a_3]}
      &\ \theta_{i,\{0,0,d\}}^{[a_3]} &\ \theta_{i,\{0,0,0\}}^{[a_3]}\\
    \end{array}
  \right),
\end{equation}
as well as
\begin{equation}
  D_{\{d_0,d_1,d_2\}}^{\alpha^{\vee}}=\Det\left(
    \begin{array}{ccc}
      \theta_{i,\{0,d,0\}}^{[a_1]}
      &\ \theta_{i,\{0,0,d\}}^{[a_1]} &\ \theta_{i,\{d_0,d_1,d_2\}}^{[a_1]}\\
      \theta_{i,\{0,d,0\}}^{[a_2]}
      &\ \theta_{i,\{0,0,d\}}^{[a_2]} &\ \theta_{i,\{d_0,d_1,d_2\}}^{[a_2]}\\
      \theta_{i,\{0,d,0\}}^{[a_3]}
      &\ \theta_{i,\{0,0,d\}}^{[a_3]} &\ \theta_{i,\{d_0,d_1,d_2\}}^{[a_3]}\\
    \end{array}
  \right).
\end{equation}
Note that $D_d=D_{\{0,0,0\}}^{\alpha^{\vee}}$ does not depend on
$\alpha^{\vee}$ since $\alpha^{\vee}=0$ when
$d_0=||\alpha^{\vee}||^2/2=0$.
Then the recursion formulas of $\IE_d$ solved from \eqref{eq:ubs} read
\begin{equation}\label{recursionZd}
  \IE_d=
  \sum_{d_0=\tfrac{1}{2}||\alpha^{\vee}||^2,\,d_{1,2}<d}^{d_0+d_1+d_2=d}
  (-1)^{|\alpha^{\vee}|}
  \frac{D^{\alpha^{\vee}}_{\{d_0,d_1,d_2\}}}{D_d}
  A_{\alpha^{\vee}}( m)
  \IE_{d_1}(m-\eq {\alpha^{\vee}},\eq,\et-\eq)
  \IE_{d_2}( m-\et {\alpha^{\vee}},\eq-\et,\et).
\end{equation}
Here the $\tau$ dependence is implied. 

Let us look at some examples. The one-string elliptic genus is given
by
\begin{equation}\label{Z1}
  \IE_1=\sum_{\substack{\alpha\in\Delta^\vee\\||\alpha^{\vee}||^2=2}}
  \frac{D^{\alpha^{\vee}}_{\{1,0,0\}}}{D_1}
  \frac{\eta^4}
  {\theta_1(m_{\alpha})
    \theta_1(m_{\alpha}-\eq)\theta_1(m_{\alpha}-\et)
    \theta_1(m_{\alpha}-\eq-\et)}
  \prod_{\displaystyle \substack{\beta\in \Delta\\
      \alpha^{\vee}\cdot \beta= 1}}
  \frac{\eta}{\theta_1(m_{\beta})}
\end{equation}
where $m_\beta =  m\cdot \beta$. In particular, for ADE type algebras, the $\IE_1$ formula can be further simplified due to the identification of roots
and coroots. Indeed, for $A_2,D_4,E_{6,7,8}$, we have the following universal formula
\begin{equation}\label{ADEZ1}
\IE_1=\sum_{\alpha\in \Delta}
\frac{D_\alpha}{D}\frac{\eta^4}{\theta_1(m_{\alpha})\theta_1(m_{\alpha}-\eq)\theta_1(m_{\alpha}-\et)\theta_1(m_{\alpha}-\eq-\et)}\prod_{\displaystyle
\substack{\beta\in \Delta\\ \alpha\cdot \beta=1}}\frac{\eta}{\theta_1(m_{\beta})}.
\end{equation}
Here $D_{\alpha}$ and $D$ are the short notations for $D^{\alpha}_{\{1,0,0\}}$ and $D_1$.

In the $Q_\tau \to 0$ limit, $\IE_1$ formula (\ref{Z1}) reduces to
the universal one-instanton partition function of 5d $\mathcal{N}=1$ pure SYM
theory \cite{Keller:2011ek,Keller:2012da}
\begin{equation}\label{eq:Z15d}
  Z_1 = \frac{1}{(1-\re^{-\eq})(1-\re^{-\et})}
  \sum_{\gamma\in\Delta_l}\frac{\re^{(h_G^\vee-1)m_\gamma/2}}
  {(1-\re^{-\eq-\et+m_\gamma})(\re^{m_\gamma/2}-\re^{-m_\gamma/2})
  \prod_{\alpha\cdot\gamma=1}(\re^{m_\alpha/2}-\re^{-m_\alpha/2})} \ ,
\end{equation}
where $\Delta_l$ denotes the set of long roots which is the same with coroots with $||\alpha^{\vee}||^2=2$.

Furthermore, the two-string elliptic genus is given by
\begin{equation}\label{Z2}
\begin{aligned}
  &\IE_2 =\frac{D^{\alpha^\vee=0}_{\{0,1,1\}}}{D_2}
    \IE_1( m,\eq,\et-\eq)
    \IE_1( m,\eq-\et,\et)+
    \sum_{||\alpha^\vee||^2 = 4} (-1)^{|\alpha^{\vee}|}
    A_{\alpha^\vee}( m) \frac{D^{\alpha^\vee}_{\{2,0,0\}}}{D_2}\\
  +&\sum_{||\alpha^\vee||^2=2} (-1)^{|\alpha^\vee|}
    A_{\alpha^\vee}( m)\bigg(\frac{D^{\alpha^\vee}_{\{1,1,0\}}}{D_2}
    \IE_1( m-\eq\alpha^\vee,\eq,\et-\eq)+
    \frac{D^{\alpha^\vee}_{\{1,0,1\}}}{D_2}
    \IE_1( m-\et\alpha^\vee,\eq-\et,\et)\bigg)   \ .
\end{aligned}
\end{equation}
Note in the bracket of the second line of (\ref{Z2}), the two terms are symmetric in $\eq\leftrightarrow\et$. In the later section, we use this formula to
compute the two-string elliptic genus of all 6d (1,0) minimal SCFTs with $G=A_2,D_4,F_4,E_{6,7,8}$. From (\ref{recursionZd}), we can also easily write
down the universal formula for three-string elliptic genus as
\begin{equation}\label{Z3}
\begin{aligned}
  \IE_3 =&\bigg[\frac{D^{\alpha^\vee=0}_{\{0,1,2\}}}{D_3}
    \IE_1( m,\eq,\et-\eq)
    \IE_2( m,\eq-\et,\et)+(\eq\leftrightarrow\et)\bigg]+
    \sum_{||\alpha^\vee||^2 = 6} (-1)^{|\alpha^{\vee}|}
    A_{\alpha^\vee}( m) \frac{D^{\alpha^\vee}_{\{3,0,0\}}}{D_3}\\
     &+\sum_{||\alpha^\vee||^2=4} (-1)^{|\alpha^\vee|}
    A_{\alpha^\vee}( m) \bigg[\frac{D^{\alpha^\vee}_{\{2,1,0\}}}{D_3}
    \IE_1( m-\eq\alpha^\vee,\eq,\et-\eq)+(\eq\leftrightarrow\et)\bigg] \\
  & +\sum_{||\alpha^\vee||^2=2} (-1)^{|\alpha^\vee|}
    A_{\alpha^\vee}( m)\bigg[\Big(\frac{D^{\alpha^\vee}_{\{1,2,0\}}}{D_3}
    \IE_2( m-\eq\alpha^\vee,\eq,\et-\eq)+(\eq\leftrightarrow\et)\Big)\\
   & \phantom{\quad\quad\quad\quad} + \frac{D^{\alpha^\vee}_{\{1,1,1\}}}{D_3}
    \IE_1( m-\eq\alpha^\vee,\eq,\et-\eq)\IE_1( m-\et\alpha^\vee,\eq-\et,\et) \bigg]   \ .
\end{aligned}
\end{equation}

From the topological string point of view, the $d$-string elliptic
genus $\IE_d$ encodes the BPS invariants $N_{j_l,j_r}^{\beta}$ with
base degree $d$ as well as multi-wrapping contributions from lower
base degree curves. Once all the elliptic genera up to certain base
degree $d$ are computed, all the BPS invariants up to base degree $d$
and arbitrary degrees along other directions can be extracted. The
recurison formulas \eqref{recursionZd} thus allow us to reproduce
the genus zero Gopakumar-Vafa invariants for the Calabi-Yau threefolds
associated to the minimal 6d SCFTs with $\fn=5,6,8,12$
\cite{DelZotto:2017mee}, and to compute the refined BPS invariants for
the first time in the literature.

Let us make a remark here concerning the validity of the recursion
formula \eqref{recursionZd}. Obviously, the recursion formula is only
well-defined when $D_d\neq 0$. We have checked that this is indeed true for all the minimal models except for the model of $G=SU(3)$ with
$d=1$, where both $D^{\alpha}_{\{1,0,0\}}$ and $D_1$ vanish. This is a
special situation since for the model with $G=SU(3)$ there are only
three choices of the characteristics $a_i$ and there may be certain symmetry enhancement for $\IE_1$ such that only two of the three unity blowup
equations are linearly independent. Note that this does
\emph{not} contradict with the fact that the universal one-instanton
partition function $Z_1$ of 5d pure SYM theories
\cite{Keller:2011ek,Keller:2012da} works perfectly for the $SU(3)$
theory which is recovered from one-string elliptic genus in the 5d
limit with $Q_\tau \to 0$. What happens in this limit is that the 6d
unity $\nound{r}$-field $\nound r=(0,0,0,3)$ splits to two 5d unity
$\nound{r}$-fields $\nound{r}_1$ and $\nound{r}_2$, and the leading $\Qtau$ order
term of $D_1$ is the difference of two contributions associated to
$\nound{r}_1$ and $\nound{r}_2$ respectively, both of which remain finite and
identical. Nevertheless, it should be emphasize that though recursion formula does not work for $\IE_1^{SU(3)}$, by assuming the refined BPS expansion,
one can still use two unity blowup equations and one vanishing blowup equation to solve out all the refined BPS invariants, which is what we have done in
\cite{Gu:2018gmy}. If further assuming the knowledge on $Z_0$, one can actually use one single unity blowup equations to solve out all refined BPS
invariants. To compute $\IE_2^{SU(3)}$ by recursion formula (\ref{Z2}), practically one can use the exact formula $\IE_1^{SU(3)}$ in \cite{Kim:2016foj}.
For all $d>1$, the recursion formula (\ref{recursionZd}) works well for $SU(3)$.

The identity $D_1^{SU(3)}=0$ here despite its simple form does
not seem so trivial.
In fact, we find it is a special case of the following series of
identities.  With $q=\re^{\pi \ri \tau}$ and $y=\re^{2 \pi \ri z}$ we define
the following double indexed $\theta$ functions
\be
\begin{array}{rl}
\theta_3^{[r/2n]}(n \tau, z)&=\displaystyle{q^\frac{r^2}{4n} y^\frac{r}{2n}\sum_{l\in \mathbb{Z}}  q^{l( nl+r)} y^l=\sum_{k=r\, (2 n)}
q^\frac{k^2}{4n}y^\frac{k}{2n}} \ ,\\ [4mm ]
\theta_4^{[r/2n]}(n \tau, z)&=\displaystyle{\re^\frac{\pi \ri r}{4n} q^\frac{r^2}{2n} y^\frac{r}{2n}\sum_{l\in \mathbb{Z}} (-1)^l  q^{l( nl+r)}
y^l=\sum_{k=r\, (2 n)} q^\frac{k^2}{4n}
(-y)^\frac{k}{2n}}\ ,
\end{array}
\ee
where the notations  $k=r\, (m)$  in the second sums, mean that $k$ runs over values of the form $r + m l $, for any  $l\in\mathbb{Z}$.
We claim that if
\be
\sum_{j=0}^{n-1} z_i=0 \quad \leftrightarrow \quad  \prod_{j=0}^{n-1} y_j=1,
\label{sumcondition}
\ee
the determinant of the $n\times n$ matrices defined from these $\theta$ functions
\begin{equation}
\begin{array}{rl}
\displaystyle{\det\left[\theta_3^{\left[\frac{i}{n}\right]}(n \tau,z_j)_{i,j=0,\ldots,n-1}\right]}&=0, \displaystyle{\quad {\rm if}\ \ \ 2\mid n
}\\[+3mm]
\displaystyle{\det\left[\theta_4^{\left[\frac{2 i+1}{2n}\right]}(n \tau,z_j)_{i,j=0,\ldots,n-1}\right]}&=0,  \displaystyle{\quad {\rm if}\ \ \ 2\nmid n}\
\\
\end{array}
\label{thetaclaim}
\end{equation}
vanish. Note the $n=1$ case is just the well-known fact
$\theta_1(\tau,0)=0$, while $n=3$ case implies the $D_1^{SU(3)}=0$
identity in our previous context. The proof proceeds in both cases in
(\ref{thetaclaim}) by showing that each term of the form
$q^m \prod_{j=1}^n y_j^{k_j}$ that occur in the expansion of the
determinant, constrained by (\ref{sumcondition}), appears once with
positive and once with negative sign\footnote{We thank Don Zagier for
  pointing this mechanism out to us.}.  The first case is notationally
simpler so we prove it explicitly. Using $2 \mid n$ and an irrelevant
rescaling\footnote{Followed by a renaming of the $\tilde z_i$ to $z_i$
  again.}  $z_i\rightarrow \tilde z_i= n z_i$ we rewrite the first
determinate in the statement in (\ref{thetaclaim}) as
\begin{equation}
\begin{array}{rl}
\ds{\det\Bigg[\bigg(\sum_{k=i\ (n)} q^\frac{k^2}{n} y_j^k\bigg)_{i,j=0,\ldots,n-1}\Bigg]}&=
\ds{ \sum_{\Pi} \sgn(\Pi)\sum_{\tiny \begin{array}{rl} k_0&= \Pi(0)\ (n)\\[-1.8mm] &\, \, \vdots \\[-.9 mm] k_{n-1}&=\Pi(n-1)\ (n)  \end{array}}}
q^\frac{k_0^2+\ldots + k_{n-1}^2}{n}
y_0^{k_0} \ldots y_{n-1}^{k_{n-1}}\\[9 mm]
&=\ds{\sum_{{\tiny \begin{array}{rl} &k_0,\ldots, k_{n-1} \in\IZ  \\[.5 mm] &k_i \neq k_l  \ {\rm mod}\ n  \end{array}}} \sgn(P_\nound{k})
q^\frac{k_0^2+\ldots + k_{n-1}^2}{n}
y_0^{k_0} \ldots y_{n-1}^{k_{n-1}}}\ .
\end{array}
\label{proof1}
\end{equation}
Here $P_\nound{k}$ is defined by the maps
$i\mapsto k_i \ {\rm mod} \ n$, for $i=0,\ldots , n-1$.  Let
$\bar k=({k_0+\ldots+k_{n-1}})/{n}$ be the average of $k_i$ and use
it to define $k_i^*=k_i- 2 \bar k$. Since $2\mid n$, $\bar k$ runs
over $\bar k=\frac{1}{2}\, (1)$ and hence $k_i^*\in \mathbb{Z}$.
Direct calculation shows that under the $*$ operation
$k_i\mapsto k_i^*$ the terms
$m_\nound{k}:=q^{(k_0^2+\ldots + k_{n-1}^2)/n} \prod_{j=0}^{n-1}
y_j^{k_j}$ in (\ref{thetaclaim}) stay invariant
$m_\nound{k}=m_{\nound{k^*}}$; the second factor due to
(\ref{sumcondition}). Let now $P^*_\nound{k}$ be defined by
$i\mapsto k^*_i \ {\rm mod} \ n$. It follows immediately that it
likewise defines a permutation of the indices $\{0,\ldots,n-1\}$,
however with the opposite parity. Hence $m_\nound{k}$ appears twice
with opposite sign and the sum (\ref{proof1}) is zero. The proof of
the second case in (\ref{thetaclaim}) proceeds analogously, with
appropriate relabelling of the indices.

\subsubsection{Uniqueness of recursion formulas}
\label{sc:uniq}

One important consequence of the universality property is that the
recursion formula \eqref{recursionZd} does not depend on the choice of
three different $\nound{r}$-fields in its construction, as it should. Let
us supppose that we already known $\IE_{<d}$ and we wish to compute
$\IE_d$ using the recursion formula \eqref{recursionZd} obtained from
three consecutive unity blowup equations forming the linear system
\begin{equation}\label{eq:lin-sys}
  \begin{pmatrix}
    \theta_{i,\{0,d,0\} }^{[a_1]} & \theta_{i,\{0,0,d\}}^{[a_1]}
    & \theta_{i,\{0,0,0\}}^{[a_1]} \\
    \theta_{i,\{0,d,0\} }^{[a_2]} & \theta_{i,\{0,0,d\}}^{[a_2]}
    & \theta_{i,\{0,0,0\}}^{[a_2]} \\
    \theta_{i,\{0,d,0\} }^{[a_3]} & \theta_{i,\{0,0,d\}}^{[a_3]}
    & \theta_{i,\{0,0,0\}}^{[a_3]}
  \end{pmatrix}
  \cdot
  \begin{pmatrix}
    \IE_d(\nound{m},\epsilon_1,\epsilon_2-\epsilon_1) \\
    \IE_d(\nound{m},\epsilon_1-\epsilon_2,\epsilon_2) \\
    - \IE_d(\nound{m},\epsilon_1,\epsilon_2)
  \end{pmatrix}
  = -
  \begin{pmatrix}
    I_d^{[a_1]}(\IE_{<d}) \\ I_d^{[a_2]}(\IE_{<d}) \\
    I_d^{[a_3]}(\IE_{<d})
  \end{pmatrix} \ ,
\end{equation}
where $a_2 - a_1 = a_3 - a_2 = 1/\fn$, and the matrix of
theta function on the l.h.s.\ is of full rank. If the
$\IE_d(\nound{m},\epsilon_1,\epsilon_2)$ solved from the linear system
is correct, so should be
$\IE_d(\nound{m},\epsilon_1,\epsilon_2-\epsilon_1)$,
$\IE_d(\nound{m},\epsilon_1-\epsilon_2,\epsilon_2)$, and the three
unity blowup equations in \eqref{eq:lin-sys} should all be correct as
well. Otherwise, the linear system could always be corrected by
\begin{equation}\label{eq:lin-sys-2}
  \begin{pmatrix}
    \theta_{i,\{0,d,0\} }^{[a_1]} & \theta_{i,\{0,0,d\}}^{[a_1]}
    & \theta_{i,\{0,0,0\}}^{[a_1]} \\
    \theta_{i,\{0,d,0\} }^{[a_2]} & \theta_{i,\{0,0,d\}}^{[a_2]}
    & \theta_{i,\{0,0,0\}}^{[a_2]} \\
    \theta_{i,\{0,d,0\} }^{[a_3]} & \theta_{i,\{0,0,d\}}^{[a_3]}
    & \theta_{i,\{0,0,0\}}^{[a_3]}
  \end{pmatrix}
  \cdot
  \begin{pmatrix}
    \IE_d(\nound{m},\epsilon_1,\epsilon_2-\epsilon_1) \\
    \IE_d(\nound{m},\epsilon_1-\epsilon_2,\epsilon_2) \\
    - \IE_d(\nound{m},\epsilon_1,\epsilon_2)
  \end{pmatrix}
  = -
  \begin{pmatrix}
    I_d^{[a_1]}(\IE_{<d}) \\ I_d^{[a_2]}(\IE_{<d}) \\
    I_d^{[a_3]}(\IE_{<d})
  \end{pmatrix}
  +
  \begin{pmatrix}
    R_1 \\ R_2 \\ R_3
  \end{pmatrix} \ .
\end{equation}
By inverting the matrix of theta functions in \eqref{eq:lin-sys} and
\eqref{eq:lin-sys-2} and subtracting the two equations from each
other, we get
\begin{equation}
    \begin{pmatrix}
    \theta_{i,\{0,d,0\} }^{[a_1]} & \theta_{i,\{0,0,d\}}^{[a_1]}
    & \theta_{i,\{0,0,0\}}^{[a_1]} \\
    \theta_{i,\{0,d,0\} }^{[a_2]} & \theta_{i,\{0,0,d\}}^{[a_2]}
    & \theta_{i,\{0,0,0\}}^{[a_2]} \\
    \theta_{i,\{0,d,0\} }^{[a_3]} & \theta_{i,\{0,0,d\}}^{[a_3]}
    & \theta_{i,\{0,0,0\}}^{[a_3]}
  \end{pmatrix}^{-1}\cdot
    \begin{pmatrix}
    R_1 \\ R_2 \\ R_3
  \end{pmatrix} = 0 \ ,
\end{equation}
which means the corrections $R_{1,2,3}$ must all vanish, as a
consequence of $D_d \neq 0$. Once the validity of the three unity
blowup equations in \eqref{eq:lin-sys} is established, using the
universality we can argue for the validity of all unity blowup
equations. The recursion formula constructed out of any three unity
blowup equations then should always gives the correct $\IE_d$ which
coincides with the solution of \eqref{eq:lin-sys}.

\subsection{Vanishing blowup equations}
\label{sc:vanishing}

We consider here the case where $\lambda \in P\backslash Q^\vee$. This
is only possible if $|P:Q^\vee| >1$, i.e.\ for the minimal 6d
$\mathcal{N}=(1,0)$ SCFTs with $G=SU(3),SO(8),E_6,E_7$. In this case,
$\phi_\lambda(Q^\vee)\neq Q^\vee$; $||\omega||^2/2$ for any
$\omega \in \phi_\lambda(Q^\vee)$ and thus $d$ is not an integer. In
fact we find
\begin{equation}\label{eq:omega-k}
  \frac{||\omega||^2}{2} = \frac{\fn-2}{\fn} + k \ ,\quad k\in
  \IZ_{\geq 0} \ ,
\end{equation}
where the minimum norm square ${(\fn-2)}/{\fn}$ is reached if and
only if $\omega$ is in a lowest dimensional irreducible
representation\footnote{This can be the fundamenal representation, the
  anti-fundamental representation, and in the case of $SO(8)$ the two
  spinor representations.} $\square_G$ of $G$. As a consequence, the
right hand side of the elliptic blowup equations \eqref{eq:uv-blowup}
vanishes
\begin{equation}\label{eq:v-blowup}
\begin{aligned}
  &\sum_{\omega\in\phi_\lambda(Q^\vee),d_{1,2}\in\IN}
  ^{\frac{1}{2}||\omega||^2+d_1+d_2 =d}
  (-1)^{|\phi_\lambda^{-1}(\omega)|} \cdot \theta_i^{[a]}
    \Big(\fn \tau,(\fn-2)(\eq+\et)-\fn\big((\tfrac{1}{2}||\omega||^2+d_1)\eq+(\tfrac{1}{2}||\omega||^2+d_2)\et
    -m_\omega\big)\Big)
  \\[-1ex]
  &\phantom{===}\times A_{\omega}( m) \cdot \IE_{d_1}(\tau,
  m-\epsilon_1\omega, \epsilon_1,\epsilon_2-\epsilon_1) \cdot
  \IE_{d_2}(\tau, m-\epsilon_2\omega, \epsilon_1
  -\epsilon_2,\epsilon_2) = 0 \ , \quad \lambda \not\in Q^\vee \,
  \text{fixed}\,d\ .
\end{aligned}
\end{equation}
Following the nomenclature of \cite{Huang:2017mis}, we call these equations of the
\emph{vanishing} type.

The number of inequivalent embeddings $\phi_\lambda$ of this kind is
$|P:Q^\vee|-1$, which happens to be the order of the automorphism
group of the Dynkin diagram $\Gamma_G$ for
$G = SU(3), SO(8), E_6, E_7$. As representatives we can choose
$\lambda$ to be a fundamental weight\footnote{Not all the fundamental
  weights generate a lowest dimensional irreducible
  representation. For instance, the fundamental weight of $SO(8)$
  corresponding to the central node in the Dynkin diagram
  $\Gamma_{SO(8)}$ generates the adjoint representation. The nodes
  associated to the $\square_G$-generating fundamental weights are
  permuted precisely by the automorphism group
  $\text{Aut}(\Gamma_G)$.} that generates $\square_G$ as the highest
weight. The total number of inequivalent vanishing blowup equations
for each of these models is $\fn(|P:Q^\vee|-1)$. Furthermore, using the property that the leading $\Qtau$ order of
$\IE_d$ is $-d h^\vee_G/6$,
the equations \eqref{eq:v-blowup} at the leading order requires that
\begin{equation}\label{vaniid}
  \sum_{\alpha\in\Delta^+} \((\omega\cdot\alpha)^2 -
  (\omega_0\cdot\alpha)^2\) = 2k h_G^\vee\ ,\quad\quad
  \omega\in \phi_\lambda(Q^\vee)\ ,\omega_0\in \square_G \ ,
\end{equation}
where the integer $k$ is associated to the weight vector $\omega$ by
\eqref{eq:omega-k}, and it is again guaranteed by the Lie algebraic
identity \eqref{eq:kk-id}.

With the elliptic genus solved from the recursion formulas
\eqref{recursionZd} plugged in, the vanishing blowup equations give
rise to infinitely many nontrivial identities of Jacobi forms. In the
lowest order, $d = (\fn-2)/\fn$ and $\phi_\lambda(Q^\vee)$ is chosen
to be one of the lowest dimensional representations
\begin{equation}\label{vanish0}
  \sum_{\omega\in\,\square_G}(-1)^{|\phi_\lambda^{-1}(\omega)|}\,
  \theta_{i}^{[a]}(\fn\tau,\fn m_\omega)
  \prod_{\displaystyle \substack{\beta\in \Delta_+\\
      \omega\cdot \beta=\pm 1}}\frac{1}{\theta_1(m_\beta)}=0 \ .
\end{equation}
This elegant formula specializing to $G=SU(3)$ and $SO(8)$ has been explicitly shown and checked in \cite{Gu:2018gmy}. Here we further checked it for
$E_6$ and $E_7$ for various characteristic $a$ to higher order of $\Qtau$. For example, for $E_6$, the relevant representation is $\bf 27$, with the
weights encoded in the character
\be
\chi_{\bf 27}^{E_6}=\sum_{i=1}^{27}\prod_{j=1}^6e^{m_jw_{ij}}.
\ee
Then for arbitrary $a\in\IZ/6$, the following identity holds:\footnote{Let us write the notations in components in case of any misunderstanding.
\be
|\phi_\lambda^{-1}(\omega_i)|={\sum_{j,k=1}^6(C_{E_6}^{-1})_{jk}\omega_{ij}},\quad m_{\omega_i}=\sum_{j,k=1}^6m_k(C_{E_6}^{-1})_{jk}\omega_{ij},\quad
{\omega_i}\cdot\beta=\sum_{j=1}^6\omega_{ij}\beta_j,\quad m_{\beta}=\sum_{j=1}^6 m_{j}\beta_j,
\ee
where $C_{E_6}$ is the Cartan matrix.}
\be\label{vanish0E6}
\sum_{i=1}^{27}(-1)^{|\phi_\lambda^{-1}(\omega_i)|}\theta_{3}^{[a]}(6\tau,6 m_{\omega_i})
  \prod_{\displaystyle \substack{\beta\in \Delta_+(E_6)\\
      \omega_i\cdot \beta=\pm 1}}\frac{1}{\theta_1(m_\beta)}=0\, .
\ee
Note there actually exist two $\bf 27$ representations due to the symmetry of Dynkin diagram of $E_6$, i.e. $|\text{Aut}(\Gamma_{E_6})|=2$, both of them
make (\ref{vanish0E6}) holds. This also explains why there are two copies of vanishing $r$ field for $E_6$ geometry, as we will see in next section in
Table \ref{tb:r-f6}. For higher base degree of the vanishing blowup equations, one can also write down some more complicated identities like
(\ref{vanish0}). We checked them for all $G=SU(3),SO(8),E_{6,7}$ in the setting of refined BPS expansion to very high orders.

We list various Lie theoretical data including the distribution of positive
roots with respect to product with any weight in $\square_G$ in
Table~\ref{tb:pr-dis}, from which one can check (\ref{vaniid}) indeed holds.
\begin{table}
  \centering
  \begin{tabular}{ccccccc}
    \toprule
    $G$  & $SU(3)$ & $SO(8)$ & $E_6$  & $E_7$\\
    \midrule
    $\mathrm{dim}(\square)$ & 3 & 8 & 27 & 56 \\
    $\mathrm{dim}(\Delta_+)$ & 3 & 12 & 36 & 63\\
    $\#\{\beta\in\Delta_+:\omega\cdot\beta=0,\forall\omega\in\square\}$
         & 1 & 6 &  20 & 36 \\
    $\#\{\beta\in\Delta_+:\omega\cdot\beta=\pm
    1,\forall\omega\in\square\}$
         & 2 & 6& 16 &  27  \\
    $\#\{\beta\in\Delta_+:\omega\cdot\beta=\pm
    2,\forall\omega\in\square\}$
         & 0 & 0 & 0 & 0 \\
    \bottomrule
  \end{tabular}
  \caption{Distribution of positive roots with respect to product with
    weights in $\square_G$.}\label{tb:pr-dis}
\end{table}

\section{Elliptic genera for 6d $(1,0)$ minimal SCFTs}
\label{sc:eg}

In this section we illustrate explicit the solution of
one-string and two-string elliptic genera of minimal 6d (1,0) SCFTs
with $G = F_4,E_6,E_7,E_8$, using the elliptic blowup equations. The
elliptic genera of the minimal theories with $G=SU(3), SO(8)$ have
been computed in \cite{Kim:2016foj,Haghighat:2014vxa}, and we also
reproduce some relevant results here. From these concrete results we
summarise some universal features of the elliptic genera, including
the expansion coefficients, the symmetric product approximation, and
some additional symmetries. They are presented immediately in the
first subsection, which one can then check in the following example
subsections.

In this and the next sections, we work with the reduced elliptic
genera which has the center-of-mass degree of freedom
removed:\footnote{We follow the notation of \cite{DelZotto:2016pvm}
  where $\mathbb{E}_{\widetilde{h}_{G}^{(k)}}$ is used to stress it is
  the RR elliptic genus of underlying 2d $(0,4)$ CFT associated to the
  $k$-strings in 6d $(1,0)$ minimal SCFT with gauge group $G$. It is
  the same with what we previously denoted as $\mathbb{E}_k$ to
  emphasize that it is coefficient of base degree $k$ in the
  topological string partition function.}  \be\label{eq:Edred}
\mathbb{E}_{h_{G}^{(k)}}=\mathbb{E}_{\widetilde{h}_{G}^{(k)}}/\mathbb{E}_{\mathrm{c.m.}},
\ee where \be
\mathbb{E}_{\mathrm{c.m.}}=-\frac{\theta_1(\eq)\theta_1(\et)}{\eta^2}.
\ee In the reduced version, elliptic genera normally obtain
simplification. For example, the reduced one-string elliptic genus is
independent from $\epsilon_-$, i.e. $SU(2)_l$, as expected.

\subsection{Universal behaviors of elliptic genera}
\label{sc:ind-gen}


\subsubsection{Universal expansion}
For all possible gauge group $G$, recall $v\equiv \exp(\pi\ri(\eq+\et))$ and $x\equiv \exp(\pi\ri(\eq-\et))$, we propose the following general ansatz for
the reduced $k$-string elliptic genera
\begin{equation}\label{gdef}
  \ehgk(v,x,\Qtau,Q_{m_i})=
  v^{k\dualCox-1}\Qtau^{-(k\dualCox-1)/6}
  \sum_{n=0}^{\infty}\Qtau^n\,  g^{(n)}_{k,G}(v,x,Q_{m_i}).
\end{equation}
Here all $g^{(n)}_{k,G}(v,x,Q_{m_i})$
are rational functions. In particular, $g^{(n)}_{1,G}$ is independent from $x$. One obvious symmetry for all $g^{(n)}_{k,G}$ is
\be\label{gsym1}
 g^{(n)}_{k,G}(v,x,Q_{m_i})=g^{(n)}_{k,G}(v,x^{-1},Q_{m_i}),
 \ee
which comes from the symmetry between $\eq$ and $\et$ in the Omega
background, and can be understood as the Weyl symmetry of
$SU(2)_x$. From on on we use $SU(2)_x$ to denote $SU(2)_l$ symmetry to stress the associated fugacity is $x$. We can further compute the $v$-expansion of
each
$g^{(n)}_{k,G}$ function where the coefficients are finite sum of
products between the characters of $SU(2)_x$ and characters of $G$
which respect Weyl symmetries of both groups. For example, $g_{k,G}^{(0)} = 1 + \ldots$ gives the Hilbert series of the reduced $k$ $G$-instanton moduli
space. In fact, we find plenty of universal coefficients for the first a few order $v$-expansion of $g^{(n)}_{k,G}$.

It is known that the Hilbert series of the reduced one-instanton moduli
space for any simple gauge group $G$ has the expansion
\cite{Benvenuti:2010pq}
\begin{equation}\label{g10G}
  g^{(0)}_{1,G}(v,Q_{m_i})=\sum_{k=0}^{\infty}\chi^G_{n\theta}v^{2n} \ ,
\end{equation}
where $\chi_{k\theta}$ is the character of the representation whose
highest weight is $k$-multiple of the longest root $\theta$; in
particular $\chi_\theta$ is the character of the adjoint
representation of $G$. In particular this is true for
$G=SU(3), SO(8), F_4, E_{6,7,8}$ when $g^{(0)}_{1,G}$ serves as the
leading contribution to one-string elliptic genus. As for sub- and
subsub-leading contributions, we find that except for $G=SU(3)$\footnote{From now on, to shorten formulas, we do not explicitly write $G$ in each
character.}
\be\label{g11G}
g^{(1)}_{1,G}(v,Q_{m_i})=1+\chi_{\theta}+\Big(1+\chi_{\theta}+\chi_{2\theta}+\chi_{\mathrm{Alt}^2 \theta}\Big)v^2
+\Big(2\chi_{2\theta}+\chi_{\mathrm{Alt}^2 \theta}+\chi_{3\theta}+B_2(G)\Big)v^4+\mathcal{O}(v^6),
\ee
while except for $G=SU(3), SO(8)$,
\begin{equation}\label{g12G}
  g^{(2)}_{1,G}(v,Q_{m_i})
  =2+2\chi_{\theta}+\chi_{\mathrm{Sym}^2\theta}+\mathcal{O}(v^2) \ .
\end{equation}
Here $B_2(G)$ are characters of some representations for which
we do not find any universal expressions, and we list them in
Table \ref{tb:g-reps}.\footnote{The bold numbers mean the character of representations with dimension of such number. Note different representations can
have the same dimension sometimes, for instance, the representations $\mathbf{35}_v$, $\mathbf{35}_s$ and $\mathbf{35}_c$ of $SO(8)$. To lighten the
notation, we do not distinguish them in the table. Nevertheless, they can be recovered by taking into account the symmetry of Dynkin diagrams.} The
exceptions of $SU(3)$ and $SO(8)$ can be
explained by the higher structures of $\IE_1$ revealed by its
intriguing relation with the Schur indices of certain rank one 4d
SCFTs discovered in \cite{DelZotto:2016pvm}, which we will review and
extend in section \ref{sec:HGk}.
\begin{table}[h]
  \centering
  \begin{tabular}{cccccccc}
    \toprule
    $G$  & $\chi_\theta$ & $\chi_{2\theta}$ & $\chi_{3\theta}$ & $B_2$
    & $C_6$ & $C_7$ & $C_8$\\
    \midrule
    $A_2$ & $\mathbf{8}$ & $\mathbf{27}$ & $\mathbf{64}$ & $2\cdot \mathbf{35}$
    & $\mathbf{1}$  &$\mathbf{27}$ & $\mathbf{8}$\\
    $D_4$ & $\mathbf{28}$ & $\mathbf{300}$ & $\mathbf{1925}$ & $\mathbf{4096}$
    & $2\cdot\mathbf{28}$ &$3\cdot\mathbf{567}$
    & $2\cdot(\mathbf{300}+\mathbf{350}+\mathbf{1})+3\cdot\mathbf{35}$\\
    $F_4$  & $\mathbf{52}$ & $\mathbf{1053}$ & $\mathbf{12376}$
     & $\mathbf{29172}$ &  $\mathbf{273}$ & $\mathbf{10829}$
     & $\mathbf{8424}+\mathbf{4096}+\mathbf{324}+\mathbf{26}$\\
    $E_6$ & $\mathbf{78}$ & $\mathbf{2430}$ & $\mathbf{43758}$
    & $\mathbf{105600}$  & $\mathbf{650}$ & $\mathbf{34749}$
    & $\mathbf{34749}+2\cdot\mathbf{5824}+\mathbf{650}+\mathbf{78}$\\
    $E_7$ & $\mathbf{133}$ & $\mathbf{7371}$ & $\mathbf{238602}$
    & $\mathbf{573440}$  & $\mathbf{1463}$ & $\mathbf{152152}$ & $\mathbf{150822}+\mathbf{40755}+\mathbf{1539}$\\
    $E_8$ & $\mathbf{248}$ & $\mathbf{27000}$ & $\mathbf{1763125}$ & $\mathbf{ 4096000}$ & 0 &  $\mathbf{779247}$ & $\mathbf{147250}$\\
    \bottomrule
  \end{tabular}
  \caption{Certain representations appearing in the expansion of
    $g_{k,G}^{(n)}$ functions.}\label{tb:g-reps}
\end{table}

Furthermore, we find the Hilbert series of reduced two-instanton modulis
space for any simple gauge group $G$ has the expansion
\be\label{g20G}
\ba
g^{(2)}_{0,G}(v,x,Q_{m_i})=&\,1 + (\chi_{\theta} + \chi_3)v^2 + \chi_{\theta}\chi_2 v^3
 +\big(\chi_{5}+\chi_{\theta}\chi_3+\chi_{\mathrm{Sym}^2\theta}\big)v^4
+\left(\chi_{\theta}\chi_4+(\chi_{2\theta}+\chi_{\mathrm{Alt}^2 \theta})\chi_2 \right) v^5\\
&+ \left(\chi_7+\chi_5\chi_{\theta}+\chi_3(\chi_{\mathrm{Sym}^2\theta}+\chi_{2\theta})+\chi_{\mathrm{Sym}^3\theta}  - C_6(G) \right) v^6\\
+\Big(\chi_{\theta}\chi_6&+(\chi_{2\theta}+\chi_{\mathrm{Alt}^2 \theta})\chi_4+(\chi_{2\theta} + \chi_{3\theta} +\chi_{\mathrm{Alt}^2
\theta}+B_2(G)+C_7(G))\chi_2\Big)v^7\\
+ \Big(\chi_9+\chi_7\chi_{\theta}+&\chi_5(\chi_{\mathrm{Sym}^2\theta}+\chi_{2\theta})+
\chi_3(\chi_{3\theta}+\chi_{2\theta}+B_2(G)+\chi_{\mathrm{Sym}^3\theta}  - C_6(G)) +\chi_{\mathrm{Sym}^4\theta}-C_8(G)\Big) v^8\\
+\Big(\chi_{\theta}\chi_8+&(\chi_{2\theta}+\chi_{\mathrm{Alt}^2 \theta})\chi_6+(\chi_{2\theta} + 2\chi_{3\theta} +\chi_{\mathrm{Alt}^2
\theta}+B_2(G)+C_7(G))\chi_4+\dots\Big)v^9+\dots
\ea
\ee

\noindent Here $\chi_n$ is the character of $n$-dimensional representation of $SU(2)_x$. The expansion coefficients up to $v^6$ were already observed
in \cite{Keller:2012da}, and we further push the observation up to
$v^8$. We have checked this expression to be consistent with all the results on Hilbert series of reduced two $G$ instanton moduli space
in \cite{Hanany:2012dm}. In particular it is true for
$G = SU(3), SO(8), F_4, E_{6,7,8}$ when $g_{2,G}^{(0)}$ is the leading
contribution to the two-string elliptic genera. Note that in this
expression, $C_6(G), C_7(G)$ are characters of certain
representations of $G$ for which universal expressions are not
found. They are collected for individual $G$ in Table \ref{tb:g-reps}. As for the subleading and subsubleading contribution to the two-string elliptic
genera, we find there exists the
following universal $v$-expansion: except for $G = SU(3)$,
\be\label{g21G}
\ba
g^{(2)}_{1,G}(v,x,Q_{m_i})=&\,1 +\chi_{\theta} + \chi_3 + (\chi_{\theta}+1)\chi_2 v
 +(\chi_{5}+(2\chi_{\theta}+3)\chi_3+(\chi_{\theta}+1)^2)v^2\\
&+\left((2\chi_{\theta}+1)\chi_4+(\chi_{2\theta}+(\chi_{\theta}+1)^2+\chi_{\mathrm{Sym}^2\theta} )\chi_2 \right) v^3\\
&+(\chi_{7}+(2\chi_{\theta}+3)\chi_5+\dots)v^4+\mathcal{O}(v^5)\, ,
\ea
\ee
while except for $G = SU(3)$ and $SO(8)$,
\be\label{g22G}
\ba
g^{(2)}_{2,G}(v,x,Q_{m_i})=&\,(\chi_5+(\chi_{\theta}+2)\chi_3+\chi_{\mathrm{Sym}^2\theta}+2\chi_{\theta}+4)
+\Big((\chi_{\theta}+1)\chi_4+((\chi_{\theta}+1)^2+2(\chi_{\theta}+1))\chi_2\Big)v\\
&+\Big(\chi_7+(2\chi_{\theta}+4)\chi_5+((\chi_{\theta}+1)^2+2\chi_{\mathrm{Sym}^2\theta}+6\chi_\theta+9)\chi_3+\dots\Big)v^2+\mathcal{O}(v^3)\, .
\ea
\ee

For the reduced three string elliptic genus $\mathbb{E}_{h_{G}^{(3)}}$, although we have not checked for all six $G$ due to the complexity of computation,
still we propose the following universal expansion:
\be\label{g30G}
\ba
g^{(3)}_{0,G}(v,x,Q_{m_i})=&\,1 + (\chi_3+\chi_{\theta})v^2 + (\chi_4+\chi_{\theta}\chi_2) v^3
 +\(\chi_{5}+\chi_{\theta}\chi_3+\chi_{\mathrm{Sym}^2\theta}+1\)v^4\\&+
 \Big(\chi_6+(2\chi_\theta+1)\chi_4+2\chi_{\mathrm{Sym}^2\theta}\Big)v^5\\
 &+\(2\chi_7+3\chi_\theta\chi_5
 +(\chi_{2\theta}+3\chi_{\mathrm{Sym}^2\theta}+1)\chi_3+\chi_{\mathrm{Sym}^3\theta}+\chi_{\mathrm{Sym}^2\theta}\)v^6+\mathcal{O}(v^7)
 \ .
\ea
\ee
We have checked this against the three-instanton Hilbert series for $SU(2)$, $G_2$, $SO(7)$, $Sp(4)$, $Sp(6)$ in \cite{Cremonesi:2014xha,Hanany:2014dia},
and against the three-string elliptic genus for $SU(3)$ \cite{Kim:2016foj}. Note the first two terms also agree with the rank three $E_6$ Hall-Littlewood
index ((A.14) in \cite{Gaiotto:2012uq}). For the subleading $\Qtau$ order, again except $SU(3)$, we propose
\be\label{g31G}
\ba
g^{(3)}_{1,G}(v,x,Q_{m_i})=&\, (\chi_3+\chi_{\theta}+1)+(\chi_4+(\chi_\theta+1)\chi_2)v\\
+(\chi_5+&(3\chi_\theta+4)\chi_3+2\chi_{\mathrm{Sym}^2\theta}+\chi_\theta+2)v^2+\mathcal{O}(v^3)
\ .
\ea
\ee
As in rank one and two cases, for $SU(3)$, the higher contributions begin to merge in at $\Qtau$ subleading order.

All above $v$-expansion coefficients can be easily obtained by setting $Q_m=1$ in $g^{(k)}_{n,G}(v,x,Q_m)$. Thus the rational functions
$g^{(k)}_{n,G}(v,x,1)$ are very useful as they encode most information. For large $k$ or $n$, such rational functions with generic $x$ turn out to be too
length. One can take the unrefined limit $x=1$ in $g^{(k)}_{n,G}(v,x,1)$ to still store meaning information on arbitrary order coefficients of
$v$-expansion. Indeed, when the fugacities of flavor as well as $SU(2)_x$ are
turned off, we find
\begin{gather}
  g_{1,G}^{(n)}(v) =
  \frac{1}{(1-v^2)^{2(h^\vee_G-1)}}\times P_{1,G}^{(n)}(v) \ ,\\
  g_{2,G}^{(n)}(v) =
  \frac{1}{(1-v^2)^{2(h^\vee_G-1)}(1+v)^{2b_G}(1+v+v^2)^{2h^\vee_G-1}}\times P_{2,G}^{(n)}(v)
  \ .
\end{gather}
The exponents $b_G$ are given by
\begin{center}
  \begin{tabular}{ccccccc}\toprule
    $G$  & $SU(3)$ & $SO(8)$ & $F_4$ & $E_6$  & $E_7$ & $E_8$ \\
    \midrule
    $b$ & 3 & 6 & 11 & 16 & 26 & 46 \\\bottomrule
  \end{tabular}\\
\end{center}
We notice that $b=5\dualCox/3-4$ except for $SU(3)$. The
numerators $P_{1,G}^{(n)}(v)$ and $P_{2,G}^{(n)}(v)$ are palindromic Laurent polynomials.
They have negative powers of $v$ when $n$ is
large. Nevertheless $P_{1,G}^{(0)}(v),P_{2,G}^{(0)}(v)$ are both
polynomials and their maximum degrees are $h_G^\vee-1$ and
$2(2h_G^\vee-1)+2b_G$ respectively. The explicit expressions of
$P_{k,G}^{(n)}(v)$ for the minimal SCFTs with $G=SU(3), SO(8),F_4,E_6,E_7,E_8$ are presented in the following example subsections and also
Appendix~\ref{ap:eg}.

\subsubsection{Symmetric product approximation}
It was noticed both in \cite{Hanany:2012dm} and \cite{Gaiotto:2012uq} that the reduced two $G$-instanton Hilbert series can be realized as certain
symmetric product of two one $G$-instantons as approximation:
\be
\frac{1}{1-vx^{\pm 1}}g^{(2)}_{0,G}(v,x,a)=\frac{1}{2}\bigg(\Big(g^{(1)}_{0,G}(v,a)\frac{1}{1-vx^{\pm
1}}\Big)^2+g^{(1)}_{0,G}(v^2,a^2)\frac{1}{1-v^2x^{\pm 2}}\bigg)+\mathcal{O}(v^4)\, .
\ee
Here we adopt their notation $a=Q_{m_G}$ to lighten the notation. It also was noticed in \cite{Cremonesi:2014xha} that the reduced three $G$-instanton
Hilbert series can be realized as certain symmetric product of three one $G$-instantons as approximation:
\be\label{threeg}
\ba
\frac{1}{1-vx^{\pm 1}}g^{(3)}_{0,G}(v,x,a)=&\frac{1}{6}\bigg(\Big(\frac{1}{1-vx^{\pm 1}}g^{(1)}_{0,G}(v,a)\Big)^3+\frac{3}{(1-vx^{\pm 1})(1-v^2x^{\pm
2})}g^{(1)}_{0,G}(v,a)g^{(1)}_{0,G}(v^2,a^2)\\
&+\frac{2}{1-v^3x^{\pm
    3}}g^{(1)}_{0,G}(v^3,a^3)\bigg)+\mathcal{O}(v^4) \ .
\ea
\ee
The above formulas have clear physical meaning. For example in (\ref{threeg}), the first term represents the configuration that three instantons are far
from each other, the second term represents the configuration that two instantons sit on the same site and the third one are far from them, while the
third term represents the configuration that all three instantons sit on the same site. Note the triple symmetric product would give the coefficient of
$v^4$ of $g^{(3)}_{0,G}$ as $\chi_{5}+\chi_{\theta}\chi_3+\chi_{2\theta}+\chi_{\mathrm{Sym}^2\theta}+1$, one can see the difference with (\ref{g30G})
begins to appear.

In fact, it is reasonable that arbitrary $k$ $G$-instanton Hilbert series can be realized as symmetric product of $k$ $G$-instantons as approximation:
\be
\ba
\frac{1}{1-vx^{\pm
    1}}g^{(k)}_{0,G}(v,x,a)
=\mathrm{Sym}^k_{\mathcal{M}_{G,1}}(v,x,a)+\mathcal{O}(v^4)\ ,
\ea
\ee
where $\mathrm{Sym}^k_{G,1}(v,x,a)$ can be obtain from generating function
\be\label{symgk}
\sum_{k=1}^\infty\mathrm{Sym}^k_{\mathcal{M}_{G,1}}(v,x,a)Q^k=\mathrm{PE}\bigg[\frac{Q}{1-vx^{\pm
1}}\,g^{(1)}_{0,G}(v,a)\bigg]\equiv\mathrm{PE}\bigg[\,\widetilde{g}^{\,(1)}_{0,G}(v,x,a)Q\bigg].
\ee
For example,
\be
\ba
\mathrm{Sym}^4_{\mathcal{M}_{G,1}}(v,x,a)=\frac{1}{24}\bigg(&\Big(\widetilde{g}^{(1)}_{0,G}(v,x,a)\Big)^4
+6\Big(\widetilde{g}^{(1)}_{0,G}(v,x,a)\Big)^2\widetilde{g}^{(1)}_{0,G}(v^2,x^2,a^2)
+3\Big(\widetilde{g}^{(1)}_{0,G}(v^2,x^2,a^2)\Big)^2\\
&+8\,\widetilde{g}^{(1)}_{0,G}(v,x,a)\widetilde{g}^{(1)}_{0,G}(v^3,x^3,a^3)+
6\,\widetilde{g}^{(1)}_{0,G}(v^4,x^4,a^4)\bigg)\\
=1 + (&\chi_3+\chi_{\theta})v^2 + (\chi_4+\chi_{\theta}\chi_2) v^3
 +\mathcal{O}(v^4) \ .
\ea
\ee
It is not hard to find that for all $k\ge 3$, the leading coefficients in $v$ expansion of symmetric product are the same:
\be
\ba
\mathrm{Sym}^k_{\mathcal{M}_{G,1}}(v,x,a)=&\,\frac{1}{k!}\bigg(\Big(\widetilde{g}^{(1)}_{0,G}(v,x,a)\Big)^k
+C_k^2\Big(\widetilde{g}^{(1)}_{0,G}(v,x,a)\Big)^{k-2}\widetilde{g}^{(1)}_{0,G}(a^2,v^2,x^2)+\dots\bigg)\\
=&\,1 + (\chi_3+\chi_{\theta})v^2 + (\chi_4+\chi_{\theta}\chi_2) v^3
 +\mathcal{O}(v^4) \ .
\ea
\ee
Here the first term represents all $k$ instantons are far from each other, while the second term represents two instantons sit at the same site and the
rest $k-2$ instanton are far from them and each other... From $v^4$, the interaction among instantons will contribute in.

We can also include $g^{(k)}_{0,G}$ into the elliptic genus to write down the above symmetric product approximation. For example in the reduced
three-string elliptic genus, since
\be
\mathbb{E}_{\widetilde{h}_{G}^{(3)}}(v,x,a,\Qtau)=\frac{v^{3\dualCox}}{1-vx^{\pm
1}}g^{(3)}_{0,G}(v,x,a)\Qtau^{-\dualCox/2}+\mathcal{O}(\Qtau^{-\dualCox/2+1}),
\ee
combining (\ref{threeg}), we obtain
\be
\ba
\mathbb{E}_{\widetilde{h}_{G}^{(3)}}(v,x,a,\Qtau)=
\frac{1}{6}\Big(&\,\mathbb{E}_{\widetilde{h}_{G}^{(1)}}(v,x,a,\Qtau)^3+
3\mathbb{E}_{\widetilde{h}_{G}^{(1)}}(v,x,a,\Qtau)\mathbb{E}_{\widetilde{h}_{G}^{(1)}}(v^2,x^2,a^2,\Qtau^2)\\
&+
2\mathbb{E}_{\widetilde{h}_{G}^{(1)}}(v^3,x^3,a^3,\Qtau^3)\Big)+\dots\, ,
\ea
\ee
which holds for the leading $\Qtau$ order and the first four $v$-expansion coefficients. For arbitrary $k$-strings elliptic genus, it is better to use
Hecke transformation. Neglecting the interaction among strings, the resulting $k$-strings elliptic genus
$\mathbb{E}_{\mathrm{Sym}_{G}^{(k)}}(v,x,a,\Qtau)$ can be generated from
\begin{equation} \label{eq:EGhecke}
\sum_{k=0}\mathbb{E}_{\mathrm{Sym}_{G}^{(k)}}(v,x,a,\Qtau)Q^k = \exp\Bigg[\sum_{n\geq 0} Q^n \frac{1}{n}
\sum_{\stackrel{cd=n}{c,d>0}}\sum_{b(\textrm{mod}~d)} \mathbb{E}_{\widetilde{h}_{G}^{(1)}}\left(\frac{c\, \tau+b}{d},c\, \epsilon_i, c\, m_G\right)
\Bigg].
\end{equation}
Note this relies on the Jacobi form nature of $\mathbb{E}_{\widetilde{h}_{G}^{(1)}}(\tau,\epsilon_i,m_G)$. Also take $d=1$ in (\ref{eq:EGhecke}), one will
go back to instanton formula (\ref{symgk}) where this is no modularity. Finally, we obtain
\be
\mathbb{E}_{\widetilde{h}_{G}^{(k)}}(v,x,a,\Qtau)=\mathbb{E}_{\mathrm{Sym}_{G}^{(k)}}(v,x,a,\Qtau)+\mathcal{O}(\Qtau^{-k\dualCox/6+1})+\mathcal{O}(v^{k\dualCox+4})
\ .
\ee
As we have checked this symmetric product approximation does not give exact subleading $\Qtau$ orders $g^{(k)}_{1,G}$ even for its leading $v$-expansion
coefficient. This means all subleading $\Qtau$ orders involves interaction among strings.

\subsubsection{Symmetries}
Besides the obvious symmetry
\begin{equation}
  \ehgk(v,x,\Qtau,Q_m)=\ehgk(v,x^{-1},\Qtau,Q_m) \ ,
\end{equation}
which comes from the symmetry between $\eq$ and $\et$ in Omega background, it was found in \cite{DelZotto:2016pvm} that the reduced one-string elliptic
genus $\mathbb{E}_{{h}_{G}^{(1)}}(v,\Qtau)$ satisfy an additional symmetry
\begin{equation}\label{E1sym}
  \mathbb{E}_{{h}_{G}^{(1)}}(\Qtau^{1/2}/v,\Qtau) = (-1)^{\fn+1}v^{2(1-{h^\vee_G}/{3})}
  Q_\tau^{({h^\vee_G}/{3}-1)/2}
  \mathbb{E}_{{h}_{G}^{(1)}}(v,\Qtau) \ .
\end{equation}
Here the dependence on $m_G$ is implicit. This symmetry was later interpreted as a spectral flow symmetry in \cite{DelZotto:2018tcj}. The left hand side
of (\ref{E1sym}) actually computes the NS-R elliptic genus, which should be equal to the R-R elliptic genus on the right hand side due to the lack of
chiral fermions in the minimal SCFT in consideration. See section 6.4 of \cite{DelZotto:2016pvm} for a detailed discussion.

We extend the symmetry (\ref{E1sym}) to arbitrary $k$-string elliptic genus $\ehgk(v,x,\Qtau)$:
\begin{equation}\label{Eksym}
  \ehgk\Big(\frac{\Qtau^{1/2}}{v},\frac{\Qtau^{1/2}}{x},\Qtau\Big)
  =(-1)^{\fn k+1}v^{\frac{k(k-5)}{6}h^\vee_G+k^2+1}
  x^{-\frac{k(k-1)}{6}h^\vee_G-k^2+1}
  Q_\tau^{{(k h^\vee_G-3)}/{6}}
  \ehgk\big(v,x,\Qtau\big)\ ,
\end{equation}
which can be derived by combining \eqref{eq:e1shift} and the modular anomaly of elliptic genera (\ref{eq:ind-Ed}). For the situation where 2d quiver
description is known, i.e. $G=SU(3)$ and $SO(8)$, the above symmetry can also be obtained by looking into the transformation of integrand of localization
with the quasi-periodicity of Jacobi theta function (\ref{thetaper1},\ref{thetaper2}). Note symmetry (\ref{Eksym}) is a nonperturbative symmetry, which
can not be seen from the $\Qtau$ expansion of the elliptic genus, except for the one-string case that is (\ref{E1sym}).\footnote{Practically, we find that
for the two-string elliptic genus, when $\Qtau$ order is enough high, for one order of $\Qtau$ goes up, the leading $v$ order goes down for 3. Thus, if
one naively does the transformation for the left hand side of (\ref{Eksym}) in $\Qtau$ expansion, one would get infinite negative order of $\Qtau$.
Similar situation also happens for three-string elliptic genus. But for one-string elliptic genus, luckily for one order of $\Qtau$ goes up, the leading
$v$ order goes down for 2, which only result in finite negative order of $\Qtau$.} This means (\ref{Eksym}) should be seen as the symmetry of the chiral
algebra associated to the underlying $(0,4)$ 2d CFT, as suggested in \cite{DelZotto:2016pvm}.

\subsection{Revisiting $G=SU(3)$ and $SO(8)$}
\label{sc:n34}

With the new understanding on the structure of ${r}$ fields
of 6d $(1,0)$ minimal SCFTs for all $G$, we now can reproduce all
${r}$ fields for $SU(3)$ and $SO(8)$ found in
\cite{Gu:2018gmy} using just the fundamental weights of the Lie
algebras. We summarize the correspondence between the ${r}$
fields given in \cite{Gu:2018gmy} and fundamental weights in Table
\ref{tb:r-a2} and \ref{tb:r-d4}.

\begin{table}
  \centering
  \begin{tabular}{*{4}{>{$}c<{$}}}\toprule
    & \nound{r} & \text{old } \nound{r}
    & \text{fundamental weights} \\\midrule
    \multirow{3}{*}{unity}
      & (0,0,0,1) & (0,0,0,1) & \multirow{3}{*}{$\omega_1+\omega_2$}\\
      & (0,0,0,3) & (0,0,0,3) & \\
      & (0,0,0,5) & (0,0,0,-1) & \\\midrule
    \multirow{3}{*}{vanishing}
      & (-2,2,0,1) & (-2,2,0,1) & \multirow{3}{*}{$\omega_1$}\\
      & (-2,2,0,3) & (0,-2,2,1) & \\
      & (-2,2,0,5) & (2,0,-2,1) & \\\midrule
    \multirow{3}{*}{vanishing}
      & (-2,0,2,1) & (-2,0,2,1) & \multirow{3}{*}{$\omega_2$}\\
      & (-2,0,2,3) & (0,2,-2,1) & \\
      & (-2,0,2,5) & (2,-2,0,1) &\\\bottomrule
  \end{tabular}
  \caption{The $\nound{r}$-fields of the $\fn=3$ model and the
    fundamental weights of ${\mf a}_2$ which induce the embedding
    $\phi: Q^\vee \hookrightarrow P$. The old $\nound r$-fields are
    from our previous paper \cite{Gu:2018gmy}. They are equivalent
    with the $\nound{r}$ in the second column by
    $2C\cdot\underline{n}$
    shift.}\label{tb:r-a2}

\end{table}

\begin{table}
  \centering
  \begin{tabular}{*{4}{>{$}c<{$}}}\toprule
    & \nound{r} & \text{old } \nound{r}
    & \text{fundamental weights} \\\midrule
    \multirow{3}{*}{unity}
      & (0,0,0,0,0,0) & (0,0,0,0,0,0) & \multirow{3}{*}{$\omega_c$}\\
      & (0,0,0,0,0,2) & (0,0,0,0,0,2) & \\
      & (0,0,0,0,0,4) & (0,0,0,0,0,4) & \\
      & (0,0,0,0,0,6) & (0,0,0,0,0,-2) & \\\midrule
    \multirow{3}{*}{vanishing}
      & (-2,2,0,0,0,0) & (-2,2,0,0,0,0) & \multirow{3}{*}{$\omega_1$}\\
      & (-2,2,0,0,0,2) & (-2,-2,0,0,2,2) & \\
      & (-2,2,0,0,0,4) & (0,0,-2,2,0,0) & \\
      & (-2,2,0,0,0,6) & (0,0,-2,-2,2,2) & \\\midrule
    \multirow{3}{*}{vanishing}
      & (-2,0,2,0,0,0) & (-2,0,2,0,0,0) & \multirow{3}{*}{$\omega_2$}\\
      & (-2,0,2,0,0,2) & (-2,0,2,0,2,2) & \\
      & (-2,0,2,0,0,4) & (0,-2,0,2,0,0) & \\
      & (-2,0,2,0,0,6) & (0,-2,0,-2,2,2) & \\\midrule
    \multirow{3}{*}{vanishing}
      & (-2,0,0,2,0,0) & (-2,0,0,2,0,0) & \multirow{3}{*}{$\omega_3$}\\
      & (-2,0,0,2,0,2) & (-2,0,0,-2,2,2) & \\
      & (-2,0,0,2,0,4) & (0,-2,2,0,0,0) & \\
      & (-2,0,0,2,0,6) & (0,-2,-2,0,2,2) &\\\bottomrule
  \end{tabular}
  \caption{The $\nound{r}$-fields of the $\fn=4$ model and the
    fundamental weights of ${\mf d}_4$ which induce the embedding
    $\phi: Q^\vee \hookrightarrow P$. The old $\nound r$-fields are
    from our previous paper \cite{Gu:2018gmy}. They are equivalent
    with the $\nound{r}$ in the second column by
    $2C\cdot\underline{n}$
    shift.}\label{tb:r-d4}
    \end{table}

The elliptic genera of 6d $(1,0)$ SCFT with $G=SU(3)$ were computed using Jeffrey-Kirwan residue in
\cite{Kim:2016foj}, and were checked to satisfy the elliptic
blowup equations \cite{Gu:2018gmy}. Following the general proposal (\ref{gdef}), the reduced two-string elliptic genus for $SU(3)$ model can be written
as
\begin{equation}
  \mathbb{E}_{{h}_{A_2}^{(2)}}(v,x,\Qtau,m_i) = v^5 Q_\tau^{-5/6}
  \sum_{n=0}^\infty Q_\tau^n g_{2,A_2}^{(n)}(v,x,Q_{m_i}) \ ,
\end{equation}
where $g_{2,A_2}^{(n)}(v,x,Q_{m_i})$ are rational functions. We
computed $g^{(n)}_{2,A_2}(v,x,Q_{m_i}=1)$ up to $n=10$. Let us turn off the fugacities of both $SU(3)$ and $SU(2)_x$, we obtain
\begin{equation}
  g^{(n)}_{2,A_2}(v,x=1,Q_{m_i}=1)
  =\frac{1}{(1-v)^{10} (1+v)^6 \left(1+v+v^2\right)^5}\times P^{(n)}_{2,A_2}(v),
\end{equation}
where all $P^{(n)}_{2,A_2}(v)$ are palindromic Laurent polynomials, in which only $P^{(0)}_{2,A_2}(v)$ is a true polynomial:
\begin{equation}
  P^{(0)}_{2,A_2}(v)=1 + v + 6 v^{2 }+ 17 v^{3 }+ 31 v^{4 }+ 52 v^{5 }+
  92 v^{6 }+ 110 v^{7 }+ 112 v^{8 }+ 110 v^{9 }+\dots+ v^{16}\ .
\end{equation}
Here the ellipsis is completed by making the expression palindromic.
This agrees with the Hilbert series of reduced two $SU(3)$-instanton
moduli space in \cite{Hanany:2012dm}. For the subleading order,
$P^{(1)}_{2,A_2}(v)$ starts with negative power of $v$, which is \emph{different}
from all the other minimal SCFTs.\footnote{This phenomenon as also occurring in one-string elliptic genus, will be discussed in detail in section
\ref{sec:HGk}.} Indeed,
\be
\ba
P^{(1)}_{2,A_2}(v)=v^{-4} \big(&1 + 3 v + 8 v^{2 }+ 11 v^{3 }+ 18 v^{4
}+ 13 v^{5 }+ 55 v^{6 }+ 238 v^{7 }+
 601 v^{8 }+ 1121 v^{9 }+ 1777 v^{10 }\\
&+ 2262 v^{11 }+ 2424 v^{12 }+ 2262 v^{13 }+\dots+v^{24}\big).
\ea
\ee

More results on $P^{(n)}_{2,A_2}(v)$ with $n>1$ can be found in Appendix~\ref{ap:eg}. Let us also show some results with generic fugacities, for example,
\be\label{g20A2}
\ba
g^{(2)}_{0,A_2}(v,x,m_i)=&\,1 + (\mathbf{8} + \chi_3)v^2 + \mathbf{8}\chi_2 v^3
 +\big(\chi_{5}+\mathbf{8}\chi_3+\mathrm{Sym}^2\mathbf{8}\big)v^4
+\left(\mathbf{8}\chi_4+(\mathbf{27}+\mathrm{Alt}^2 \mathbf{8})\chi_2 \right) v^5\\
&+ \left(\chi_7+\mathbf{8}\chi_5+(\mathrm{Sym}^2\mathbf{8}+\mathbf{27})\chi_3+\mathrm{Sym}^3\mathbf{8}  -\mathbf{1} \right) v^6+\mathcal{O}(v^7),
\ea
\ee
\be
\ba
v^4g^{(2)}_{1,A_2}(v,x,m_i)=&1+\chi_2v+(\chi_3+\mathbf{8})v^2+(\chi_4+\mathbf{8}\chi_2)v^3+(\chi_5+(\mathbf{8}+\mathbf{1})\chi_3+\mathrm{Sym}^2\mathbf{8})v^4\\
&+(\chi_6+\mathbf{8}\chi_4+\mathrm{Sym}^2\mathbf{8}\chi_2)v^5+\mathcal{O}(v^6).
\ea
\ee
Note (\ref{g20A2}) agree with our universal expansion formula (\ref{g20G}).

Similarly, the reduced three-string elliptic genus for $SU(3)$ model can be written as
\begin{equation}
  \mathbb{E}_{{h}_{A_2}^{(2)}}(v,x,\Qtau,m_i) = v^8 Q_\tau^{-4/3}
  \sum_{n=0}^\infty Q_\tau^n g_{3,A_2}^{(n)}(v,x,Q_{m_i}),
\end{equation}
where all $g_{3,A_2}^{(n)}(v,x,Q_{m_i})$ are rational functions. We
computed $g^{(n)}_{3,A_2}(v,x,Q_{m_i}=1)$ up to $n=6$. Turning off the fugacities of both $SU(3)$ and $SU(2)_x$, we obtain
\begin{equation}
  g^{(n)}_{3,A_2}(v,x=1,Q_{m_i}=1)
  =\frac{1}{(1-v)^{16} (1+v)^{10} \left(1+v^2\right)^5 \left(1+v+v^2\right)^6}\times P^{(n)}_{3,A_2}(v),
\end{equation}
where all $P^{(n)}_{3,A_2}(v)$ are palindromic Laurent polynomials, in which only $P^{(0)}_{3,A_2}(v)$ is a true polynomial:
\begin{equation}
\ba\label{P03A2}
  P^{(0)}_{3,A_2}(v)=&1 + 6 v^2 + 14 v^3 + 40 v^4 + 82 v^5 + 213 v^6 + 388 v^7 + 772 v^8 +
 1260 v^9 + 2079 v^{10} + 2986 v^{11}\\
 & + 4226 v^{12} + 5226 v^{13} +
 6384 v^{14} + 6940 v^{15} + 7334 v^{16} + 6940 v^{17}+\ldots+ v^{32},
 \ea
\end{equation}
\begin{equation}
\ba
  P^{(1)}_{3,A_2}(v)=&v^{-4}(1+v^2)(1 + 2 v + 8 v^{2 }+ 24 v^{3 }+ 62 v^{4 }+ 114 v^{5 }+ 242 v^{6 }+ 456 v^{7 }+
 964 v^{8 }+ 1926 v^{9 }\\
 &+ 4225 v^{10 }+ 8448 v^{11 }+ 16317 v^{12 }+
 28038 v^{13 }+ 44954 v^{14 }+ 64960 v^{15 }+ 87437 v^{16 }\\
 &+ 106636 v^{17 }+
 121046 v^{18 }+ 125368 v^{19 }+ 121046 v^{20}+\ldots+v^{38}).
 \ea
\end{equation}
Note $g^{(0)}_{3,A_2}$ agrees with our universal expansion formula (\ref{g30G}). More higher $P^{(n)}_{3,A_2}(v)$ can be find in Appendix ~\ref{ap:eg}.

The elliptic genera for the 6d $(1,0)$ SCFT with $G = SO(8)$ were
computed using Jeffrey-Kirwan residue in \cite{DelZotto:2017mee}, and they were checked to satisfy elliptic blowup equations in
\cite{Gu:2018gmy}. Let us write the reduced two-string elliptic genus as
\begin{equation}
  \mathbb{E}_{{h}_{D_4}^{(2)}}(v,x,\Qtau,m_i)
  =v^{11}\Qtau^{-11/6}\sum_{n=0}^{\infty}\Qtau^n
  g^{(n)}_{2,D_4}(v,x,m_i)\ .
\end{equation}
We computed $g^{(2)}_{n,D_4}(v,x,m_i=0)$ up to $n=6$. In particular, $g^{(2)}_{n,D_4}$ for $n=0,1,2$ agree with our universal expansion formulas
(\ref{g20G}, \ref{g21G}). Turning off the $SU(2)_x$ fugacity, we have
\begin{equation}
  g^{(n)}_{2,D_4}(v,x=1,m_i=0)
  =\frac{1}{(1-v)^{22} (1+v)^{12} \left(1+v+v^2\right)^{11}}\times
  P^{(n)}_{2,D_4}(v) \ .
\end{equation}
Here $P^{(n)}_{2,D_4}(v)$ are palindromic Laurent polynomials. In particular, only for $n=0,1$, they are true polynomials:
\be\label{P02D4}
\ba
P^{(0)}_{2,D_4}(v)=& 1+v+20 v^2+65 v^3+254 v^4+841v^5+2435 v^6  +6116
v^7+14290 v^8+29700 v^9\\&+55947 v^{10}+96519 v^{11}+152749
v^{12}+220408 v^{13}   +293226 v^{14}+359742 v^{15}\\&+406014
v^{16}+421960 v^{17}+406014 v^{18}+ \dots+v^{34} .
\ea
\ee
\begin{equation}
\ba
P^{(1)}_{2,D_4}(v)=&\,(1+v^2)\big(32 + 90 v + 697 v^2 + 2913 v^3 + 10582 v^4 + 34415 v^5 + 97961 v^6 +
 242492 v^7\\& + 540749 v^8 + 1085137 v^9 + 1958185 v^{10} +
 3205774 v^{11} + 4789888 v^{12} + 6522178 v^{13}\\& + 8110633 v^{14} +
 9248825 v^{15} + 9668450 v^{16} + 9248825 v^{17}+\dots+v^{32}\big) \ .
 \ea
\end{equation}
\noindent Note the above $g^{(2)}_{0,D_4}$ agrees with the $SO(8)$ two-instanton Hilbert series in
\cite{Hanany:2012dm}. More results on $P^{(n)}_{2,D_4}(v)$ with $n>1$ can be found in Appendix~\ref{ap:eg}.

\subsection{$G = F_4$}
\label{sc:n5}


The divisors and curves of the non-compact $\fn=5$ geometry are explained in
\cite{DelZotto:2017pti}. There are five compact divisors, all of which
are Hirzebruch surfaces $\IF_{n_i}$ of various degrees $n_i$. We
denote them by $\mf D_I$ ($I=0,1,\ldots,5$). They intersect with each
other like the affine dynkin diagram of $\mf f_4$
\begin{center}
  \includegraphics[width=0.5\linewidth]{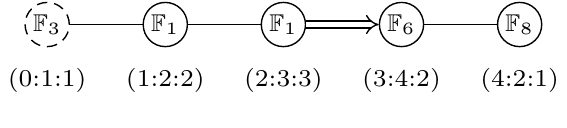}
\end{center}\vspace{-3ex}
where each node corresponds to a Hirzebruch surface, and two nodes are
connected if the corresponding Hirzebruch surfaces intersect at a
$\IP^1$ normal to their respective $\IP^1$ fibers. In the diagram
above we also give the ordering of the nodes $I$ and the associated
marks/comarks $a_I/a_I^\vee$ with the notation $(I:a_I:a_I^\vee)$
following \cite{DiFrancesco:1997nk}. The $\IF_3$ denoted by a dashed
circle corresponds to the affine node and it intersects with the base
at the $\IP^1$ with normal bundle
$\mc O(-5)\oplus \mc O(3) \to \IP^1$. The arrow with double line means
the $\IF_1$ and $\IF_6$ intersect at a $\IP^1$ which is the double
cover of the $(+1)$ curve in $\IF_1$.  See the illustration
in Figure~\ref{fg:F4}. There are six linearly independent curves,
which we choose for the moment to be the $\IP^1$ fibers $\Sigma_I$ of
the divisors $\mf D_I$ and the $(-5)$ curve in the base denoted by
$\Sigma_B$.  Denoting their complexified K\"ahler moduli by $t_I$ and
$t_B$, the linear combination
\begin{equation}\label{eq:ttau-n5}
  \sum_{I=0}^5 a_I t_I = \tau \ ,
\end{equation}
with $a_I$ the marks of $\mf f_4$, is the volume of the elliptic
fiber. Since we will be interested in the extraction of BPS invariants
from the partition function, we would like to also identify among the
compact curves the Mori cone generators. They include the $\IP^1$
fibers $\Sigma_I$ ($I=0,1,\ldots,4$), as well as the $\IP^1$ base of
the $\IF_1$ surface that intersects with $\IF_6$ (see the Dynkin
diagram above and the Figure~7 in \cite{DelZotto:2017pti}, which we
reproduce in Figure~\ref{fg:F4}). We denote the latter curve by
$\Sigma_b$, and it is related to $\Sigma_B$ by
\begin{equation}\label{eq:SBSb-n5}
  [\Sigma_B] = [\Sigma_b] + 3[\Sigma_0] + [\Sigma_1] \ .
\end{equation}
This implies the relation between their K\"ahler moduli
\begin{equation}\label{eq:tBtb-n5}
  t_B = t_b + 3t_0 + t_1 \ .
\end{equation}

The $C$-matrix of intersection between $\Sigma_I,\Sigma_b$ and
$D_I$\footnote{Note that here we use $\Sigma_b$
instead of $\Sigma_B$, which is why the matrix $C$ does not follow
exactly the pattern \ref{eq:matC}.}
\begin{equation}
   C =
  \begin{pmatrix}
    -2 & 1 & 0 & 0 & 0 \\
    1 & -2 & 1 & 0 & 0 \\
    0 & 1 & -2 & 2 & 0 \\
    0 & 0 & 1 & -2 & 1 \\
    0 & 0 & 0 & 1 & -2 \\
    0 & -1 & -1 & 0 & 0 \\
  \end{pmatrix}  \ .
\end{equation}
The semiclassical components of the partition function can be computed
using the prescription in section~\ref{sc:Zpert} with the
normalisation scheme in Appendix~\ref{ap:norm}. We obtain
\begin{equation}
  \begin{aligned}
    -F_{(0,0)}^{\text{cls}} =&
    \Big(\frac{t_{0}}{10}+\frac{t_{1}}{5}+\frac{3 t_{2}}{10}+\frac{2
        t_{3}}{5}+\frac{t_{4}}{5}\Big) t_{b}^2
    +\Big(\frac{3t_{0}^2}{10} +\frac{t_{0} t_{1}}{5}
      +\frac{t_{1}^2}{5}+\frac{3 t_{2}^2}{10}+\frac{6
        t_{3}^2}{5}+\frac{4 t_{4}^2}{5}+\frac{4t_{2} t_{3}}{5}\\
    &
      +\frac{2t_{2} t_{4}}{5} +\frac{6t_{3}
        t_{4}}{5} \Big) t_{b} +\frac{3t_{0}^3}{10} +\frac{3 t_{0}^2t_{1}}{10}
    +\frac{t_{0}t_{1}^2}{10}
    +\frac{t_{1}^3}{15}+\frac{t_{2}^3}{10}+\frac{6
      t_{3}^3}{5}+\frac{16 t_{4}^3}{15}+\frac{6t_{2} t_{3}^2}{5}\\
      &
    +\frac{4t_{2} t_{4}^2}{5} +\frac{12t_{3} t_{4}^2}{5}
    +\frac{2t_{2}^2 t_{3}}{5} +\frac{t_{2}^2 t_{4}}{5} +\frac{9t_{3}^2
      t_{4}}{5} +\frac{6t_{2} t_{3}t_{4}}{5},
  \end{aligned}
\end{equation}

\noindent which is consist with the universal formula \eqref{eq:pertF0-g}. Furthermore, using the relations
\eqref{eq:telltB},\eqref{eq:tBtb-n5} and \eqref{eq:ttau-n5},
$F_{(0,0)}^{\text{cls}}$ can be more succinctly written as
\begin{equation}
  -F^{(0,0)} = \frac{1}{10} t_{\rm ell}^2\tau
  + \frac{1}{2} t_{\rm ell} ( m, m) - \frac{3}{4} \tau ( m, m) + \ldots
\end{equation}
up to $\tau^3$ and terms cubic in $m_i$, which agrees with the
universal formula \eqref{eq:pertF0-uni}. Therefore the analysis in
section~\ref{sc:adm} goes through. Here for $F_4$,
\begin{equation}
   m = \sum_{i=1}^4 m_i \omega_i^\vee \ ,
\end{equation}
and
\begin{equation}
  ( m, m)=
  m_1^2 + 3 m_1 m_2 + 3 m_2^2 +
  4 m_1 m_3 + 8 m_2 m_3 + 6 m_3^2 + 2 m_1 m_4 +
  4 m_2 m_4 +
  6 m_3 m_4 + 2 m_4^2 \ .
\end{equation}
We also find
\begin{equation}
  F_{(1,0)}^{\text{cls}} =
  \frac{18}{5}t_0+\frac{16}{5}t_1+6t_2+12t_3+8t_4+\frac{36}{5}t_b  \ .
\end{equation}

\begin{table}
  \centering
  \begin{tabular}{*{3}{>{$}c<{$}}}\toprule
    & \nound{r}
    & \text{fundamental weights} \\\midrule
    \multirow{5}{*}{unity}
    & (0,0,0,0,0,1) & \multirow{5}{*}{$\omega_i \;(i=1,\ldots,4)$}\\
    & (0,0,0,0,0,3) & \\
    & (0,0,0,0,0,5) & \\
    & (0,0,0,0,0,7) & \\
    & (0,0,0,0,0,9) & \\\bottomrule
  \end{tabular}
  \caption{The $\nound{r}$-fields of the $\fn=5$ model and the fundamental
    weights of $\mf f_4$ which induce the same embedding
    $\phi: Q^\vee \hookrightarrow P$. All the $\nound{r}$-fields and all
    the fundamental weight induce the same embedding as $Q^\vee = P$
    for $\mf f_4$.}\label{tb:r-f5}
\end{table}

Imposing the admissibility condition \eqref{eq:rtau0} and the BPS
checkerboard pattern condition \eqref{eq:rn}, which
specialises to
\begin{equation}
  \nound{r} \equiv (0,0,0,0,0,1)\quad \text{mod}\;2 \ ,
\end{equation}
there are only
five inequivalent $\nound{r}$-fields, and we list their representatives
in Table~\ref{tb:r-f5}. According to the discussion in
section~\ref{sc:ebp}, we should classify them according to the
embedding $\phi_\lambda: Q^\vee \hookrightarrow P$ induced by the
reduced $\nound{r}$-vector $\lambda$ defined in
\eqref{eq:lambda-alpha}. In the case of the $\fn=5$ model, all the
$\nound{r}$-fields have the same reduced $\lambda = (0,0,0,0)$, which
induces the unique embedding
$\phi_{(0,0,0,0)}: Q^\vee \hookrightarrow P = Q^\vee$. As a
consequence, this model has no blowup equation of the vanishing
type. We notice that all the fundamental weights $\omega_i$ also
induce the same embedding (which is not the case in all the other
models.)

We use \eqref{Z1} and \eqref{Z2} to compute the one-string and two-string
elliptic genus.
The one-string elliptic genus does not depend on $SU(2)_x$. Its expansion
in $Q_\tau$ reads
\begin{equation}
  \mathbb{E}_{{h}_{F_4}^{(1)}}(v,\Qtau,m_i)=
  v^{8}\Qtau^{-4/3}\sum_{n=0}^{\infty}g^{(n)}_{1,F_4}(v,Q_{m_i})\Qtau^n,
\end{equation}
where $g^{(n)}_{1,F_4}(v,Q_{m_i})$ are rational
functions. Turning off all flavor fugacities,
\begin{equation}
  g^{(n)}_{1,F_4} (v,Q_{m_i}=1) = \frac{1}{(1-v^2)^{16}}\times
  P^{(n)}_{1,F_4}(v)~,
\end{equation}
where
\begin{equation}
  \ba
  P^{(0)}_{1,F_4}(v)
  &=1+36 v^2+341 v^4+1208 v^6+1820 v^8+1208 v^{10}+341 v^{12}+36v^{14}+v^{16},\\
  P^{(1)}_{1,F_4}(v)
  &=(1 + v^2)^2\(53 + 1478 v^2 + 9419 v^4 + 18036 v^6 + 9419 v^8 +
  1478 v^{10} + 53 v^{12}\),\\
  P^{(2)}_{1,F_4}(v)
  &=1484 + 36252 v^2 + 241608 v^4 + 663716 v^6 + 909400 v^8 +\dots + 1484 v^{16}.
  \ea
\end{equation}
The ellipsis in $P^{(2)}_{1,F_4}(v)$ is completed by making the
expression palindromic. Here the leading order expression
$g^{(0)}_{1,F_4}$ agrees with the Hilbert series of the reduced moduli
space of one $F_4$-instanton in \cite{Benvenuti:2010pq}, which is not
surprising since the one-string formula \eqref{Z1} reduces to the
one-instanton partition function \eqref{eq:Z15d} in the $Q_\tau \to 0$
limit. Furthermore, higher order expressions agree with
\cite{DelZotto:2016pvm}.

The $Q_\tau$ expansion of the two-string elliptic genus reads
\begin{equation}
  \mathbb{E}_{{h}_{F_4}^{(2)}}(v,x,\Qtau,m_i)=
  v^{17}\Qtau^{-17/6}\sum_{n=0}^{\infty}g^{(n)}_{2,F_4}(v,x,Q_{m_i})\Qtau^n,
\end{equation}
where $g^{(n)}_{2,F_4}(v,x,Q_{m_i})$ are rational functions. Turning
of flavor fugacities and $SU(2)_x$, we find
\begin{equation}
  g^{(n)}_{2,F_4} (v,x=1,Q_{m_i} =1)
  = \frac{1}{(1-v)^{34} (1+v)^{22} \left(1+v+v^2\right)^{17}}\times P^{(n)}_{2,F_4}(v)~,
\end{equation}
where
\begin{equation}
  \ba
  P^{(0)}_{2,F_4}(v)
  &=\,1+5 v+48 v^2+287 v^3+1560 v^4+7503 v^5+32316 v^6+125355 v^7+444325
  v^8\\
  &+1443572 v^9+4322993 v^{10}  +11989241 v^{11}+30913094 v^{12}+74321701 v^{13}+167106519
  v^{14}\\
  &+352245510 v^{15}+697557618 v^{16}+1300152932 v^{17} +2284606168 v^{18}+3790004228 v^{19}\\
  &+5943020899
  v^{20}+8818128233 v^{21}+12392104012 v^{22}+16505926853 v^{23}\\
  &  +20851379873 v^{24}+24994963144
  v^{25}+28442119825 v^{26}+30731161887 v^{27}\\
  &+31533797982 v^{28}  +30731161887 v^{29} + \dots+v^{56} .\\
  \ea
\end{equation}
\begin{equation}
  \ba
  P^{(1)}_{2,F_4}(v)
  &=\,(1+v^{2}) (56+386 v+3217 v^{2}+20295 v^{3}+110327 v^{4}+529286
  v^{5}+2266151 v^{6}\\
  & +8718327 v^{7}+30479449 v^{8}+97433532 v^{9}+286304088 v^{10}+777049966 v^{11}\\
  &+1956035588 v^{12}
  +4581942186 v^{13}+10017235514 v^{14}+20492637094 v^{15}\\
  &+39315499928 v^{16}+70871529676 v^{17}
  +120240591034 v^{18}+192278945658 v^{19}\\
  &+290168035137 v^{20}+413676858801 v^{21} +557641624668
  v^{22}+711294838217 v^{23}\\
  &+859008747683 v^{24}+982638991174 v^{25} +1065069893896 v^{26}
  +1094033908456 v^{27}\quad\quad\\
  &+1065069893896 v^{28}+\dots+v^{56}).
  \ea
\end{equation}
and the ellipses are again completed by making the expressions
palindromic.  Here the leading order expression $g^{(0)}_{2,F_4}$
agrees with the Hilbert series of the reduced moduli space of two
$F_4$-instanton in \cite{Hanany:2012dm}. Some polynomials
$P_{2,F_4}^{(n)}(v)$ of higher order $n$ can be found in the
appendix~\ref{ap:eg}.

We also use the expressions of $\IE_1,\IE_2$ to extract the BPS
invariants $N_{j_l,j_r}^{\beta}$. For this purpose, we need to use
instead the K\"ahler moduli $t_b,t_I$ ($I=0,1,\ldots,4$) associated to
the Mori cone generators. The results are tabulated in
appendix~\ref{ap:BPS}. They display the proper checkerboard pattern, and
reproduce the known genus 0 Gopakumar-Vafa invariants
\cite{DelZotto:2017mee}. At base degree one, we also notice a pattern that the only non-vanishing BPS invariants for the
curve classes $\beta = (0,k,0,0,0,1)$, $(0,0,k,0,0,1)$ are
\begin{equation}
  N^{(0,k,0,0,0,1)}_{0,k} = N^{(0,0,k,0,0,1)}_{0,k} = 1 \ ,\quad
  k=0,1,\ldots \ ,
\end{equation}
which in fact can be proved from the one-string formula \eqref{Z1} with the help of (\ref{clusterformula}), (\ref{clusterformula2}) and
(\ref{clusterformula3}).

\subsection{$G = E_6$}
\label{sc:n6}


The divisors and curves of the non-compact $\fn=6$ geometry is explained in
\cite{DelZotto:2017pti}. There are seven compact divisors, which are
Hirzebruch surfaces $\IF_{n_i}$ of various degrees $n_i$. We denote
them by $D_I$ ($I=0,1,\ldots,6$). They intersect with each other like
the affine dynkin diagram of $\mf e_6$
\begin{center}
  \includegraphics[width=0.5\linewidth]{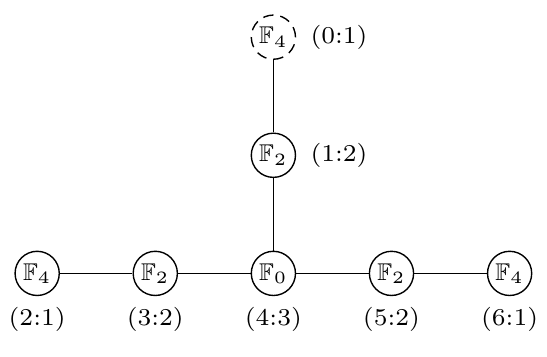}
\end{center}\vspace{-3ex}
where each node corresponds to a Hirzebruch surface and two nodes are
connected if the corresponding Hirzebruch surfaces intersect (see
Figure~5 in \cite{DelZotto:2017pti}).  In the diagram above we also give
the ordering of the nodes $I$ and the associated marks $a_I$ with the
notation $(I:a_I)$ following \cite{DiFrancesco:1997nk}. The $\IF_4$
denoted by a dashed circle corresponds to the affine node and it
intersects with the base at the $\IP^1$ with normal bundle
$\mc O(-6)\oplus \mc O(4) \to \IP^1$. There are eight linearly
independent curves, which we choose for the moment to be the $\IP^1$
fibers $\Sigma_I$ of the divisors $D_I$ and the $(-6)$ curve in the
base denoted by $\Sigma_B$. Denoting their complexified K\"ahler moduli
by $t_I$ and $t_B$, the linear combination
\begin{equation}\label{eq:ttau-n6}
  \sum_{I=0}^6 a_I t_I = \tau
\end{equation}
with $a_I$
the marks of $\mf e_6$, is the volume of the elliptic fiber. We also
identify the Mori cone generators. They include the $\IP^1$ fibers
$\Sigma_I$ ($I=0,1,\ldots,6$), as well as the $\IP^1$ base of the
$\IF_0$ surface in the center. We denote the last curve by $\Sigma_b$,
and it is related to $\Sigma_B$ by
\begin{equation}
  [\Sigma_B] = [\Sigma_b] + 4[\Sigma_0] + 2[\Sigma_1] \ .
\end{equation}
This implies the following relation of their K\"ahler moduli
\begin{equation}\label{eq:tBtb-n6}
  t_B = t_b + 4t_0 + 2t_1 \ .
\end{equation}

The $ C$-matrix of intersection between $\Sigma_I,\Sigma_b$ and
$D_I$ is
\begin{equation}
   C =
  \begin{pmatrix}
    -2 & 1 & 0 & 0 & 0 & 0 & 0 \\
    1 & -2 & 0 & 0 & 1 & 0 & 0 \\
    0 & 0 & -2 & 1 & 0 & 0 & 0 \\
    0 & 0 & 1 & -2 & 1 & 0 & 0 \\
    0 & 1 & 0 & 1 & -2 & 1 & 0 \\
    0 & 0 & 0 & 0 & 1 & -2 & 1 \\
    0 & 0 & 0 & 0 & 0 & 1 & -2 \\
    0 & 0 & 0 & 0 & -2 & 0 & 0 \\
  \end{pmatrix}  \ .
\end{equation}
The semiclassical components of the partition function can be computed
using the prescription in section~\ref{sc:Zpert} with the
normalisation scheme in Appendix~\ref{ap:norm}. We obtain
\begin{equation}
\begin{aligned}
  &-F_{(0,0)}^{\text{cls}}
  =
    \left(\frac{t_{0}}{12}+\frac{t_{1}}{6}+\frac{t_{2}}{12}+\frac{t_{3}}{6}
    +\frac{t_{4}}{4}+\frac{t_{5}}{6}+\frac{t_{6}}{12}\right) t_{b}^2\\
  &\phantom{=}\left(\frac{t_{0}^2}{3}
    +\frac{t_{1}^2}{3}+\frac{t_{2}^2}{3}+\frac{t_{3}^2}{3}+\frac{t_{5}^2}{3}
    +\frac{t_{6}^2}{3}+\frac{t_{0}
    t_{1}}{3} +\frac{t_{2} t_{3}}{3} +\frac{t_{5}
    t_{6}}{3} \right) t_{b} \\
  &\phantom{=}+\frac{4t_{0}^3}{9} +\frac{2
    t_{1}^3}{9}+\frac{4 t_{2}^3}{9}+\frac{2 t_{3}^3}{9}+\frac{2 t_{5}^3}{9}+\frac{4
    t_{6}^3}{9}
    +\frac{2t_{0}^2t_{1}}{3} +\frac{t_{0}t_{1}^2}{3}
    +\frac{t_{2} t_{3}^2}{3}  +\frac{2t_{2}^2
    t_{3}}{3} +\frac{2t_{5}
    t_{6}^2}{3}+\frac{t_{5}^2 t_{6}}{3},
\end{aligned}
\end{equation}

\noindent which is consistent with the universal formula
\eqref{eq:pertF0-uni}. Using the relations
\eqref{eq:telltB}, \eqref{eq:tBtb-n6}, \eqref{eq:ttau-n6}, we can
express $F^{(0,0)}(\nound t)$ in terms of the K\"ahler moduli
$t_{\text{ell}}, \tau, m_i$ $(i=1,\ldots,6)$ and find
\begin{equation}
  -F^{(0,0)} = \frac{1}{12} t_{\rm ell}^2\tau
  + \frac{1}{2} t_{\rm ell} ( m, m) - \tau ( m, m) + \ldots
\end{equation}
up to $\tau^3$ and terms cubic in $m_i$, where
\begin{equation}
   m = \sum_{i=1}^6 m_i \omega_i^\vee \ .
\end{equation}
It is in agreement with the universal expression
\eqref{eq:pertF0-uni}, and thus the analysis in section~\ref{sc:adm}
goes through.
We also find
\begin{equation}
  F_{(1,0)}^{\text{cls}} =
  \frac{9}{2}t_0+5t_1+\frac{9}{2}t_2+5t_3+\frac{3}{2}t_4+5t_5+\frac{9}{2}t_6+8t_b \ .
\end{equation}

\begin{table}
  \centering
  \begin{tabular}{*{4}{>{$}c<{$}}}\toprule
    & \multicolumn{2}{c}{$\nound{r}$}
    & \text{fundamental weights} \\\midrule
    \multirow{3}{*}{unity}
    & (0,0,0,0,0,0,0,0) & (0,0,0,0,0,0,0,2) & \multirow{3}{*}{$\omega_1,\omega_4$}\\
    & (0,0,0,0,0,0,0,4) & (0,0,0,0,0,0,0,6) & \\
    & (0,0,0,0,0,0,0,8) & (0,0,0,0,0,0,0,10) &\\\midrule
    \multirow{3}{*}{vanishing}
    & (-2,0,2,0,0,0,0,0) & (-2,0,2,0,0,0,0,2) & \multirow{3}{*}{$\omega_2,\omega_5$}\\
    & (-2,0,2,0,0,0,0,4) & (-2,0,2,0,0,0,0,6) & \\
    & (-2,0,2,0,0,0,0,8) & (-2,0,2,0,0,0,0,10) &\\\midrule
    \multirow{3}{*}{vanishing}
    & (-2,0,0,0,0,0,2,0) & (-2,0,0,0,0,0,2,2) & \multirow{3}{*}{$\omega_3,\omega_6$}\\
    & (-2,0,0,0,0,0,2,4) & (-2,0,0,0,0,0,2,6) & \\
    & (-2,0,0,0,0,0,2,8) & (-2,0,0,0,0,0,2,10) & \\\bottomrule
  \end{tabular}
  \caption{The $\nound{r}$-fields of the $\fn=6$ model and the
    fundamental weights of $\mf e_6$ which induce the same embedding
    $\phi: Q^\vee \hookrightarrow P$. They can be divided into three
    groups; inside each group $\nound{r}$-fields or fundamental weights
    induce the same embedding.}\label{tb:r-f6}
\end{table}

Imposing the admissibility condition \eqref{eq:rtau0} and the BPS
checkerboard pattern condition \eqref{eq:rn}, which
specialises to
\begin{equation}
  \nound{r} \equiv (0,0,0,0,0,0,0,0)\quad \text{mod}\;2 \ ,
\end{equation}
there are in
total 18 inequivalent $\nound{r}$-fields, and we list their
representatives in Table~\ref{tb:r-f6}. We classify them according to
the embeddings $\phi_\lambda: Q^\vee \hookrightarrow P$ induced by the
reduced $\nound{r}$-field $\lambda$ defined in
\eqref{eq:lambda-alpha}, \eqref{eq:QvP}. We also list in the table the
fundamental weights which induce the same embedding.

We use \eqref{Z1} and \eqref{Z2} to compute the one-string and
two-string elliptic genera. The results are again presented in terms
of the reduced elliptic genera defined in \eqref{eq:Edred}.

The reduced one-string elliptic genus does not depend on $SU(2)_x$. The expansion
in $Q_\tau$ reads
\begin{equation}
  \mathbb{E}_{{h}_{E_6}^{(1)}}(v,\Qtau,m_i) = v^{11}Q_{\tau}^{-11/6}
  \sum_{n=0}^\infty g_{1,E_6}^{(n)}(v,Q_{m_i}) Q_\tau^n \ ,
\end{equation}
where $g_{1,E_6}^{(n)}(v,Q_{m_i})$ are rational
functions. Turning off all flavor fugacities
\begin{equation}
  g^{(n)}_{1,E_6} (v,Q_{m_i}=1) = \frac{1}{(1-v^2)^{22}}\times P^{(n)}_{1,E_6}(v)~,
\end{equation}
where the first few orders are
\begin{equation}
\ba
P^{(0)}_{1,E_6}(v)&=1+56 v^2+945 v^4+6776 v^6+23815 v^8+43989 v^{10}+\dots+v^{22},\\
P^{(1)}_{1,E_6}(v)&=79 + 3774 v^2 + 54206 v^4 + 337457 v^6 + 1067286 v^8 + 1862806 v^{10} +
 \dots +
 79 v^{22},\\
P^{(2)}_{1,E_6}(v)&=3239 + 130034 v^2 + 1603334 v^4 + 8798601 v^6 + 25393522 v^8 +
 42223058 v^{10} + \dots+ 3239 v^{22}.
\ea
\end{equation}

\noindent The ellipses are completed by making the expression
palindromic. Here $g^{(0)}_{1,E_6}$ agrees with the Hilbert series of
reduced one $E_6$-instanton moduli space \cite{Benvenuti:2010pq},
while higher order contributions agree with \cite{DelZotto:2016pvm}.

The $Q_\tau$ expansion of the two-string elliptic genus reads
\begin{equation}
  \mathbb{E}_{{h}_{E_6}^{(2)}}(v,x,\Qtau,m_i) = v^{23}Q_{\tau}^{-23/6}
  \sum_{n=0}^\infty g_{2,E_6}^{(n)}(v,x,Q_{m_i}) Q_\tau^n \ .
\end{equation}
Turning off flavor fugacities and $SU(2)_x$, we obtain
\begin{equation}
  g^{(n)}_{2,E_6} (v,x,Q_{m_i}=1)
  = \frac{1}{(1-v)^{46} (1+v)^{32} \left(1+v+v^2\right)^{23}}\times P^{(n)}_{2,E_6}(v)~,
\end{equation}
where the first few orders are
\begin{equation}
\ba
P^{(0)}_{2,E_6}&(v) = 1+9 v+94 v^2+739 v^3+5121 v^4+31432 v^5+173895 v^6+874485 v^7+4036298 v^8\\
+&17200367 v^9   +68039474 v^{10}+250943933 v^{11}+866242068 v^{12}+2807705547 v^{13}\\
+&8569454706 v^{14}+24690503239 v^{15}   +67304396959 v^{16}+173919980352 v^{17}\\
+&426790882149 v^{18}+996158535441 v^{19}+2214670938701 v^{20}   +4695878015170 v^{21}\\
+&9507297417908 v^{22}+18398716114730 v^{23}+34066083855696 v^{24}+60399840583490 v^{25} \\
  +&102628223553496 v^{26}+167232472484542 v^{27}+261500117384417 v^{28}+392614934492341 v^{29} \\
  +&566271723784347 v^{30}+784947220008032 v^{31}+1046126546231772 v^{32}+1340924322289616 v^{33} \\
  +&1653587141756229 v^{34}+1962268356880815 v^{35}+2241216639463322 v^{36}+2464163123099051 v^{37}  \\
 +&2608327634962043 v^{38}+2658213934310966 v^{39}+ \dots+v^{78}~.
\ea
\end{equation}

\noindent Note that $g^{(0)}_{2,E_6}$ agrees with the Hilbert series
of reduced two $E_6$-instanton moduli space \cite{Hanany:2012dm}.

We also use the expressions of $\IE_1,\IE_2$ to extract the BPS
invariants $N_{j_l,j_r}^{\beta}$. For this purpose, we need to use the
K\"ahler moduli $t_I,t_b$ ($I=0,1,\ldots,6$) associated to the Mori cone
generators. The results are tabulated in appendix~\ref{ap:BPS}. They
display the proper checkerboard pattern, and reproduce the known genus
0 Gopakumar-Vafa invariants \cite{DelZotto:2017mee}. At base degree one,
we notice the interesting pattern that the only non-vanishing BPS
invariants for the curve classes
$\beta = (0,k,0,0,0,0,0,1)$, $(0,0,0,k,0,0,0,1)$, $(0,0,0,0,0,k,0,1)$,
$(0,0,0,0,k,0,0,1)$ are
\begin{equation}
  N^{(0,k,0,0,0,0,0,1)}_{0,k-1/2} = N^{(0,0,0,k,0,0,0,1)}_{0,k-1/2} =
  N^{(0,0,0,0,0,k,0,1)}_{0,k-1/2} = N^{(0,0,0,0,k,0,0,1)}_{0,k+1/2} = 1 \ ,\quad
  k=1,2,\ldots \ ,
\end{equation}
which in fact can be proved from the one-string formula \eqref{Z1} with the help of (\ref{clusterformula}), (\ref{clusterformula2}) and
(\ref{clusterformula3}).

\subsection{$G = E_7$}
\label{sc:n8}


The divisors and curves of the non-compact $\fn=8$ geometry is explained in
\cite{DelZotto:2017pti}. There are eight compact divisors, which are
Hirzebruch surfaces $\IF_{n_i}$ of various degrees $n_i$. We denote
them by $D_I$ ($I=0,1,\ldots,7$). They intersect with each other like
the affine dynkin diagram of $\mf e_7$
\begin{center}
  \includegraphics[width=0.7\linewidth]{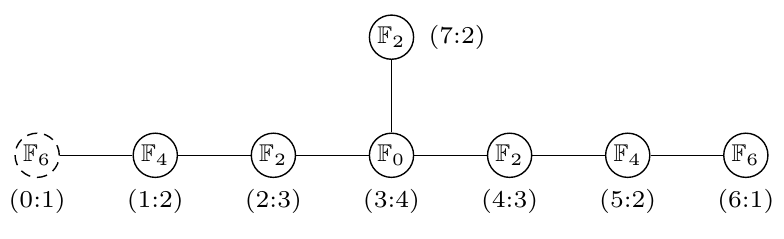}
\end{center}\vspace{-3ex}
In the diagram above we also give the ordering of the nodes $I$ and
the associated marks $a_I$ with the notation $(I:a_I)$. The $\IF_6$
denoted by a dashed circle corresponds to the affine node and it
intersects with the base at the $\IP^1$ with normal bundle
$\mc O(-8)\oplus \mc O(6) \to \IP^1$. There are nine linearly
independent curves, which we choose for the moment to be the $\IP^1$
fibers $\Sigma_I$ of the divisors $D_I$ and the $(-8)$ curve in the
base denoted by $\Sigma_B$. Denoting their complexified K\"ahler moduli
by $t_I$ and $t_B$, the linear combination
\begin{equation}\label{eq:ttau-n8}
  \sum_{I=0}^7 a_I t_I = \tau
\end{equation}
with $a_I$
the marks of $\mf e_7$, is the volume of the elliptic fiber. We
identify the Mori cone
generators.
They include the $\IP^1$ fibers $\Sigma_I$ ($I=0,1,\ldots,7$), as well
as the $\IP^1$ base of the $\IF_0$ surface in the middle. We denote
the last curve by $\Sigma_b$, which is related to $\Sigma_B$ by
\begin{equation}
  [\Sigma_B] = [\Sigma_b] + 6[\Sigma_0] + 4[\Sigma_1] + 2[\Sigma_2] \ .
\end{equation}
Their K\"ahler moduli are consequently related by
\begin{equation}\label{eq:tBtb-n8}
  t_B = t_b + 6t_0 + 4t_1 + 2t_2 \ .
\end{equation}

The $ C$-matrix of intersection between $\Sigma_I,\Sigma_b$ and
$D_I$ is
\begin{equation}
   C =
  \begin{pmatrix}
    -2 & 1 & 0 & 0 & 0 & 0 & 0 & 0\\
    1 & -2 & 1 & 0 & 0 & 0 & 0 & 0\\
    0 & 1 & -2 & 1 & 0 & 0 & 0 & 0\\
    0 & 0 & 1 & -2 & 1 & 0 & 0 & 1\\
    0 & 0 & 0 & 1 & -2 & 1 & 0 & 0\\
    0 & 0 & 0 & 0 & 1 & -2 & 1 & 0\\
    0 & 0 & 0 & 0 & 0 & 1 & -2 & 0\\
    0 & 0 & 0 & 1 & 0 & 0 & 0 & -2\\
    0 & 0 & 0 & -2 & 0 & 0 & 0 & 0\\
  \end{pmatrix}  \ .
\end{equation}

The semiclassical components of the partition function can be computed
using the prescription in section~\ref{sc:Zpert} with the
normalisation scheme in Appendix~\ref{ap:norm}. We obtain
\begin{equation}
  \begin{aligned}
   -F_{(0,0)}^{\text{cls}} = &
    \Big(\frac{t_{0}}{16}+\frac{t_{1}}{8}+\frac{3
        t_{2}}{16}+\frac{t_{3}}{4}+\frac{3
        t_{4}}{16}+\frac{t_{5}}{8}+\frac{t_{6}}{16}+\frac{t_{7}}{8}\Big)
    t_{b}^2
    +\Big(\frac{3t_{0}^2}{8} +\frac{t_{1}^2}{2}+\frac{3
        t_{2}^2}{8}+\frac{3 t_{4}^2}{8}+\frac{t_{5}^2}{2}+\frac{3
        t_{6}^2}{8}\\&+\frac{t_{7}^2}{4} +\frac{t_{0}t_{1}}{2}
      +\frac{t_{0} t_{2}}{4} +\frac{t_{1} t_{2}}{2} +\frac{t_{4}
        t_{5}}{2} +\frac{t_{4} t_{6}}{4} +\frac{t_{5}
        t_{6}}{2} \Big) t_{b} +\frac{3t_{0}^3}{4} +\frac{2
      t_{1}^3}{3}+\frac{t_{2}^3}{4}+\frac{t_{4}^3}{4}+\frac{2
      t_{5}^3}{3}\\&+\frac{3
      t_{6}^3}{4}+\frac{t_{7}^3}{6}+\frac{3t_{0}^2t_{1}}{2}
    +\frac{3t_{0}^2 t_{2}}{4} + t_{0}t_{1}^2 +\frac{t_{0} t_{2}^2}{4}
    +t_{0}t_{1} t_{2}
    +\frac{t_{1} t_{2}^2}{2} +t_{4} t_{5}^2+\frac{3t_{4} t_{6}^2}{4} +\frac{3t_{5}
      t_{6}^2}{2} \\&+t_{1}^2 t_{2} +\frac{t_{4}^2 t_{5}}{2}
    +\frac{t_{4}^2 t_{6}}{4} +t_{5}^2 t_{6}+t_{4} t_{5} t_{6},
  \end{aligned}
\end{equation}

\noindent which is consistent with the universal formula
\eqref{eq:pertF0-uni}. Using the relations
\eqref{eq:telltB}, \eqref{eq:tBtb-n8} and \eqref{eq:ttau-n8}, we can
express $F^{(0,0)}(\nound t)$ in terms of the K\"ahler moduli
$t_{\text{ell}}, \tau, m_i$ $(i=1,\ldots,7)$ and find
\begin{equation}
  -F^{(0,0)} = \frac{1}{16} t_{\rm ell}^2\tau
  + \frac{1}{2} t_{\rm ell} ( m, m) - \frac{3}{2}\tau ( m, m) + \ldots
\end{equation}
up to $\tau^3$ and terms cubic in $m_i$, where
\begin{equation}
   m = \sum_{i=1}^7 m_i \omega_i^\vee \ .
\end{equation}
It is in agreement with the universal expression
\eqref{eq:pertF0-uni}, and therefore the analysis in
section~\ref{sc:adm} goes through, which then leads to the elliptic
blowup equations \eqref{eq:uv-blowup}.
We also find
\begin{equation}
  F_{(1,0)}^{\text{cls}} =
  \frac{51}{8}t_0+\frac{35}{4}t_1+\frac{57}{8}t_2+\frac{3}{2}t_3+\frac{57}{8}t_4
  +\frac{35}{4}t_5+\frac{51}{8}t_6+\frac{11}{4}t_b \ .
\end{equation}

\begin{table}
  \centering
  \begin{tabular}{*{4}{>{$}c<{$}}}\toprule
    &   \multicolumn{2}{c}{$\nound{r}$}
    & \text{fundamental weights} \\\midrule
    \multirow{4}{*}{unity}
    & (0,0,0,0,0,0,0,0,0)  & (0,0,0,0,0,0,0,0,2)
    &\multirow{4}{*}{$\omega_1,\omega_2,\omega_3,\omega_5$}\\
    & (0,0,0,0,0,0,0,0,4)  & (0,0,0,0,0,0,0,0,6) & \\
    & (0,0,0,0,0,0,0,0,8)  & (0,0,0,0,0,0,0,0,10) & \\
    & (0,0,0,0,0,0,0,0,12) & (0,0,0,0,0,0,0,0,14) & \\\midrule
    \multirow{4}{*}{vanishing}
    & (-2,0,0,0,0,0,2,0,0)  & (-2,0,0,0,0,0,2,0,2)
    &\multirow{4}{*}{$\omega_4,\omega_6,\omega_7$}\\
    & (-2,0,0,0,0,0,2,0,4)  & (-2,0,0,0,0,0,2,0,6) & \\
    & (-2,0,0,0,0,0,2,0,8)  & (-2,0,0,0,0,0,2,0,10) & \\
    & (-2,0,0,0,0,0,2,0,12) & (-2,0,0,0,0,0,2,0,14) & \\\bottomrule
  \end{tabular}
  \caption{The $\nound{r}$-fields of the $\fn=8$ model and the
    fundamental weights of $\mf e_7$ which induce the same embedding
    $\phi: Q^\vee \hookrightarrow P$. They can be divided into two
    groups; inside each group $\nound{r}$-fields or fundamental weights
    induce the same embedding.}\label{tb:r-f8}
\end{table}

Imposing the admissibility condition \eqref{eq:rtau0} and the BPS
checkerboard pattern condition \eqref{eq:rn}, which
specialises to
\begin{equation}
  \nound{r} \equiv (0,0,0,0,0,0,0,0,0)\quad \text{mod}\;2 \ ,
\end{equation}
there are in
total 16 inequivalent $\nound{r}$-fields, and we list their
representatives in Table~\ref{tb:r-f8}.
We classify them according to the embeddings
$\phi_\lambda: Q^\vee \hookrightarrow P$ induced by the reduced
$\nound{r}$-field $\lambda$ defined in \eqref{eq:lambda-alpha},
\eqref{eq:QvP}.  We also list in the table the fundamental weights
which induce the same embedding.

We use \eqref{Z1} and \eqref{Z2} to compute the one-string and two-string
elliptic genera and convert them to reduced versions. The one string
elliptic genus when expanded in $Q_\tau$ reads
\begin{equation}
  \mathbb{E}_{{h}_{E_7}^{(1)}}(v,\Qtau,m_i) = v^{17} Q_\tau^{-17/6}
  \sum_{n=0}^\infty g^{(n)}_{1,E_7}(v,Q_{m_i}) Q_\tau^n \ ,
\end{equation}
where $g^{(n)}_{1,E_7}(v,Q_{m_i})$ are rational
functions. When all flavor fugacities are turned off
\begin{equation}
  g^{(n)}_{1,E_7} (v,Q_{m_i} = 1) = \frac{1}{(1-v^2)^{34}}\times P^{(1)}_{n,E_7}(v)~,
\end{equation}
where the leading order contributions are
\begin{equation}
\ba
P^{(0)}_{1,E_7}(v)=&\,(1+v^2)(1 + 98 v^{2 }+ 3312 v^{4 }+ 53305 v^{6 }+ 468612 v^{8 }+ 2421286 v^{10 }+
 7664780 v^{12 }\\&+ 15203076 v^{14 }+ 19086400 v^{16 }+ 15203076 v^{18 }+ \dots+v^{32}
),\\
P^{(1)}_{1,E_7}(v)=&\,(1+v^2) (134 + 11593 v^{2 }+ 345521 v^{4 }+ 4931707 v^{6 }+ 38850151 v^{8 }+
 182614170 v^{10 }\\&+ 536726278 v^{12 }+ 1014596958 v^{14 }+
 1252490096 v^{16 }+ 1014596958 v^{18 }+ \dots+v^{32}),\\
P^{(2)}_{1,E_7}(v)=&\,(1+v^2)(9179 + 693316 v^{2 }+ 18210733 v^{4 }+ 231525774 v^{6 }+ 1645739978 v^{8 }\\&+
 7093827388 v^{10 }+ 19507715662 v^{12 }+ 35350906224 v^{14 }+
 43009574252 v^{16 }\\&+ 35350906224 v^{18 }\dots+v^{32}).
\ea
\end{equation}

\noindent where the ellipses are completed by palindrome. Here
$g^{(0)}_{1,E_7}$ agrees with the Hilbert series of reduced one
$E_7$-instanton moduli space in \cite{Benvenuti:2010pq}, while higher
order contributions agree with \cite{DelZotto:2016pvm}.

The $Q_\tau$ expansion of the two-string elliptic genus reads
\begin{equation}
  \mathbb{E}_{{h}_{E_7}^{(2)}}(v,x,\Qtau,m_i)
  = v^{35} Q_\tau^{-35/6} \sum_{n=0}^\infty
  g_{2,E_7}^{(n)}(v,x,Q_{m_i}) Q_\tau^n \ .
\end{equation}
Turning off $SU(2)_x$ and flavours, we have
\begin{equation}
  g_{2,E_7}^{(n)}(v,x,Q_{m_i}=1)
  = \frac{1}{(1-v)^{70} (1+v)^{52} \left(1+v+v^2\right)^{35}}\times P^{(2)}_{n,E_7}(v)~.
\end{equation}
We have computed $P^{(2)}_{n,E_7}(v)$ for $n=0,1$ which we put in the
appendix~\ref{ap:eg}. Indeed, our $g^{(0)}_{2,E_7}$ agrees with the
Hibert series of reduced two $E_7$-instanton moduli space in
\cite{Hanany:2012dm}.

We also use the expressions of $\IE_1,\IE_2$ to extract the BPS
invariants $N_{j_l,j_r}^{\beta}$. For this purpose, we need to use the
K\"ahler moduli $t_I,t_b$ ($I=0,1,\ldots,7$) associated to the Mori cone
generators. The results are tabulated in appendix~\ref{ap:BPS}. They
display the proper checkerboard pattern, and reproduce the known genus
zero Gopakumar-Vafa invariants \cite{DelZotto:2017mee}. At base degree one,
we notice the interesting pattern that the only non-vanishing BPS
invariants for the curve classes
$\beta = (0,0,k,0,0,0,0,0,1)$, $(0,0,0,0,k,0,0,0,1)$, $(0,0,0,0,0,0,0,k,1)$,
$(0,0,0,k,0,0,0,0,1)$ are
\begin{equation}
  N^{(0,0,k,0,0,0,0,0,1)}_{0,k-1/2} = N^{(0,0,0,0,k,0,0,0,1)}_{0,k-1/2} =
  N^{(0,0,0,0,0,0,0,k,1)}_{0,k-1/2} = N^{(0,0,0,k,0,0,0,0,1)}_{0,k+1/2} = 1, \ ,\quad
  k=1,2,\ldots \ ,
\end{equation}
which in fact can be proved from the one-string formula \eqref{Z1} with the help of (\ref{clusterformula}), (\ref{clusterformula2}) and
(\ref{clusterformula3}).

\subsection{$G = E_8$}
\label{sc:n12}


The divisors and curves of the non-compact $\fn=12$ geometry is explained in
\cite{DelZotto:2017pti}. There are nine compact divisors, which are
Hirzebruch surfaces of various degrees. We denote them by $D_I$
($I=0,1,\ldots,8$). They intersect with each other like the affine
dynkin diagram of $\mf e_8$
\begin{center}
  \includegraphics[width=0.8\linewidth]{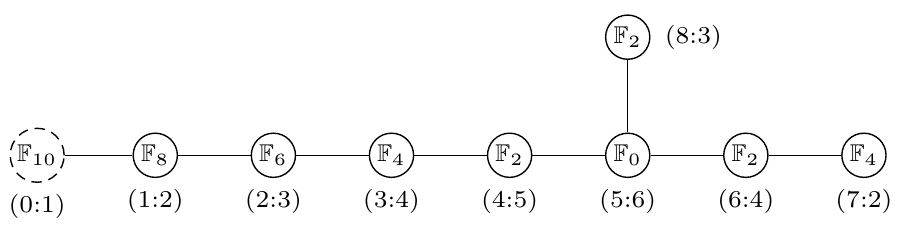}
\end{center}\vspace{-3ex}
In the diagram above we also give the ordering of the nodes $I$ and
the associated marks $a_I$ with the notation $(I:a_I)$. The
$\IF_{10}$ denoted by a dashed circle corresponds to the affine node
and it intersects with the base at the $\IP^1$ with normal bundle
$\mc O(-12)\oplus \mc O(10) \to \IP^1$. There are ten linearly
independent curves, which we choose for the moment to be the $\IP^1$
fibers $\Sigma_I$ of the divisors $D_I$ and the $(-12)$ curve in the
base denoted by $\Sigma_B$. Let $t_I$ and $t_B$ be their complexified
K\"ahler moduli. The linear combination
\begin{equation}\label{eq:ttau-n12}
  \sum_{I=0}^8 a_I t_I = \tau
\end{equation}
with $a_I$
marks of $\mf e_8$, is the volume of the elliptic fiber. We identify
the Mori cone generators. They include $\IP^1$ fibers $\Sigma_I$
($I=0,1,\ldots,8$) as well as the $\IP^1$ base of the $\IF_0$ surface
in the center. We denote the latter by $\Sigma_b$, which is related to
$\Sigma_B$ by
\begin{equation}
  [\Sigma_B] =
  [\Sigma_b] + 10[\Sigma_0] + 8[\Sigma_1] + 6[\Sigma_2] + 4[\Sigma_3] + 2[\Sigma_4] \ ,
\end{equation}
which implies
\begin{equation}\label{eq:tBtb-n12}
  t_B = t_b + 10t_0 + 8t_1 + 6t_2 + 4t_3 + 2t_4 \ .
\end{equation}

The $C$-matrix of intersection between $\Sigma_I,\Sigma_b$ and
$D_I$ is
\begin{equation}
   C =
  \begin{pmatrix}
    -2 & 1 & 0 & 0 & 0 & 0 & 0 & 0 & 0\\
    1 & -2 & 1 & 0 & 0 & 0 & 0 & 0 & 0\\
    0 & 1 & -2 & 1 & 0 & 0 & 0 & 0 & 0\\
    0 & 0 & 1 & -2 & 1 & 0 & 0 & 0 & 0\\
    0 & 0 & 0 & 1 & -2 & 1 & 0 & 0 & 0\\
    0 & 0 & 0 & 0 & 1 & -2 & 1 & 0 & 1\\
    0 & 0 & 0 & 0 & 0 & 1 & -2 & 1 & 0\\
    0 & 0 & 0 & 0 & 0 & 0 & 1 & -2 & 0\\
    0 & 0 & 0 & 0 & 0 & 1 & 0 & 0 & -2\\
    0 & 0 & 0 & 0 & 0 & -2 & 0 & 0 & 0\\
  \end{pmatrix}  \ .
\end{equation}

The semiclassical components of the partition function can be computed
using the prescription in section~\ref{sc:Zpert} with the
normalisation scheme in Appendix~\ref{ap:norm}. We obtain

\begin{equation}
\begin{aligned}
  -F_{(0,0)}^{\text{cls}}
  &=
    \Big(\frac{t_{0}}{24}+\frac{t_{1}}{12}+\frac{t_{2}}{8}+\frac{t_{3}}{6}+
    \frac{5
    t_{4}}{24}+\frac{t_{5}}{4}+\frac{t_{6}}{6}+\frac{t_{7}}{12}+\frac{t_{8}}{8}\Big)
    t_{b}^2\\
  &+\Big(\frac{5t_{0}^2}{12}
    +\frac{2t_{1}^2}{3}+\frac{3
    t_{2}^2}{4}+\frac{2 t_{3}^2}{3}+\frac{5
    t_{4}^2}{12}+\frac{t_{6}^2}{3}+\frac{t_{7}^2}{3}+\frac{t_{8}^2}{4}
    +\frac{2t_{0}t_{1}}{3}
    +\frac{t_{0}t_{2}}{2} +\frac{t_{0}t_{3}}{3}  +\frac{t_{0}t_{4}}{6}\\
  &+t_{1}t_{2}+\frac{2t_{1} t_{3}}{3} +t_{2} t_{3}+\frac{t_{1}
    t_{4}}{3} +\frac{t_{2}
    t_{4}}{2} +\frac{2t_{3} t_{4}}{3} +\frac{t_{6} t_{7}}{3} \Big) t_{b} \\
  &+\frac{25t_{0}^3}{18} +\frac{16 t_{1}^3}{9}+\frac{3 t_{2}^3}{2}
    +\frac{8 t_{3}^3}{9}+\frac{5 t_{4}^3}{18}+\frac{2 t_{6}^3}{9}+\frac{4
    t_{7}^3}{9}+\frac{t_{8}^3}{6}
    +\frac{10t_{0}^2t_{1}}{3}
    +\frac{5t_{0}^2t_{2}}{2}
    +\frac{5t_{0}^2t_{3}}{3}  +\frac{5t_{0}^2t_{4}}{6}\\
  &+\frac{8t_{0}t_{1}^2}{3}
    +\frac{3t_{0}t_{2}^2}{2}  +\frac{2t_{0}t_{3}^2}{3}
    +\frac{t_{0}t_{4}^2}{6}
    +4 t_{0} t_{1} t_{2} +\frac{8t_{0}t_{1} t_{3}}{3}
    +2 t_{0} t_{2} t_{3} +\frac{4}{3}
    t_{0}t_{1} t_{4} +t_{0} t_{2} t_{4} +\frac{2t_{0}t_{3} t_{4}}{3}\\
  &+3 t_{1} t_{2}^2+\frac{4t_{1}
    t_{3}^2}{3} +2 t_{2} t_{3}^2+\frac{t_{1} t_{4}^2}{3} +\frac{t_{2}
    t_{4}^2}{2} +\frac{2t_{3} t_{4}^2}{3} +\frac{2t_{6}
    t_{7}^2}{3} +4 t_{1}^2 t_{2}+\frac{8t_{1}^2 t_{3}}{3} +3 t_{2}^2 t_{3}+4 t_{1} t_{2}
    t_{3}\\
  &+\frac{4t_{1}^2 t_{4}}{3} +\frac{3t_{2}^2 t_{4}}{2} +\frac{4t_{3}^2
    t_{4}}{3} +2 t_{1} t_{2} t_{4}+\frac{4t_{1} t_{3} t_{4}}{3} +2 t_{2}
    t_{3} t_{4}+\frac{t_{6}^2 t_{7}}{3},
\end{aligned}
\end{equation}

\noindent which is consistent with \eqref{eq:pertF0-uni}. Using the
relations \eqref{eq:telltB}, \eqref{eq:tBtb-n12}, \eqref{eq:ttau-n12},
we can express $F^{(0,0)}(\underline t)$ in terms of the K\"ahler moduli
$t_{\text{ell}}, \tau, m_i$ $(i=1,\ldots,8)$ and find
\begin{equation}
  -F^{(0,0)} = \frac{1}{24} t_{\rm ell}^2\tau
  + \frac{1}{2} t_{\rm ell} ( m, m) - \frac{5}{2}\tau ( m, m) + \ldots
\end{equation}
up to $\tau^3$ and terms cubic in $m_i$, where
\begin{equation}
   m = \sum_{i=1}^8 m_i \omega_i^\vee \ .
\end{equation}
It is in agreement with the universal expression
\eqref{eq:pertF0-uni}, and therefore the analysis in
section~\ref{sc:adm} goes through leading to the elliptic blowup
equations \eqref{eq:uv-blowup}.
We also find
\begin{equation}
  F_{(1,0)}^{\text{cls}} =
  \frac{41}{4}t_0+\frac{33}{2}t_1+\frac{75}{4}t_2+17t_3+\frac{45}{4}t_4
  +\frac{3}{2}t_5+5t_6+\frac{9}{2}t_7+\frac{11}{4}t_8+10t_b \ .
\end{equation}

\begin{table}
  \centering
  \begin{tabular}{c *{4}{>{$}c<{$}}}\toprule
    & \multicolumn{2}{c}{$\nound{r}$}
    & \text{fundamental weights} \\\midrule
    \multirow{6}{*}{unity}
    & (0,0,0,0,0,0,0,0,0,0) & (0,0,0,0,0,0,0,0,0,2)
    & \multirow{6}{*}{$\omega_i \;(i=1,\ldots,8)$}\\
    & (0,0,0,0,0,0,0,0,0,4) & (0,0,0,0,0,0,0,0,0,6) & \\
    & (0,0,0,0,0,0,0,0,0,8) & (0,0,0,0,0,0,0,0,0,10) & \\
    & (0,0,0,0,0,0,0,0,0,12) & (0,0,0,0,0,0,0,0,0,14) & \\
    & (0,0,0,0,0,0,0,0,0,16) & (0,0,0,0,0,0,0,0,0,18) & \\
    & (0,0,0,0,0,0,0,0,0,20) & (0,0,0,0,0,0,0,0,0,22) &\\\bottomrule
  \end{tabular}
  \caption{The $\nound{r}$-fields of the $\fn=12$ model and the fundamental
    weights of $\mf e_8$ which induce the same embedding
    $\phi: Q^\vee \hookrightarrow P$. All the $\nound{r}$-fields and all
    the fundamental weight induce the same embedding as $Q^\vee = P$
    for $\mf e_8$.}\label{tb:r-f12}
\end{table}

Imposing the admissibility condition \eqref{eq:rtau0} and the BPS
checkerboard pattern condition \eqref{eq:rn}, which
specialises to
\begin{equation}
  \nound{r} \equiv (0,0,0,0,0,0,0,0,0,0)\quad \text{mod}\;2 \ ,
\end{equation}
there are in
total 12 inequivalent $\nound{r}$-fields, and we list their
representatives in Table~\ref{tb:r-f12}.
All of them have the
same reduced $\lambda$ which induces the same embedding
$\phi_\lambda: Q^\vee \hookrightarrow P = Q^\vee$. In this special
case, there is no vanishing blowup equations.

We use \eqref{Z1} and \eqref{Z2} to compute the one-string and two-string
elliptic genera and convert them to reduced versions. The one-string
reduced elliptic genus in $Q_\tau$ expansion reads
\begin{equation}
  \mathbb{E}_{{h}_{E_8}^{(1)}}(v,\Qtau,m_i) = v^{29}Q_\tau^{-29/6}
  \sum_{n=0}^\infty g_{1,E_8}^{(n)}(v,Q_{m_i}) Q_{\tau}^n,
\end{equation}
where $g_{1,E_8}^{(n)}$ are rational functions. Turning off flavor
fugacities
\begin{equation}
  g^{(n)}_{1,E_8} (v,Q_{m_i}=1)
  = \frac{1}{(1-v^2)^{58}}\times P^{(1)}_{n,E_8}(v)~,
\end{equation}
where the leading orders are
\begin{equation}
\ba
P^{(1)}_{0,E_8}(v) &=\,(1+v^2)(1+189 v^{2}+14080 v^{4}+562133 v^{6}+13722599 v^{8}+220731150 v^{10}\\&
 +2454952400 v^{12}+19517762786 v^{14}+113608689871 v^{16}+492718282457 v^{18}\\&
 +1612836871168 v^{20}+4022154098447 v^{22}+7692605013883 v^{24}+11332578013712 v^{26}\\&
 +12891341012848 v^{28}+11332578013712 v^{30}+\dots+v^{56}),
\ea
\end{equation}
\begin{equation}
\ba
P^{(1)}_{1,E_8}(v) &=\,249 + 43435 v^{2 }+ 2998484 v^{4 }+ 111587988 v^{6 }+ 2558096217 v^{8 }+
 38985250263 v^{10 }\\&
  + 415090167480 v^{12 }+ 3197400818096 v^{14 }+
 18281159666407 v^{16 }+ 79099752469353 v^{18 }\\&
  + 262872507223458 v^{20 }+
 678620928038790 v^{22 }+ 1372471431431505 v^{24 }\\&
  +
 2187800775100695 v^{26 }+ 2759575276449180 v^{28 }+ 2759575276449180 v^{30}+\dots+v^{58},
 \ea
\end{equation}
\begin{equation}
\ba
   P^{(1)}_{2,E_8}(v) &=\,31374 + 4996185 v^{2 }+ 316301853 v^{4 }+ 10844316461 v^{6 }+
 230109165319 v^{8 }\\&
  + 3262175735364 v^{10 }+ 32482207865920 v^{12 }+
 235331998114532 v^{14 }\\&
  + 1273365718136904 v^{16 }+
 5249113972780491 v^{18 }+ 16738824444898167 v^{20 }\\&
  +
 41781447040327605 v^{22 }+ 82360817736515085 v^{24 }+
 129037047832755990 v^{26 }\\&
  + 161349436368883950 v^{28 }+ 161349436368883950 v^{30}+\dots+v^{58}.
\ea
\end{equation}

\noindent where the ellipses are completed by palindome. Note
$g^{(0)}_{1,E_8}$ indeed agrees with the Hilbert series of reduced one
$E_8$-instanton moduli space in \cite{Benvenuti:2010pq}. Higher order
contributions agree with \cite{DelZotto:2016pvm}.

The two-string reduced elliptic genus in $Q_\tau$ expansion reads
\begin{equation}
  \mathbb{E}_{{h}_{E_8}^{(2)}}(v,x,\Qtau,m_i)
  = v^{59}Q_\tau^{-59/6} \sum_{n=0}^\infty
  g^{(n)}_{2,E_8}(v,x,Q_{m_i}) Q_\tau^n \ ,
\end{equation}
where $g^{(n)}_{2,E_8}(v,x,Q_{m_i})$ are rational functions. Turning
off flavor fugacities and $SU(2)_x$
\begin{equation}
  g^{(n)}_{2,E_8} (v,x,Q_{m_i}=1) = \frac{1}{(1-v)^{118} (1+v)^{92}
    (1+v+v^2)^{59}}\times P^{(n)}_{2,E_8}(v)~.
\end{equation}

We have computed $g^{(0)}_{2,E_8}$ which indeed agrees with
the Hilbert series of two $E_8$-instanton reduced moduli space in
\cite{Hanany:2012dm}.

We also use the expressions of $\IE_1,\IE_2$ to extract the BPS
invariants $N_{j_l,j_r}^{\beta}$. For this purpose, we need to use the
K\"ahler moduli $t_I,t_b$ ($I=0,1,\ldots,8$) associated to the Mori cone
generators. The results are tabulated in appendix~\ref{ap:BPS}. They
display the proper checkerboard pattern, and reproduce the known genus
zero Gopakumar-Vafa invariants \cite{DelZotto:2017mee}. At base degree one,
we notice a pattern that the only non-vanishing BPS
invariants for the curve classes
$\beta = (0,0,0,0,k,0,0,0,0,1)$, $(0,0,0,0,0,0,k,0,0,1)$,
$(0,0,0,0,0,0,0,0,k,1)$, $(0,0,0,0,0,k,0,0,0,1)$ are
\begin{equation}
  N^{(0,0,0,0,k,0,0,0,0,1)}_{0,k-1/2} = N^{(0,0,0,0,0,0,k,0,0,1)}_{0,k-1/2} =
  N^{(0,0,0,0,0,0,0,0,k,1)}_{0,k-1/2} = N^{(0,0,0,0,0,k,0,0,0,1)}_{0,k+1/2} = 1 \ ,\quad
  k=1,2,\ldots \ ,
\end{equation}
which in fact can be proved from the one-string formula \eqref{Z1} with the help of (\ref{clusterformula}), (\ref{clusterformula2}) and
(\ref{clusterformula3}).

\section{On the relation with 4d SCFTs of type $H^{(k)}_G$}
\label{sec:HGk}

The purpose of this section is to connect the $k$-string elliptic
genera $\mathbb{E}_{h^{(k)}_G}$ for the minimal ${\cal N}=(1,0)$ 6d
SCFTs with $G=A_2,D_4,F_4,E_{6,7,8}$ discussed above to the
superconformal indices of the $\mathcal{N}=2$ 4d SCFTs of rank $k$
denoted by $H_G^{(k)}$.  The simplest series of $\mathcal{N}=2$ SCFTs
namely $H_G^{(1)}$ can be obtained by geometric engineering on
non-compact del Pezzo geometries and contains the Minahan-Nemeschansky
theories.  The main result is an extension of a surprising conjecture
by Del Zotto-Lockhart from the rank one case~\cite{DelZotto:2016pvm} to
the higher rank cases. To be precise~\cite{DelZotto:2016pvm}
recognised that the one-string elliptic genus
$\mathbb{E}_{{h}_{G}^{(1)}}(\Qtau,v)$ can be decomposed in terms of a
seemingly more fundamental function $L_G(\Qtau,v)$, which for special
choices of $\Qtau$ and $v$ specialises to the Hall-Littlewood index or
the Schur index of the $H_G^{(1)}$ theories. With the two string
elliptic genera computed in our previous sections, we are able to study
this conjectural relation at rank two and in principle at arbitrary
rank, and find indeed that similar striking relations exist.

We first review some basic properties of 4d rank $k$ type $H_G^{(k)}$
-- and ${\widetilde H}_G^{(k)}$ theories, including their class
$\mathcal{S}$ theory construction, and then review the superconformal
indices of 4d SCFTs in various physically motivated limits as well as
the methods to computed them. Next we state the conjectural relation
at rank one from~\cite{DelZotto:2016pvm}, and explain in some detail
the new relations at rank two for all $G$. We also extend the analysis
to some rank three cases. For all choices of rank and $G$ we analysed,
the surprising relation between elliptic genera and superconformal
indices exists. We define an intermediate function at rank $k$ called
$L_G^{(k)}$\footnote{The $L_G$ function in \cite{DelZotto:2016pvm}
  becomes $L_G^{(1)}$ here}. This function is on the one hand the
ingredient of $k$-string elliptic genus, on the other hand gives the
Hall-Littelwood index and Schur index of $H_G^{(k)}$ theories at
special choices of parameters. This general structure allows us to
calculate the latter indices efficiently from the
$\mathbb{E}_{h^{(k)}_G}$ that are determined from the elliptic blowup
equations.

\subsection{Rank $k$ $H_G$ theories}
The 4d $\mathcal{N}=2$ SCFTs $H_G^{(k)}$ are well known to exist for
$G=\emptyset,A_1,A_2,D_4,E_{6,7,8}$ and $k=1,2,3\dots$
\cite{Argyres:1995xn,Banks:1996nj,Douglas:1996js,Minahan:1996fg,Minahan:1996cj}\footnote{The
  $G=\emptyset,A_1,A_2$ type theories are also traditionally denoted
  as $H_{0,1,2}$ theories. Here we follow the notations in
  \cite{DelZotto:2016pvm}.}.  In type IIB superstring theory, they are
realized as the worldvolume theory for $k$ multiple D3-branes probing
a stack of exotic seven-branes.  Such seven-branes in F-theory are
defined as codimension one singularities with Kodaire type
$II,III,IV,I_0^*,IV^*,III^*,$ and $II^* $, which give the gauge
symmetries $G$ for the low energy 8d SYM theories. The number $k$ is
usually called the rank of $H_G$ theories.  For example, the rank one
$H_{\emptyset,A_1,A_2}$ theories appear as certain limit of $SU(2)$
gauge theory with $N_f=1,2,3$ respectively \cite{Argyres:1995xn}.  The
rank one $H_{D_4}$ theory is well known to be the $SU(2)$ gauge theory
with $N_f=4$, while the higher rank cases with $k>1$ are equivalent to
$USp(2k)$ gauge theories with four fundmental hypermultiplets and one
antisymmetric hypermultiplet, which are all Lagrangian theories. The
rank one $H_{E_{6,7,8}}$ are also known as the Minahan-Nemeschansky
theories \cite{Minahan:1996fg,Minahan:1996cj}, where the simplest
example rank one $E_6$ theory is in S-duality with $SU(3)$, $N_f=6$
theory \cite{Argyres:2007cn}.

All $H_G^{(k)}$ theory can be coupled with a free hypermultiplet
associated to the center of mass motion of the instantons. We follow
\cite{DelZotto:2016pvm} and denote these theories as
$\widetilde{H}_G^{(k)}$. As was observed in \cite{Gaiotto:2012uq}, for
higher rank cases, $\widetilde{H}_G$ are sometimes more natural than
$H_G$ theories. One major difference between rank one and higher rank
$H_G$ theories is the flavour symmetry.  Besides the flavour $G$ given
by the strings stretched between D3-branes and exotic seven-brane, for
$k>1$ there is one more $SU(2)$ symmetry coming from the transverse
space in the seven-brane. By coupling a free hypermultiplet, all
$\widetilde{H}_G^{(k)}$ theories share flavour symmetry
$G\times SU(2)$.

The $H_G^{(k)}$ theories of interest in this paper are $G=A_2,D_4,E_{6,7,8}$ as they are directly related to 6d minimal
$(1,0)$ SCFTs with corresponding gauge group $G$. To be precise, the RR elliptic genus is identified as the $\beta$-twisted
$T^2\times S^2$ partition function of the 4d SCFTs:
\begin{equation}\label{twistHG}
\mathbb{E}_{ h_{G}^{(k)}} = Z_{(T^2 \times S^2)_\beta}\big({H}_G^{(k)}\big),
\end{equation}
Adding the ``tildes'', one can also obtain the equality with the free hypermultiplet coupled. Here the $\beta$-twist was introduced by Kapustin in
\cite{Kapustin:2006hi} to preserve half of the supersymmetries on the backgrounds such as $T^2\times S^2$. See a good description of such twist in for
example section 3.2 of \cite{DelZotto:2016pvm}. The identification (\ref{twistHG}) makes it sometimes possible to compute the elliptic genus from 4d
setting, in which cases the S-duality with a Lagrangian theory is invoked and one can use certain analogy of Spiridonov-Warnaar inverse formula
\cite{spiridonov2006inversions} to compute the $T^2\times S^2$ partition function. This was indeed achieved for one string elliptic genus with
$G=D_4,E_{6,7}$ \cite{Putrov:2015jpa,DelZotto:2016pvm,Gadde:2015xta,Agarwal:2018ejn}. For example, the elliptic genus of one $E_7$ instanton string was
obtained in \cite{Agarwal:2018ejn} via $SU(4)$ gauge theory $N_f=8$ and appropriate Higgsing as
\be
\ba
 Z_{(T^2 \times S^2)_\beta}\big({H}_{E_7}^{(1)}\big) &= 1 + \chi^{E_7}_{133} v^2 + \chi^{E_7}_{7371}v^4 + \chi^{E_7}_{238602} v^6 + \chi^{E_7}_{5248750}
 v^8+ \ldots \\
 &\quad+ \Qtau \Big( 1 + \chi^{E_7}_{133} + \big( 1 + 2 \chi^{E_7}_{133} + \chi^{E_7}_{7331} + \chi^{E_7}_{8645}  \big) v^2 \\
  &\quad\quad\quad + \big( \chi^{E_7}_{133} + 2  \chi^{E_7}_{7371} + \chi^{E_7}_{8645} + \chi^{E_7}_{238602} + \chi^{E_7}_{573440} \big) v^4+ \ldots \Big)
  \\
  &\quad+ \Qtau^2 \left( 3 + 2  \chi^{E_7}_{133} + \chi^{E_7}_{1539} + \chi^{E_7}_{7371} + \ldots \right)+ \CO(\Qtau^3) \ ,
\ea
\ee
which completely agrees with our universal expansion formula (\ref{g10G}), (\ref{g11G}) and (\ref{g12G}).\footnote{In the coefficients of $\Qtau^2$, one
also need to use the Joseph relation $\mathrm{Sym}^2{\bf 133}=1+{\bf 133}+{\bf 7371}$ to obtain the identification.} We also checked for $D_4$ and $E_6$,
where the agreement holds to all known orders.

Another important feature of $H_{G}^{(k)}$ theories is that they all admit 6d construction. It is well known all rank $k$ $H_{D_4,E_{6,7,8}}$ theories
can be realized by compactifying a 6d $A_{N-1}$ (2,0) SCFT on some punctured sphere with regular singularities \cite{Benini:2009gi},
i.e. they are class $\mathcal{S}$ theories. The regular singularities are classified by embeddings of $SU(2)$ in $SU(N)$, thus can be denoted as Young
diagrams.
Such punctures with associated Young diagram represent how the $SU(N)$ decomposes and what is the residual flavour symmetry.
For example, the rank one $H_{SO(8)}$ theory is obtained by compactifying 6d $A_{1}$ (2,0) SCFT on a sphere with four full punctures
$\{1^2\}$, i.e. the residual flavour symmetry is $SU(2)$. Thus the resulting 4d theory has gauge symmetry $SU(2)$ and four fundamentals, as was mentioned
already above.
We summarize the gauge algebras and punctures for the 6d construction of all $H_{G}^{(k)}$ theories with $G=D_4,E_{6,7,8} $ in Table \ref{punctures}.
\begin{table}[h]
 \begin{center}
\begin{tabular}{c| c|c }$G$& 6d $(2,0)$ $A_{N-1}$ & punctures $\Lambda_i$ \\ \hline
$D_4$ & $A_{2k-1}$ & four $\{k^2\}$ \\  \hline
$E_6$ & $A_{3k-1}$ & three $\{k^3\}$ \\ \hline
$E_7$ & $A_{4k-1}$ & $\{(2k)^2\}$ and two $\{k^4\}$\\ \hline
$E_8$ & $A_{6k-1}$ & $\{(3k)^2\}$,$\{(2k)^3\}$ and $\{k^6\}$
\end{tabular}
\caption{6d construction for rank $k$ $H_G$ theory}\label{punctures}
\end{center}
\end{table}
The 6d construction for rank $k$ $H_{A_2}$ theories however involves
irregular punctures. For example, they can be realized by
compactifying 6d $A_{2k-1}$ theory on a sphere with one regular
puncture with Young diagram $\{k^2\}$ and one irregular puncture of
form
\be
\Phi=\frac{1}{z^3}\mathrm{diag}(1,\dots,1_{k_{\mathrm{th}}},-1,\dots,-1_{k_{\mathrm{th}}})+\dots
\ee
with the coefficient of $z^{-2}$ and $z^{-1}$ have the same type
of matrix \cite{Xie:2012hs}. In particular, the rank
one $H_{A_2}$ theory coincides with $(A_1,D_4)$ Argyres-Douglas
theory. See also the 6d construction involving irregular punctures in \cite{Bonelli:2011aa}.

Class $\mathcal{S}$ 4d SCFTs are also known to be connected to 2d
vertex operator algebra, i.e. chiral algebra
\cite{Beem:2013sza,Beem:2014rza}. This correspondence relies directly
on the class $\mathcal{S}$ construction and can be understood from
certain generalized TQFT structure on the punctured Riemann
surface. This relation sometimes gives a new approach to compute the
indices of 4d SCFT by realizing them as the vacuum character of
associated chiral algebra. For example, the chiral algebras associated
to rank one $H_{D_4}$ and $H_{E_6}$ theories are identified as
$\mathfrak{so}(8)$ affine Lie algebra at level $k_{2d}=-2$ and
$\mathfrak{e}(6)$ affine Lie algebra at level $k_{2d}=-3$ in
\cite{Beem:2013sza}. See some recent works trying to explain VOA/SCFT
correspondence
\cite{Pan:2017zie,Pan:2019bor,Oh:2019bgz,Dedushenko:2019yiw,Jeong:2019pzg}. Besides,
the rank one $H_{D_4,E_6,E_7}$ theories are also connected with the
curved $\beta\gamma$ systems on cones over the complex Grassmannian
$\mathrm{Gr}(2,4)$, the complex orthogonal Grassmannian
$\mathrm{OG}^+(5,10)$, and the complex Cayley plane $\mathbb{OP}^2$
respectively in \cite{Eager:2019zrc}.

\subsection{Hall-Littlewood and Schur indices}
The superconformal index of 4d $\mathcal{N}=2$ SCFT is defined as \cite{Kinney:2005ej,Romelsberger:2005eg}
\be\label{indqpt}
{\mathcal I}(p, q, t) = \mathrm{Tr}\,(-1)^F\Big(\frac{t}{pq}\Big)^r\,p^{j_{12}} \,
q^{j_{34}}\,
t^{R}\,
\prod_i a_i^{f_i}\, ,
\ee
where $j_{12}=j_2+j_1$ and $j_{34}=j_2-j_1$ denote the rotation generators in $\mathbb{C}^2$ with $j_{1,2}$ representing each $SU(2)$ Lorentz symmetry,
and $r$ and $R$ denote the $U(1)_r$ and $SU(2)_R$ generators respectively. Besides, $a_i$ are the fugacities for the flavour generators $f_i$ which
sometimes are set to be zero for simplicity.
For generic 4d SCFT, the full superconformal indices with $(p,q,t)$ are difficult to compute. For example, among all
$H_{G}^{(k)}$ theories, the full superconformal indices to our knowledge are only computable so far
for $H_{SO(8)}$ with arbitrary rank owing to their Lagrangian nature and $H_{E_6,E_7}$ for
rank one owing to the existence of certain $\mathcal{N}=1$ Lagrangian flow \cite{Gadde:2015xta,Agarwal:2018ejn}.

Certain limits of superconformal index are particularly interesting due to symmetry enhancement. The name of limit comes from the observation
that the resulting indices involve corresponding symmetric polynomial known in mathematics literature.
Following \cite{Gadde:2011uv}, we list three of them here:
\bitem
\item(Macdonald) $p\to 0$. Superconformal index when taking the Macdonald limit is computable for all class $\mathcal{S}$ theory with regular punctures.
    For a genus ${\frak g}$ theory with $s$ punctures compactified from 6d $A_{N-1}$ (2,0) SCFT, the Macdonald index is given in \cite{Gadde:2011uv} as
\be\label{macdonald}
{\mathcal I}^{\rm M}_{{\frak g},s} ({\bf a},q,t)=\prod_{j=2}^{N}(t^j;q)^{2{\frak g}-2+s}\frac{(t;q)^{(k-1)(1-{\frak g})+s}}
{(q;q)^{(k-1)(1-{\frak g})}}\;
\sum_{\lambda}
\frac{\prod_{i=1}^s \hat {\mathcal K}_{\Lambda_i}({ {\bf a}_i})\;P^\lambda({{\bf a}_i}(\Lambda_i)|q,t)}
{\left[P^\lambda(t^{\frac{k-1}{2}},t^{\frac{k-3}{2}},\dots,t^{\frac{1-k}{2}}|q,t)\right]^{2{\frak g}-2+s}}\,.\\
\ee
Here $P^\lambda({ {\bf a}_i}(\Lambda_i)|q,t)$ are Macdonald polynomials and the summation is over all possible Young diagrams
$\lambda=\{\lambda_1,\lambda_2,\dots,\lambda_{N-1},0\}$. The Pochhammer symbol $(a;b)$ is defined by
\be
(a;b)=\prod_{i=0}^\infty(1-ab^i)\,.
\ee
The $\hat {\mathcal K}_{\Lambda_i}$ factors are defined by
\be
\hat {\mathcal K}_{\Lambda}({\bf a})=
\prod_{i=1}^{row(\Lambda)}\prod_{j, k=1}^{l_i}\mathrm{PE}\bigg[\frac{{a}^i_j\bar {a}^i_k}{1-q}\bigg]_{{\frak a}_i,q}\, ,
\ee
with the coefficients ${a}^i_k$ associated to the Young diagram
as 
\be
{a}^i_j=c_jv^{\lambda_j+1-i}\quad\textrm{and}\quad \bar
{a}^i_k=c_k^{-1}v^{\lambda_k+1-i} \ ,
\ee
with $v^2=t$. Here these $c_j$ parameterize the residual flavour symmetry and are subject to constrain $\prod_{i=1}^{row(\Lambda)}\prod_{j}^{l_i}c_j=1$ to preserve the
traceless condition of $SU(N)$. The association of the flavour fugacities for a puncture $a(\Lambda)$ in Macdonald polynomial is defined similiarly as
$c_j v^{-\lambda_j-1+2i}$. Some good figures to visualize these definitions can be found in \cite{Gadde:2011uv,Gaiotto:2012uq}.
\item(Hall-Littlewood) $p,q\to 0$. By taking limit in (\ref{macdonald}), it is easy to obtain the Hall-Littlewood index for all class $\mathcal{S}$
    theories. As only genus zero theories are of concern in this paper, we only write down the formulas with ${\frak g}=0$. For example, the
    Hall-Littlewood index of 4d SCFT compactified from 6d $A_{N-1}$ theory is
\be\label{hlall}
{\mathcal I}^{\rm HL}=\mathcal{N}_{N,s}\sum_\lambda\frac{\prod_{i=1}^s \hat {\mathcal K}_{\Lambda_i}({\mathbf a_i})\;\psi^\lambda({\mathbf
a_i}(\Lambda_i)|v)}
{\left[\psi^\lambda(v^{N-1},v^{N-3},\dots,v^{1-N}|v)\right]^{s-2}}\, \ ,
\ee
where
\be
\mathcal{N}_{N,s}=(1-v^2)^{N-1+s}\prod_{j=2}^N(1-v^{2j})^{s-2},
\ee
and $\psi^\lambda$ is the Hall-Littlewood polynomials defined as
\be\label{HLdef}
\psi^{\lambda}(x_1,\dots,x_N|v)=
{\mathcal N}_\lambda(v)\;\sum_{\sigma \in S_N}
x_{\sigma(1)}^{\lambda_1} \dots x_{\sigma(N)}^{\lambda_N}
\prod_{i<j}   \frac{  x_{\sigma(i)}-v^2 x_{\sigma(j)} } {x_{\sigma(i)}-x_{\sigma(j)}}\,,
\ee
with
\be\label{normHL1}
{\mathcal N}_{\lambda}(v)=\prod_{i=0}^\infty \prod_{j=1}^{m(i)}\,
 \left(\frac{1-v^{2j}}{1-v^2}\right)^{-1/2}\, ,
\ee where $m(i)$ is the number of rows in the Young diagram $\lambda=(\lambda_1,\dots,\lambda_N)$ of length $i$. Here we have made the substitution
$t=v^2$ for convenience.

It is argued in \cite{Gadde:2011uv} that for linear quiver theories
the HL index is equivalent to the Hilbert series of the Higgs
branch. In particular, this is true for all $H_G$ theories. It is
well-known the Higgs branch of $H_G^{(k)}$ theories are the reduced
moduli space of $k$ $G$-instantons, which can be understood from the
probing picture that the $k$ D3-branes dissolving into the
seven-branes resemble $k$ instantons in the transverse space. Thus the
HL index of $H_G^{(k)}$ theory are supposed to be equal to the Hilbert
series of reduced moduli space of $k$ $G$-instantons. On the other
hand, the Hilbert series can also be obtained from the 5d Nekrasov
partition function with pure gauge group $G$, which are just the 5d
limit of elliptic genus of 6d minimal $(1,0)$ SCFT with type
$G$. Therefore, we arrive at the relation:
\be\label{tri} {\mathcal
  I}^{\rm HL}_{H_G^{(k)}}=\mathrm{Hilb}_{G}^{k}=g^{(0)}_{k,G},
\ee
where $g^{(0)}_{k,G}$ as we defined previously in (\ref{gdef}) is the
coefficient of leading $\Qtau$ order of $k$-string elliptic genus
$\mathbb{E}_{h_{G}^{(k)}}$. One can also add ``tildes'' to get the
equality with a free hypermultiplet coupled, in which situation one
encounters the full Hilbert series other than the reduced. We have
checked relation (\ref{tri}) for $k=1,2$ for all possible $G$ and
$k=3$ for $SU(3)$.\footnote{For $SU(3)$ and $F_4$, we are not aware
  how to compute the HL indices directly. Still, the Hilbert series
  are well-defined and computed in
  \cite{Benvenuti:2010pq,Hanany:2012dm}, which are in perfect
  agreement with our computation for elliptic genus from blowup
  equations.}
\item(Schur) $q=t$ with $p$ arbitrary. In fact, it can be shown in such specialization the index is independent of $p$. Thus, taking $\,p\to 0$,
    Schur index is actually a limit of Macdonald index. Using (\ref{macdonald}), the Schur index for a class $\mathcal{S}$ theory is given by
\be\label{schurall}
{\mathcal I}^{\rm Schur}=\hat{\mathcal{N}}_{N,s}\frac{\prod_{i=1}^s \hat {\mathcal K}_{\Lambda_i}({\mathbf a_i})\;\chi^\lambda({\mathbf a_i}(\Lambda_i))}
{\left[\chi^\lambda(v^{N-1},v^{N-3},\dots,v^{1-N})\right]^{s-2}}\,,
\ee
where\footnote{As in this paper we only deal with the cases with three or four punctures, we also shorten $\mathcal{N}_{N,3}$ as $\mathcal{N}_{N}$ and
$\mathcal{N}_{N,4}$ as $\mathcal{N}_{N}'$ in the latter subsections, and same for those with hat.}
\be
\hat{\mathcal{N}}_{N,s}=(v^2;v^2)^{s}\prod_{j=2}^N(v^{2j};v^2)^{s-2},
\ee
and $\chi^\lambda$ is the Schur polynomials defined as
\be
\chi_{{\lambda}}(\mathbf a)=\frac{\det(a_i^{\lambda_j+k-j})}{\det (a_i^{k-j})}\, .
\ee
At last, one replaces back $v^2\to q$.

The Schur indices in some sense are more interesting than the
Hall-Littlewood indices. For instance, for class $\mathcal{S}$
theories, Schur indices equal the $q$-deformed topological 2d
Yang-Mills parition function on the punctured Riemann surface
\cite{Gadde:2011ik}, and also equal the vacuum character of the
associated chiral algebra
\cite{Beem:2013sza,Beem:2014rza}. Furthermore, Schur indices can be
computed in IR via wall crossing for theories even beyond class
$\mathcal{S}$, such as certain Argyres-Douglas theories
\cite{Cordova:2015nma} including rank one $H_{A_2}$ theory.
\eitem

The full superconformal indices of rank one $H_{D_4,E_{6,7}}$ theories
have been computed in
\cite{Gadde:2010te,Putrov:2015jpa,Agarwal:2018ejn}. The Schur index of
rank one $H_{E_8}$ was given in \cite{DelZotto:2016pvm} and the Schur
index of rank one $H_{A_2}$ was given in \cite{Cordova:2015nma}. To
compute the Hall-Littlewood indices and Schur indices of higher rank
$H_{D_4,E_{6,7,8}}$ theories one will encounter certain subtle
issues. Directly using the general formulas (\ref{hlall}) and
(\ref{schurall}) fails to give correct results, because at a given
order of $v$ infinite number of Young diagrams $\lambda$ contribute
in. To cure such divergence, it was suggested in \cite{Gaiotto:2012uq}
that one reduce the flavor symmetry ``one box at a time'', that is to
change one specific puncture by moving one box down in the associated
Young diagram. The physical meaning of such operation is interpreted
as coupling a free hypermultiplet to $H_{G}^{(k)}$ theory, which in
our notation is just $\widetilde{H}_{G}^{(k)}$ theory. In the
terminology of \cite{Gaiotto:2012uq}, $H_{G}^{(k)}$ are ``bad''
theories, while $\widetilde{H}_{G}^{(k)}$ are ``good'' theories. One
can directly use (\ref{hlall}) and (\ref{schurall}) to compute the
indices of $\widetilde{H}_{G}^{(k)}$, then divide by the index of a
free hypermultiplet which is well defined, finally one will obtain the
finite indices of $H_{G}^{(k)}$. Following this procedure, the
Hall-Littlewood indices of rank two $H_{D_4,E_{6,7,8}}$ theories was
computed in \cite{Gaiotto:2012uq}. Similarly, we computed the Schur
indices of rank two and three $H_{D_4,E_{6,7,8}}$ theories which will
be shown in details in later sections. For higher rank $H_{A_2}$ we
are not aware how to compute its Schur indices due to the irregular
punctures of 6d construction.  Although there exist no $H_G^{k}$
theory for $G=F_4$, we suspect certain analogy can be constructed such
that Hall-Littlewood indices still make sense as the Hilbert series of
moduli space of $k$ $F_4$ instantons, and the Schur indices can be
associated with affine $\mathfrak{f}_4$ algebra. One support for such
speculation is that the Hilbert series for arbitrary $k$ $F_4$
instantons has been constructed from certain folding from $E_6$
\cite{Cremonesi:2014xha}. Thus we sometimes informally denote the
analogy as $H_{F_4}^{(k)}$ theories.

\subsection{Rank one: Del Zotto-Lockhart's conjecture}
In \cite{DelZotto:2016pvm}, Del Zotto-Lockhart found an intriguing
structure of one string elliptic genera of 6d mininal $(1,0)$ SCFTs
and a surprising relation between the elliptic genera and the
supersymmetric indices of rank one $H_G$ theories. Let us rephrase
their conjecture here:
\begin{conj}[Del Zotto-Lockhart]
There exists a function $L_{G}^{(1)}(v,m_G,\Qtau)=\sum_{i,j=0}^\infty b^G_{i,j}\Qtau^iv^j$ such that
\begin{enumerate}
\item $b^G_{i,j}$ can be written as the sum of characters of
  irreducible representation of $G$ with integral coefficients.
\item $ L_{G}^{(1)}(v,m_G,0) $ is the Hilbert series of the reduced
  moduli space of one $ G $-instanton, i.e. the Hall-Littlewood index
  of the $ H^{(1)}_{G} $ theory.
\item $ L_{G}^{(1)}(q^{1/2},m_G,q^{2}) $ is the Schur index of the
  $ H^{(1)}_{G} $ theory.
\item The reduced one-string elliptic genus
  $\mathbb{E}_{h_{G}^{(1)}}(v)$ can be generated from $L_{G}^{(1)}(v)$
  by the following formula in which the symmetry (\ref{E1sym}) is
  manifest:\footnote{Here the dependence on $\Qtau$ and $Q_m$ are
    implied.} 
  \be \ba
  \mathbb{E}_{h_{G}^{(1)}}(v)=&v^{2h-1}\Qtau^{1/6}\sum_{n\geq 0}\Qtau^{2n}\bigg[u^{4h}L_G(\Qtau^n\,v)-(-1)^{2h} u^{-4h}L_G(\Qtau^{n+1/2}/v)\\[-1mm]
  &+(1+(-1)^{2h})\Qtau^{h+1/2}\left(u^2L_G(\Qtau^{n+1/2}\,v)-u^{-2} L_G(\Qtau^{n+1}/v)\right)\\
  &+\Qtau^2\left((-1)^{2h}u^{4(1-h)}L_G(\Qtau^{n+1}\, v)-
    u^{-4(1-h)}L_G(\Qtau^{n+3/2}/v)\right)\bigg] \label{eq:conjLZ} \ea
  \ee where $h=h^\vee_G/6$, $u=v/\Qtau^{1/4}$.
\end{enumerate}
\end{conj}
The conjectural formula (\ref{eq:conjLZ}) is quite intricate. Roughly
speaking, it means the coefficient matrix of reduced one-string
elliptic genus contains several ``blocks'', overlapping or
non-overlapping, and each block contains infinite copies of the
$L_G^{(1)}$ function. The number of blocks turns out to be 2 for
$SU(3)$, $4$ for $F_4$ and 6 for the other $G$. In the following we
show the coefficient matrix of one-string elliptic genus of $SO(8)$ in
a way consistent with our later higher rank discussion. The
coefficient matrix of elliptic genus and the $L_G^{(1)}$ functions for
other $G$ can be found in \cite{DelZotto:2016pvm}. Let us denote \be
\mathbb{E}_{h_{SO(8)}^{(1)}}(v,
\Qtau,m_i=0)=v^{5}\Qtau^{-5/6}\sum_{i,j=0}^{\infty}c^{SO(8)}_{i,j}
v^j(\Qtau v^{-4})^i.  \ee Then we have Table \ref{tb:col} for the
coefficients $c^{SO(8)}_{i,j}$ where each ``block'' is colored
differently: the coefficients coming from the first term in the square
bracket in (\ref{eq:conjLZ}) is colored red, the second black, the
third blue, the forth orange, the fifth cyan and the last magenta.
\begin{table}[h]
\begin{center}\begin{small}
\begin{tabular}{c| cccccccccc }$i,j$ &0&2&4&6&8&$10$&$12$& $14$\\
\hline
0& \color{red}1&\color{red} 28&\color{red} 300&\color{red} 1925&\color{red} 8918 & \color{red}32928 & \color{red}102816 &\color{red} 282150\\
1& 0& 0& \color{red} 29&\color{red} 707& \color{red}6999& \color{red}42889 &\color{red}193102  &\color{red} 699762\\
2& $-1$& 0& 0&\color{blue} $2\cdot 1$& \color{red}$463+1$& \color{red}9947&\color{red}
92391 & \color{red} 544786\\
3&0 & $-28$& $-29$&
\color{orange}$-2\cdot1$&{\color{cyan}1}{\color{magenta}$\,- 1$}& \color{blue} $2 \cdot 29$& {\color{red}$5280+
29$}\color{blue}$\,+2\cdot 28$& \color{red}101850\\
4& 0& 0&
$-300$& $-707$& $-463-1$& \color{orange}$-2\cdot29$& {\color{cyan}$29$}\color{magenta}$\,- 29$& \color{blue}$2\cdot 463+2\cdot1$ \\
5& 0 &0 &0 &
$-1925$& $-6999$& $-9947$& $-5280-29$\color{orange}$\,-2\cdot 28$&\color{orange}$-2\cdot 463-2\cdot 1$\\
6& 0 &0 &0 & 0& $-8918$& $-42889$&
$-92391$& $-101850$\\
7& 0 &0 &0 & 0& 0& $-32928$& $-193102$& $-544786$\\
8& 0 &0 &0 & 0& 0& 0& $-102816$& $-699762$\\
9& 0 &0 &0 & 0& 0& 0& 0& $-282150$
\end{tabular}\end{small}
\end{center}
\caption{Expansion coefficients $ c^{SO(8)}_{i, j} $ for one $SO(8)$ instanton string.}\label{tb:col}
\end{table}
As we can see from the table, the reduced one-string elliptic indeed depends on $v^2$. One can also see the symmetry (\ref{E1sym}) on the two sides of the
ray with slop $-1/2$. Here the $L_{SO(8)}^{(1)}(v,\Qtau)$ function can be defined by all the red number in Table \ref{tb:col} with the red ${\color{red}
+1}$ and ${\color{red}+29}$ moving out, as they come from $n=1$ term in the summation. Thus we have\be
\ba
L^{(1)}_{SO(8)}(v,\Qtau)=&  (1\! +\! 28 v^2\! +\! 300 v^4\! +\! 1925 v^6\! +\! 8918 v^8\! +\! 32928 v^{10}\! +\!
   102816 v^{12}\! +\! 282150 v^{14}\!  +\! \mathcal{O}(v^{16}))\\[-0.5mm]
&+(29\! +\! 707 v^2\! +\! 6999 v^4\! +\! 42889 v^6\! +\!
    193102 v^8\! +\! 699762 v^{10}\!  +\mathcal{O}(v^{12}))\Qtau\\
&+(463 +
    9947 v^2 + 92391 v^4 + 544786 v^6 +  \mathcal{O}(v^{8}))\Qtau^2\\
 & +  (5280 + 101850 v^2 +   +\mathcal{O}(v^{4}))\Qtau^3+ \mathcal{O}(\Qtau^4).
\label{eq:so8L1}
\ea
\ee
Clearly, the first row in Table \ref{tb:col} gives the well-known
Hilbert series for the reduced moduli space for one $SO(8)$ instanton, i.e. the
Hall-Littlewood index for rank one $H_{SO(8)}$ theory:
\be
L^{(1)}_{SO(8)}(v,0)=\sum_{n=0}^\infty\chi_{n\theta}^{SO(8)}v^{2n}=1\! +\! 28v^2\! +\! 300v^4\! +\! 1925 v^6\! +\! 8918 v^8\! +\! 32928 v^{10}\! +\!
   102816 v^{12}\!   +\! \mathcal{O}(v^{14})
\ee
Adding the red numbers from $L^{(1)}_{SO(8)}(v,\Qtau)$ in each column of Table \ref{tb:col} together, one expects to obtain the Schur index of rank one
$H_{SO(8)}$ theory. Indeed, by making $v\to q^{1/2}$ to make contact with the literature, we obtain
\be L_{SO(8)}(q^{1/2},q^2) = 1 + 28 q + 329 q^2 + 2632 q^{3} + 16380 q^4\! + 85764 q^{5} +\!
 393589 q^6 +\! 1628548 q^{7}\! +\mathcal{O}(q^{8}).
\ee
Such series was actually already obtained by a lot of methods. For example, from the viewpoint of VOA/SCFT correspondece, it equals the vacuum character
of affine Lie algebra $\mathfrak{so}(8)_{k=-2}$ \cite{Beem:2013sza}. From the nature that rank one $H_{SO(8)}$ theory is actually just $SU(2)$ gauge
theory with $N_f=4$, the Schur index can be computed both from UV Lagrangian and IR wall-crossing formula \cite{Cordova:2016uwk}. See the Schur series
from vacuum character up to $q^{14}$ in the end of the appendix of \cite{Cordova:2016uwk}.

Such comparison between the reduced elliptic genus and Schur index for all other rank one $H_G$ theory except $G=F_4$ has been done in
\cite{DelZotto:2016pvm}. In particular, all $L_G^{(1)}(v,\Qtau,m_G=0)$ functions are identified, and the conjectural formula (\ref{eq:conjLZ}) holds to
substantial orders. Similarly, one can also couple a free hypermultiplet to establish the relation between original one-string elliptic genus
$\mathbb{E}_{\tilde{h}_{G}^{(1)}}(v)$ and the Hall-Littlewood and Schur indices of $\widetilde{H}_G^{(k)}$ theory. Indeed, the Schur index of a 4d
hypermultiplet is known to be \cite{Cordova:2016uwk}
\be
{\cal{I}}^{\rm Schur}_{h.m.}=\,\mathrm{PE}\bigg[\frac{q^{1/2}}{1-q}\big(x+x^{-1}\big)\bigg],
\ee
which can also be obtained by taking limit $\mathbb{E}_{\rm c.m.}(v, x,\Qtau)\to \mathbb{E}_{\rm cm}(q^{1/2}, x,q^2)$. The Hall-Littlewood index of a 4d
hypermultiplet i.e. the Hilbert series of $\mathbb{C}^2$ is well-known to be
\be
{\cal{I}}^{\rm HL}_{h.m.}=\frac{1}{(1-vx^{\pm1})},
\ee
which can also be obviously obtained by taking limit $\mathbb{E}_{\rm c.m.}(v, x,\Qtau\to 0)$, with a factor $v\Qtau^{-1/6}$ absorbed into the overall
factor of (\ref{gdef}). This makes the whole story consistent.

In the viewpoint of pure 4d, this intriguing conjecture indicates
there exists certain precise relation between the $\beta$-twisted
partition function on $T^2\times S^2$ and the partition function on
$S^3\times S^1$. We suspect the connection may be established by
realizing one $S^1$ of $T^2$ as the Hopf fibration over $S^2$ to get
$S^3\times S^1$. To find the consequence of such realization one has
to go into the details of localization which is beyond the scope of
current paper.\footnote{Guglielmo Lockhart came up independently with
  a similar idea (private communication).}

\subsection{Rank two}
We would like to generalize Del Zotto-Lockhart's conjecture to the rank two cases, where there exist more flavour symmetry that is $SU(2)_x$ in $H_G$
theories. To be precise, we want to find some functions $L_{G}^{(2)}(v,x,m_G,\Qtau)=\sum_{i,j=0}^\infty b^G_{i,j}\Qtau^iv^j$ such that
\begin{enumerate}
\item $b^G_{i,j}$ can be written as the sum of products between the character of irreducible representation of $SU(2)_x$ and the character of
    irreducible representation of $G$ with integral coefficients.
\item $ L_{G}^{(2)}(v,x,m_G,0) $ is the Hilbert series of the reduced moduli space of two $ G $-instanton, i.e. the Hall-Littlewood index of the $
    H^{(2)}_{G} $ theory.
\item $ L_{G}^{(2)}(q^{1/2},x,m_G,q^{2}) $ is the Schur index of the $ H^{(2)}_{G} $ theory.
\item The reduced two-string elliptic genus $\mathbb{E}_{h_{G}^{(2)}}(v,x,m_G,\Qtau)$ can be generated from $L_{G}^{(2)}(v,x,m_G,\Qtau)$ and
    $L_{G}^{(1)}(v,x,m_G,\Qtau)$ functions.
\end{enumerate}
It turns out the rank two cases are much more complicated than the rank one cases, one reason for which is that we can not rely on the additional symmetry
(\ref{Eksym}). Although we have not achieved an exact formula to generate the two string elliptic genus, we successfully manage to identify the
$L_{G}^{(2)}$ functions to substantial orders, which we will elaborate
on later for each example. In fact, the leading and subleading $\Qtau$ order of
$L_{G}^{(2)}(v,x,m_G,\Qtau)$ are just given by $g^{(0)}_{2,G}(v,x,m_G)$ in (\ref{g20G}) and $g^{(1)}_{2,G}(v,x,m_G)$ in (\ref{g21G}), while the
subsubleading order is given by
\be\label{g22L}
(\chi_5+(\chi_{\theta}+2)\chi_3+\chi_{\mathrm{Sym}^2\theta}+2\chi_{\theta}+3)
+\Big((\chi_{\theta}+1)\chi_4+((\chi_{\theta}+1)^2+(2\chi_{\theta}+1))\chi_2\Big)v+\dots
\ ,
\ee
which differs from $g^{(2)}_{2,G}(v,x,m_G)$ in (\ref{g22G}) by $1+\chi_2 v+\dots$. Such difference is recognized as what we call ``blue'' series in
contrast to the red $L_G^{(2)}$ functions. Indeed, the reason we also include $L_{G}^{(1)}$ in the last condition is that we observe a ``blue'' series
appearing multiple times in the coefficient matrix of $\mathbb{E}_{h_{G}^{(2)}}$:\\[-1mm]
\be\label{blue2}
\ba
M_G^{(2),\rm blue}(v,x)=&\sum_{n=0}^{\infty}v^n\sum_{i+2j=n+1}\chi_i\chi_{j\theta}=\frac{1}{(1-vx)(1-v/x)}
g^{(0)}_{1,G}(v)\\
=1+\chi_2v+(\chi_3+\chi_\theta)v^2+&(\chi_4+\chi_\theta\chi_2)v^3+(\chi_5+\chi_\theta\chi_3+\chi_{2\theta})v^4
+(\chi_6+{\chi_{\theta}}\chi_4+\chi_{2\theta}\chi_2)v^5+\dots \ .
\ea
\ee
For example, the blue series always appear at $\Qtau$ order $\dualCox/3$ with leading $v$ order $-2\dualCox/3$ (comparing to the leading $\Qtau$ order).
The reason for such phenomenon is yet not clear to us.

On the other hand, from the technique of class $\mathcal{S}$ theory, we can compute the Schur index of ${H}^{(2)}_G$ theories for $G=D_4,E_{6,7,8}$. All
of them are in agreement with our expectation from elliptic genera up to quite high orders. For example, from the $L_{G}^{(2)}$ functions, we are able to
write down the following general formula for the Schur indices up to $q^{7/2}$:
\be\label{SchurG}
\ba
{\cal I}_{H^{(2)}_G}^{\rm Schur}&={\cal I}_{\widetilde{H}^{(2)}_G}^{\rm Schur}/{\cal{I}}^{\rm
Schur}_{h.m.}=\,1+(\chi_3+{\chi_{\theta}})q+\chi_{\theta}\chi_2q^{3/2}
+\Big(\chi_5+(\chi_{\theta}+1)\chi_3+\chi_{\mathrm{Sym}^2\theta}+\chi_{\theta}+1\Big)q^2\\[-1 mm]
&+\Big(\chi_\theta\chi_4+(\chi_{2\theta}+\chi_{\mathrm{Sym}^2\theta}+1)\chi_2\Big)q^{5/2}+\Big(\chi_7+(\chi_{\theta}+1)\chi_5
+(\chi_{2\theta}+\chi_{\mathrm{Sym}^2\theta}+2\chi_{\theta}+3)\chi_3\\
&+\chi_{\mathrm{Sym}^3\theta}+(\chi_\theta+1)^2-C_6(G)\Big)q^3+\Big(\chi_{\theta}\chi_6+(\chi_{2\theta}+\chi_{\mathrm{Alt}^2
\theta}+2\chi_{\theta}+1)\chi_4\\
&+(\chi_{3\theta} +2\chi_{2\theta}
+(\chi_{\theta}+1)^2+\chi_{\mathrm{Sym}^2\theta}+
\chi_{\mathrm{Alt}^2\theta}+B_2(G)+C_7(G))\chi_2\Big)q^{7/2}+\dots \ .
\ea
\ee
In the following, we show the striking comparison between elliptic genus and indices at rank two for all symmetry group $G$.
\newline
\subsubsection*{$\bf SU(3)$}
For $SU(3)$, let us denote the two-string elliptic genus as
\be
\mathbb{E}_{h_{A_2}^{(2)}}(v,x, \Qtau,Q_{m})=v^{5}\Qtau^{-5/6}\sum_{i,j=0}^{\infty} c^{A_2}_{i,j}(x,Q_{m}) v^j(\Qtau v^{-4})^i.
\ee
Then we have the unrefined coefficients $c^{A_2}_{i,j}(x=1,Q_m=1)$ listed in Table \ref{tb:A2red2}. Keeping in mind that all such numbers can be refined to
incorporate $SU(2)_x$, we show the unrefined coefficients just to make them look clearer. The red numbers give the definition of $L_{G}^{(2)}$ functions.
In particular, they are in agreement with the universal expansion (\ref{g20G}), (\ref{g21G}) and (\ref{g22L}). Note the red numbers in the first row
agrees with the Hilbert series for reduced moduli space of two $A_2$ instantons in \cite{Hanany:2012dm}. The two red numbers in the $i=2$ rows are
predicted from (\ref{g22L}). Besides, the blue numbers agree with our proposal (\ref{blue2}). Adding the red numbers in each column together, we expect to
obtain a series that is equal to the Schur index of rank two $H_{A_2}$ 4d SCFT.
\begin{table}[h]
 \begin{center}
 \begin{small}
\begin{tabular}{c| ccccccccccccc }$i,j$&$-2$ & $-1$ &0&1&2 & 3 & 4 & 5& 6 &7 &8  \\
\hline
0& 0 & 0&\color{red}1& \color{red}0& \color{red}{11}& \color{red}{16} & \color{red}65 & \color{red}142 &\color{red}335 & \color{red}700 &\color{red}1542
\\
1& 0& 0& {\color{blue}1} &\color{blue} 2 & \color{blue}11 &\color{blue} 20 & {\color{red}12}{\color{blue}+56} & {\color{red}18}{\color{blue}+92}&
{\color{red}143}{\color{blue}+192} & {\color{red}356}{\color{blue}+292} &{\color{red}1091}{\color{blue}+517} \\
2&$-1$ &0 & 1& $-2$ &  2 & 0 & 51& 150 & 473 & 1032 & {\color{red}90}+2225 \\
\hline
\end{tabular}
\end{small}
 \begin{small}
\begin{tabular}{c| cccc}$i,j$&9 &10 & 11 & 12 \\
\hline
0& \color{red}2788 &\color{red}5350 & \color{red}9288 &\color{red}16184\\
1&  {\color{red}2676}{\color{blue}+742} &{\color{red}6387}{\color{blue}+1183} & {\color{red}13476}{\color{blue}+1624}
&{\color{red}28204}{\color{blue}+2408}\\
2& {\color{red}232}+4024 & 8589 & 15552 &30469\\
\end{tabular}
\end{small}
\caption{Unrefined coefficients $ c^{A_2}_{i,j} $ for the elliptic genus of two $ SU(3) $  instanton strings.}\label{tb:A2red2}
\end{center}
\end{table}

The construction of $H_{A_2}^{(2)}$ theory from 6d involves irregular punctures. We are not aware how to directly compute its indices. We write our
prediction from elliptic genus here: the Hall-Littlewood index of rank two $H_{A_2}$ theory is
\be
\ba
{\cal I}^{\rm HL}_{H_{A_2}^{(2)}}&=1+(\chi_3+8)q+8\chi_2q^{3/2}+(\chi_5+8\chi_3+36)q^2+(8\chi_4+55\chi_2)q^{5/2}\\[-2mm]
&+(\chi_7+8\chi_5+63\chi_3+119)q^3+(8\chi_6+55\chi_4+216\chi_2)q^{7/2}\\
+(\chi_9&+8\chi_7+63\chi_5+280\chi_3+322)q^{4}+(8\chi_8+55\chi_6+280\chi_4+637\chi_2)q^{9/2}
+\mathcal{O}(q^{5})\, ,
\ea
\ee
which agrees with the Hilbert series of reduced moduli space of two $SU(3)$ instantons \cite{Hanany:2012dm}, and the Schur index of rank two $H_{A_2}$
theory is
\be\label{rank2A2schur}
\ba
{\cal I}^{\rm Schur}_{H_{A_2}^{(2)}}&=1+(\chi_3+8)q+8\chi_2q^{3/2}+(\chi_5+9\chi_3+45)q^2+(8\chi_4+64\chi_2)q^{5/2}\\[-2mm]
&+(\chi_7+9\chi_5+82\chi_3+200)q^3+(8\chi_6+72\chi_4+360\chi_2)q^{7/2}\\
+(\chi_9&+9\chi_7+83\chi_5+479\chi_3+799)q^{4}+(8\chi_8+72\chi_6+496\chi_4+1608\chi_2)q^{9/2}
+\mathcal{O}(q^{5})\, .
\ea
\ee
Taking $x=1$ in (\ref{rank2A2schur}), we have the unrefined Schur index as
\be
1+11 q + 16 q^{3/2} + 77 q^2 + 160 q^{5/2} + 498 q^3 + 1056 q^{7/2} +
2723 q^4 + 5696 q^{9/2}+ \mathcal{O}(q^{5}) \ .
\ee
This is in complete agreement with Beem-Rastalli's to appear computation from chiral algebra!\footnote{We thank Beem and Rastelli for providing us their
unpublished results on the unrefined Schur index of rank two $SU(3)$ theory.}
\newline
\subsubsection*{$\bf SO(8)$}
The $H_{D_4}^{(2)}$ theory can be constructed by compactifying $A_3$ $(2,0)$ 6d SCFT on a sphere with four square punctures $\{2^2\}$, i.e. 2222 theory,
which is expected to be a $usp(4)$ gauge theory with four fundamental hypermultiplets and one anti-fundamental. On the other hand, the
$\widetilde{H}_{D_4}^{(2)}$ theory can be constructed as a $222L$ theory, i.e. we replaces one $\{2^2\}$ puncture to $\{2,1^2\}$. In
\cite{Gaiotto:2012uq}, the Hall-Littlewood indices of both $222L$ theory and $usp(4)+4f+1a$ theory was computed, which are in relation
\be
{\cal I}_{222L}(v,x,m_{i})=\frac{1}{1-vx^{\pm1}}\,
{\cal I}_{usp(4)+4f+1a}(v,x,m_{i})\,.
\ee
We expect and indeed checked to high orders
\be
{\cal I}_{usp(4)+4f+1a}(v,x,m_{i})=g^{(2)}_{0,D_4}(\tau,a,m_i).
\ee
For example, one can directly see the series coefficients in (A.12) of \cite{Gaiotto:2012uq} agree with the $\Qtau^0$ entries in Table \ref{tb:so8red}.

The Schur index of $222L$ theory can be obtained in a similar manner. Following the general formula in \cite{Gadde:2011uv}, we obtain
\be
\ba
{\cal I}^{\rm Schur}_{222L}&(c,d,e; a,b)=\hat{\cal N}_4'\, \hat{\cal K}_1(c)\,\hat{\cal K}_1(d)\,\hat{\cal K}_1(e)\,\hat{\cal K}_2(a,b)\,
\sum_\lambda\frac{\chi_\lambda(v b,v^{-1} b,b^{-1}a,b^{-1}a^{-1})}{\chi_\lambda^2(v^{-3},v^{-1},v,v^{3})}\,\\\times
\chi_\lambda(&v c,v^{-1} c,v c^{-1},v^{-1} c^{-1})
\chi_\lambda(v d,v^{-1} d,v d^{-1},v^{-1} d^{-1})
\chi_\lambda(v e,v^{-1} e,v e^{-1},v^{-1} e^{-1})\, ,
\ea
\ee
where ($b_1=b,\;b_2=1/b$). The summation is over Young diagrams $\lambda=(\lambda_1,\lambda_2,\lambda_3,0)$. The $\hat {\mathcal N}$ and $\hat {\mathcal
K}$ factors are given by
\be
\ba
\hat{\cal N}_4' &=(v^2;v^2)^4\prod_{j=2}^4(v^{2j};v^2)^2\,,\\
\hat{\cal K}_1(b)&=\mathrm{PE}\left[\frac{(v^{2}+v^4)(b^2+b^{-2}+2)}{1-v^2}\right]\,,\\
\hat{\cal K}_2(a,b)&=\mathrm{PE}\left[\frac{3v^2+v^4+v^3 b^{\pm2}a^{\pm1}+v^2 a^{\pm2}}{1-v^2}\right]\,.
\ea
\ee
At last, one usually replaces $v\to q^{1/2}$ to make contact with literatures. From the above formula, we computed the Schur index up to $v^{20}$ as
\be
\ba
{\cal I}^{\rm Schur}_{222L}=&1+\chi_2v+(2\chi_3+{
  28})v^2+(2\chi_4+58\chi_2)v^3+(3\chi_5+87\chi_3+465)v^4+\dots \ .
\ea
\ee
Decoupling the free hypermultiplet, we obtain the Schur index of $H_{D_4}^{(2)}$ theory
\be\label{so8rank2schur}
{\cal I}^{\rm Schur}_{H_{D_4}^{(2)}}={\cal I}^{\rm
  Schur}_{usp(4)+4f+1a}={\cal I}^{\rm Schur}_{222L}/{\cal{I}}^{\rm
  Schur}_{h.m.}
\ee
up to $q^{10}$. The first $12$ terms with full $SU(2)_x$ fugacity are
\be
\ba
{\cal I}^{\rm Schur}_{H_{D_4}^{(2)}}&=1+(\chi_3+28)q+28\chi_2q^{3/2}+(\chi_5+29\chi_3+435)q^2+(28\chi_4+707\chi_2)q^{5/2}\\[-2.5 mm]
&+(\chi_7+29\chi_5+765\chi_3+4845)q^3+(28\chi_6+735\chi_4+9947\chi_2)q^{7/2}\\
&+(\chi_9+29\chi_7+766\chi_5+12337\chi_3+43353)q^4\\
&+(28\chi_8+735\chi_6+12607\chi_4+101878\chi_2)q^{9/2}\\
&+(\chi_{11}+29\chi_9+766\chi_7+12667\chi_5+141518\chi_3+330360)q^5\\
&+(28\chi_{10}+735\chi_8+12635\chi_6+155449\chi_4+845225\chi_2)q^{11/2}+\mathcal{O}(q^{6})
\ .
\ea
\ee
We can compare this with elliptic genus up to $q^{11/2}$. Let us denote the $SO(8)$ two-string elliptic genus as
\be
\mathbb{E}_{h_{D_4}^{(2)}}(v,x, \tau,m_i=0)=v^{11}\Qtau^{-11/6}\sum_{i,j=0}^{\infty}c^{SO(8)}_{i,j}(x) v^j(\Qtau v^{-4})^i.
\ee
Then we have Table \ref{tb:so8red} for the coefficients $c^{SO(8)}_{i,j}(x=1)$.
\begin{table}[h]
 \begin{center}
 \begin{small}
\begin{tabular}{c| ccccccccc }$i,j$&0&1&2&3&4&5&6&7&8\\
\hline
0&\color{red}1& \color{red}0& \color{red}{31}& \color{red}{56} & \color{red}495 & \color{red}1468 &\color{red}6269 & \color{red}19680&\color{red}64768\\
1&0& 0& 0& 0& \color{red}{32} & \color{red}58 & \color{red}1023 & \color{red}3322 & \color{red}19078 \\
2&  \color{blue}$-1$ & \color{blue}$-2$ &\color{blue} $-31$ & \color{blue}$-60$ & \color{blue}$-389$ & \color{blue}$-718$& \color{blue}$-2972+2\cdot1$
&\color{blue} $-5226+2\cdot2$ & \color{red}{560}\color{blue}$-16398+2\cdot 31+1$\\
\hline
\end{tabular}
\end{small}
 \begin{small}
\begin{tabular}{c| ccc }$i,j$&9 & 10 & 11 \\
\hline
0 & \color{red}187792 & \color{red}537021 &\color{red}1424526\\
1 & \color{red}69114& \color{red}266799 & \color{red}886104 \\
2 & \color{red}{1912}\color{blue}$-27570+2\cdot 60+2$ &{\color{red}{20063}}\color{blue}$-71670+2\cdot 389+31$ &
{\color{red}{83586}}\color{blue}$-115770+2\cdot 718+60$ \\
\end{tabular}
\end{small}
\caption{Series coefficients $ c^{SO(8)}_{i,j} $ for the elliptic genus of two $ SO(8) $ instanton strings.}\label{tb:so8red}
\end{center}
\end{table}
Here the red numbers are from the $L_{D_4}^{(2)}$ series. Add the red numbers in each column together, we expect to obtain a series that is equal to the
Schur index of rank two $H_{D_4}$ 4d SCFT. Indeed, we have
\be
\ba
L_{D_4}^{(2)}(q^{1/2},x=1,&m_{D_4}=0,q^2)=1+31 q+56 q^{3/2}+527 q^2+1526 q^{5/2}+7292 q^3+23002 q^{7/2}\\
&+84406 q^8+258818 q^{9/2}+823883 q^5+2394216 q^{11/2}+\dots \ .
\ea
\ee
On the other hand, by taking the unrefined limit $x=1$ in (\ref{so8rank2schur}), we obtain the unrefined Schur series
\be
\ba
1&+31 q+56 q^{3/2}+527 q^2+1526 q^{5/2}+7292 q^3+23002 q^{7/2}+84406 q^8+258818 q^{9/2}+823883 q^5\\
&+2394216 q^{11/2}+6943434 q^{6}+19082748 q^{13/2}+51665849 q^7+134888730 q^{15/2}+345764537 q^8\\
&+862482876 q^{17/2}+2112344321 q^9+5061362222 q^{19/2}+11921262927
q^{10}+\mathcal{O}(q^{21/2}) \ .
\ea
\ee
One can see the two series match perfectly up to $q^{11/2}$!
\newline
\subsubsection*{$\bf F_4$}
Let us denote the two-string elliptic genus with gauge symmetry $F_4$ as
\be
\mathbb{E}_{h_{F_4}^{(2)}}(v,x=1, \tau,m_i=0)=v^{17}\Qtau^{-17/6} \sum_{i,j=0}^{\infty}c^{F_4}_{i,j} v^j(\Qtau v^{-4})^i.
\ee
Then we have Table \ref{tb:f4red} for the unrefined coefficients $c^{F_4}_{i,j}$. The red numbers in the first row agrees with the Hilbert series for
reduced moduli space of two $F_4$ instantons in \cite{Hanany:2012dm}.
\begin{table}[h]
 \begin{center}
 \begin{small}
\begin{tabular}{c| cccccccccccc }$i,j$&0&1&2 & 3 & 4 & 5& 6 &7 &8 &9  \\
\hline
0&\color{red}1& \color{red}0& \color{red}{55}& \color{red}{104} & \color{red}1539 & \color{red}4966 &\color{red}32091 & \color{red}119340
&\color{red}542109  & \color{red}1973088  \\
1& 0& 0& 0& 0& \color{red}{56} & \color{red}106 & \color{red}3135 & \color{red}10900 & \color{red}97125 & \color{red}405480\\
2&0 &0 & 0& 0 &  0 & 0 & 0& 0 & {\color{red}1652}\color{blue}+1 & {\color{red}6040}\color{blue}+2 \\
3 & \color{blue}1& \color{blue}2 &\color{blue} 55 &\color{blue} 108 &\color{blue} 1214 &\color{blue} 2320 &\color{blue} 15802 &\color{blue} 29284 &
\color{blue}143542$-1$ & \color{blue}257800$-2$   \\ \hline
\end{tabular}
\begin{tabular}{c| ccc }$i,j$ &10 &11 &12 \\
\hline
0&\color{red}7460100 &\color{red}25288640 & \color{red}84766812 \\
1&\color{red}2210027 & \color{red}9075756 &\color{red}38900537\\
2& {\color{red}99611}\color{blue}+55& {\color{red}466860}\color{blue}+108 & {\color{red}3399668}\color{blue}+1214\\
3    & \color{blue}999970$-55$ & \color{blue}1742140$-108$ &5704242\\
\end{tabular}
\end{small}
\caption{Series coefficients $ c^{F_4}_{i,j} $ for the elliptic genus of two $ F_4 $  instanton strings.}\label{tb:f4red}
\end{center}
\end{table}
By summing over the red numbers in each column, we obtain certain analogy of Schur index of rank two $H_G$ theory for $F_4$ up to $q^{11/2}$. The
unrefined version is
\be
\ba
1 &+ 55 q + 104 q^{3/2} + 1595 q^2 + 5072 q^{5/2} + 35226 q^3 + 130240 q^{7/2} + 640886 q^4 + 2384608 q^{9/2}\\
& + 9769738 q^5 + 34831256 q^{11/2}+\mathcal{O}(q^6) \ .
\ea
\ee
This is in complete agreement with Beem-Rastalli's to appear computation from chiral algebra!\footnote{We thank Beem and Rastelli for providing us their
unpublished results on the unrefined Schur index of rank two $F_4$ theory.}
\newline
\subsubsection*{$\bf E_6$}
The $H_{E_6}^{(2)}$ theory can be constructed by compactifying $A_5$ $(2,0)$ 6d SCFT on a sphere with three $\{2^3\}$ punctures, which is a ``bad''
theory. One can change one of the punctures to $\{2^2,1^2\}$ to add a decoupled hypermultiplet, i.e. the $\widetilde{H}_{E_6}^{(2)}$ theory. The
Hall-Littlewood index of this theory was computed in \cite{Gaiotto:2012uq}. We expect and indeed checked
\be
{\cal I}^{\rm HL}_{{H}_{E_6}^{(2)}}(v,x,m_{E_6})=\,g^{(2)}_{0,E_6}(v,x,m_{E_6}).
\ee

The Schur index can be obtained in a similar manner. Following the general formula in \cite{Gadde:2011uv}, we obtain
\be
\ba
{\cal I}^{\rm Schur}_{\widetilde{H}_{E_6}^{(2)}}&=\hat{\cal N}_6\, \hat{\cal K}_1(a_1,a_2)\,\hat{\cal K}_1(a_3,a_4)\,
\hat{\cal K}_2(a_5,a_6,x)\,\sum_\lambda
\frac{\chi_\lambda(v a_5,v^{-1} a_5,v a_6,v^{-1} a_6,
\frac{x}{a_5a_6},\frac{x^{-1}}{a_5a_6})}{\chi_\lambda(v^{-5},v^{-3},v^{-1},v^{1},v^{3},v^{5})}\,\\[-2mm]
&\times
\chi_\lambda(v a_1,v^{-1} a_1,v a_2,v^{-1} a_2,v \frac{1}{a_1a_2},v^{-1} \frac{1}{a_1a_2})
\chi_\lambda(v a_3,v^{-1} a_3,v a_4,v^{-1} a_4,v \frac{1}{a_3a_4},v^{-1} \frac{1}{a_3a_4})\,.
\ea
\ee
Here $\lambda=(\lambda_1,\cdots,\lambda_5,0)$ and ($b_3\equiv\frac{1}{b_1b_2}$)
\be
\ba
\hat{\cal N}_6&=(v^2;v^2)^3\prod_{j=2}^6(v^{2j};v^2)\,,\\[-1mm]
\hat{\cal K}_1(b_1,b_2)&=\prod_{\ell=1}^2\prod_{i,j=1}^3\mathrm{PE}\left[\frac{v^{2\ell} b_i/b_j}{1-v^2}\right],\\[-1mm]
\hat{\cal K}_2(b_1,b_2,x)&=\prod_{\ell=1}^2\prod_{i,j=1}^2\mathrm{PE}\left[\frac{v^{2\ell}
b_i/b_j}{1-v^2}\right]\times\mathrm{PE}\left[\frac{2v^2+v^2x^{\pm2}+v^3(b_1^2b_2x^{\pm1})^{\pm1}+
v^3(b_1b_2^2x^{\pm1})^{\pm1}}{1-v^2}\right]\,.
\ea
\ee
At last, one need to replace $v\to q^{1/2}$. We computed the Schur index up to $q^{7}$:
\be\label{SchurE6}
\ba
{\cal I}^{\rm Schur}_{H_{E_6}^{(2)}}&=\,1+(\chi_3+78)q+78\chi_2q^{3/2}+(\chi_5+79\chi_3+3160)q^2+(78\chi_4+5512\chi_2)q^{5/2}\\[-2mm]
&+(\chi_7+79\chi_5+5670\chi_3+87751)q^3
+(78\chi_6+5590\chi_4+201292\chi_2)q^{7/2}+(\chi_9+79\chi_7\\
&+5671\chi_5+248290\chi_3+1871196)q^4
+(78\chi_8+5590\chi_6+250640\chi_4+5048654\chi_2)q^{9/2}\\
&+(\chi_{11}+79\chi_9+5671\chi_7+250400\chi_5+7248975\chi_3+32615793)q^5\\
&+(78\chi_{10}+5590\chi_8+250718\chi_6+7900243\chi_4+97665932\chi_2)q^{11/2}\\
&+(\chi_{13}+79\chi_{11}+5671\chi_9+250801\chi_7+7949911\chi_5+157280287\chi_3+483480405)q^{6}\\
&+(78\chi_{12}+5590\chi_{10}+250718\chi_8+7949591\chi_6+186447755\chi_4+1552411211\chi_2)q^{13/2}\\
&+(\chi_{15}+79\chi_{13}+5671\chi_{11}+250801\chi_9+7952421\chi_7+193661181\chi_5+2725694921\chi_3\\
&+6263699772)q^{7}+\dots \ .
\ea
\ee
Note the leading terms up to $q^{7/2}$ agree with our general proposal (\ref{SchurG}).

Let us denote the two-string elliptic genus as
\be
\mathbb{E}_{h_{E_6}^{(2)}}(v,x=1, \tau,m_i=0)=v^{23}\Qtau^{-23/6}\sum_{i,j=0}^{\infty} c^{E_6}_{i,j} v^j(\Qtau v^{-4})^i.
\ee
Then we have Table \ref{tb:E6red} for the coefficients $c^{E_6}_{i,j}$.
\begin{table}[h]
 \begin{center}
\begin{small}
\begin{tabular}{c| cccccccccc }$i,j$&0&1&2&3&4&5&6&7&8&9\\
\hline
0& \color{red}1&\color{red}0 & \color{red}81 & \color{red}156 & \color{red} 3320 & \color{red}11178 & \color{red} 98440 & \color{red}
 401280 & \color{red} 2344619 &\color{red}9785226  \\
1&0& 0& 0&0&\color{red} 82 &\color{red}158 &\color{red} 6723 &\color{red} 24132 &\color{red}296879 &\color{red} 1335694  \\
2&0 &0 & 0& 0 & 0 & 0 & 0 &0 & {\color{red}3485}\color{blue}$+1$ & {\color{red}13112}\color{blue}$+2$ \\
3&0 &0 & 0& 0 & 0 & 0 & 0 & 0 & \color{blue}$-1$ & \color{blue}$-2$ \\
4&\color{blue}$-1$ & \color{blue}$-2$ & \color{blue}$-81$ & \color{blue}$-160$ & \color{blue}$-2669$ & \color{blue}$-5178$ & \color{blue}$-51445$ &
\color{blue}$-97712$ & \color{blue}$-681945$& \color{blue}$-1266178$\\ \hline
\end{tabular}
\end{small}
\begin{small}
\begin{tabular}{c| ccccc }$i,j$&10&11&12& 13 \\
\hline
0& \color{red}45870686 &\color{red}182872426 &\color{red}746229150 &\color{red}2782158570  \\
1 &\color{red}
 9484963 & \color{red}44112702&\color{red}236141466 &\color{red}1042037420  \\
2 & {\color{red}301488}\color{blue}$+81$ & {\color{red}1497516}\color{blue}$+160$ & {\color{red}14405643}\color{blue}$+2669$ &
{\color{red}75613998}\color{blue}$+5178$ \\
3& \color{blue}$-81$ & \color{blue}$-160$ & {\color{red}102090}{\color{blue}$-2669$}$+83$ & {\color{red}563580}{\color{blue}$-5178$}$+322$ \\
4& \color{blue}$-6819518+2\cdot1$& \color{blue}$-12372858+2\cdot2$& {\color{blue}$-54611704+2\cdot81$}$-83$& {\color{blue}$-96850550+2\cdot 160$}$-322$
\\
\end{tabular}
\end{small}
\begin{small}
\begin{tabular}{c| cc }$i,j$&14& 15\\
\hline
0&\color{red}10261780870 &\color{red}35695088906\\
1 &\color{red}4709271558 &\color{red}19202312882\\
2 &{\color{red}486421964}\color{blue}$+51445$ &{\color{red}2415319754}\color{blue}$+97712$\\
3 & {\color{red}9603627}{\color{blue}$-51445$}$+7039$  &{\color{red}58071366}{\color{blue}$-97712$}$+24620$\\
4&{\color{blue}$-365050846+2\cdot 2669$}$-7039$ &{\color{blue}$-633251142$}{\color{blue}$+2\cdot 5178$}$-24620$\\
\end{tabular}
\end{small}
\caption{Series coefficients $ c^{E_6}_{i,j} $ for the unrefined elliptic genus of two $ E_6 $ instanton strings.}\label{tb:E6red}
\end{center}
\end{table}
Here the red numbers are from the $L_{E_6}^{(2)}$ series. Add the red numbers in each column together, we expect to obtain a series that is equal to the
Schur index of rank two $H_{E_6}$ 4d SCFT. Indeed, we have
\be
\ba
L_{E_6}^{(2)}(q^{1/2},x=1,&m_{E_6}=0,q^2)=1+81 q+156 q^{3/2}+3402 q^2+11336 q^{5/2}+105163 q^3+425412 q^{7/2}\\
&+2644983 q^4+11134032 q^{9/2}+55655137 q^5+228482644 q^{11/2}+996878349 q^6\\
&+3900373568 q^{13/2}+15467078019 q^7+57370792908q^{15/2}+\dots \ .
\ea
\ee
On the other hand, taking $x=1$ in (\ref{SchurE6}), the unrefined Schur index is
\be
\ba
1&+81 q+156 q^{3/2}+3402 q^2+11336 q^{5/2}+105163 q^3+425412 q^{7/2}+2644983 q^4+11134032 q^{9/2}\\
&+55655137 q^5+228482644 q^{11/2}+996878349 q^6+3900373568
q^{13/2}+15467078019 q^7+\dots \ .
\ea
\ee
We can see the two series match perfectly up to $q^7$!
\newline

\subsubsection*{$\bf E_7$}
The $H_{E_7}^{(2)}$ theory can be constructed by compactifying $A_7$ $(2,0)$ 6d SCFT on a sphere with one $\{4^2\}$ puncture and two $\{2^4\}$ punctures,
which is a ``bad'' theory. One can change one of the $\{2^4\}$ punctures to $\{2^3,1^2\}$ to add a decoupled hypermultiplet, i.e. the
$\widetilde{H}_{E_7}^{(2)}$ theory. The Hall-Littlewood index of this theory was computed in \cite{Gaiotto:2012uq}. We find it agrees with our computation
for $g^{(2)}_{0,E_7}(\tau,x,m_{E_7})$. The Schur index can be obtained in a similar manner. Following the general formula in \cite{Gadde:2011uv}, we
obtain
\be
\ba
{\cal I}^{\rm Schur}_{\widetilde{H}_{E_7}^{(2)}}&=\hat{\cal N}_8\, \hat{\cal K}_1(a_1,a_2,a_3)\,\hat{\cal K}_2(a_4,a_5,a_6,x)\,
\hat{\cal K}_3(a_7)\,\\[-2mm]
&\times\sum_\lambda
\frac{
\chi_\lambda(v^3 a_7,v^{-3} a_7,v a_7,v^{-1} a_7,v^3 a_7^{-1},v^{-3} a_7^{-1},v a_7^{-1},v^{-1} a_7^{-1})
}
{\chi_\lambda(v^{-7},v^{-5},v^{-3},v^{-1},v,v^{3},v^{5},v^{7})}\,\\[-2mm]
&\times
\chi_\lambda(v a_1,v^{-1} a_1,v a_2,v^{-1} a_2,v^{} a_3,v^{-1} a_3,v
\frac{1}{a_1a_2a_3},v^{-1} \frac{1}{a_1a_2a_3})
\\[-2mm]
&\times\chi_\lambda(v a_4,v^{-1} a_4,v a_5,v^{-1} a_5,v^{} a_6,v^{-1} a_6,\frac{x}{a_4a_5a_6},\frac{x^{-1}}{a_4a_5a_6})\,.
\ea
\ee
Here
\be
\ba
&\hat{\cal K}_3(b)=\mathrm{PE}\left[\frac{2(v^2+v^4+v^6+v^8)}{1-v^2}\right]\prod_{\ell=1}^4\mathrm{PE}\left[\frac{v^{2\ell} b^{\pm2}}{1-v^2}\right]\,,\\
\hat{\cal K}_2(b_1,b_2,b_3,x)&=\mathrm{PE}\left[\frac{2v^2+v^2 x^{\pm
2}}{1-v^2}\right]\bigg[\prod_{\ell=1}^2\prod_{i,j=1}^3\mathrm{PE}\left[\frac{v^{2\ell} b_i/b_j}{1-v^2}\right]\bigg],
\prod_{i=1}^3\left[\frac{v^3 (x^{-1}\,b_{i}/b_4)^{\pm1}+v^3
    (x^{}\,b_{i}/b_4)^{\pm1}}{1-v^2}\right] \ ,
\\
&\hat{\cal K}_1(b_1,b_2,b_3)=\prod_{\ell=1}^2\prod_{i,j=1}^4\mathrm{PE}\left[\frac{v^{2\ell} b_i/b_j}{1-v^2}\right]\, .
\ea
\ee
At last, one need to replace $v\to q^{1/2}$. We computed the Schur index up to $q^2$ order. %

After decoupling the free hypermultiplet, the Schur index of $H_{E_7}^{(2)}$ theory is given by
\be\label{SchurE7}
{\cal I}^{\rm Schur}_{{H}_{E_7}^{(2)}}={\cal I}^{\rm Schur}_{\widetilde{H}_{E_7}^{(2)}}/{\cal{I}}^{\rm Schur}_{h.m.}=1+(\chi_3+{\bf 133})q+{\bf
133}\chi_2q^{3/2}+(\chi_5+({\bf 133}+1)\chi_3+\mathrm{Sym}^2{\bf
133}+{\bf 133}+1)q^2+\dots \ .
\ee

Let us denote the two-string elliptic genus as
\be
\mathbb{E}_{h_{E_7}^{(2)}}(v,x=1,
\tau,m_i=0)=v^{35}\Qtau^{-35/6}\sum_{i,j=0}^{\infty} c^{E_7}_{i,j}
v^j(\Qtau v^{-4})^i \ .
\ee
Then we have Table \ref{tb:E7red} for the coefficients $c^{E_7}_{i,j}$.
\begin{table}[h]
 \begin{center}
 \begin{small}
\begin{tabular}{c| ccccccccccc }$i,j$&0&1&2&3&4&5&6&7&8&9\\
\hline
0& \color{red}1&\color{red}0 & \color{red}136 & \color{red}266 & \color{red} 9315 & \color{red}32830 & \color{red} 449050 & \color{red}
 2026080 & \color{red} 17179899 &\color{red}84195608  \\
1&0& 0& 0&0&\color{red} 137 &\color{red}268 &\color{red} 18768 &\color{red} 69544 &\color{red} 1349005 &\color{red} 6575250  \\
2&0 &0 & 0& 0 & 0 & 0 & 0 &0 & {\color{red}9590}\color{blue}$+1$ & {\color{red}36982}\color{blue}$+2$  \\
\end{tabular}
\end{small}
\caption{Series coefficients $ c^{E_7}_{i,j} $ for the unrefined elliptic genus of two $ E_7 $  instanton.}\label{tb:E7red}
\end{center}
\end{table}
Here the red numbers are from the $L_{E_7}^{(2)}$ series. Add the red numbers in each column together, we expect to obtain a series that is equal to the
Schur index of rank two $H_{E_7}$ 4d SCFT. Thus, we predict the unrefined Schur index as
\be
\ba
1&+136 q+266 q^{3/2}+9452 q^2+33098 q^{5/2}+467818 q^3+2095624 q^{7/2}+18538494 q^4\\
&+90807840 q^{9/2}+\mathcal{O}(q^5) \ .
\ea
\ee
Indeed, taking $x=1$ in (\ref{SchurE7}), the unrefined Schur index is given by
\be
1+136 q+266 q^{3/2}+9452 q^2+\mathcal{O}(q^{5/2}) \ .
\ee
We can see the two series match perfectly!
\newline
\subsubsection*{$\bf E_8$}
The $H_{E_8}^{(2)}$ theory can be constructed by compactifying $A_{11}$ $(2,0)$ 6d SCFT on a sphere with three $\{6^2\}$, $\{4^3\}$ and $\{2^6\}$, which
is a ``bad'' theory. One can change the $\{2^6\}$ puncture to $\{2^5,1^2\}$ to add a decoupled free hypermultiplet, i.e. the $\widetilde{H}_{E_8}^{(2)}$
theory. Following the general formula in \cite{Gadde:2011uv}, we obtain its Schur index as
\be
\ba
{\cal I}^{\rm Schur}_{\widetilde{H}_{E_8}^{(2)}}&=\hat{\cal N}_{12}\, \hat{\cal K}_1(a_1,a_2,a_3,a_4,a_5,x)\,\hat{\cal K}_2(a_6,a_7)\,
\hat{\cal K}_3(a_8)\sum_\lambda
\frac{
\chi_\lambda(v a_1,v^{-1} a_1,\dots,v a_5,v^{-1} a_5,\frac{x}{a_1\cdots a_5},\frac{x^{-1}}{a_1\cdots a_5})
}
{\chi_\lambda(v^{-11},v^{-9},\dots,v^{9},v^{11})}\,\\[-1mm]
&\times
\chi_\lambda(v^3 a_6,v^{-3} a_6,v a_6,v^{-1} a_6,v^3 a_7,v^{-3} a_7,v a_7,v^{-1} a_7,v^3 \frac{1}{a_6a_7},v^{-3} \frac{1}{a_6a_7},v
\frac{1}{a_6a_7},v^{-1} \frac{1}{a_6a_7})
\\
&\times\chi_\lambda(v^{-5} a_8,v^{-3} a_8,\dots, v^3 a_8,v^5 a_8,v^{-5} a_8^{-1},v^{-3} a_8^{-1},\dots, v^3 a_8^{-1},v^5 a_8^{-1})\,.
\ea
\ee
Here $\lambda=(\lambda_1,\cdots,\lambda_{11},0)$ and
\be
\ba
\hat{\cal K}_3(b)=&\prod_{\ell=1}^6\mathrm{PE}\left[\frac{2v^{2\ell}+v^{2\ell} b^{\pm2}}{1-v^2}\right]\,,\\
\hat{\cal K}_2(b_1,b_2)=&\prod_{\ell=1}^4\prod_{i,j=1}^3\mathrm{PE}\left[\frac{v^{2\ell} b_i/b_j}{1-v^2}\right]\,,\\
\hat{\cal K}_1(c_1,c_2,c_3,c_4,c_5,x)=&\,\mathrm{PE}\left[\frac{2v^2+v^2\,x^{\pm2}}{1-v^2}\right]
\left[\prod_{\ell=1}^2\prod_{i,j=1}^5\mathrm{PE}\left[\frac{v^{2\ell} c_i/c_j}{1-v^2}\right]\right]\\&\times
\prod_{i=1}^5\mathrm{PE}\left[\frac{v^3 (x^{-1}\,c_{i}/c_6)^{\pm1}+v^3 (x^{}\,c_{i}/c_6)^{\pm1}}{1-v^2}\right]
\,
,
\ea
\ee
where $b_3\equiv\frac{1}{b_1b_2}$ and $c_6\equiv\frac{1}{c_1c_2c_3c_4c_5}$. At last, one need to replace $v\to q^{1/2}$. As the leading terms up to
$q^{3/2}$ are contributed from rank one theory, the Schur index is given by
\be
{\cal I}^{\rm
  Schur}_{\widetilde{H}_{E_8}^{(2)}}=1+\chi_2q^{1/2}+(2\chi_3+{\bf
  248})q+(2\chi_4+2({\bf 248}+1)\chi_2)q^{3/2}+\dots \ .
\ee

After decoupling the free hypermultiplet, the Schur index of $H_{E_8}^{(2)}$ theory is
\be\label{SchurE8}
{\cal I}^{\rm Schur}_{{H}_{E_8}^{(2)}}={\cal I}^{\rm Schur}_{\widetilde{H}_{E_7}^{(2)}}/{\cal{I}}^{\rm Schur}_{h.m.}=1+(\chi_3+{\bf 248})q+{\bf
248}\chi_2q^{3/2}+\dots \ .
\ee

Let us denote the two-string elliptic genus as
\be
\mathbb{E}_{h_{E_8}^{(2)}}(v,x=1, \tau,m_i=0)=v^{59}\Qtau^{-59/6}\sum_{i,j=0}^{\infty} c^{E_8}_{i,j} v^j(\Qtau v^{-4})^i.
\ee
Then we have Table \ref{tb:E8red} for the coefficients $c^{E_8}_{i,j}$.
\begin{table}[h]
 \begin{center}
 \begin{small}
\begin{tabular}{c| ccccccccc }$i,j$&0&1&2&3&4&5&6&7\\
\hline
0& \color{red}1&\color{red}0 & \color{red}251 & \color{red}496 & \color{red} 31625 & \color{red}116248 & \color{red} 2747875 & \color{red}
 13624000  \\
1&0& 0& 0&0&\color{red} 252 &\color{red}498 &\color{red} 63503 &\color{red} 241742   \\
\end{tabular}
\end{small}
\caption{Series coefficients $ c^{E_8}_{i,j} $ for the unrefined elliptic genus of two $ E_8 $ instanton strings.}\label{tb:E8red}
\end{center}
\end{table}
Here the red numbers are from the $L_{E_8}^{(2)}$ series. Add the red numbers in each column together, we expect to obtain a series that is equal to the
Schur index of rank two $H_{E_8}$ 4d SCFT. Thus, we predict from the general formula (\ref{SchurG}) for the unrefined Schur index as
\be
1+251 q+496 q^{3/2}+31877 q^2+116746 q^{5/2}+2811378 q^3+13865742
q^{7/2}+\mathcal{O}(q^4) \ .
\ee
Indeed, taking $x=1$ in (\ref{SchurE8}), the unrefined Schur index is given by
\be
1+251 q+496 q^{3/2}+\mathcal{O}(q^{2}) \ .
\ee
Indeed, the two series match perfectly!

\subsection{Rank three and higher}
We expect the Del Zotto-Lockhart's conjecture can be generalized to rank three and higher. From the universal leading expansion for three-string elliptic
genus (\ref{g30G}) and (\ref{g31G}), we are able to predict the Schur index of rank three $ H_{G} $ SCFT up to order $q^3$:
\be\label{schurrank3}
\ba
{\cal I}^{\rm Schur}_{{H}_{G}^{(3)}}=&\,1 + (\chi_3+\chi_{\theta})q + (\chi_4+\chi_{\theta}\chi_2) q^{3/2}
 +\(\chi_{5}+(\chi_{\theta}+1)\chi_3+\chi_{\mathrm{Sym}^2\theta}+\chi_{\theta}+2\)q^2\\[-1mm]
 &+
 \Big(\chi_6+(2\chi_\theta+2)\chi_4+2\chi_{\mathrm{Sym}^2\theta}+\chi_{\theta}+1\Big)q^{5/2}\\
 &+\(2\chi_7+(3\chi_\theta+1)\chi_5
 +(\chi_{2\theta}+3\chi_{\mathrm{Sym}^2\theta}+3\chi_\theta+5)\chi_3+\chi_{\mathrm{Sym}^3\theta}+3\chi_{\mathrm{Sym}^2\theta}+\chi_\theta+2\)q^3\\
 &+\mathcal{O}(q^{7/2}) \ .
\ea
\ee
This is actually because (\ref{g30G}) and (\ref{g31G}) are also the
definition of leading and subleading $\Qtau$ order of $L_G^{(3)}$
functions. Besides, we observe in the coefficient matrix of reduced
three string elliptic genus, other than the $L^{(3)}_{G}$ function
that appears as expected, the blue series also appears as in the rank
two. The difference is that here the blue series are generated from
the leading $\Qtau$ order of \emph{two} string elliptic genus!
\be\label{bule3}
M_G^{(3),\rm blue}(v,x)=\frac{1}{(1-vx)(1-v/x)}g^{(2)}_{0,G}(v,x).
\ee
Note $g^{(2)}_{0,G}(v,x)$ is also the leading $\Qtau$ order of $L^{(2)}_{G}$. In the following, we show the relation between reduced elliptic genus of
three strings and the Schur index of $H_G^{(3)}$ theories for each $G$.
\newline
\subsubsection*{$\bf SU(3)$}
The formula for the elliptic genus of three $SU(3)$ string has been written down via Jeffrey-Kirwan residues in \cite{Kim:2016foj}, using which we
computed $\mathbb{E}_{h_{A_2}^{(3)}}$ up to $\Qtau^6$ order. Denote
\be
\mathbb{E}_{h_{A_2}^{(3)}}(v,x, \tau,m_i=0)=v^{8}\Qtau^{-4/3}\sum_{i,j=0}^{\infty}c^{SU(3)}_{i,j}(x) v^j(\Qtau v^{-4})^i.
\ee
Then the unrefined $L_G^{(3)}$ function is shown red in the coefficient matrix of $\mathbb{E}_{h_{A_2}^{(3)}}$ in Table \ref{tb:3su3red}.
\begin{table}[h]
 \begin{center}
\begin{small}
\begin{tabular}{c| cccccccc }$i,j$&0&1&2&3&4&5&6&7\\
\hline
0& \color{red}1&\color{red}0 & \color{red}11 & \color{red}20 & \color{red} 90  & \color{red}218 & \color{red} 698 &\color{red}1618\\
1&\color{blue}1& \color{blue}2& \color{blue}14&\color{blue}22&{\color{blue}135+}\color{red} 12 &{\color{blue}370+}\color{red}22
&{\color{blue}960+}\color{red}171 & {\color{blue}2250+}\color{red}502\\
\end{tabular}
\end{small}
\caption{Coefficients $ c^{SU(3)}_{i,j} $ for the unrefined elliptic genus of three $ SU(3) $ instanton strings.}\label{tb:3su3red}
\end{center}
\end{table}
Note the red numbers are in agreement with our universal expansion (\ref{g30G}) and (\ref{g31G}), while the blue numbers are in agreement with our
proposal (\ref{bule3}).

The construction for rank three $H_{A_2}$ theory from 6d involves certain irregular punctures as the rank two case. We are not aware how to compute its
indices directly. We write down our prediction for the Schur index of rank three $H_{A_2}$ theory here:
\be
\ba
{\cal I}^{\rm Schur}_{{H}_{A_2}^{(3)}}&=1+(\chi_3+8)q+(\chi_4+8\chi_2)q^{3/2}+
(\chi_5+17\chi_3+46)q^2+(\chi_6+18\chi_4+81\chi_2)q^{5/2}\\[-2mm]
&+
(2\chi_7+25\chi_5+164\chi_3+248)q^3+(\chi_8+27\chi_6+209\chi_4+557\chi_2)q^{7/2}+\mathcal{O}(q^4)\, .
\ea
\ee
The unrefined limit is
\be
\ba
{\cal I}^{\rm Schur}_{{H}_{A_2}^{(3)}}(x=1)=&\,1+11 q +  20q^{3/2} +  102 q^2 + 240  q^{5/2} + 869q^3 + 2120q^{7/2} +\mathcal{O}(q^4)\, .
\ea
\ee
\newline
\subsubsection*{$\bf SO(8)$}
We can use class $\mathcal{S}$ theory technique to compute the HL and Schur index of rank three $H_{D_4}$ 4d SCFT. The $H_{D_4}^{(3)}$ theory can be
constructed by compactifying $A_5$ $(2,0)$ 6d SCFT on a sphere with four $\{3^2\}$ punctures, which is a ``bad'' theory. We need instead to consider
$\widetilde{H}_{D_4}^{(3)}$ theory obtained from three $\{3^2\}$ punctures and one $\{3,2,1\}$ puncture.
We compute the Schur index as
\be
\ba
{\cal I}^{\rm Schur}_{\widetilde{H}_{D_4}^{(3)}}(c_1,c_2,c_3; x,b)=&\hat{\cal N}_6'\, \hat{\cal K}_1(c)\,\hat{\cal K}_1(d)\,\hat{\cal K}_1(e)\,\hat{\cal
K}_2(x,b)\,
\sum_\lambda\frac{\chi_\lambda(v^2 b,b,v^{-2} b,v b^{-1}x,v^{-1}b^{-1}x,b^{-1}x^{-2})}{\chi_\lambda^2(v^{-5},v^{-3},v^{-1},v,v^{3},v^{5})}\,\\
&\times
\prod_{i=1,2,3}\chi_\lambda(v^2 c_i,c_i,v^{-2} c_i,v^2 c_i^{-1},c_i^{-1},v^{-2} c_i^{-1})\, ,
\ea
\ee
with ($b_1=b,\;b_2=1/b$)
\be
\ba
\hat{\cal N}_6' &=(v^2;v^2)^4\prod_{j=2}^6(v^{2j};v^2)^2\,,\\[-1mm]
\hat{\cal K}_1(b)&=\mathrm{PE}\left[\frac{(v^{2}+v^4+v^6)(b^2+b^{-2}+2)}{1-v^2}\right]\,,\\
\hat{\cal K}_2(a,b)&=\mathrm{PE}\left[\frac{3v^2+2v^4+v^6+(v^3+v^5) (b^{2}a^{-1})^{\pm 1}+v^3 a^{\pm3}+v^4 (b^{2}a^{2})^{\pm 1}}{1-v^2}\right]\,.
\ea
\ee
From the above formula, we compute the Schur index up to $q^{11/2}$. After decoupling the free hypermultiplet, we obtain
\be
\ba
{\cal I}^{\rm Schur}_{{H}_{D_4}^{(3)}}&={\cal I}^{\rm Schur}_{\widetilde{H}_{D_4}^{(3)}}/{\cal{I}}^{\rm
Schur}_{h.m.}=1+(\chi_3+28)q+(\chi_4+28\chi_2)q^{3/2}+
(\chi_5+57\chi_3+436)q^2\\[-2mm]
&+(\chi_6+58\chi_4+841\chi_2)q^{5/2}+
(2\chi_7+85\chi_5+1607\chi_3+5308)q^3\\
&+(\chi_8+87\chi_6+2042\chi_4+14135\chi_2)q^{7/2}+(2\chi_9+115\chi_7+2806\chi_5+29042\chi_3+55871)q^4\\
&+(2\chi_{10}+115\chi_8+3242\chi_6+43166\chi_4+177896\chi_2)q^{9/2}\\
&+(2\chi_{11}+144\chi_9+4008\chi_7+60673\chi_5+392233\chi_3+527217)q^5\\
&+(2\chi_{12}+145\chi_{10}+4441\chi_8+75128\chi_6+649112\chi_4+1857119\chi_2)q^{11/2}+\mathcal{O}(q^6)
\ .
\ea
\ee
The unrefined limit is
\be
\ba
{\cal I}^{\rm Schur}_{{H}_{D_4}^{(3)}}(x=1)=&1+31 q + 60 q^{3/2} + 612 q^2 + 1920 q^{5/2} + 10568 q^3 + 36968 q^{7/2}+157850 q^4\\[-2mm]
 &+ 548848 q^{9/2} + 2036655 q^5+6798456 q^{11/2}+\mathcal{O}(q^6) \ .
\ea
\ee

Let us denote the reduced three-string elliptic genus as
\be
\mathbb{E}_{h_{D_4}^{(3)}}(v,x, \tau,m_i=0)=v^{17}\Qtau^{-17/6}\sum_{i,j=0}^{\infty}c^{SO(8)}_{i,j} v^j(\Qtau v^{-4})^i.
\ee
Then from (\ref{g30G}) and (\ref{g31G}), we expect to have Table \ref{tb:so8red3} for the unrefined coefficients $c^{SO(8)}_{i,j}$.
\begin{table}[h]
 \begin{center}
 \begin{small}
\begin{tabular}{c| ccccccc }$i,j$&0&1&2&3&4&5&6\\
\hline
0&\color{red}1& \color{red}0& \color{red}{31}& \color{red}{60} & \color{red}580 & \color{red}1858 &\color{red}9457 \\
1&0& 0& 0& 0& \color{red}{32} & \color{red}62 & \color{red}1111  \\
\end{tabular}
\end{small}
\caption{Expected coefficients $ c^{SO(8)}_{i,j} $ for the unrefined elliptic genus of three $ SO(8) $ instanton strings.}\label{tb:so8red3}
\end{center}
\end{table}
Here the red numbers are from the $L_{D_4}^{(3)}$ series. Add the red numbers in each column together, we expect to obtain a series that is equal to the
Schur index of rank three $H_{D_4}$ 4d SCFT. Indeed, we have
\be
\ba
L_{D_4}^{(3)}(q^{1/2},x=1,Q_m=1,q^2)=1+31 q + 60 q^{3/2} + 612 q^2 +
1920 q^{5/2} + 10568 q^3+\dots \ .
\ea
\ee
One can see the two series match perfectly up to $q^{3}$!
\newline
\subsubsection*{$\bf E_6$}
The formula to compute the Hall-Littlewood index of rank three $H_{E_6}$ SCFT has been written down in \cite{Gaiotto:2012uq}. Similarly, we compute the
Schur index as
\be
\ba
&{\cal I}^{\rm Schur}_{\widetilde{H}_{E_6}^{(3)}}=\hat{\cal N}_9\, \hat{\cal K}_1(a_1,a_2)\,\hat{\cal K}_1(a_3,a_4)\,
\hat{\cal K}_2(a_5,a_6,x)\,\times\\
&\qquad\sum_\lambda
\frac{
\chi_\lambda(v^2 a_5,v^{-2} a_5,a_5,v^2 a_6,v^{-2} a_6,a_6,
 v \frac{x}{a_5a_6}, v^{-1}\frac{x}{a_5a_6},\frac{x^{-2}}{a_5a_6})
}
{\chi_\lambda(v^{-8},v^{-6},v^{-4},v^{-2},1,v^2,v^{4},v^{6},v^{8})}\,\times\nonumber\\
&\qquad\times
\chi_\lambda(v^2 a_1,v^{-2} a_1,a_1,v^2 a_2,v^{-2} a_2,a_2,v^2 \frac{1}{a_1a_2},v^{-2} \frac{1}{a_1a_2},\frac{1}{a_1a_2})
\nonumber\\
&\qquad\times
\chi_\lambda(v^2 a_3,v^{-2} a_3,a_3,v^2 a_4,v^{-2} a_4,a_4,v^2 \frac{1}{a_3a_4},v^{-2} \frac{1}{a_3a_4}, \frac{1}{a_3a_4})\,.
\nonumber
\ea
\ee
with
\be
\ba
\hat{\cal K}_1(b_1,b_2)=\mathrm{PE}&\bigg[\sum_{i,j=1}^3\frac{(v^{2}+v^4+v^6)b_i/b_j}{1-v^2}\bigg]\,,\\[-1mm]
\hat{\cal K}_2(b_1,b_2,x)=\mathrm{PE}&\bigg[\frac{1}{1-v^2}\Big((v^{2}+v^4+v^6)\Big(\sum_{i,j=1}^2
b_i/b_j\Big)+2v^2+v^4+v^3x^{\pm3}\\[-1mm]&+(b_1+b_2)((v^3+v^5)(b_3x)^{\pm 1}+v^4(b_3x^{-2})^{\pm 1})\Big)\bigg]
\, ,\nonumber
\ea
\ee
where $b_1b_2b_3=1$. From the above formula, we computed the Schur index up to $q^{2}$. After decoupling the free hypermultiplet, we obtain
\be
\ba
{\cal I}^{\rm Schur}_{{H}_{E_6}^{(3)}}&={\cal I}^{\rm Schur}_{\widetilde{H}_{E_6}^{(3)}}/{\cal{I}}^{\rm
Schur}_{h.m.}=1+(\chi_3+78)q+(\chi_4+78\chi_2)q^{3/2}+
(\chi_5+157\chi_3+3161)q^2\\[-2mm]
&\phantom{00000000}+(\chi_6+158\chi_4+6241\chi_2)q^{5/2}+
(2\chi_7+235\chi_5+11912\chi_3+91483)q^3\\
&\phantom{00000000}+(\chi_8+237\chi_6+15072\chi_4+260821\chi_2)q^{7/2}+\mathcal{O}(q^{4})
\ .
\ea
\ee
The unrefined limit is
\be\label{schurE63}
\ba
{\cal I}^{\rm Schur}_{{H}_{E_6}^{(3)}}(x=1)=&1+81 q + 160 q^{3/2} +
3637 q^2+13120 q^{5/2} + 128408 q^3 + 583360 q^{7/2} +
\mathcal{O}(q^{4}) \ .
\ea
\ee

On the other hand, the universal leading expansion (\ref{g30G}) and (\ref{g31G}) indicate the following Table \ref{tb:E6red3} for the coefficients of
reduced three-string elliptic genus for $E_6$.
\begin{table}[h]
 \begin{center}
\begin{small}
\begin{tabular}{c| ccccccc }$i,j$&0&1&2&3&4&5&6\\
\hline
0& \color{red}1&\color{red}0 & \color{red}81 & \color{red}160 & \color{red} 3555  & \color{red}12958 & \color{red} 121447 \\
1&0& 0& 0&0&\color{red} 82 &\color{red}162 &\color{red} 6961 \\
\end{tabular}
\end{small}
\caption{Coefficients $ c^{E_6}_{i,j} $ for the unrefined elliptic genus of three $ E_6 $ instanton strings.}\label{tb:E6red3}
\end{center}
\end{table}
By adding the red numbers in each column together, one can indeed obtain the same unrefined Schur series as (\ref{schurE63}) up to $q^3$.
\newline
\subsubsection*{$\bf F_4,E_7,E_8$}
The Schur indices with generic $SU(2)_x$ fugacity for rank three $H_G$ theories can be predicted from (\ref{schurrank3}) up to $q^3$ order. Let us just
mark the unrefined series here:
\be
\ba
{\cal I}^{\rm Schur}_{{H}_{F_4}^{(3)}}&=1 + 55 q + 108 q^{3/2} + 1752
q^2 + 6048 q^{5/2} + 45835 q^3+\mathcal{O}(q^{7/2})\ ,\\
{\cal I}^{\rm Schur}_{{H}_{E_7}^{(3)}}&=1 + 136 q + 270 q^{3/2} + 9852
q^2 + 36990 q^{5/2} + 533401 q^3+\mathcal{O}(q^{7/2}) \ ,\\
{\cal I}^{\rm Schur}_{{H}_{E_8}^{(3)}}&=1 + 251 q + 500 q^{3/2} +
32622 q^2 + 126000 q^{5/2} + 3030748 q^3+\mathcal{O}(q^{7/2}) \ .\\
\ea
\ee
Note for $F_4$, we always mean the analogy for $H_G$ theories.\\
\newline
In summary, we arrive at the final conjecture for arbitrary rank:
\begin{conj}
There exists an infinite series of functions $L_{G}^{(n)}(v,x,m_G,\Qtau)=\sum_{i,j=0}^\infty b^{G,n}_{i,j}\Qtau^iv^j$, $n=1,2,\dots$ such that
\begin{enumerate}
\item $b^{G,n}_{i,j}$ can be written as the sum of products between the character of irreducible representation of $SU(2)_x$ and the character of
    irreducible representation of $G$ with integeral coefficients.
\item $ L_{G}^{(n)}(v,x,m_G,0) $ is the Hilbert series of the reduced moduli space of $n$ $ G $-instanton, i.e. the Hall-Littlewood index of the $
    H^{(n)}_{G} $ theory.
\item $ L_{G}^{(n)}(q^{1/2},x,m_G,q^{2}) $ is the Schur index of the $ H^{(n)}_{G} $ theory.
\item The $n$-string elliptic genus $\mathbb{E}_{h_{G}^{(n)}}(v,x,m_G,\Qtau)$ can be generated from the first $n$ $L_G$ functions, i.e.
    $L_{G}^{(r)}(v,x,m_G,\Qtau)$, $r=1,2,\dots,n$.
\end{enumerate}
\end{conj}

\section{Conclusion and outlook}
\label{se:outlook}

In this paper we study the elliptic blowup equations for minimal 6d
$(1,0)$ SCFTs with all six possible gauge groups
$G=SU(3),SO(8),F_4,E_{6,7,8}$. The study is twofold, topological
string partition on elliptic non-compact Calabi-Yau and elliptic genera
for 6d $(1,0)$ SCFTs. From the viewpoint of Calabi-Yau, we use the
geometric construction in \cite{Haghighat:2014vxa} and the generalized
blowup equations in \cite{Huang:2017mis} to solve the refined BPS
invariants to high base degrees, which in turn serve as numerous
nontrivial checks for the blowup equations, both unity and vanishing
ones. From the viewpoint of 6d SCFTs, we use the de-affinisation
procedure to derive some elegant functional equations for the elliptic
genera, from which we obtain an exact and universal recursion formula
for the elliptic genera of arbitrary number of strings and arbitrary
gauge group. In particular, we explicitly compute the one and
two-string elliptic genera for all $G$, which recover all previous
partial results from refined topological string, modular bootstrap,
Hilbert series, 2d quiver gauge theories and the $\beta$-twisted
partition function of $\mathcal{N}=2$ superconformal $H_{G}$
theories. We also prove the modularity of the elliptic blowup
equations which is a strong support that they hold for arbitrary
number of strings.

The elliptic genera we solved out from blowup equations could be
useful in many aspects. For example, they would help to identify the
2d quiver description of the 6d minimal SCFT with exceptional gauge
symmetry, see some attempts for $G=E_7$ in \cite{Kim:2018gjo}. They
also serve as the calibration to determine modular ansatz for
higher-string elliptic genus and the web of topological vertex for the
associated non-toric Calabi-Yau threefolds \cite{Hayashi:2017jze}. The
elliptic genera of 2d $(0,4)$ SCFTs we studied also play a role in the
context of certain compact elliptic Calabi-Yau threefolds
\cite{Haghighat:2015ega,Hayashi:2019fsa}. For example, the $SO(8)$ and
$E_6$ minimal SCFT serve as the constituents of the 2d quiver gauge
theories associated to the $T^6/\IZ_2\times\IZ_2$ and
$T^6/\IZ_3\times\IZ_3$ geometries respectively, and their elliptic
genera are useful to compute the degeneracies of 5d spinning BPS black
holes in the dual gravity picture, as suggested in
\cite{Hayashi:2019fsa}. We hope our exact formulas of the elliptic
genera for exceptional minimal SCFTs would contribute to this subject.

It is also interesting to investigate the K-theoretic blowup equations
for all possible 5d SYM theories. The K-theoretic blowup equations
are quite different from the elliptic ones in that they exists for all
simple Lie groups. Part of the unity K-theoretic blowup equations were
already conjectured by Nakajima-Yoshioka \cite{Nakajima:2003pg} and
explicitly checked by Keller-Song \cite{Keller:2012da}. One may
suggest to use dimensional reduction i.e. $\Qtau\to 0$ to obtain the
K-theoretic blowup equations for the six gauge groups $G$. However,
intriguingly we find that the 5d reduction of elliptic blowup
equations does \emph{not} produce all non-equivalent K-theoretic
blowup equations for these six gauge groups. We leave these issues and
the complete set of K-theoretic blowup equations for all simple Lie
group to a separate paper \cite{GKSW2}.

The elliptic blowup equations for 6d $(1,0)$ minimal
SCFTs allow us to further study more complicated examples. Some
immediate models are the non-Higgsable clusters with matters which we
will investigate in a subsequent paper \cite{GKSW3}. There are four of
them: one belongs to minimal (1,0) SCFT with $\fn=7$, and the other
three have more than one dimensional tensor branch. See some primary
results on the elliptic genera for such theories in
\cite{Haghighat:2014vxa,DelZotto:2018tcj,Kim:2018gjo}. One can also
use blowup equations to study elliptic Calabi-Yau with multi-sections.
See some discussion on such geometries in
\cite{Braun:2014oya,Morrison:2014era}. Our final
goal is to find an exact, explicit and universal formula for the
elliptic genera of all 6d (1,0) SCFTs in the atomic classification
\cite{Heckman:2015bfa}, see also a good review on the classification in \cite{Heckman:2018jxk}. It is known there are two approaches to
classifying 6d (1,0) SCFTs: top down and bottom up. We expect blowup
equations make sense in both settings. In the top down approach, given
the explicit description of a elliptic non-compact Calabi-Yau,
i.e. base and elliptic fibration, one should be able to use the
generalized blowup equations in \cite{Huang:2017mis} to solve the
partition function of refined topological strings. On the other hand,
in the bottom up approach, given the explicit content of 6d multiplets
which satisfy the anomaly cancellations, one should also be able to
directly write down the elliptic blowup equations for the elliptic
genera of such 6d SCFT, as a generalization of the current paper. The
two pictures are related by geometric engineering, as the two
formalism of blowup equations are related by de-affinisation.

The K-theoretic Nekrasov partition function inspired the study on
K-theoretic invariants for general 4-manifolds, specially complex
surfaces \cite{Gottsche:2006bm}. Since the elliptic genus of 6d $(1,0)$
SCFTs in 5d limit gives K-theoretic Nekrasov partition function,
naturally one wonders if elliptic genus can be used to construct some
elliptic version of 4-manifold invariants, such as Donaldson
invariants. Besides, the Nekrasov partition function is known to
relate to W-algebras. In 4d, the equality between the universal
one-instanton Nekrasov partition function and the norm of
Gaiotto-Whittaker vectors in W-algebra has been checked in
\cite{Keller:2011ek}. See proof in \cite{Braverman:2014xca}. In 5d,
the relation between K-theoretic $SU(N)$ Nekrasov partition function
and $q$-deformed W$_N$-algebra was also studied in
\cite{Taki:2014fva}. It seems natural to extend such relation to 6d
where the elliptic genus should be related to the elliptic
W-algebras. The elliptic W-algebra associated to general Lie algebras
is very difficult to study. We hope our exact formula on the elliptic
genus could shed some new light. For example, it would be nice to see
if one can use elliptic W-algebra to make comparison with our
universal one-string elliptic genus formula (\ref{Z1}) like those
comparison done in \cite{Keller:2011ek}. One can even ask whether the
structure of blowup equations itself can find some origin in pure
algebras. See a possible direction \cite{Fukuda}.

One major remaining question is of course how to prove the elliptic
blowup equations, or more general, the blowup equations for all local
Calabi-Yau in \cite{Grassi:2016nnt,Gu:2017ccq,Huang:2017mis}. As the
refined BPS invariants for non-compact Calabi-Yau threefolds have been
rigorously defined via refined stable pairs in \cite{Choi:2012jz},
these functional equations for the partition functions are indeed
well-formulated mathematical conjectures. See also the definition of
refined invariants in \cite{MR3504535}. The proof of
G\"{o}ttsche-Nakajima-Yoshioka K-theoretic blowup equations
\cite{Nakajima:2009qjc} relies deeply on the structure of gauge
theories, which may not be exactly suitable for Calabi-Yau setting, as
the latter does not necessarily engineer a gauge theory.  As
emphasized before, the formalism of generalized blowup equations is
not sensitive to additional structures of non-compact Calabi-Yau
threefolds, be they toric or elliptic. Let us also point out there is
even no physical proof for the generalized blowup equations in
\cite{Grassi:2016nnt,Gu:2017ccq,Huang:2017mis}. In particular, it
would be good to see if one can connect the blowup equations with
refined holomorphic anomaly equations. Specializing to elliptic blowup
equations studied in this paper, we suspect by using the Kac-Weyl
character formulas and following the 4d derivation in
\cite{Keller:2011ek}, one may be able to derive the universal $\IE_1$
formulas (\ref{Z1}) and the identities from the leading degree of
vanishing blowup equations (\ref{vanish0}). We leave these for future
studies.

Another major question is how to explain the surprising relation
between the elliptic genera of 6d (1,0) SCFT and the Schur indices of
4d $\mathcal{N}=2$ $H_G$ theories. Despite the striking relation for
rank two cases and even some rank three cases shown in this paper, we
do not find the exact formulas connecting the two and three-string
elliptic genera and those $L_G$ functions like (\ref{eq:conjLZ}) in
rank one cases found in \cite{DelZotto:2016pvm}. To obtain such
fascinating formulas for arbitrary rank, it seems one has to answer
some questions first. For example, what is the physical meaning for
the $L_G^{(k)}$ functions?\footnote{Naively one may tempt to identify
  $L_G$ functions as Macdonald indices, since they both have two
  parameters, and both serve as an unification of Hall-Littlewood
  indices and Schur indices. However, this seems not ture. For
  example, the Macdonald index of rank one $H_{E_7}$ can be easily
  obtained by taking limit in the full superconformal index in
  \cite{Agarwal:2018ejn} as
  \begin{equation}
    \mathcal{I}^{M}_{H_{E_7}^{(1)}}(q,t)
    =1+133t+(134tq+7371t^2)+(134tq^2+16149t^2q+238602
    t^3)+(134tq^3+25193t^2q^2+819413t^3q+5248750t^4)+\dots
  \end{equation}
  While the $L_{E_7}^{(1)}$ function is determined in
  \cite{DelZotto:2016pvm} as \be
  L_{E_7}^{(1)}(\Qtau,v)=1+133v^2+7371v^4+238602
  v^6+5248750v^8+\dots+\Qtau(134+16283v^2+835562v^4+\dots)+\Qtau^2(31373+\dots)+\dots
  \ee One can see they are indeed not the same, albeit
  $\mathcal{I}^{M}_{H_{E_7}^{(1)}}(0,v^2)=L_{E_7}^{(1)}(0,v)$ and
  $\mathcal{I}^{M}_{H_{E_7}^{(1)}}(q,q)=L_{E_7}^{(1)}(q^2,q^{1/2})$.
} How to interpret and make use of those nonpertubative symmetries
(\ref{Eksym})? In 4d SCFT $H_G$, precisely how should the
$\beta$-twisted partition function on $T^2\times S^2$ be related to
the superconformal indices on $S^3\times S^1$? And how should the
SCFT/VOA correspondence be put in this picture? One possible direction
is to look into the localization on the 4d backgrounds following the
recent works
\cite{Pan:2017zie,Pan:2019bor,Oh:2019bgz,Dedushenko:2019yiw,Jeong:2019pzg}.

\section*{Acknowledgement}
We thank Babak Haghighat for the early participation of this
project. We also thank Christopher Beem, Giulio Bonelli, Michele Del Zotto, Lothar
G\"{o}ttsche, Min-xin Huang, Guglielmo Lockhart, Joonho Kim, Seok Kim,
Hiraku Nakajima, Leonardo Rastelli, Alessandro Tanzini, Wenbin Yan,
Don Zagier and Rui-Dong Zhu for useful
discussion. JG is supported by the Fonds National Suisse,
subsidiary 200020-175539 (project ``Quantum mechanics, geometry and
strings'').

\appendix

\section{Lie algebraic convention}
\label{ap:Lie}

We collect some definitions in (affine) Lie algebras and fix our
convention used throughout the paper. Given a simple Lie algebar $\fg$
of rank $\mathrm{rk}(G)$, there are four $\mathrm{rk}(G)$-dimensional lattices of importance,
the root and coroot lattices $Q, Q^{\vee}$, the weight and coweight
lattices $P, P^{\vee}$. They satisfy
\begin{gather}
  Q^\vee \subset P \subset \fh_{\IC} \ ,\\
  Q \subset P^\vee \subset \fh^*_{\IC} \ .
\end{gather}
Here $\fh_{\IC}, \fh_{\IC}^* \cong \IC^\mathrm{rk}$ are the complexified Cartan
subalgebra and its dual. They are isomorphic to each other via the
natural inner product
\begin{equation}
  \vev{\bullet,\bullet}: \fh_{\IC}^* \times \fh_{\IC} \to \IC \ .
\end{equation}
The Cartan matrix is then defined by
\begin{equation}
  A_{ij} = \vev{\alpha_i,\alpha_j^\vee}  \ ,
\end{equation}
where $\alpha_i$ are simple roots. Consider the invariant bilinear
form $(\bullet,\bullet)$ on the coroot lattice $Q^\vee$ normalized so
that the norm square of the shortest coroot $\theta^\vee$ is two. It
can be generalized to a bilinear form on $\fh_{\IC}$ in which $Q^\vee$
is embedded. By the isomorphism between $\fh_{\IC}$ and $\fh_{\IC}^*$,
it induces also an invariant bilinear form with the same notation on
the latter vector space. With our normalization, the bilinear form
satisfies
\begin{equation}\label{eq:kk-id}
  (k,k) = \frac{1}{2\hg}\sum_{\alpha\in\Delta}\vev{\alpha,k}^2 \ ,\quad
  k\in\fh\ ,
\end{equation}
where $h^\vee_\fg$ is the dual Coxeter number, and $\Delta$ the set of
all roots.

We also define the fundamental weights $\omega_i\in P$ and fundamental
coweights $\omega_i^\vee\in P^\vee$ ($i=1,\ldots,\mathrm{rk}$) through
\begin{equation}
  \vev{\alpha_i,\omega_j} = \vev{\omega_i^\vee, \alpha_j^\vee} = \delta_{ij}\ .
\end{equation}
They are related to roots and coroots by
\begin{equation}\label{eq:al-omega}
  \alpha_i = \sum_{j=1}^\mathrm{rk} A_{ij}\omega_j^\vee \ ,\quad
  \alpha_i^\vee = \sum_{j=1}^\mathrm{rk} \omega_j A_{ji} \ .
\end{equation}

Most of these definitions can be generalized to the affine Lie algebra
$\hat{\fg}$. We add an additional simple root $\alpha_0$ satisfying
\begin{equation}
  \alpha_0 = \alpha_0^\vee \ ,\quad
  (\alpha_0,\alpha_0) = (\alpha_0^\vee,\alpha_0^\vee) = 2 \ .
\end{equation}
The affine Cartan matrix is defined to be
\begin{equation}
  \hat{A}_{IJ} = \vev{\alpha_I, \alpha_J^\vee} \ ,\quad
  I,J=0,1,\ldots,\mathrm{rk} \ ,
\end{equation}
where $a_I, a_J^\vee$ are the marks and the comarks of the affine Lie
algebra $\hat{\fg}$ respectively. The affine Cartan matrix satisfies
\begin{equation}\label{eq:aA0}
  \sum_{I=0}^\mathrm{rk} a_I \hat{A}_{IJ} = \sum_{J=0}^\mathrm{rk} \hat{A}_{IJ} a_J^\vee =
  0 \ .
\end{equation}
Note that $\alpha_0$ can be written in terms of the longest root
$\theta$ and the imaginary root $\delta$, which annihilates anything
in $\fh$ or $\fh^*$ and has a vanishing norm square, by
\begin{equation}
  \alpha_0 = \delta -\theta \, . \label{eq:alpha0}
\end{equation}

Similarly we can also define the fundamental weights
$\widehat{\omega}_I$ and coweights $\widehat{\omega}_I^\vee$ in the
affine Lie algebra $\hat{\fg}$ by
\begin{equation}
  \vev{\alpha_I,\widehat{\omega}_J} =
  \vev{\widehat{\omega}_I^\vee,\alpha_J^\vee} =
  \delta_{IJ} \ ,\quad
  I,J=0,\ldots,\mathrm{rk} \ .
\end{equation}
For $i=1,\ldots,\mathrm{rk}$ the fundamental (co-)weights in $\hat{\fg}$ are
related those in $\fg$ by
\begin{gather}
  \widehat{\omega}_i = \omega_i + a_i \widehat{\omega}_0 \ ,\\
  \widehat{\omega}_i^\vee = \omega_i^\vee + a_i^\vee \widehat{\omega}_0^\vee \ ,
\end{gather}
while $\widehat{\omega}_0 = \widehat{\omega}_0^\vee$ is imaginary, and
it satisfies
\begin{equation}
  \vev{\alpha_i,\widehat{\omega}_0} = \vev{\widehat{\omega}_0,\alpha_i^\vee} =
  0\ ,\quad (\delta,\widehat{\omega}_0) = 1 \ .
\end{equation}
Using these relations together with \eqref{eq:al-omega}, we find the
affine version of \eqref{eq:al-omega}
\begin{equation}
  \alpha_I = \sum_{J=0}^\mathrm{rk} \hat{A}_{IJ} \widehat{\omega}^\vee_J
  +\delta \cdot \delta_{I,0}\ ,\quad
  \alpha_I^\vee = \sum_{J=0}^\mathrm{rk} \widehat{\omega}_J \hat{A}_{JI}
  + \delta \cdot \delta_{I,0}\ .
\end{equation}

\section{Mirror symmetry for elliptic non-compact Calabi-Yau
  three-folds}
\label{ap:norm}
The prescription in section~\ref{sc:Zpert} can determine all the
triple intersection numbers in the non-compact Calabi-Yau $X$ associated to
a minimal 6d SCFT except for the number $\kappa_{\tau\tau\tau}$. Given
the non-compactness of $X$ we do not expect all the triple
intersection numbers to be computable, and the number
$\kappa_{\tau\tau\tau}$ is irrelevant for the blowup equations in any
case. Nevertheless we propose here a reasonable normalisation scheme
for $\kappa_{\tau\tau\tau}$, which involves a local version of mirror
symmetry. We use this normalisation scheme in the example
section~\ref{sc:eg}.

For the compact Calabi-Yau $\hat{X}$ where $X$ is embedded, one can
define for every toric charge $l^{(i)}$ a Picard-Fuchs operator
$\hat{\mc L}_i$ which annihilates the homogeneous periods
$\hat{\omega}_0,\hat{\Pi}^{(1)}_i, \hat{\Pi}^{(2)}_i$. In the
decompactification limit $z_{\text{de}} \to 0$, the Picard-Fuchs (PF)
operator associated to the $(0)$-curve in the base vanishes, while the
other operators $\mc L_I$ remain well-defined and non-trivial, and
they annihilate all the finite homogeneous periods\footnote{For these
  elliptic Calabi-Yau threefolds decompactified in the horizontal
  direction, the fundamental period $\hat{\omega}_0$ does not become a
  constant but remains a non-trivial holomorphic function $\omega_0$.}
$\omega_0,\omega_0 t_I, \omega_0 F_J$ of the resulting local
Calabi-Yau.  These operators, however, do not form a PF complete
system, in the sense that they have extra independent solutions. To
cure this problem, we define in addition the PF operator $\mc L_\tau$
from the toric charge of the elliptic fiber
$\tau: (-6,2,3,0,\ldots,1,\ldots)$\footnote{The charge $1$ corresponds
  to the zero section of the elliptic fibration.}. It annihilates all
the finite homogeneous periods, but not the other superfluous
solutions, thus making the PF system complete.

The number $\kappa_{\tau\tau\tau}$ is contained in the homogeneous
B-period $\omega_0 F_\tau = \omega_0\pd_\tau F_{(0,0)}$ which
corresponds to the zero section. We find that it is completely fixed
by the normalisation condition
\begin{equation}
  \mc L_\tau (\omega_0 \pd_\tau F_{(0,0)}) = 0 \ .
\end{equation}
Note that the resulting homogeneous B-period is not a solution to the
complete PF system. When it is acted upon by the other PF operators it
does not vanish but produces $\omega_0$ up to a scaling factor, which
may have some open string interpretation.

There are several ways to understand this normalisation scheme. Once
all the triple intersection numbers are known, the normalised Euler
characteristic can be computed by\footnote{Here it is understood that
  we omit the toric charge of the $(0)$-curve in the summation. If one
  wishes the charge entry associated to the pullback of the base curve
  (denoted by $S$ in Table~\ref{tb:PD4}) can also be ignored as it is
  only nonzero for the $(0)$-curve. See for instance Table~\ref{tb:PD4}
  for the model with $G=F_4$.}
\begin{equation}\label{eq:euler}
  \chi = \int_{X}c_3(X) = \frac{1}{3}\sum_{ijk} \kappa_{ijk}
  l^{(i)}_nl^{(j)}_nl^{(k)}_n \ ,
\end{equation}
we list the results of all the minimal 6CFTs except for the cases of
$\fn=3,7$ in Table~\ref{tb:euler}. Note that the calculation of the
normalised Euler characteristic for the case $\fn >3$ is different
from that for the first three cases. For $\fn> 2$ the elliptic
singularity is constant over the $-\fn$ curve in the base. Except for
the $\fn=3$ case, the resolved geometry can be described as
configuration of Hirzebruch surfaces inside the compact CY-3-fold with
only even Betti numbers and $\chi(X)=1+b_2+b_4$. The case of $\fn=3$,
$G=SU(3)$ is special as the reduction from the compact geometry to the
non-compact geometry also involves a flop operation.  The geometry of
$\fn=2$ has more supersymmetry and odd Betti numbers $b_0=b_4=1$,
$b_1=b_2=b_3=2$. In the case $\fn=1$ we normalise the Euler
characteristic using the formula for the compact
threefold~\cite{Klemm:1996ts} with $E_8$ elliptic fibre type
$\chi(X_{comp})=-2 \cdot 30 \times \int_B c_1^2(M)$. The effect of
blowing up a $C^2=-\fn=1$ curve decreases $\int_B c_1^2(M)$ by
one. Hence the contribution of the non-compact geometry should be
$\chi(M)=60$.

The Euler characteristics thus computed for the theories with a pure gauge
bulk agree with the naive definition in terms of the numbers $b_n$ of
compact $n$-cycles $\chi = \sum_n (-1)^n b_n$. Furthermore, by
integrating the B-period $\pd_\tau F_{(0,0)}$ we can compute the genus
0 GW invariants in the $\tau$ direction, which should be the same as
the Euler characteristics. We checked this for the $\fn=1,2,5,6$
models.\footnote{For the remaining models the first non-vanishing
  invariants appear at very high degree and we fail to obtain them
  within a reasonable period of time.}
\begin{table}
  \centering
  \begin{tabular}{*{8}{>{$}c<{$}}}\toprule
    \fn & 1& 2& 4 & 5 & 6 & 8 & 12\\\midrule
    G &-&-& SO(8) & F_4 & E_6 & E_7 & E_8 \\
    b_2&1& & 6 & 6 & 8 & 9 & 10 \\
    b_4 &10&& 5 & 5 & 7 & 8 & 9 \\
    \chi(X) & 60 & 0 & 12 & 12 & 16 & 18 & 20 \\\bottomrule
  \end{tabular}
  \caption{The normalised Euler characteristics of the non-compact
    elliptic Calabi-Yau threefolds associated to minimal 6d SCFTs.}
  \label{tb:euler}
\end{table}

In the following, we illustrate this idea with two examples.

\subsection*{$\bf \fn=1$}

\begin{equation} \label{F1}
 \begin{array}{crrrr|rrrr|}
   D &\multicolumn{4}{c}{  \nu_i^*}
   &l^{(1)}& l^{(2)}& l^{(3)} & \\
   D_0    &  0&   0&   0&   0\phantom{\ }&-6 & 0 & 0  &\\
   D_1    & -1&   0&   0&   0\phantom{\ }&2 & 0 &0&  \\
   D_2    &  0&  -1&   0&   0\phantom{\ }& 3 & 0 & 0 &  \\
   S'     &  2&   3&   0&  -1\phantom{\ } &0 & -1 & 1 & \\
   K      &  2&   3&   0&   0 \phantom{\ }&1 & -1& -2 &  \\
   F      &  2&   3&  -1&  -1\phantom{\ } &0 & 1 & 0 &  \\
   S      &  2&   3&   0&   1\phantom{\ }&0 & 0& 1 & \\
   F      &  2&   3&   1&   0\phantom{\ }&0 & 1 & 0 &   \\
   \end{array} \
\end{equation}
The Picard-Fuchs operators of the compact geometry are
\begin{equation}
\begin{split}
\hat{\mathcal{L}}_1&=\theta _1 \left(\theta _1-\theta _2-2 \theta _3\right)-12 z_1 \left(6 \theta _1+1\right)
   \left(6 \theta _1+5\right) ,\\
\hat{\mathcal{L}}_2&=  \theta _2^2-z_2\left(\theta _1-\theta _2-2 \theta _3\right) \left(\theta _3-\theta
   _2\right), \\
\hat{\mathcal{L}}_3&= \theta _3 \left(\theta _3-\theta _2\right)-z_3\left(\theta _1-\theta _2-2 \theta
   _3-1\right) \left(\theta _1-\theta _2-2 \theta _3\right),
\end{split}
\end{equation}
where $\theta_i:=z_i\frac{\partial}{\partial z_i}$.  Denote
$\hat{F}_0$ the compact genus zero free energy, we have the periods:
$$
\hat{X}_0 =\hat{\omega}_0, \hat{X}_1=\hat{\omega}_0 t_1, \hat{X}_2= \hat{\omega}_0 t_2,
$$
\begin{equation}
\begin{split}
\hat{X}_3&=( \frac{\partial}{\partial \rho_2 \partial \rho_ 3}+ \frac{1}{2}\frac{\partial}{\partial \rho_3^2} )\hat{\omega}_0(\rho)|_{\rho=0}, \\
\hat{X}_4&= (-\frac{1}{2} \frac{\partial}{\partial \rho_1^2 }- \frac{\partial}{\partial \rho_1\partial \rho_ 2} )\hat{\omega}_0(\rho)|_{\rho=0}, \\
\hat{X}_5&= ( - \frac{\partial}{\partial \rho_1^2 }-\frac{\partial}{\partial \rho_1\partial \rho_3} )\hat{\omega}_0(\rho)|_{\rho=0}, \\
\end{split}
\end{equation}
and another period with triple logarithmic singularity. Here
$\hat{\omega}_0(\rho)$ is the deformed fundamental period
\begin{small}
\be
\hat{\omega}_0(\rho)=\sum_{\substack{n_1,n_2,n_3\in \mathbb{Z}_{\ge 0} \\
x_i=n_i+\rho_i}}\frac{\Gamma(x_1+1)}{\Gamma(x_1+1)\Gamma(x_1+1)\Gamma(x_1-x_2-2x_3+1)\Gamma(-x_2+x_3+1)\Gamma(x_3+1)\Gamma(x_2+1)^2}z_1^{x_1}z_2^{x_2}z_3^{x_3},
\ee
\end{small}
\noindent and $\hat{\omega}_0=\hat{\omega}_0({0})$. The
non-compact geometry is related to the compact geometry by setting
$z_{\text{dc}}:=z_3 \rightarrow 0$. Then the Picard-Fuchs operator
$\hat{\mc L}_3$ vanishes while the other two become:
\begin{equation}
\begin{split}
\mathcal{L}_1&=\theta _1 \left(\theta _1-\theta _2\right)-12
z_1\left(6 \theta _1+1\right) \left(6
   \theta _1+5\right), \\
 \mathcal{L}_2&= \theta _2^2+z_2\left(\theta _1-\theta _2\right)
 \theta _2. \\
\end{split}
\end{equation}
The deformed fundamental period $\hat{\omega}_0(\rho)$
becomes
$\omega_0(\rho)=\hat{\omega}_0(\rho)|_{z_3\rightarrow
  0}$.  There are one period
$X_0=\omega_0=\hat{\omega}_0|_{z_3\rightarrow 0}$ without singularity,
two periods with logarithmic singularities
\begin{equation}
X_1=\hat{X}_1|_{z_3 \rightarrow 0},\ \ \  X_2=\hat{X}_2|_{z_3
  \rightarrow 0},
\end{equation}
one with double logarithmic singularities
\begin{equation}
  X_3= ( - \frac{1}{2}\frac{\partial}{\partial \rho_1^2
  }-\frac{\partial}{\partial \rho_1\partial \rho_ 2}
  )\omega_0(\rho)|_{\rho=0}=\omega_0 \frac{\partial}{\partial
    t_2}F_0=\hat{X}_3|_{z_3 \rightarrow 0},
\end{equation}
and no solution with triple logarithmic singularities.

The reason we cannot fix the $\tau$ terms is because we do not know
the $\tau=t_1$ derivative of the free energy, however, there is an
interesting ``period''
$X_4=\omega_0 \frac{\partial}{\partial \tau}F_{(0,0)}$ which satisfies (up
to all the orders we have checked)
\begin{equation}
  \mathcal{L}_1 X_4=0,\ \ \ \mathcal{L}_2 X_4=2\omega_0.
\end{equation}
Note that $\mc L_1$ is precisely the PF operator in the
$\tau$-direction. If $X_4$ is a special period, we can integrate both
$X_3, X_4$ and fix the full triple intersection ring
\begin{equation}
  \mathcal{R}=-J_1^3-J_2 J_1^2-J_2^2 J_1 \  .
\end{equation}
and then proceed to compute the Euler number $\chi=60$ as well as
$\int c_2 J_1=-10, \int c_2 J_1=-12$ using
\eqref{eq:euler} and \eqref{eq:bGV-comp}.

Note that $X_4$ indeed descends from a period of the compact geometry.
In the non-compact geometry it is the properly normalised integral
over a non-compact cycle in the mirror Calabi-Yau.

\subsection*{$\bf \fn=5$}

The computation for the geometry with $\fn=1$ is kind of trivial. Let
us now consider a more complicated model with $\fn=5$. Notice in the
compact cases, we embed our elliptic Calabi-Yau 3-fold into a toric
variety described by a reflexive polytope and its star
triangulation. Then the Mori cone generators $l^{(i)}$, which are also
known as toric charges, are related to the star triangulation
directly. In the de-compactification limit, a point of the polytope is
missing, leaving a non-reflexive polytope. The dual polytope now in
principle have infinite size. As depicted in section \ref{sc:Zpert},
the limit happens to take the variable $z_{\text{dc}}$ to $0$. This is
equivalent to deleting one Mori generator, and keeping the
others. From the polytope point of view, we delete a sub-polytope from
it, and keep the same triangulation on the remaining part. We may
assume that the standard method of mirror symmetry for a compact
hyper-surface embedded in a compact toric variety still holds.

For now, we try to triangulate the non-reflexive polytope, it has 16
star triangulations. For one of them, the toric charges are
$l^{(i)},i=1,2,3,4,6,7$ in Table~\ref{tb:PD4}, we say the associated
curves form the toric basis of compact curves. One can in principle
write down the Picard-Fuchs equations, and then try to find
solutions. For this model, it is possible to change the variables of
complex structure parameters $z_i$ so that the solutions do not
change, and the mirror maps have expansions with positive powers of
$z_i$. The charges of the new basis $l^{(i)}_{F_4}$ can be found in
Table~\ref{tb:PD4}, which correspond exactly to nodes in the Dynkin
diagram. Then the complete Picard-Fuchs operators are given in
(\ref{picardfuchsn5}). There are five B-periods solved from these
operators, and an extra one $X_0 \frac{\partial}{\partial \tau}F_{(0,0)}$
annihilated by all the operators except for $\mathcal{L}_b$, with
$\mathcal{L}_b (\omega_0 \frac{\partial}{\partial \tau}F_{(0,0)})\sim
\omega_0$. The Euler number can be predicted from the $\tau$ direction
genus zero invariant as 12.

\begin{equation}\label{picardfuchsn5}
\begin{split}
\mathcal{L}_0=&\theta _1 \left(\theta _1-2 \theta _2+\theta _3-\theta _6\right)-z_1\left(2 \theta _1-\theta _2\right) \left(2 \theta _1-\theta
_2+1\right) \ ,\\
\mathcal{L}_1=&-\left(2 \theta _1-\theta _2\right) \left(\theta _2-2 \theta _3+\theta _4-\theta _6\right)-z_2\left(\theta _1-2 \theta _2+\theta _3-\theta
_6-1\right) \left(\theta _1-2 \theta _2+\theta _3-\theta _6\right)\ ,\\
\mathcal{L}_2=&\left(2 \theta _3-2 \theta _4+\theta _5-1\right) \left(2 \theta _3-2 \theta _4+\theta _5\right) \left(\theta _1-2 \theta _2+\theta
_3-\theta _6\right) \ ,\\
&-2z_3\left(2 \theta _3+1\right) \left(\theta _2-2 \theta _3+\theta
  _4-\theta _6-1\right) \left(\theta _2-2 \theta _3+\theta _4-\theta
  _6\right) \ ,\\
\mathcal{L}_3=&\left(\theta _4-2 \theta _5\right) \left(\theta _2-2 \theta _3+\theta _4-\theta _6\right)-z_4\left(2 \theta _3-2 \theta _4+\theta
_5-1\right) \left(2 \theta _3-2 \theta _4+\theta _5\right)\ ,\\
\mathcal{L}_4=&\theta _5 \left(2 \theta _3-2 \theta _4+\theta
  _5\right)-z_5\left(\theta _4-2 \theta _5-1\right) \left(\theta _4-2
  \theta _5\right) \ ,\\
\mathcal{L}_b=&\theta _6^2-z_6\left(\theta _1-2 \theta _2+\theta
  _3-\theta _6\right) \left(\theta _2-2 \theta _3+\theta _4-\theta
  _6\right)\ ,\\
\mathcal{L}_\tau=&\theta _1 \left(\theta _5-1\right) \theta _5-8z_1 z_2^2 z_3^3 z_4^4 z_5^2\left(2 \theta _3+1\right) \left(2 \theta _3+3\right) \left(2
\theta _3+5\right) \ .
\end{split}
\end{equation}

\section{Geometric data}
\label{ap:geodata}

We express here the Mori cone generators of the elliptic non-compact
Calabi-Yau threefolds $X$ associated to the minimal 6d SCFTs in terms
of the Mori cone generators $l^{(i)}$ of the compact Calabi-Yau
$\hat{X}$ given in \cite{Haghighat:2014vxa}.

\subsection*{$G=SO(8)$}

$l^{(5)}$ is the direction of decompactification.
\begin{equation}
  \Sigma_b = l^{(4)},\;
  \Sigma_0 = l^{(6)},\;
  \Sigma_1=2l^{(1)} + l^{(6)}+2l^{(7)},\;
  \Sigma_2 = 2l^{(1)}+l^{(6)},\;
  \Sigma_3 = l^{(3)}+l^{(6)},\;
  \Sigma_4 = l^{(1)}+l^{(2)}+2l^{(7)}\ .
\end{equation}

\subsection*{$G=F_4$}

$l^{(5)}$ is the direction of decompactification.
\begin{equation}
  \Sigma_b = l^{(4)},\;\;
  \Sigma_0 = l^{(3)},\;\;
  \Sigma_1 = l^{(2)}+l^{(6)}+2l^{(7)},\;\;
  \Sigma_2 = l^{(1)},\;\;
  \Sigma_3 = l^{(6)},\;\;
  \Sigma_4 = l^{(7)}\ .
\end{equation}

\subsection*{$G=E_6$}

$l^{(3)}$ is the direction of decompactifiction.
\begin{equation}
  \begin{gathered}
    \Sigma_b = l^{(5)},\;\; \Sigma_0 = l^{(4)},\;\;
    \Sigma_1 = l^{(1)}+l^{(6)}+2l^{(7)}+2l^{(9)},\;\; \Sigma_2 = l^{(8)}+l^{(9)}, \\
    \Sigma_3 = l^{(6)}+2l^{(7)}+l^{(8)},\;\;
    \Sigma_4 = l^{(2)}+l^{(7)}+l^{(8)},\;\; \Sigma_5 = l^{(6)},\;\;
    \Sigma_6 = l^{(7)}+l^{(9)} \ .
  \end{gathered}
\end{equation}

\subsection*{$G=E_7$}

$l^{(4)}$ is the direction of decompactification.
\begin{equation}
\begin{gathered}
  \Sigma_b = l^{(3)},\;\;
  \Sigma_0 = l^{(5)},\;\;
  \Sigma_1 = l^{(6)},\;\;
  \Sigma_2 = l^{(2)}+l^{(7)}+2l^{(8)}+l^{(9)},\;\;
  \Sigma_3 = l^{(1)}+l^{(7)}+l^{(10)},\;\;\\
  \Sigma_4 = l^{(7)},\;\;
  \Sigma_5 = l^{(8)}+l^{(10)},\;\;
  \Sigma_6 = l^{(9)},\;\;
  \Sigma_7 = l^{(7)}+2l^{(8)}+l^{(9)} \ .
\end{gathered}
\end{equation}

\subsection*{$G=E_8$}

$l^{(3)}$ is the direction fo decompactification.
\begin{equation}
\begin{gathered}
  \Sigma_b = l^{(8)},\;\;
  \Sigma_0 = l^{(4)},\;\;
  \Sigma_1 = l^{(5)},\;\;
  \Sigma_2 = l^{(6)},\;\;
  \Sigma_3 = l^{(7)},\;\;
  \Sigma_4 = l^{(1)}+l^{(2)}+l^{(10)}+2l^{(11)},\;\;\\
  \Sigma_5 = l^{(9)},\;\;
  \Sigma_6 = l^{(10)},\;\;
  \Sigma_7 = l^{(11)},\;\;
  \Sigma_8 = l^{(1)}+l^{(10)}+2l^{(11)} \ .
\end{gathered}
\end{equation}

\section{Useful identities}
\label{ap:ids}

Jacobi theta functions with characteristics are defined
as
\be \ba
\theta_1^{[a]}(\tau,z)=&-\ri\sum_{k\in\IZ}(-1)^{k+a}\Qtau^{(k+1/2+a)^2/2}Q_z^{k+1/2+a},\\
\theta_2^{[a]}(\tau,z)=&\sum_{k\in\IZ}\Qtau^{(k+1/2+a)^2/2}Q_z^{k+1/2+a},\\
\theta_3^{[a]}(\tau,z)=&\sum_{k\in\IZ}\Qtau^{(k+a)^2/2}Q_z^{k+a},\\
\theta_4^{[a]}(\tau,z)=&\sum_{k\in\IZ}(-1)^{k+a}\Qtau^{(k+a)^2/2}Q_z^{k+a},
\ea \ee which satisfy the well-known addition formulas
\be\label{addformulas} \ba
&\theta_3^{[a_1]}(\tau,z_1)\theta_3^{[a_2]}(\tau,z_2)=\sum_{i=2,3}\theta_i^{[\frac{a_1+a_2}{2}]}(2\tau,z_1+z_2)\theta_i^{[\frac{a_1-a_2}{2}]}(2\tau,z_1-z_2),\\
&\theta_4^{[a_1]}(\tau,z_1)\theta_4^{[a_2]}(\tau,z_2)=\sum_{i=1,4}\theta_i^{[\frac{a_1+a_2}{2}]}(2\tau,z_1+z_2)\theta_i^{[\frac{a_1-a_2}{2}]}(2\tau,z_1-z_2).
\ea \ee

Jacobi theta function $\theta_1$ can be defined as triple products
\be
\theta_1(\tau,z)=\ri\Qtau^{\frac{1}{12}}Q_z^{-\frac{1}{2}}\eta(\tau)\prod_{n=1}^{\infty}\(1-Q_z\Qtau^{n-1}\)\(1-\frac{\Qtau^n}{Q_z}\),
\ee
which satisfies the quasi-periodicity
\be
\theta_1(\tau,z+1)=-\theta_1(\tau,z),
\ee
\be\label{thetaper1}
\theta_1(\tau,z+\tau)=-\Qtau^{-1/2}Q_z^{-1}\theta_1(\tau,z),
\ee
\be\label{thetaper2}
\theta_1(\tau,z-\tau)=-\Qtau^{-1/2}Q_z\theta_1(\tau,z).
\ee

For a cluster of refined BPS invariants $N_{(0,k)}^k=1$ for all $k\ge 0$, the total contribution to BPS partition function is
\be\label{clusterformula}
\sum_{k=0}^{\infty}\frac{Q^k\chi_{2k+1}\((q_1q_2)^{1/2}\)}{\(q_1^{1/2}-q_1^{-1/2}\)\(q_2^{1/2}-q_2^{-1/2}\)}=\frac{1+Q}{\(q_1^{1/2}-q_1^{-1/2}\)\(q_2^{1/2}-q_2^{-1/2}\)\(1-Qq_1q_2\)\(1-Qq_1^{-1}q_2^{-1}\)}.
\ee
Similiarly, for a cluster of refined BPS invariants $N_{(0,k+1/2)}^k=-1$ for all $k\ge 0$, the total contribution to BPS partition function is
\be\label{clusterformula2}
\sum_{k=0}^{\infty}\frac{Q^k\chi_{2k+2}\((q_1q_2)^{1/2}\)}{\(q_1^{1/2}-q_1^{-1/2}\)\(q_2^{1/2}-q_2^{-1/2}\)}=\frac{(q_1q_2)^{1/2}+(q_1q_2)^{-1/2}}{\(q_1^{1/2}-q_1^{-1/2}\)\(q_2^{1/2}-q_2^{-1/2}\)\(1-Qq_1q_2\)\(1-Qq_1^{-1}q_2^{-1}\)}.
\ee

We also often encounter the case where a cluster of refined BPS invariants $N_{(0,k-1/2)}^k=-1$ for all $k\ge 0$ are combined with a ``zero'' degree
invariants $N_{(0,1/2)}^0=-1$. In such case, the total contribution is
\be\label{clusterformula3}
\ba
&\frac{1}{\(q_1^{1/2}-q_1^{-1/2}\)\(q_2^{1/2}-q_2^{-1/2}\)}+\sum_{k=1}^{\infty}\frac{Q^k\chi_{2k}\((q_1q_2)^{1/2}\)}{\(q_1^{1/2}-q_1^{-1/2}\)\(q_2^{1/2}-q_2^{-1/2}\)}\\
=&\frac{\Big((q_1q_2)^{1/2}+(q_1q_2)^{-1/2}\Big)(1+Q+Q^2-\(q_1q_2+q_1^{-1}q_2^{-1}\)Q)}{\(q_1^{1/2}-q_1^{-1/2}\)\(q_2^{1/2}-q_2^{-1/2}\)\(1-Qq_1q_2\)\(1-Qq_1^{-1}q_2^{-1}\)}.
\ea
\ee

In the computation of vector multiplets, we often encounter the following expressions:
\be
\mathrm{PE}\[\frac{Q}{1-\Qtau}\]=\prod_{n=0}^{\infty}\frac{1}{1-Q\Qtau^n}
\ ,
\ee
and
\be\label{petheta}
\mathrm{PE}\[\Big(Q_z+{\Qtau\over Q_z}\Big)\Big(\frac{1}{1-\Qtau}\Big)\]=\frac{\ri \Qtau^{\frac{1}{12}}Q_z^{-\frac{1}{2}}\eta(\tau)}{\theta_1(\tau,z)}.
\ee
In counting the total index quadratic form of the contribution from vector multiplets, we often encounter the following expression:
\be\label{brevedef}
\mathrm{PE}\(-\Big(Bl_{(0,1/2,R)}(q_1,q_2)Q_z+Bl_{(0,1/2,-R)}(q_1,q_2){\Qtau\over Q_z}\Big)\Big({1\over {1-\Qtau}}\Big)\).
\ee
Here
\be\label{Blfunc}
Bl_{(j_l,j_r,R)}(q_1,q_2)=f_{(j_l,j_r)}(q_1,q_2/q_1)q_1^R+f_{(j_l,j_r)}(q_1/q_2,q_2)q_2^R-f_{(j_l,j_r)}(q_1,q_2),
\ee
and
\be
f_{(j_l,j_r)}(q_1,q_2)=\frac{\chi_{j_l}(q_L)\chi_{j_r}(q_R)}{\(q_1^{1/2}-q_1^{-1/2}\)\(q_2^{1/2}-q_2^{-1/2}\)}.
\ee
%
Supposing $R\geq2$, the expression (\ref{brevedef}) can be written as
\be
\ba
& \prod_{\displaystyle \substack{m,n\geq 0\\m+n\leq R-1}}\frac{\ri\Qtau^{1/12}\eta(Q_zq_1^mq_2^n)^{-1/2}}{\theta_1(z+m\eq+n\et)}\prod_{\displaystyle
\substack{m,n\geq 0\\m+n\leq R-2}}\frac{\ri\Qtau^{1/12}\eta(Q_zq_1^{m+1}q_2^{n+1})^{-1/2}}{\theta_1(z+(m+1)\eq+(n+1)\et)}\\
=\, \big(\ri\Qtau^{1/12}& Q_z^{-1/2}\big)^{R^2}(q_1q_2)^{-\frac{(R-1)R(R+1)}{6}}\prod_{\displaystyle \substack{m,n\geq 0\\m+n\leq
R-1}}\frac{\eta}{\theta_1(z+m\eq+n\et)}\prod_{\displaystyle
\substack{m,n\geq 0\\m+n\leq
  R-2}}\frac{\eta}{\theta_1(z+(m+1)\eq+(n+1)\et)} \ .
\ea
\ee
We normally denote the modular part of (\ref{brevedef}), i.e. those $\theta_1$ and $\eta$ functions in the above expression together as $\breve{\theta}$.
The index quadratic form of $\breve{\theta}$ can be computed as
\be\label{indbreve}
\mathrm{Ind}_{\breve{\theta}}(z,R)=-{R^2z^2\over
2}-\frac{(R-1)R(R+1)}{3}z(\eq+\et)-\frac{(R-1)R^2(R+1)}{12}(\epsilon_1^2+\epsilon_1\epsilon_2+\epsilon_2^2),
\ee
which actually holds for all $R\in\IZ$. See more details in the Appendix A in \cite{Gu:2018gmy}.

\section{Relation with modular ansatz}
\label{ap:mod-ansatz}

In this appendix, we show how the modular ansatz, or its denominator to
be specific, for the elliptic genus emerges from our exact
formulas. Simply speaking, the denominator of the modular ansatz comes
from suming over all $\alpha^{\vee}$ with a fixed norm square in the
recursion formula.

It was proposed in \cite{DelZotto:2016pvm,DelZotto:2017mee} the
$k$-string elliptic genus satifies the following ansatz
\begin{equation}
  \mathbb{E}_{\widetilde
    h_{G}^{(k)}}(\tau,\eq,\et,m_\alpha) =
  \frac{\mathcal{N}_{G,k}(\tau,\eq,\et,m_\alpha)}
    {\mathcal{D}_{G,k}(\tau,\eq,\et,m_\alpha)}\ ,
  \label{eq:ansatzG}
\end{equation}
where both the numerator and the denominator are Weyl invariant Jacobi
forms. Furthermore the denominator has the following unique structure
as a Weyl invariant Jacobi form which reproduces the poles of the
Hilbert series of the moduli space of $k$ $G$-gauge instantons
\cite{Cremonesi:2014xha} and the correct leading order of $Q_\tau$:
\begin{equation}\label{eq:Dk-0}
  \mc D_{G,k} = \eta(\tau)^{4kh^\vee_G}
  \prod_{i=1}^k
  \varphi_{-1,\tfrac{1}{2}}(i\eq)
  \varphi_{-1,\tfrac{1}{2}}(i\et) \tilde{D}_{G,k}
\end{equation}
with the gauge group related factor
\begin{equation}\label{eq:Dk-1}
  \tilde{\mc D}_{G,k} =  \prod_{\alpha\in\Delta}\prod_{i=1}^k\prod_{\ell=0}^{i-1}
  \varphi_{-1,\tfrac{1}{2}}((i+1)\epsilon_++(i-1-2\ell)\epsilon_-+m_\alpha)
  \ ,
\end{equation}
multiplying over the set of roots. Later \cite{Kim:2018gak} claims
that $\IE_k$ has actually fewer poles and as a consequence the
denominator is smaller (see also \cite{Keller:2011ek}). It can be
written as\footnote{Here ``red'' means ``reduced'', i.e. the number of poles reduces. The ``tilde'' and ``reduced'' in this section should not be confused
with them in the main text where them mean a free hypermultiplet is coupled or decoupled.}
\begin{equation}\label{eq:Drk-0}
  \mc D_{G,k}^{\text{red}} = \eta(\tau)^{4kh^\vee_G}
  \prod_{i=1}^k
  \varphi_{-1,\tfrac{1}{2}}(i\eq)
  \varphi_{-1,\tfrac{1}{2}}(i\et) \tilde{\mc D}_{G,k}^{\text{red}}
\end{equation}
with the gauge group related factor
\begin{equation}\label{eq:Drk-1}
  \tilde{\mc D}_{G,k}^{\text{red}}=
  \prod_{\alpha\in\Delta^l}
  \mc D_{k,\alpha}^{\text{SU(2)}}(\tau,\nound{m})
  \prod_{\alpha\in\Delta^s}
  \mc D_{\lfloor k/c_\alpha
    \rfloor,\alpha}^{\text{SU(2)}}(\tau,\nound{m}) \ ,
\end{equation}
where $\Delta^l,\Delta^s$ are the set of long roots and short roots
respectively, the constants $c_\alpha$ are
\begin{equation}
  \begin{aligned}
    c_\alpha =2 \quad &\text{if}\; G= Sp(N), SO(2N+1), F_4 \ ,\\
    c_\alpha=3 \quad &\text{if}\; G=G_2 \ ,
  \end{aligned}
\end{equation}
and that
\begin{equation}
  \mc D_{k,\alpha}^{\text{SU(2)}}(\tau,\nound{m}) =
  \prod_{\substack{a,b\leq k\\a,b>0}}
  \varphi_{-1,\tfrac{1}{2}}(a\eq+b\et + m_\alpha) \ .
\end{equation}
To see that $\mc D_{G,k}^{\text{red}}$ is actually smaller than
$\mc D_{G,k}$, we spell out explicitly the components of
$\tilde{\mc D}_{G,k}, \tilde{\mc D}_{G,k}^{\text{red}}$ for some small
values of $k$. To be concrete, we take the model of $\fn=5$ with
$G = F_4$.  On the one hand,
\begin{align}
  \tilde{\mc D}_1 =
  \prod_{\alpha\in\Delta}
  &\varphi_{-1,\tfrac{1}{2}}(\eq+\et+m_\alpha) \ , \label{eq:tD1}\\
  \tilde{\mc D}_2 =
  \prod_{\alpha\in\Delta}
  &\varphi_{-1,\tfrac{1}{2}}(\eq+\et+m_\alpha)
    \varphi_{-1,\tfrac{1}{2}}(2\eq+\et+m_\alpha)
    \varphi_{-1,\tfrac{1}{2}}(\eq+2\et+m_\alpha) \ , \label{eq:tD2}\\
  \tilde{\mc D}_3 =
  \prod_{\alpha\in\Delta}
  &\varphi_{-1,\tfrac{1}{2}}(\eq+\et+m_\alpha)
    \varphi_{-1,\tfrac{1}{2}}(2\eq+\et+m_\alpha)
    \varphi_{-1,\tfrac{1}{2}}(\eq+2\et+m_\alpha)\nn
  &\varphi_{-1,\tfrac{1}{2}}(3\eq+\et+m_\alpha)
    \varphi_{-1,\tfrac{1}{2}}(2\eq+2\et+m_\alpha)
    \varphi_{-1,\tfrac{1}{2}}(\eq+3\et+m_\alpha) \ . \label{eq:tD3}
\end{align}
On the other hand,
\begin{align}
  \tilde{\mc D}_1^{\text{red}} = \prod_{\alpha\in\Delta^l}
  &\varphi_{-1,\tfrac{1}{2}}(\eq+\et+m_\alpha) \ , \label{eq:tDr1}\\
  \tilde{\mc D}_2^{\text{red}} =\prod_{\alpha\in\Delta^s}
  &\varphi_{-1,\tfrac{1}{2}}(\eq+\et+m_\alpha) \times\nn
    \prod_{\alpha\in\Delta^l}
  &\varphi_{-1,\tfrac{1}{2}}(\eq+\et+m_\alpha)
    \varphi_{-1,\tfrac{1}{2}}(2\eq+\et+m_\alpha)
    \varphi_{-1,\tfrac{1}{2}}(\eq+2\et+m_\alpha) \ , \label{eq:tDr2}\\
  \tilde{\mc D}_3^{\text{red}} =\prod_{\alpha\in\Delta^s}
  &\varphi_{-1,\tfrac{1}{2}}(\eq+\et+m_\alpha) \times\nn
    \prod_{\alpha\in\Delta^l}
  &\varphi_{-1,\tfrac{1}{2}}(\eq+\et+m_\alpha)
    \varphi_{-1,\tfrac{1}{2}}(2\eq+\et+m_\alpha)
    \varphi_{-1,\tfrac{1}{2}}(\eq+2\et+m_\alpha)\nn
  &\varphi_{-1,\tfrac{1}{2}}(3\eq+\et+m_\alpha)
    \varphi_{-1,\tfrac{1}{2}}(\eq+3\et+m_\alpha) \ . \label{eq:tDr3}
\end{align}
We will demonstrate that our recursion formulas \eqref{recursionZd}
are consistent with \eqref{eq:Drk-0} and \eqref{eq:Drk-1} rather than
\eqref{eq:Dk-0},\eqref{eq:Dk-1}.

Let us first take a look at the case of $k=1$ where the recursion
formulas \eqref{recursionZd} simply read
\begin{equation}\label{eq:E1}
  \IE_1 = \sum_{||\alpha^\vee||^2= 2} (-1)^{|\alpha^\vee}
  \frac{D^{\alpha^\vee}_{\{1,0,0\}}}{D_1} A_{\alpha^\vee} \ ,
\end{equation}
We consider the poles contributed by each component.  Suppose we
choose three unity $\nound{r}$ fields with $a_{1,2,3}$, which differ from
each other by $a_i-a_j=s_{ij}/\fn$. According to the requirement for
$a_{1,2,3}$, we know that all $s_{ij}$ are intergers and
$0< |s_{ij}|< n$. Using (\ref{addformulas}), it is not difficult to
show that both $D^{\alpha^\vee}_{1,0,0}$ and $D_1$ contain the zero
$\eq-\et=0$ of order $\mathrm{min}( |s_{ij}|, n- |s_{ij}|)$. For
example, the minor
\begin{equation}
  \Delta_{1,2}=\det\left(
\begin{array}{cc}
  \theta^{[a_1]}_3(n\tau,-2\eq+(n-2)\et)
  &\ \theta^{[a_1]}_3(n\tau,(n-2)\eq-2\et)  \\
  \theta^{[a_2]}_3(n\tau,-2\eq+(n-2)\et)
  & \theta^{[a_2]}_3(n\tau,(n-2)\eq-2\et) \\
\end{array}
\right),
\end{equation}
can be rewritten as
\begin{equation}
  \Delta_{1,2}=
  \sum_{i=2,3}\theta^{[\frac{a_1-a_2}{2}]}_i(2n\tau,n(\et-\eq))-
  \theta^{[\frac{a_1-a_2}{2}]}_i(2n\tau,n(\eq-\et)).
\end{equation}
which clearly contain zeros $\eq-\et=0$ of order
$\mathrm{min}(|s_{ij}|, n- |s_{ij}|)$. Now by the universality
argument in section~\ref{sc:unv}, we can choose arbitrary three
$a_{1,2,3}$ in recursion formulas. Let us choose three successive ones
with $a_3 -a_2 = a_2 -a_1 = 1/\fn$. Both $D^{\alpha^\vee}_{1,0,0}$ and
$D_1$ have the simple zero $\eq-\et=0$. In fact, more is true, the
determinant
\begin{equation}
  D(z_1,z_2,z_3) =\det
  \begin{pmatrix}
    \theta_i^{[a_1]}(\fn \tau, \fn z_1) &
    \theta_i^{[a_1]}(\fn \tau, \fn z_2) &
    \theta_i^{[a_1]}(\fn \tau, \fn z_3) \\
    \theta_i^{[a_2]}(\fn \tau, \fn z_1) &
    \theta_i^{[a_2]}(\fn \tau, \fn z_2) &
    \theta_i^{[a_2]}(\fn \tau, \fn z_3) \\
    \theta_i^{[a_3]}(\fn \tau, \fn z_1) &
    \theta_i^{[a_3]}(\fn \tau, \fn z_2) &
    \theta_i^{[a_3]}(\fn \tau, \fn z_3) \\
  \end{pmatrix}
\end{equation}
with characteristics $a_{1,2,3}$ chosen as above and $i=3,4$ always
has simple zeros at $z_1 - z_2 = z_2 - z_3 = z_3-z_1 = 0$. Therefore
$D^{\alpha^\vee}_{1,0,0}/D_1$ has zeros/poles
\begin{equation}
  \frac{(\eq-\et)(m_\alpha-\eq)(m_\alpha-\et)}{(\eq-\et)\eq \et} \ ,
\end{equation}
which can be boosted to the modular object
\begin{equation}
  \frac{\theta_1(m_\alpha-\eq)\theta_1(m_\alpha-\et)}
  {\theta_1(\eq)\theta_1(\et)} \ .
\end{equation}

The other component in \eqref{eq:E1} for $\IE_1$ is
\begin{equation}
  A_{\alpha^\vee}(\nound{m}y) =
  \frac{\eta^4}{\theta_1(m_\alpha)\theta_1(m_\alpha- \eq)
    \theta_1(m_\alpha- \et)
  \theta_1(m_\alpha- \eq- \et)}
  \prod_{\substack{\beta\in\Delta\\\beta\cdot\alpha^\vee=1}}
  \frac{\eta}{\theta_1(m_\beta)} \ ,
\end{equation}
where $\alpha = \alpha^\vee\cdot 2/||\alpha^\vee||^2$ in the root
associated to the coroot $\alpha^\vee$.  The components
$\theta_1(m_\alpha-\eq)\theta_1(m_\alpha-\et)$ cancel with
corresponding components in the numerator of
$D_{1,0,0}^{\alpha^\vee}/D_1$. Thus naively $A_{\alpha^\vee}$ should
contribute poles\footnote{There are also poles at positions shifted by
  $1$ or $\tau$ of course.}  $m_\beta = 0$ for $\beta\in\Delta$ and
$m_\alpha-\eq-\et = 0$ for $\alpha\in\Delta^l$ to $\IE_1$. The former
poles, however, get canceled with some factors in the numerator after
performing the summation over short coroots in \eqref{eq:E1}, which
can either be seen in explicit calculations\footnote{This cancellation
  can be checked in all the minimal models, not only in the $\fn=5$
  model.}, or be argued formally from the Hilbert series of gauge
instanton moduli space \cite{Cremonesi:2014xha} as well as from
W-algebra via AGT correspondence \cite{Keller:2011ek}. The remaining
genuine poles are $m_\alpha-\eq-\et = 0$ for $\alpha\in\Delta^l$,
which are consistent with the slimmer expression \eqref{eq:tDr1}
rather than \eqref{eq:tD1}.

The discussion above indicates that the recursion formulas
\eqref{recursionZd} for $\IE_k$ naively contain both genuine poles and
spurious poles, while the cancellation of the latter is not
obvious. Nevertheless, all true poles should already be visible in the
recursion formulas, which allows us to distinguish \eqref{eq:Drk-0}
from \eqref{eq:Dk-0}. In the following, we will argue in favor of
\eqref{eq:Drk-0} by pointing out that extra poles indicated in
\eqref{eq:Dk-0} are not present in the recursion formulas
\eqref{recursionZd}, and assume along the way that the spurious poles
not consistent with either \eqref{eq:Dk-0} or \eqref{eq:Drk-0} cancel
automatically. With this comment in mind, let us now look at the
denominators of $\IE_2,\IE_3$ from the recursion formulas, and
consider only \emph{relevant} poles, the poles which appear also in
either \eqref{eq:Dk-0} or \eqref{eq:Drk-0}.

The right hand side of \eqref{recursionZd} for $\IE_2$ has three
sectors with
\begin{equation}
  (||\alpha^\vee||^2,d_1,d_2) = (4,0,0), (2,1,0)/(2,0,1), (0,1,1)
\end{equation}
respectively. With an explicit calculation in the $\fn=5$ models, we
find the following relevant poles in each of the three sectors
\footnote{We also suppress the poles $\eq=\et=0$, which are guaranteed
  by the $D^{\alpha^\vee}_{\{d_0,d_1,d_2\}}/D_d$ structure similar to
  the previous discussion.}
\begin{center}
  \begin{tabular}{cp{25em}}\toprule
    $(||\alpha^\vee||^2,d_1,d_2)$
    & relevant poles\\\midrule
    (4,0,0)
    & $m_\alpha-\eq-\et,\; \alpha\in\Delta$ \\
    (2,1,0)/(2,0,1)
    &$m_\alpha-\eq-\et,m_\alpha-2\eq-\et,m_\alpha-\eq-2\et,\;\alpha\in\Delta^l$\\
    (0,1,1)
    & $-$ \\\bottomrule
  \end{tabular}
\end{center}
They are clearly consistent with \eqref{eq:tDr2}. Likewise, for
$\IE_3$ there are five sectors on the right hand sie of
\eqref{recursionZd}, and we find the relevant poles as follows
\begin{center}
  \begin{tabular}{cp{25em}}\toprule
    $(||\alpha^\vee||^2,d_1,d_2)$
    & relevant poles\\\midrule
    (6,0,0)
    & $m_\alpha-\eq-\et,\; \alpha\in\Delta$;\\
    & $m_\alpha-2\eq-\et,m_\alpha-\eq-2\et,\;\alpha\in\Delta^l$\\
    (4,1,0)/(4,0,1)
    &$m_\alpha-\eq-\et,m_\alpha-2\eq-\et,m_\alpha-\eq-2\et,\;\alpha\in\Delta^l$\\
    (2,2,0)/(2,0,2)
    & $m_\alpha-\eq-\et,\; \alpha\in\Delta;$\\
    & $m_\alpha-2\eq-\et,m_\alpha-\eq-2\et,$ \\
    & $m_\alpha-3\eq-\et,m_\alpha-\eq-3\et,\;\alpha\in\Delta^l$\\
    (2,1,1)
    & $m_\alpha-\eq-\et,m_\alpha-2\eq-\et,m_\alpha-\eq-2\et,\;\alpha\in\Delta^l$\\
    (0,2,1)
    & $m_\alpha-\eq-\et,\;\alpha\in\Delta^l$ \\\bottomrule
  \end{tabular}
\end{center}
which are consistent with \eqref{eq:tDr3}.

\section{Elliptic genera}
\label{ap:eg}
We record some high $\Qtau$ order results for the reduced elliptic genus here. Recall the $k$-string elliptic genus when expanded with respect to $\Qtau$
can be written as
\begin{equation*}
  \ehgk(v,x,\Qtau,Q_{m_i})=
  v^{k\dualCox-1}\Qtau^{-(k\dualCox-1)/6}
  \sum_{n=0}^{\infty}\Qtau^n\,  g^{(n)}_{k,G}(v,x,Q_{m_i}).
\end{equation*}
We are interested in the $v$-expansion of $g^{(n)}_{k,G}(v,x,Q_{m_i})$. Usually, the leading $v$ power becomes more and more negative when $\Qtau$ order
$n$ goes up. When $n$ is enough high, we observe some patterns for the leading $v$-expansion behavior. For two-strings elliptic genus, we observe for
$i\ge {2(\fn-2)}$,
\be
 g^{(2)}_{i,G}(v,x,Q_{m_i})=-v^{-3(\fn+i)-5}\big(\chi_{(i-2(\fn-2))\theta}\chi_{i-(\fn-2)}\big)-\mathcal{O}(v^{-3(\fn+i)-4})\dots
 \ .
\ee
As high as the $\Qtau$ order we have reached, this is true. It is nice to see how to explain this phenomenon.
\subsection*{$\bf A_2$}
For the reduced two-string elliptic genus for $G=SU(3)$ model, recall
\begin{equation*}
  g^{(n)}_{2,A_2}(v,x=1,Q_{m_i}=1)
  =\frac{1}{(1-v)^{10} (1+v)^6 \left(1+v+v^2\right)^5}\times P^{(n)}_{2,A_2}(v).
\end{equation*}
We have
{\footnotesize{
\begin{equation*}
\ba
P^{(2)}_{2,A_2}(v)&=v^{-10} \big(-1 - v + 6 v^2 + 9 v^3 - 10 v^4 - 33
v^5 + 41 v^6 + 256 v^7 +
 428 v^8 + 220 v^9 - 347 v^{10} - 823 v^{11} - 131 v^{12}\\& + 2652 v^{13} +
 7721 v^{14} + 14419 v^{15} + 21826 v^{16} + 27125 v^{17} + 28966 v^{18} +
 27125 v^{19}+\dots+v^{36}\big),\\
 \ea
\end{equation*}
\begin{equation*}
\ba
P^{(3)}_{2,A_2}(v)&=v^{-13} \big(-16-28 v+50 v^2+198 v^3+138 v^4-399
v^5-963 v^6-419 v^7+1716 v^8+4316 v^9+5014 v^{10}\\& +2174 v^{11}-4110
v^{12}-10701 v^{13}-12583 v^{14}-2128 v^{15}+27073 v^{16}+75426
v^{17}+136089 v^{18}+198723 v^{19}\\& +244336 v^{20}+260628
v^{21}+244336 v^{22}+\dots+v^{42}\big),\\
\ea
\end{equation*}
\begin{equation*}
\ba
P^{(4)}_{2,A_2}(v)&=v^{-16} \big(-81 - 223 v - 5 v^{2 }+ 880 v^{3 }+
1695 v^{4 }+ 577 v^{5 }- 3110 v^{6 }-
 6735 v^{7 }- 5803 v^{8 }+ 2582 v^{9 }+ 17019 v^{10 }\\& + 31735 v^{11 }+
 38096 v^{12 }+ 25783 v^{13 }- 11462 v^{14 }- 68637 v^{15 }- 118109 v^{16 }-
 116300 v^{17 }- 11102 v^{18 }\\& + 231810 v^{19 }+ 605425 v^{20 }+
 1054375 v^{21 }+ 1497688 v^{22 }+ 1816145 v^{23 }+ 1930514 v^{24 }+
 \dots+v^{48}\big),\\
\ea
\end{equation*}
\begin{equation*}
\ba
P^{(5)}_{2,A_2}(v)&=v^{-19} \big(-256 -874 v -810  v^{2 }+1432  v^{3
}+5510  v^{4 }+7553  v^{5 }+2483  v^{6 }-10671  v^{7 }-26644  v^{8
}-34874  v^{9 }\\&-23600  v^{10 }+16620  v^{11 }+90316  v^{12 }+185964
v^{13 }+260250  v^{14 }+241846  v^{15 }+62988  v^{16 }-279728  v^{17
}\\&-692866  v^{18 }-963287  v^{19 }-815649  v^{20 }+30457  v^{21
}+1729760  v^{22 }+4189517  v^{23 }+7044615  v^{24 }+9777494  v^{25
}\\&+11727151  v^{26 }+12428414  v^{27 }+11727151  v^{28 }+
 \dots+v^{54}\big),\\
\ldots\\
\ea
\end{equation*}
\begin{equation*}
\ba
P^{(10)}_{2,A_2}(v)(v)&=v^{-34} \big(-6561-31129 v-72335 v^{2}-120018
v^{3}-168868 v^{4}-206886 v^{5}-222250 v^{6}-170936 v^{7}+98441
v^{8}\\& +773013 v^{9}+1959712 v^{10}+3612563 v^{11}+5410138
v^{12}+6633691 v^{13}+5994478 v^{14}+1510639 v^{15}\\& -9064153
v^{16}-27314880 v^{17}-52091127 v^{18}-77137569 v^{19}-89646012
v^{20}-69861741 v^{21}+5587892 v^{22}\\& +154737705 v^{23}+377669387
v^{24}+641219584 v^{25}+868425362 v^{26}+940214412 v^{27}+718317837
v^{28}\\& +93293293 v^{29}-951170969 v^{30}-2275980208
v^{31}-3545719630 v^{32}-4250592736 v^{33}-3785597548 v^{34}\\&
-1612094548 v^{35}+2569710910 v^{36}+8659446500 v^{37}+16075828305
v^{38}+23808502207 v^{39}\\& +30625221095 v^{40}+35302136340
v^{41}+36965592032 v^{42}+35302136340 v^{43}+\dots+v^{84}\big).\\
\ea
\end{equation*}}}

\noindent Note that the leading
power of $v$ in $P^{(n)}_{2,A_2}(v)$ becomes more and more negative as
$n$ increases. In fact, we notice for $n\ge 2$,
\begin{equation}
\ba
P^{(n)}_{2,A_2}(v)=v^{-3n-4} \big(&-(n-1)^4-\dots+\text{palindrome up
  to $v^{6(n+4)}$}\big).
\ea
\end{equation}
With generic $x$, we also observe the following general expansion: for $n\ge 2$,
\begin{equation}
  \begin{aligned}
    v^{3n+4} g^{(n)}_{2,A_2}(v,x,Q_{m_i}) =
    &-\chi^{A_2}_{(n-2)\theta}\chi_{n-1}
      -\big(\chi^{A_2}_{(n-3)\theta}\chi_{n}+\dots\big)v
      -\big(\chi^{A_2}_{(n-4)\theta}\chi_{n+1}+\dots\big)v^2+\dots \ .
  \end{aligned}
\end{equation}

For the reduced elliptic genus of three-strings, recall
\begin{equation}
  g^{(n)}_{3,A_2}(v,x=1,Q_{m_i}=1)
  =\frac{1}{(1-v)^{16} (1+v)^{10} \left(1+v^2\right)^5 \left(1+v+v^2\right)^6}\times P^{(n)}_{3,A_2}(v).
\end{equation}
We obtain
{\footnotesize{
\begin{equation*}
\ba
P^{(2)}_{3,A_2}(v)&=v^{-10}(-1 - 2 v - 5 v^{2 }- 4 v^{3 }+ 11 v^{4 }+ 30 v^{5 }+ 74 v^{6 }+ 84 v^{7 }+
 267 v^{8 }+ 880 v^{9 }+ 2718 v^{10 }+ 6130 v^{11 }\\&+ 11512 v^{12 }+
 17594 v^{13 }+ 24774 v^{14 }+ 35306 v^{15 }+ 64404 v^{16 }+ 133960 v^{17 }+
 290609 v^{18 }+ 569846 v^{19 }\\&+ 1020364 v^{20 }+ 1628376 v^{21 }+
 2376984 v^{22 }+ 3145582 v^{23 }
 + 3848955 v^{24 }+ 4318298 v^{25 }+
 4501676 v^{26 }\\&+ 4318298 v^{27}+\dots+v^{52}),\\
 \ea
\end{equation*}
\begin{equation*}
\ba
P^{(3)}_{3,A_2}(v)&=v^{-17}(2 - 10 v^{2 }- 9 v^{3 }+ 12 v^{4 }+ 44 v^{5 }- 48 v^{6 }- 367 v^{7 }- 622 v^{8 }-
 183 v^{9 }+ 1740 v^{10 }+ 4336 v^{11 }+ 4488 v^{12 }\\&- 1116 v^{13 }-
 12504 v^{14 }- 17424 v^{15 }+ 4576 v^{16 }+ 77007 v^{17 }+ 199630 v^{18 }+
 337366 v^{19 }+ 417362 v^{20 }\\&+ 382080 v^{21 }+ 255946 v^{22 }+
 265962 v^{23 }+ 839446 v^{24 }+ 2662413 v^{25 }+ 6368124 v^{26 }+
 12524957 v^{27 }\\&+ 21034734 v^{28 }+ 31405824 v^{29 }+ 42159674 v^{30 }+
 51803451 v^{31 }+ 58351834 v^{32 }+ 60785174 v^{33 }\\&+ 58351834 v^{34}+\ldots+v^{66}),\\
 \ea
\end{equation*}
\begin{equation*}
\ba
P^{(4)}_{3,A_2}(v)&=v^{-21}(32 + 40 v - 84 v^{2 }- 339 v^{3 }- 426 v^{4 }+ 487 v^{5 }+ 2242 v^{6 }+
 2876 v^{7 }- 1500 v^{8 }- 12817 v^{9 }- 23440 v^{10 }\\&- 16831 v^{11 }+
 25630 v^{12 }+ 97493 v^{13 }+ 150494 v^{14 }+ 103012 v^{15 }- 112582 v^{16 }-
 448583 v^{17 }- 700938 v^{18 }\\&- 495362 v^{19 }+ 512476 v^{20 }+
 2411835 v^{21 }+ 4741886 v^{22 }+ 6461123 v^{23 }+ 6254586 v^{24 }+
 3310035 v^{25 }\\&- 1672684 v^{26 }- 5447913 v^{27 }- 2195270 v^{28 }+
 16177917 v^{29 }+ 57171170 v^{30 }+ 126228382 v^{31 }+ 222042686 v^{32 }\\&+
 337620393 v^{33 }+ 457167512 v^{34 }+ 563129124 v^{35 }+ 635332178 v^{36 }+
 661625872 v^{37 }+ 635332178 v^{38}+\dots +v^{74}),\\
 \ea
\end{equation*}
\begin{equation*}
\ba
P^{(5)}_{3,A_2}(v)&=v^{-25}(162 + 412 v + 276 v^{2 }- 1295 v^{3 }- 4518 v^{4 }- 5771 v^{5 }+ 1292 v^{6 }+
 21793 v^{7 }+ 44712 v^{8 }+ 38030 v^{9 }\\&- 38688 v^{10 }- 192767 v^{11 }-
 341004 v^{12 }- 309898 v^{13 }+ 101910 v^{14 }+ 927411 v^{15 }+
 1862626 v^{16 }+ 2189163 v^{17 }\\&+ 1001340 v^{18 }- 2210183 v^{19 }-
 6887108 v^{20 }- 10727974 v^{21 }- 10030208 v^{22 }- 703145 v^{23 }+
 19133352 v^{24 }\\&+ 46719280 v^{25 }+ 72876216 v^{26 }+ 83251991 v^{27 }+
 63444248 v^{28 }+ 7529926 v^{29 }- 72005488 v^{30 }- 136950773 v^{31 }\\&-
 124639134 v^{32 }+ 43752233 v^{33 }+ 438669558 v^{34 }+ 1103016208 v^{35 }+
 2019390168 v^{36 }+ 3111882801 v^{37 }\\&+ 4234930428 v^{38 }+
 5220665928 v^{39 }+ 5892000612 v^{40 }+ 6133530828 v^{41 }+ 5892000612 v^{42}+\dots+v^{82}),\\
 \ea
\end{equation*}
\begin{equation*}
\ba
P^{(6)}_{3,A_2}(v)&=v^{-29}(512 + 1690 v + 3044 v^{2 }+ 766 v^{3 }- 10596 v^{4 }- 31191 v^{5 }-
 46930 v^{6 }- 19288 v^{7 }+ 89188 v^{8 }+ 274511 v^{9 }\\&+ 419176 v^{10 }+
 288140 v^{11 }- 374722 v^{12 }- 1614464 v^{13 }- 2963132 v^{14 }-
 3347473 v^{15 }- 1280918 v^{16 }\\&+ 4247141 v^{17 }+ 12688484 v^{18 }+
 20686636 v^{19 }+ 22001562 v^{20 }+ 9184210 v^{21 }- 22497646 v^{22 }-
 69062433 v^{23 }\\&- 113553734 v^{24 }- 124631077 v^{25 }- 65395112 v^{26 }+
 91413273 v^{27 }+ 339731228 v^{28 }+ 622477442 v^{29 }\\&+ 828007584 v^{30 }+
 813952151 v^{31 }+ 463211226 v^{32 }- 241280313 v^{33 }- 1140297436 v^{34 }-
 1857927117 v^{35 }\\&- 1829759686 v^{36 }- 398271211 v^{37 }+
 2994640544 v^{38 }+ 8636915664 v^{39 }+ 16332876522 v^{40 }+
 25397471432 v^{41 }\\&+ 34642293136 v^{42 }+ 42690290417 v^{43 }+
 48157593258 v^{44 }+ 50110832268 v^{45 }+ 48157593258 v^{46}+\dots+v^{90}).\\
 \ea
\end{equation*}
}}
For $n\ge 3$, we notice the following universal leading behavior
\be
g^{(3)}_{n,A_2}(v,x)=v^{-4n-5}(\chi_{2(n-2)}\chi_{(n-3)\theta}+\mathcal{O}(v))
\ .
\ee

\subsection*{$\bf D_4$}
For the reduced two-string elliptic genus of $SO(8)$ 6d SCFT, recall
\begin{equation*}
  g^{(n)}_{2,D_4}(v,x=1,Q_{m_i}=1)
  =\frac{1}{(1-v)^{22} (1+v)^{12} \left(1+v+v^2\right)^{11}}\times
  P^{(n)}_{2,D_4}(v) \ .
\end{equation*}
We have
{\footnotesize{
\begin{equation*}
\ba
P^{(2)}_{2,D_4}(v)&=-v^{-8} (1+3 v+22 v^{2}+47 v^{3}+108 v^{4}+29 v^{5}-184 v^{6}-861 v^{7}-1762 v^{8}-3305 v^{9}-13965 v^{10}-65210 v^{11}\\&-260932
v^{12}-884324 v^{13}-2589008 v^{14}-6645978 v^{15}-15280924 v^{16}-31673046 v^{17}-59506626 v^{18}\\&-101871752 v^{19}-159566189 v^{20}-229273231
v^{21}-303095099 v^{22}-369460130 v^{23}-415787266 v^{24}\\&-432409780 v^{25}-415787266 v^{26}+\dots+v^{50}),\\
 \ea
\end{equation*}
\begin{equation*}
\ba
  P^{(3)}_{2,D_4}(v)&=v^{-11} (1+v^{2}) (2+3 v-23 v^{2}-281 v^{3}-1338 v^{4}-3786 v^{5}-5794 v^{6}-384 v^{7}+22410 v^{8}+57106 v^{9}\\&+65200 v^{10}-8878
  v^{11}-148110 v^{12}-82859 v^{13}+911821 v^{14}+4473554 v^{15}+14488449 v^{16}+39847612 v^{17}\\&+97908991 v^{18}+217119419 v^{19}+434067674
  v^{20}+782158341 v^{21}+1274332193 v^{22}+1885916954 v^{23}\\&+2545465687 v^{24}+3144719815 v^{25}+3566407238 v^{26}+3718703248 v^{27}+3566407238
  v^{28}+\dots+v^{54}),\\
  \ea
\end{equation*}
\begin{equation*}
\ba
P^{(4)}_{2,D_4}(v)&= -v^{-19} (2+2 v-22 v^{2}-42 v^{3}+88 v^{4}+329 v^{5}-11 v^{6}-1411 v^{7}-1930 v^{8}+1877 v^{9}+12149 v^{10}\\&+38843 v^{11}+113904
v^{12}+240220 v^{13}+226932 v^{14}-364906 v^{15}-1770622 v^{16}-3067485 v^{17}-1974439 v^{18}\\&+3262797 v^{19}+10272490 v^{20}+10304550 v^{21}-10767498
v^{22}-74227039 v^{23}-228511951 v^{24}\\&-601571798 v^{25}-1479273049 v^{26}-3376185023 v^{27}-7028420238 v^{28}-13251946148 v^{29}\\&-22656311766
v^{30}-35281600040 v^{31}-50298462814 v^{32}-65951082224 v^{33}-79826800694 v^{34}\\&-89424037262 v^{35}-92856436862 v^{36}-89424037262
v^{37}+\dots+v^{72}),\\
\ea
\end{equation*}
\begin{equation*}
\ba
P^{(5)}_{2,D_4}(v)&=-v^{-22} (1+v^{2}) (84+146 v-831 v^{2}-2495 v^{3}+2027 v^{4}+16874 v^{5}+12844 v^{6}-55230 v^{7}-118023 v^{8}\\&+45251 v^{9}+412658
v^{10}+306961 v^{11}-747606 v^{12}-1085075 v^{13}+2514527 v^{14}+8955862 v^{15}+7677132 v^{16}\\&-13638464 v^{17}-48153754 v^{18}-58843606 v^{19}-3166625
v^{20}+114419603 v^{21}+211444789 v^{22}\\&+164548302 v^{23}-98377407 v^{24}-542911427 v^{25}-1148224939 v^{26}-2306012154 v^{27}-5445277331
v^{28}\\&-13585439053 v^{29}-31372682848 v^{30}-64179900027 v^{31}-116221363135 v^{32}-188200853472 v^{33}\\&-275432340276 v^{34}-367411722852
v^{35}-449506711376 v^{36}-506517773112 v^{37}-526950799964 v^{38}\\&-506517773112 v^{39}+\dots+v^{76}), \\
 \ea
\end{equation*}
\begin{equation*}
\ba
 P^{(6)}_{2,D_4}(v)&=-v^{-25} (1200+3374 v-6724 v^{2}-40296 v^{3}-31436 v^{4}+159328 v^{5}+410682 v^{6}+15592 v^{7}-1467224 v^{8}\\&-2279805 v^{9}+1088019
 v^{10}+8007211 v^{11}+8314028 v^{12}-9090084 v^{13}-32918254 v^{14}-19726713 v^{15}\\&+67719233 v^{16}+186559606 v^{17}+183703487 v^{18}-107579194
 v^{19}-661966959 v^{20}-1107115384 v^{21}\\&-832956121 v^{22}+557841530 v^{23}+2672286699 v^{24}+4109758618 v^{25}+3046056345 v^{26}-1483241349
 v^{27}\\&-8623040702 v^{28}-16826056321 v^{29}-28634625983 v^{30}-59027415291 v^{31}-144463390006 v^{32}\\&-347331281609 v^{33}-749797256633
 v^{34}-1433909622002 v^{35}-2450540661913 v^{36}-3786292890317 v^{37}\\&-5341633935638 v^{38}-6932828077068 v^{39}-8322655072044 v^{40}-9274316155458
 v^{41}-9612992092064 v^{42}\\&-9274316155458 v^{43}+\dots+v^{84}). \\
\ea
\end{equation*}}}
For $n\ge 4$, we observe the following general leading order behavior
\begin{equation}
  v^{3n+7} g^{(n)}_{2,D_4}(v,x,Q_{m_i})=-\chi^{D_4}_{(n-4)\theta}\chi_{n-2}(x)
  +\cO(v).
\end{equation}

\subsection*{$\bf F_4$}

For the reduced two-string elliptic genus of $F_4$ 6d SCFT, recall

\begin{equation}
  g^{(n)}_{2,F_4} (v,x=1,Q_{m_i}=1)
  = \frac{1}{(1-v)^{34} (1+v)^{22} \left(1+v+v^2\right)^{17}}\times
  P^{(n)}_{2,F_4}(v)~.
\end{equation}
We have
{\footnotesize{
\begin{equation*}
\ba
P^{(2)}_{2,F_4}(v)&=\, 1653+14307 v+118305 v^{2}+770928 v^{3}+4293754 v^{4}+20938534 v^{5}+90874761 v^{6} +354378137 v^{7}\\&+1254181600 v^{8}+4057153817
v^{9}+12068000241 v^{10}+33173318084 v^{11} +84638789902 v^{12}+201171570880 v^{13}\\&+446852450528 v^{14}+930177353330 v^{15} +1818950197662
v^{16}+3348446417048 v^{17}+5813458948881 v^{18}\\&+9534395062259 v^{19} +14791974953829 v^{20}+21734938030680 v^{21}+30278882934513 v^{22}+40026745086453
v^{23}\\& +50246339190488 v^{24}+59931545438647 v^{25}+67951300040637 v^{26}+73260270949890 v^{27}\\& +75118982210308 v^{28}+73260270949890
v^{29}+\dots+v^{56},\\
\ea
\end{equation*}
\begin{equation*}
\ba
P^{(3)}_{2,F_4}(v)&=\,v^{-12} (1+7 v+58 v^{2}+277 v^{3}+1071 v^{4}+2976 v^{5}+5964 v^{6}+5832 v^{7}-9266 v^{8}-57418 v^{9} -135552 v^{10}\\&-169110
v^{11}+55432 v^{12}+957484 v^{13}+4373263 v^{14}+21374265 v^{15}+111764209 v^{16} +546538072 v^{17}\\&+2392703794 v^{18}+9374157248 v^{19}+33193123730
v^{20}+107176864396 v^{21} +317784470985 v^{22}\\&+870042414425 v^{23}+2209544249477 v^{24}+5224968453408 v^{25} +11543089455336 v^{26}+23893104753132
v^{27}\\&+46455227531026 v^{28}+85029824340612 v^{29} +146800695692191 v^{30}+239466759024769 v^{31}\\&+369631201685833 v^{32}+540578567158590 v^{33}
+749891181508407 v^{34}+987634558444181 v^{35}\\&+1235929372818009 v^{36}+1470501793161916 v^{37} +1664270696390466 v^{38}+1792334859311106
v^{39}\\&+1837134386523548 v^{40}+1792334859311106 v^{41}
+\dots+v^{80})\ ,\\
\ea
\end{equation*}
\begin{equation*}
\ba
P^{(4)}_{2,F_4}(v)&=\,-v^{-15} (2+11 v-9 v^{2}-729 v^{3}-6740 v^{4}-37612 v^{5}-146242 v^{6}-411322 v^{7}-786424 v^{8} -683123 v^{9}\\&+1519991
v^{10}+7863641 v^{11}+17299644 v^{12}+19396808 v^{13}-7574716 v^{14} -81039308 v^{15}-177701238 v^{16}\\&-228893241 v^{17}-328644969 v^{18}-1704624732
v^{19} -10239254139 v^{20}-48437902734 v^{21}-193165328741 v^{22}\\&-683657580163 v^{23} -2198459381202 v^{24}-6491341536008 v^{25}-17700027945928
v^{26}-44757889196479 v^{27} \\&-105345430184255 v^{28}-231557373968430 v^{29}-476759190688969 v^{30}-921916213048446 v^{31}\\& -1678228390978831
v^{32}-2881832101851776 v^{33}-4676542813241321 v^{34}-7182956156002759 v^{35}\\& -10456822688468996 v^{36}-14445439874041454 v^{37}-18955600691329352
v^{38} -23647746652092629 v^{39}\\&-28066464498494385 v^{40}-31707630617443838 v^{41} -34110133094328607 v^{42}-34949875241183086
v^{43}\\&-34110133094328607 v^{44}+\dots+v^{86})\ ,\\
\ea
\end{equation*}
\begin{equation*}
\ba
P^{(5)}_{2,F_4}(v)&=\,v^{-18} (1+v-187 v^{2}-1942 v^{3}-8588 v^{4}+388 v^{5}+256583 v^{6}+1875127 v^{7}+8097680 v^{8}+23995511 v^{9}\\&+47743204
v^{10}+45403031 v^{11}-77639509 v^{12}-441768092 v^{13}-990656365 v^{14}-1141697391 v^{15}\\&+314912356 v^{16}+4322862537 v^{17}+9413229443
v^{18}+9932344114 v^{19}-1369097445 v^{20}-22028614537 v^{21}\\&-15689824140 v^{22}+143994570705 v^{23}+853983234794 v^{24}+3391440788051
v^{25}+11730377038184 v^{26}\\&+37232002426385 v^{27}+109311773773871 v^{28}+297084435895112 v^{29}+748700764659463 v^{30}\\&+1754971776388633
v^{31}+3839341467548316 v^{32}+7864675977003847 v^{33}+15128265661342010 v^{34}\\&+27394204368470677 v^{35}+46797420093685421 v^{36}+75560202097716654
v^{37}+115501696318370322 v^{38}\\&+167393226212698984 v^{39}+230299469534789175 v^{40}+301112810847585121 v^{41}+374497086121044331
v^{42}\\&+443382800033202476 v^{43}+500006416441373208 v^{44}+537304885313273980 v^{45}+550330680323572096 v^{46}\\&+537304885313273980
v^{47}+\dots+v^{92}).
\ea
\end{equation*}}}
For $n\ge 6$, we observe the following general leading order behavior
\begin{equation}
  v^{3n+10} g^{(n)}_{2,F_4}(v,x,Q_{m_i})=-\chi^{F_4}_{(n-6)\theta}\chi_{n-3}(x)
  +\cO(v).
\end{equation}
\subsection*{$\bf E_6$}
For the reduced two-string elliptic genus of $E_6$ 6d SCFT, recall
\begin{equation*}
  g^{(n)}_{2,E_6} (v,x,Q_{m_i}=1)
  = \frac{1}{(1-v)^{46} (1+v)^{32} \left(1+v+v^2\right)^{23}}\times P^{(n)}_{2,E_6}(v)~.
\end{equation*}
We have
{\footnotesize
\begin{equation*}
\ba
P^{(1)}_{2,E_6}(v) &= (1+v^2)(82 + 896 v + 9129 v^{2 }+ 73825 v^{3 }+ 515477 v^{4 }+ 3176394 v^{5 }+
  17567385 v^{6 }+ 88082527 v^{7 }\\&+ 404122599 v^{8 }+ 1707996910 v^{9 }+
  6687039606 v^{10 }+ 24365673656 v^{11 }+ 82957003626 v^{12 }\\&+
  264812209428 v^{13 }+ 794925309293 v^{14 }+ 2249848989493 v^{15 }+
  6017588149603 v^{16 }+ 15241390482586 v^{17 }\\&+ 36623148751459 v^{18 }+
  83623554563863 v^{19 }+ 181712020504595 v^{20 }+ 376267731853770 v^{21 }\\&+
  743340720549339 v^{22 }+ 1402570753853399 v^{23 }+
  2530053857442778 v^{24 }+ 4367001323365453 v^{25 }\\&+
  7218179887542376 v^{26 }+ 11433257908228549 v^{27 }+
  17365401325615558 v^{28 }+ 25305594210396759 v^{29 }\\&+
  35398201343930359 v^{30 }+ 47551931562200552 v^{31 }+
  61367940071565626 v^{32 }\\&+ 76109936363599780 v^{33 }+
  90737018750916024 v^{34 }+ 104007721490984500 v^{35 }\\&+
  114645634265369518 v^{36 }+ 121537998101131452 v^{37 }+
  123925354694394472 v^{38 }+
  \dots+v^{76}).
\ea
\end{equation*}
\begin{equation*}
\ba
P^{(2)}_{2,E_6}(v) &= 3486 + 44488 v + 464913 v^{2 }+ 3873323 v^{3 }+ 27606333 v^{4 }+
 172731506 v^{5 }+ 966630727 v^{6 }+ 4894721995 v^{7 }\\&+ 22642609833 v^{8 }+
 96385144324 v^{9 }+ 379809898432 v^{10 }+ 1392335683050 v^{11 }+
 4768406721146 v^{12 }\\&+ 15311706805952 v^{13 }+ 46244599549903 v^{14 }+
 131729726893973 v^{15 }+ 354773291170080 v^{16 }\\&+ 905322588772629 v^{17 }+
 2193217132886674 v^{18 }+ 5052883393549785 v^{19 }+
 11088020153427871 v^{20 }\\&+ 23207949295452654 v^{21 }+
 46391596457503471 v^{22 }+ 88666687652950697 v^{23 }+
 162200225646883046 v^{24 }\\&+ 284262935556020849 v^{25 }+
 477678248920949928 v^{26 }+ 770246305708025175 v^{27 }\\&+
 1192619227946678339 v^{28 }+ 1774271549538256254 v^{29 }+
 2537602212267590587 v^{30 }\\&+ 3490792045588793343 v^{31 }+
 4620696016056616197 v^{32 }+ 5887558970641741644 v^{33 }\\&+
 7223464858978994614 v^{34 }+ 8535984817703595260 v^{35 }+
 9717463741648588196 v^{36 }\\&+ 10658977913348459838 v^{37 }+
 11266596547576742504 v^{38 }+ 11476648364287371362 v^{39 }\\&+
 11266596547576742504 v^{40 }+
  \dots+v^{78}.
\ea
\end{equation*}
\begin{equation*}
\ba
P^{(3)}_{2,E_6}(v) &=v^{-4}(-1 - 11 v - 112 v^{2 }- 769 v^{3 }+ 97959 v^{4 }+ 1465146 v^{5 }+
 15949836 v^{6 }+ 136827854 v^{7 }+ 992270036 v^{8 }\\&+ 6276463714 v^{9 }+
 35347238184 v^{10 }+ 179602878296 v^{11 }+ 831787432544 v^{12 }+
 3538777989264 v^{13 }\\&+ 13918341585911 v^{14 }+ 50873087148945 v^{15 }+
 173571288630315 v^{16 }+ 554884515982156 v^{17 }\\&+
 1667591219224745 v^{18 }+ 4724849885190467 v^{19 }+
 12653182331604747 v^{20 }+ 32099832942977106 v^{21 }\\&+
 77298235563217848 v^{22 }+ 177003542946746952 v^{23 }+
 386049290828201197 v^{24 }\\&+ 803129445851193549 v^{25 }+
 1595815215747637914 v^{26 }+ 3032155653955325045 v^{27 }\\&+
 5515192004170661557 v^{28 }+ 9612446692952872376 v^{29 }+
 16067738341715188425 v^{30 }\\&+ 25779182779041415519 v^{31 }+
 39727277739048609244 v^{32 }+ 58842809938209124305 v^{33 }\\&+
 83817103245449704330 v^{34 }+ 114875661886259702901 v^{35 }+
 151556399981844604507 v^{36 }\\&+ 192548720900049944088 v^{37 }+
 235652392733234663776 v^{38 }+ 277900332002367574858 v^{39 }\\&+
 315856987228959670753 v^{40 }+ 346060446910493617699 v^{41 }+
 365533661213463473501 v^{42 }\\&+ 372262200765577638648 v^{43}+
  \dots+v^{86}).
\ea
\end{equation*}
\begin{equation*}
\ba
P^{(4)}_{2,E_6}(v) &=v^{-16}(-1 - 11 v - 112 v^{2 }- 769 v^{3 }- 4214 v^{4 }- 18313 v^{5 }- 64197 v^{6 }-
 177594 v^{7 }- 364431 v^{8 }- 421607 v^{9 }\\&+ 420751 v^{10 }+ 3722444 v^{11 }+
 10737460 v^{12 }+ 18191652 v^{13 }+ 11206753 v^{14 }- 38487665 v^{15 }\\&-
 147892027 v^{16 }- 235567050 v^{17 }+ 204232919 v^{18 }+ 4014329887 v^{19 }+
 28747116555 v^{20 }\\&+ 177469181418 v^{21 }+ 994624050267 v^{22 }+
 5061754204737 v^{23 }+ 23475175955326 v^{24 }\\&+ 99852192764195 v^{25 }+
 392058843196059 v^{26 }+ 1428966813600884 v^{27 }+
 4857570355921361 v^{28 }\\&+ 15462381593811917 v^{29 }+
 46247090390273044 v^{30 }+ 130358984643118795 v^{31 }\\&+
 347206472133377093 v^{32 }+ 875864441137943176 v^{33 }+
 2096960972906450647 v^{34 }\\&+ 4773710145181408321 v^{35 }+
 10350471755632365352 v^{36 }+ 21407072068973867721 v^{37 }\\&+
 42290120178103689822 v^{38 }+ 79898824171145741885 v^{39 }+
 144526191934600175231 v^{40 }\\&+ 250551336753503603832 v^{41 }+
 416666541458353995971 v^{42 }+ 665245395666408218767 v^{43 }\\&+
 1020471334680763094257 v^{44 }+ 1505001425501016329934 v^{45 }+
 2135260459421424616916 v^{46 }\\&+ 2915910779649516764294 v^{47 }+
 3834519574899765264582 v^{48 }+ 4857787518331415856898 v^{49 }\\&+
 5930740363027659407709 v^{50 }+ 6979905103120498497197 v^{51 }+
 7920702637212573264754 v^{52 }\\&+ 8668249061573197611391 v^{53 }+
 9149752299658818747593 v^{54 }+ 9316044470623822160548 v^{55}+
  \dots+v^{110}).
\ea
\end{equation*}
}

\subsection*{$\bf E_7$}
For the reduced two-string elliptic genus of $E_7$ 6d SCFT, recall
\begin{equation}
  g_{2,E_7}^{(n)}(v,x,Q_{m_i}=1)
  = \frac{1}{(1-v)^{70} (1+v)^{52} \left(1+v+v^2\right)^{35}}\times P^{(2)}_{n,E_7}(v)~.
\end{equation}
We have
 {\footnotesize
 \begin{equation*}
 \ba
 P^{(2)}_{0,E_7}(v) &= 1+17 v+237 v^2+2628 v^3+25193 v^4+213819 v^5+1638666 v^6+11476871 v^7+74152233 v^8   \\
 & +445070980 v^9+2495671432 v^{10}+13133928036 v^{11}+65121712327 v^{12}+305215505275 v^{13}   \\
 & +1356033968529 v^{14}+5725284334978 v^{15}+23021851542594 v^{16}+88338636956104 v^{17}   \\
 & +324035139906700 v^{18}+1138031848052668 v^{19}+3832341391241046 v^{20}+12390621413785440 v^{21}   \\
 & +38509222288582663 v^{22}+115175603408208175 v^{23}+331836472263902521 v^{24}+921861932483495244 v^{25}   \\
 & +2471530433876763846 v^{26}+6399961693050532054 v^{27}+16018745367471142680 v^{28}   \\
 & +38781560068496818142 v^{29}+90876821066275028695 v^{30}+206242719899419463791 v^{31}   \\
 & +453576963793872584712 v^{32}+967171231109021529977 v^{33}+2000571291562232590513 v^{34}   \\
 & +4016126507767354504238 v^{35}+7828073649219480743672 v^{36}+14820947289312246349740 v^{37}   \\
 & +27267076918737091016348 v^{38}+48764087264312469202730 v^{39}+84802326792798968389732 v^{40}   \\
 & +143449590902653729399624 v^{41}+236104043071240448693797 v^{42}+378216261606533139497461 v^{43}   \\
 & +589822792928957883073617 v^{44}+895677339869346647226824 v^{45}+1324728639658651633703727 v^{46}  \\
 & +1908697079658876873038411 v^{47}+2679565476854052143878502 v^{48}+3665936157860425562998541 v^{49}   \\
 & +4888414479465062757831170 v^{50}+6354435158683924634396271 v^{51}+8053206553397859455383003 v^{52}   \\
 & +9951646269406905770095206 v^{53}+11992251412402642586454948 v^{54}\\
 &+14093734406768042617860546 v^{55}    +16154939755233169917249815 v^{56}\\
 &+18062065264884658609927825 v^{57}+19698620890606501833935055 v^{58}   \\
 & +20956986683280640928389866 v^{59}+21750009714684524653667914 v^{60}\\
 &+22020920210850484561094012 v^{61}    +21750009714684524653667914 v^{62}+ \dots+v^{122}~.
 \ea
 \end{equation*}}
 {\footnotesize
 \begin{equation*}
 \ba
 P^{(2)}_{1,E_7}&(v) =(1 + v^2) (137 + 2597 v + 37024 v^{2 }+ 419921 v^{3 }+ 4077137 v^{4 }+
    34901534 v^{5 }+ 268811177 v^{6 }+ 1887255497 v^{7 }\\
  +& 12196657853 v^{8 }+
    73094300214 v^{9 }+ 408614442098 v^{10 }+ 2140990474296 v^{11 }+
    10556715862964 v^{12 }\\
  +& 49151597538306 v^{13 }+ 216730904533865 v^{14 }+
    907396069059573 v^{15 }+ 3615374924636545 v^{16 } \\
 +&
    13736293007916068 v^{17 }+ 49857926256318138 v^{18 }+
    173164585658174276 v^{19 }+ 576354715341079126 v^{20 }\\
  +&
    1840835604225541174 v^{21 }+ 5649018624246617909 v^{22 }+
    16674709176092326437 v^{23 }\\
  +& 47394303096706259811 v^{24 }+
    129836595656234291790 v^{25 }+ 343131359707453293583 v^{26 }\\
  +&
    875542039936399623515 v^{27 }+ 2158650362542725175948 v^{28 }+
    5146221002718346735055 v^{29 }\\
  +& 11870970192394860758359 v^{30 }+
    26512271515436823962474 v^{31 }+ 57361948999125457686102 v^{32 } \\
 +&
    120296712068566252009120 v^{33 }+ 244657061843538883914723 v^{34 }+
    482772800850075541856889 v^{35 }\\
  +& 924703269912581018952608 v^{36 }+
    1719956874446161848789295 v^{37 }+ 3107832107172475492688890 v^{38 }\\
  +&
    5457334614794588632799143 v^{39 }+ 9316106764452824593797657 v^{40 }+
    15465256039202958794051688 v^{41 }\\
  +&
    24973370386295921380753761 v^{42 }+
    39238673033178891558314265 v^{43 }+
    60003949644181883287996554 v^{44 } \\
 +&
    89325800153382434388949763 v^{45 }+
    129479490500199449940430040 v^{46 } +
    182784785431420765743008945 v^{47 }\\
    +&
    251347720581234682951528991 v^{48 } +
    336728327510605097378060508 v^{49 }+
    439563294320255738140535927 v^{50 } \\+&
    559192119429259737303598283 v^{51 }+
    693350791574808124298361559 v^{52 } +
    838003231433873171234645238 v^{53 }\\+&
    987373409206497489976018270 v^{54 } +
    1134218396741161783456978908 v^{55 }\\+&
    1270346407634715071774123344 v^{56 } +
    1387339824356009883032251758 v^{57 }\\+&
    1477400106011059794784293700 v^{58 } +
    1534199301083129786878770830 v^{59 }\\+&
    1553610262702054425407310320 v^{60 }  + \dots + v^{120})~.
 \ea
 \end{equation*}}
 Note $g^{(2)}_{0,E_7}$ agree with the two-instanton $E_7$ Hilbert
 series in \cite{Hanany:2012dm}.

\section{Refined BPS invariants}
\label{ap:BPS}

The refined BPS invariants are solved from the generalised blowup equations
with the following initial input: the triple intersection numbers
$\kappa_{ijk}$ of divisors, the intersection numbers $b_i^{\text{GV}}$
of divisors with $c_2(X)$ (these are two ingredients of
$Z^{\text{cls}}$), the curve-divisor intersection matrix $C$, one
unity $\nound{r}$ field with nonzero $r_b$, as well as the one-loop partition function
$Z^{\text{1-loop}}$ and the bounds $j_{l,r}^{\text{max}}$.

In the case of $\fn=5,12$ models, there is no vanishing $\nound{r}$-fields, the unity $\nound{r}$ do not have enough constraints on $Z^{\text{1-loop}}$.
On the other hand, as seen in \eqref{eq:1-loop}, $Z^{\text{1-loop}}$ is easily computed, we simply input $Z^{\text{1-loop}}$ for all the models.

The input of $j_{l,r}^{\text{max}}$ is strictly speaking also not
necessary, as the bounds can be generated from the blowup equations
with the other input data, but the inclusion of the bounds in the
program makes the computation much faster. In any case, for $d_b\geq1$,
we observe an experimental formula for $j_{l,r}^{\text{max}}$.
For $F_4$, we observed for $d+d_b \leq 14$
\be
\begin{split}
j_r^{\text{max}}(d,d_b)&= d (d_b + 1)-d_b(d_b-1)/2,\\
j_l^{\text{max}}(d,d_b)&=(d - 1) (d_b - 1) -d_b(d_b-1)/2+  \delta_{d_b,1}\lfloor d/(1+d_\tau) \rfloor,\\
\end{split}
\ee
where $d$ is the total degree of the fibers, $d_b$ is the degree of the base, and $d_\tau$ is the first total degree of fibers when $\tau$ appears.
For $E_{6,7,8}$, we observed for $d+d_b \leq 12 \leq d_\tau$
\be
\begin{split}
j_r^{\text{max}}(d,d_b)&= d_b+d (d_b + 1),\\
j_l^{\text{max}}(d,d_b)&=(d - 1) (d_b - 1).\\
\end{split}
\ee
For $F_4$ model, we compute all the refined BPS invariants up to total degree 14, with $4777$ non-vanishing. For $E_{6,7,8}$, we compute all the BPS
invariants up to total degree 12, with $10383$, $10491$, $10068$ non-vanishing respectively. We list part of the refined BPS invariants in the affine Lie
algebra bases in the following tables, for complete lists, one can find them at \cite{kl}. It is worthwhile to point out that unlike the elliptic genus
\cite{DelZotto:2017mee}, the Weyl symmetry of gauge group $G$ is not manifest in the refined BPS invariants. This is simply because the Weyl symmetry will
change the sign of some K\"ahler parameters, while the refined BPS expansion is always in positive degrees. Note this should not be confused with the
situation where the refined BPS invariants of E-strings do have manifest $E_8$ symmetry, in which case the $E_8$ is a global symmetry other than the gauge
symmetry $G$ we considered in this paper.

\footnotesize{

\tablehead{Header of first column & Header of second column \\}
 \tablefirsthead{%
   \hline
   \multicolumn{1}{|c}{ ${\beta}$} &
   \multicolumn{1}{|c||}{$\oplus N^{{\beta}}_{j_l,j_r}(j_l,j_r)$} &
   ${\beta}$ &
   \multicolumn{1}{c|}{$\oplus N^{{\beta}}_{j_l,j_r}(j_l,j_r)$} \\
   \hline}
 \tablehead{%
   \hline
   \multicolumn{1}{|c}{ ${\beta}$} &
   \multicolumn{1}{|c||}{$\oplus N^{{\beta}}_{j_l,j_r}(j_l,j_r)$} &
   ${\beta}$ &
   \multicolumn{1}{c|}{$\oplus N^{{\beta}}_{j_l,j_r}(j_l,j_r)$} \\
   \hline}
 \tabletail{%
   \hline
   \multicolumn{4}{r}{\small\emph{continued on next page}}\\
   }
 \tablelasttail{\hline}
 \bottomcaption{Refined BPS invariants of 6d $F_4$ minimal SCFT.}
\center{
 \begin{supertabular}{|c|p{5.7cm}||c|p{5.7cm}|}\label{tb:F_4-BPS}$(0, 0, 0, 0, 0, 1)$&$(0,0)$&$(0, 0, 1, 0, 0, 1)$&$(0,1)$\\ \hline 
 $(0, 0, 1, 1, 0, 1)$&$(0,0)\oplus(0,1)$&$(0, 0, 1, 1, 1, 1)$&$(0,0)\oplus(0,1)$\\ \hline 
 $(0, 0, 1, 2, 0, 1)$&$(0,1)$&$(0, 0, 1, 2, 1, 1)$&$(0,0)\oplus(0,1)$\\ \hline 
 $(0, 0, 1, 2, 2, 1)$&$(0,1)$&$(0, 0, 2, 0, 0, 1)$&$(0,2)$\\ \hline 
 $(0, 0, 2, 1, 0, 1)$&$(0,1)\oplus(0,2)$&$(0, 0, 2, 1, 1, 1)$&$(0,1)\oplus(0,2)$\\ \hline 
 $(0, 0, 2, 2, 0, 1)$&$(0,0)\oplus(0,1)\oplus(0,2)$&$(0, 0, 2, 2, 1, 1)$&$(0,0)\oplus2(0,1)\oplus(0,2)$\\ \hline 
 $(0, 0, 2, 3, 0, 1)$&$(0,1)\oplus(0,2)$&$(0, 0, 3, 0, 0, 1)$&$(0,3)$\\ \hline 
 $(0, 0, 3, 1, 0, 1)$&$(0,2)\oplus(0,3)$&$(0, 0, 3, 1, 1, 1)$&$(0,2)\oplus(0,3)$\\ \hline 
 $(0, 0, 3, 2, 0, 1)$&$(0,1)\oplus(0,2)\oplus(0,3)$&$(0, 0, 4, 0, 0, 1)$&$(0,4)$\\ \hline 
 $(0, 0, 4, 1, 0, 1)$&$(0,3)\oplus(0,4)$&$(0, 0, 5, 0, 0, 1)$&$(0,5)$\\ \hline 
 $(0, 1, 0, 0, 0, 1)$&$(0,1)$&$(0, 1, 1, 0, 0, 1)$&$(0,0)\oplus(0,1)$\\ \hline 
 $(0, 1, 1, 1, 0, 1)$&$(0,0)\oplus(0,1)$&$(0, 1, 1, 1, 1, 1)$&$(0,0)\oplus(0,1)$\\ \hline 
 $(0, 1, 1, 2, 0, 1)$&$(0,0)\oplus(0,1)$&$(0, 1, 1, 2, 1, 1)$&$(0,0)\oplus(0,1)$\\ \hline 
 $(0, 1, 2, 0, 0, 1)$&$(0,1)\oplus(0,2)$&$(0, 1, 2, 1, 0, 1)$&$(0,0)\oplus2(0,1)\oplus(0,2)$\\ \hline 
 $(0, 1, 2, 1, 1, 1)$&$(0,0)\oplus2(0,1)\oplus(0,2)$&$(0, 1, 2, 2, 0, 1)$&$2(0,0)\oplus3(0,1)\oplus(0,2)$\\ \hline 
 $(0, 1, 3, 0, 0, 1)$&$(0,2)\oplus(0,3)$&$(0, 1, 3, 1, 0, 1)$&$(0,1)\oplus2(0,2)\oplus(0,3)$\\ \hline 
 $(0, 1, 4, 0, 0, 1)$&$(0,3)\oplus(0,4)$&$(0, 2, 0, 0, 0, 1)$&$(0,2)$\\ \hline 
 $(0, 2, 1, 0, 0, 1)$&$(0,1)\oplus(0,2)$&$(0, 2, 1, 1, 0, 1)$&$(0,1)\oplus(0,2)$\\ \hline 
 $(0, 2, 1, 1, 1, 1)$&$(0,1)\oplus(0,2)$&$(0, 2, 1, 2, 0, 1)$&$(0,1)\oplus(0,2)$\\ \hline 
 $(0, 2, 2, 0, 0, 1)$&$(0,0)\oplus(0,1)\oplus(0,2)$&$(0, 2, 2, 1, 0, 1)$&$(0,0)\oplus2(0,1)\oplus(0,2)$\\ \hline 
 $(0, 2, 3, 0, 0, 1)$&$(0,1)\oplus(0,2)\oplus(0,3)$&$(0, 3, 0, 0, 0, 1)$&$(0,3)$\\ \hline 
 $(0, 3, 1, 0, 0, 1)$&$(0,2)\oplus(0,3)$&$(0, 3, 1, 1, 0, 1)$&$(0,2)\oplus(0,3)$\\ \hline 
 $(0, 3, 2, 0, 0, 1)$&$(0,1)\oplus(0,2)\oplus(0,3)$&$(0, 4, 0, 0, 0, 1)$&$(0,4)$\\ \hline 
 $(0, 4, 1, 0, 0, 1)$&$(0,3)\oplus(0,4)$&$(0, 5, 0, 0, 0, 1)$&$(0,5)$\\ \hline 
 $(1, 1, 0, 0, 0, 1)$&$(0,0)\oplus(0,1)$&$(1, 1, 1, 0, 0, 1)$&$(0,0)\oplus(0,1)$\\ \hline 
 $(1, 1, 1, 1, 0, 1)$&$(0,0)\oplus(0,1)$&$(1, 1, 1, 1, 1, 1)$&$(0,0)\oplus(0,1)$\\ \hline 
 $(1, 1, 1, 2, 0, 1)$&$(0,0)\oplus(0,1)$&$(1, 1, 2, 0, 0, 1)$&$(0,1)\oplus(0,2)$\\ \hline 
 $(1, 1, 2, 1, 0, 1)$&$(0,0)\oplus2(0,1)\oplus(0,2)$&$(1, 1, 3, 0, 0, 1)$&$(0,2)\oplus(0,3)$\\ \hline 
 $(1, 2, 0, 0, 0, 1)$&$(0,1)\oplus(0,2)$&$(1, 2, 1, 0, 0, 1)$&$(0,0)\oplus2(0,1)\oplus(0,2)$\\ \hline 
 $(1, 2, 1, 1, 0, 1)$&$(0,0)\oplus2(0,1)\oplus(0,2)$&$(1, 2, 2, 0, 0, 1)$&$(0,0)\oplus2(0,1)\oplus(0,2)$\\ \hline 
 $(1, 3, 0, 0, 0, 1)$&$(0,2)\oplus(0,3)$&$(1, 3, 1, 0, 0, 1)$&$(0,1)\oplus2(0,2)\oplus(0,3)$\\ \hline 
 $(1, 4, 0, 0, 0, 1)$&$(0,3)\oplus(0,4)$&$(2, 1, 0, 0, 0, 1)$&$(0,1)$\\ \hline 
 $(2, 2, 0, 0, 0, 1)$&$(0,0)\oplus(0,1)\oplus(0,2)$&$(2, 2, 1, 0, 0, 1)$&$(0,0)\oplus2(0,1)\oplus(0,2)$\\ \hline 
 $(2, 3, 0, 0, 0, 1)$&$(0,1)\oplus(0,2)\oplus(0,3)$&$(3, 1, 0, 0, 0, 1)$&$(0,2)$\\ \hline 
 $(3, 2, 0, 0, 0, 1)$&$(0,1)\oplus(0,2)$&$(4, 1, 0, 0, 0, 1)$&$(0,3)$\\ \hline 
 $(0, 0, 2, 0, 0, 2)$&$(0,5/2)$&$(0, 0, 2, 1, 0, 2)$&$(0,3/2)\oplus(0,5/2)$\\ \hline 
 $(0, 0, 2, 1, 1, 2)$&$(0,3/2)\oplus(0,5/2)$&$(0, 0, 2, 2, 0, 2)$&$(0,1/2)\oplus(0,3/2)\oplus(0,5/2)$\\ \hline 
 $(0, 0, 3, 0, 0, 2)$&$(0,5/2)\oplus(0,7/2)\oplus(1/2,4)$&$(0, 0, 3, 1, 0, 2)$&$(0,3/2)\oplus3(0,5/2)\oplus2(0,7/2)\oplus(1/2,3)\oplus(1/2,4)$\\ \hline 
 $(0, 0, 4, 0, 0, 2)$&$(0,5/2)\oplus(0,7/2)\oplus2(0,9/2)\oplus(1/2,4)\oplus(1/2,5)\oplus(1,11/2)$&$(0, 1, 2, 0, 0, 2)$&$(0,3/2)\oplus(0,5/2)$\\ \hline 
 $(0, 1, 2, 1, 0, 2)$&$(0,1/2)\oplus2(0,3/2)\oplus(0,5/2)$&$(0, 1, 3, 0, 0, 2)$&$(0,3/2)\oplus3(0,5/2)\oplus2(0,7/2)\oplus(1/2,3)\oplus(1/2,4)$\\ \hline 
 $(0, 2, 0, 0, 0, 2)$&$(0,5/2)$&$(0, 2, 1, 0, 0, 2)$&$(0,3/2)\oplus(0,5/2)$\\ \hline 
 $(0, 2, 1, 1, 0, 2)$&$(0,3/2)\oplus(0,5/2)$&$(0, 2, 2, 0, 0, 2)$&$2(0,1/2)\oplus2(0,3/2)\oplus2(0,5/2)\oplus(0,7/2)$\\ \hline 
 $(0, 3, 0, 0, 0, 2)$&$(0,5/2)\oplus(0,7/2)\oplus(1/2,4)$&$(0, 3, 1, 0, 0, 2)$&$(0,3/2)\oplus3(0,5/2)\oplus2(0,7/2)\oplus(1/2,3)\oplus(1/2,4)$\\ \hline 
 $(0, 4, 0, 0, 0, 2)$&$(0,5/2)\oplus(0,7/2)\oplus2(0,9/2)\oplus(1/2,4)\oplus(1/2,5)\oplus(1,11/2)$&$(1, 1, 2, 0, 0, 2)$&$(0,3/2)\oplus(0,5/2)$\\ \hline 
 $(1, 2, 0, 0, 0, 2)$&$(0,3/2)\oplus(0,5/2)$&$(1, 2, 1, 0, 0, 2)$&$(0,1/2)\oplus2(0,3/2)\oplus(0,5/2)$\\ \hline 
 $(1, 3, 0, 0, 0, 2)$&$(0,3/2)\oplus3(0,5/2)\oplus2(0,7/2)\oplus(1/2,3)\oplus(1/2,4)$&$(2, 2, 0, 0, 0, 2)$&$(0,1/2)\oplus(0,3/2)\oplus(0,5/2)$\\ \hline 
 $(0, 0, 3, 0, 0, 3)$&$(0,3)\oplus(1/2,9/2)$&$(0, 3, 0, 0, 0, 3)$&$(0,3)\oplus(1/2,9/2)$\\ \hline 
  \end{supertabular}}
}  
\footnotesize{

\tablehead{Header of first column & Header of second column \\}
 \tablefirsthead{%
   \hline
   \multicolumn{1}{|c}{ ${\beta}$} &
   \multicolumn{1}{|c||}{$\oplus N^{{\beta}}_{j_l,j_r}(j_l,j_r)$} &
   ${\beta}$ &
   \multicolumn{1}{c|}{$\oplus N^{{\beta}}_{j_l,j_r}(j_l,j_r)$} \\
   \hline}
 \tablehead{%
   \hline
   \multicolumn{1}{|c}{ ${\beta}$} &
   \multicolumn{1}{|c||}{$\oplus N^{{\beta}}_{j_l,j_r}(j_l,j_r)$} &
   ${\beta}$ &
   \multicolumn{1}{c|}{$\oplus N^{{\beta}}_{j_l,j_r}(j_l,j_r)$} \\
   \hline}
 \tabletail{%
   \hline
   \multicolumn{4}{r}{\small\emph{continued on next page}}\\
   }
 \tablelasttail{\hline}
 \bottomcaption{Refined BPS invariants of 6d $E_6$ minimal SCFT.}
\center{
 \begin{supertabular}{|c|p{5.3cm}||c|p{5.3cm}|}\label{tb:E_6-BPS}$(0, 0, 0, 0, 0, 0, 0, 1)$&$(0,1/2)$&$(0, 1, 0, 0, 0, 0, 0, 1)$&$(0,1/2)$\\ \hline 
 $(0, 2, 0, 0, 0, 0, 0, 1)$&$(0,3/2)$&$(0, 3, 0, 0, 0, 0, 0, 1)$&$(0,5/2)$\\ \hline 
 $(0, 4, 0, 0, 0, 0, 0, 1)$&$(0,7/2)$&$(0, 5, 0, 0, 0, 0, 0, 1)$&$(0,9/2)$\\ \hline 
 $(0, 0, 0, 0, 0, 1, 0, 1)$&$(0,1/2)$&$(0, 0, 0, 0, 0, 1, 1, 1)$&$(0,1/2)$\\ \hline 
 $(0, 0, 0, 0, 0, 2, 0, 1)$&$(0,3/2)$&$(0, 0, 0, 0, 0, 2, 1, 1)$&$(0,1/2)\oplus(0,3/2)$\\ \hline 
 $(0, 0, 0, 0, 0, 2, 2, 1)$&$(0,1/2)\oplus(0,3/2)$&$(0, 0, 0, 0, 0, 2, 3, 1)$&$(0,3/2)$\\ \hline 
 $(0, 0, 0, 0, 0, 3, 0, 1)$&$(0,5/2)$&$(0, 0, 0, 0, 0, 3, 1, 1)$&$(0,3/2)\oplus(0,5/2)$\\ \hline 
 $(0, 0, 0, 0, 0, 3, 2, 1)$&$(0,1/2)\oplus(0,3/2)\oplus(0,5/2)$&$(0, 0, 0, 0, 0, 4, 0, 1)$&$(0,7/2)$\\ \hline 
 $(0, 0, 0, 0, 0, 4, 1, 1)$&$(0,5/2)\oplus(0,7/2)$&$(0, 0, 0, 0, 0, 5, 0, 1)$&$(0,9/2)$\\ \hline 
 $(0, 0, 0, 0, 1, 0, 0, 1)$&$(0,3/2)$&$(0, 1, 0, 0, 1, 0, 0, 1)$&$(0,1/2)\oplus(0,3/2)$\\ \hline 
 $(0, 2, 0, 0, 1, 0, 0, 1)$&$(0,1/2)\oplus(0,3/2)$&$(0, 3, 0, 0, 1, 0, 0, 1)$&$(0,3/2)\oplus(0,5/2)$\\ \hline 
 $(0, 4, 0, 0, 1, 0, 0, 1)$&$(0,5/2)\oplus(0,7/2)$&$(0, 0, 0, 0, 1, 1, 0, 1)$&$(0,1/2)\oplus(0,3/2)$\\ \hline 
 $(0, 1, 0, 0, 1, 1, 0, 1)$&$2(0,1/2)\oplus(0,3/2)$&$(0, 2, 0, 0, 1, 1, 0, 1)$&$(0,1/2)\oplus(0,3/2)$\\ \hline 
 $(0, 3, 0, 0, 1, 1, 0, 1)$&$(0,3/2)\oplus(0,5/2)$&$(0, 0, 0, 0, 1, 1, 1, 1)$&$(0,1/2)\oplus(0,3/2)$\\ \hline 
 $(0, 1, 0, 0, 1, 1, 1, 1)$&$2(0,1/2)\oplus(0,3/2)$&$(0, 2, 0, 0, 1, 1, 1, 1)$&$(0,1/2)\oplus(0,3/2)$\\ \hline 
 $(0, 0, 0, 0, 1, 2, 0, 1)$&$(0,1/2)\oplus(0,3/2)$&$(0, 1, 0, 0, 1, 2, 0, 1)$&$(0,1/2)\oplus(0,3/2)$\\ \hline 
 $(0, 0, 0, 0, 1, 2, 1, 1)$&$2(0,1/2)\oplus(0,3/2)$&$(0, 1, 0, 0, 1, 2, 1, 1)$&$2(0,1/2)\oplus(0,3/2)$\\ \hline 
 $(0, 0, 0, 0, 1, 2, 2, 1)$&$(0,1/2)\oplus(0,3/2)$&$(0, 0, 0, 0, 1, 3, 0, 1)$&$(0,3/2)\oplus(0,5/2)$\\ \hline 
 $(0, 1, 0, 0, 1, 3, 0, 1)$&$(0,3/2)\oplus(0,5/2)$&$(0, 0, 0, 0, 1, 3, 1, 1)$&$(0,1/2)\oplus2(0,3/2)\oplus(0,5/2)$\\ \hline 
 $(0, 0, 0, 0, 1, 4, 0, 1)$&$(0,5/2)\oplus(0,7/2)$&$(0, 0, 0, 0, 2, 0, 0, 1)$&$(0,5/2)$\\ \hline 
 $(0, 1, 0, 0, 2, 0, 0, 1)$&$(0,3/2)\oplus(0,5/2)$&$(0, 2, 0, 0, 2, 0, 0, 1)$&$(0,1/2)\oplus(0,3/2)\oplus(0,5/2)$\\ \hline 
 $(0, 3, 0, 0, 2, 0, 0, 1)$&$(0,1/2)\oplus(0,3/2)\oplus(0,5/2)$&$(0, 0, 0, 0, 2, 1, 0, 1)$&$(0,3/2)\oplus(0,5/2)$\\ \hline 
 $(0, 1, 0, 0, 2, 1, 0, 1)$&$(0,1/2)\oplus2(0,3/2)\oplus(0,5/2)$&$(0, 2, 0, 0, 2, 1, 0, 1)$&$2(0,1/2)\oplus2(0,3/2)\oplus(0,5/2)$\\ \hline 
 $(0, 0, 0, 0, 2, 1, 1, 1)$&$(0,3/2)\oplus(0,5/2)$&$(0, 1, 0, 0, 2, 1, 1, 1)$&$(0,1/2)\oplus2(0,3/2)\oplus(0,5/2)$\\ \hline 
 $(0, 0, 0, 0, 2, 2, 0, 1)$&$(0,1/2)\oplus(0,3/2)\oplus(0,5/2)$&$(0, 1, 0, 0, 2, 2, 0, 1)$&$2(0,1/2)\oplus2(0,3/2)\oplus(0,5/2)$\\ \hline 
 $(0, 0, 0, 0, 2, 2, 1, 1)$&$(0,1/2)\oplus2(0,3/2)\oplus(0,5/2)$&$(0, 0, 0, 0, 2, 3, 0, 1)$&$(0,1/2)\oplus(0,3/2)\oplus(0,5/2)$\\ \hline 
 $(0, 0, 0, 0, 3, 0, 0, 1)$&$(0,7/2)$&$(0, 1, 0, 0, 3, 0, 0, 1)$&$(0,5/2)\oplus(0,7/2)$\\ \hline 
 $(0, 2, 0, 0, 3, 0, 0, 1)$&$(0,3/2)\oplus(0,5/2)\oplus(0,7/2)$&$(0, 0, 0, 0, 3, 1, 0, 1)$&$(0,5/2)\oplus(0,7/2)$\\ \hline 
 $(0, 1, 0, 0, 3, 1, 0, 1)$&$(0,3/2)\oplus2(0,5/2)\oplus(0,7/2)$&$(0, 0, 0, 0, 3, 1, 1, 1)$&$(0,5/2)\oplus(0,7/2)$\\ \hline 
 $(0, 0, 0, 0, 3, 2, 0, 1)$&$(0,3/2)\oplus(0,5/2)\oplus(0,7/2)$&$(0, 0, 0, 0, 4, 0, 0, 1)$&$(0,9/2)$\\ \hline 
 $(0, 1, 0, 0, 4, 0, 0, 1)$&$(0,7/2)\oplus(0,9/2)$&$(0, 0, 0, 0, 4, 1, 0, 1)$&$(0,7/2)\oplus(0,9/2)$\\ \hline 
 $(0, 0, 0, 0, 5, 0, 0, 1)$&$(0,11/2)$&$(0, 0, 0, 1, 0, 0, 0, 1)$&$(0,1/2)$\\ \hline 
 $(0, 0, 0, 1, 1, 0, 0, 1)$&$(0,1/2)\oplus(0,3/2)$&$(0, 1, 0, 1, 1, 0, 0, 1)$&$2(0,1/2)\oplus(0,3/2)$\\ \hline 
 $(0, 2, 0, 1, 1, 0, 0, 1)$&$(0,1/2)\oplus(0,3/2)$&$(0, 3, 0, 1, 1, 0, 0, 1)$&$(0,3/2)\oplus(0,5/2)$\\ \hline 
 $(0, 0, 0, 1, 1, 1, 0, 1)$&$2(0,1/2)\oplus(0,3/2)$&$(0, 1, 0, 1, 1, 1, 0, 1)$&$3(0,1/2)\oplus(0,3/2)$\\ \hline 
 $(0, 2, 0, 1, 1, 1, 0, 1)$&$(0,1/2)\oplus(0,3/2)$&$(0, 0, 0, 1, 1, 1, 1, 1)$&$2(0,1/2)\oplus(0,3/2)$\\ \hline 
 $(0, 1, 0, 1, 1, 1, 1, 1)$&$3(0,1/2)\oplus(0,3/2)$&$(0, 0, 0, 1, 1, 2, 0, 1)$&$(0,1/2)\oplus(0,3/2)$\\ \hline 
 $(0, 1, 0, 1, 1, 2, 0, 1)$&$(0,1/2)\oplus(0,3/2)$&$(0, 0, 0, 1, 1, 2, 1, 1)$&$2(0,1/2)\oplus(0,3/2)$\\ \hline 
 $(0, 0, 0, 1, 1, 3, 0, 1)$&$(0,3/2)\oplus(0,5/2)$&$(0, 0, 0, 1, 2, 0, 0, 1)$&$(0,3/2)\oplus(0,5/2)$\\ \hline 
 $(0, 1, 0, 1, 2, 0, 0, 1)$&$(0,1/2)\oplus2(0,3/2)\oplus(0,5/2)$&$(0, 2, 0, 1, 2, 0, 0, 1)$&$2(0,1/2)\oplus2(0,3/2)\oplus(0,5/2)$\\ \hline 
 $(0, 0, 0, 1, 2, 1, 0, 1)$&$(0,1/2)\oplus2(0,3/2)\oplus(0,5/2)$&$(0, 1, 0, 1, 2, 1, 0, 1)$&$4(0,1/2)\oplus4(0,3/2)\oplus(0,5/2)$\\ \hline 
 $(0, 0, 0, 1, 2, 1, 1, 1)$&$(0,1/2)\oplus2(0,3/2)\oplus(0,5/2)$&$(0, 0, 0, 1, 2, 2, 0, 1)$&$2(0,1/2)\oplus2(0,3/2)\oplus(0,5/2)$\\ \hline 
 $(0, 0, 0, 1, 3, 0, 0, 1)$&$(0,5/2)\oplus(0,7/2)$&$(0, 1, 0, 1, 3, 0, 0, 1)$&$(0,3/2)\oplus2(0,5/2)\oplus(0,7/2)$\\ \hline 
 $(0, 0, 0, 1, 3, 1, 0, 1)$&$(0,3/2)\oplus2(0,5/2)\oplus(0,7/2)$&$(0, 0, 0, 1, 4, 0, 0, 1)$&$(0,7/2)\oplus(0,9/2)$\\ \hline 
 $(0, 0, 0, 2, 0, 0, 0, 1)$&$(0,3/2)$&$(0, 0, 0, 2, 1, 0, 0, 1)$&$(0,1/2)\oplus(0,3/2)$\\ \hline 
 $(0, 1, 0, 2, 1, 0, 0, 1)$&$(0,1/2)\oplus(0,3/2)$&$(0, 0, 0, 2, 1, 1, 0, 1)$&$(0,1/2)\oplus(0,3/2)$\\ \hline 
 $(0, 1, 0, 2, 1, 1, 0, 1)$&$(0,1/2)\oplus(0,3/2)$&$(0, 0, 0, 2, 1, 1, 1, 1)$&$(0,1/2)\oplus(0,3/2)$\\ \hline 
 $(0, 0, 0, 2, 2, 0, 0, 1)$&$(0,1/2)\oplus(0,3/2)\oplus(0,5/2)$&$(0, 1, 0, 2, 2, 0, 0, 1)$&$2(0,1/2)\oplus2(0,3/2)\oplus(0,5/2)$\\ \hline 
 $(0, 0, 0, 2, 2, 1, 0, 1)$&$2(0,1/2)\oplus2(0,3/2)\oplus(0,5/2)$&$(0, 0, 0, 2, 3, 0, 0, 1)$&$(0,3/2)\oplus(0,5/2)\oplus(0,7/2)$\\ \hline 
 $(0, 0, 0, 3, 0, 0, 0, 1)$&$(0,5/2)$&$(0, 0, 0, 3, 1, 0, 0, 1)$&$(0,3/2)\oplus(0,5/2)$\\ \hline 
 $(0, 1, 0, 3, 1, 0, 0, 1)$&$(0,3/2)\oplus(0,5/2)$&$(0, 0, 0, 3, 1, 1, 0, 1)$&$(0,3/2)\oplus(0,5/2)$\\ \hline 
 $(0, 0, 0, 3, 2, 0, 0, 1)$&$(0,1/2)\oplus(0,3/2)\oplus(0,5/2)$&$(0, 0, 0, 4, 0, 0, 0, 1)$&$(0,7/2)$\\ \hline 
 $(0, 0, 0, 4, 1, 0, 0, 1)$&$(0,5/2)\oplus(0,7/2)$&$(0, 0, 0, 5, 0, 0, 0, 1)$&$(0,9/2)$\\ \hline 
 $(0, 0, 1, 1, 0, 0, 0, 1)$&$(0,1/2)$&$(0, 0, 1, 1, 1, 0, 0, 1)$&$(0,1/2)\oplus(0,3/2)$\\ \hline 
 $(0, 1, 1, 1, 1, 0, 0, 1)$&$2(0,1/2)\oplus(0,3/2)$&$(0, 2, 1, 1, 1, 0, 0, 1)$&$(0,1/2)\oplus(0,3/2)$\\ \hline 
 $(0, 0, 1, 1, 1, 1, 0, 1)$&$2(0,1/2)\oplus(0,3/2)$&$(0, 1, 1, 1, 1, 1, 0, 1)$&$3(0,1/2)\oplus(0,3/2)$\\ \hline 
 $(0, 0, 1, 1, 1, 1, 1, 1)$&$2(0,1/2)\oplus(0,3/2)$&$(0, 0, 1, 1, 1, 2, 0, 1)$&$(0,1/2)\oplus(0,3/2)$\\ \hline 
 $(0, 0, 1, 1, 2, 0, 0, 1)$&$(0,3/2)\oplus(0,5/2)$&$(0, 1, 1, 1, 2, 0, 0, 1)$&$(0,1/2)\oplus2(0,3/2)\oplus(0,5/2)$\\ \hline 
 $(0, 0, 1, 1, 2, 1, 0, 1)$&$(0,1/2)\oplus2(0,3/2)\oplus(0,5/2)$&$(0, 0, 1, 1, 3, 0, 0, 1)$&$(0,5/2)\oplus(0,7/2)$\\ \hline 
 $(0, 0, 1, 2, 0, 0, 0, 1)$&$(0,1/2)\oplus(0,3/2)$&$(0, 0, 1, 2, 1, 0, 0, 1)$&$2(0,1/2)\oplus(0,3/2)$\\ \hline 
 $(0, 1, 1, 2, 1, 0, 0, 1)$&$2(0,1/2)\oplus(0,3/2)$&$(0, 0, 1, 2, 1, 1, 0, 1)$&$2(0,1/2)\oplus(0,3/2)$\\ \hline 
 $(0, 0, 1, 2, 2, 0, 0, 1)$&$(0,1/2)\oplus2(0,3/2)\oplus(0,5/2)$&$(0, 0, 1, 3, 0, 0, 0, 1)$&$(0,3/2)\oplus(0,5/2)$\\ \hline 
 $(0, 0, 1, 3, 1, 0, 0, 1)$&$(0,1/2)\oplus2(0,3/2)\oplus(0,5/2)$&$(0, 0, 1, 4, 0, 0, 0, 1)$&$(0,5/2)\oplus(0,7/2)$\\ \hline 
 $(0, 0, 2, 2, 0, 0, 0, 1)$&$(0,1/2)\oplus(0,3/2)$&$(0, 0, 2, 2, 1, 0, 0, 1)$&$(0,1/2)\oplus(0,3/2)$\\ \hline 
 $(0, 0, 2, 3, 0, 0, 0, 1)$&$(0,1/2)\oplus(0,3/2)\oplus(0,5/2)$&$(0, 0, 3, 2, 0, 0, 0, 1)$&$(0,3/2)$\\ \hline 
 $(1, 1, 0, 0, 0, 0, 0, 1)$&$(0,1/2)$&$(1, 2, 0, 0, 0, 0, 0, 1)$&$(0,1/2)\oplus(0,3/2)$\\ \hline 
 $(1, 3, 0, 0, 0, 0, 0, 1)$&$(0,3/2)\oplus(0,5/2)$&$(1, 4, 0, 0, 0, 0, 0, 1)$&$(0,5/2)\oplus(0,7/2)$\\ \hline 
 $(1, 1, 0, 0, 1, 0, 0, 1)$&$(0,1/2)\oplus(0,3/2)$&$(1, 2, 0, 0, 1, 0, 0, 1)$&$2(0,1/2)\oplus(0,3/2)$\\ \hline 
 $(1, 3, 0, 0, 1, 0, 0, 1)$&$(0,1/2)\oplus2(0,3/2)\oplus(0,5/2)$&$(1, 1, 0, 0, 1, 1, 0, 1)$&$2(0,1/2)\oplus(0,3/2)$\\ \hline 
 $(1, 2, 0, 0, 1, 1, 0, 1)$&$2(0,1/2)\oplus(0,3/2)$&$(1, 1, 0, 0, 1, 1, 1, 1)$&$2(0,1/2)\oplus(0,3/2)$\\ \hline 
 $(1, 1, 0, 0, 1, 2, 0, 1)$&$(0,1/2)\oplus(0,3/2)$&$(1, 1, 0, 0, 2, 0, 0, 1)$&$(0,3/2)\oplus(0,5/2)$\\ \hline 
 $(1, 2, 0, 0, 2, 0, 0, 1)$&$(0,1/2)\oplus2(0,3/2)\oplus(0,5/2)$&$(1, 1, 0, 0, 2, 1, 0, 1)$&$(0,1/2)\oplus2(0,3/2)\oplus(0,5/2)$\\ \hline 
 $(1, 1, 0, 0, 3, 0, 0, 1)$&$(0,5/2)\oplus(0,7/2)$&$(1, 1, 0, 1, 1, 0, 0, 1)$&$2(0,1/2)\oplus(0,3/2)$\\ \hline 
 $(1, 2, 0, 1, 1, 0, 0, 1)$&$2(0,1/2)\oplus(0,3/2)$&$(1, 1, 0, 1, 1, 1, 0, 1)$&$3(0,1/2)\oplus(0,3/2)$\\ \hline 
 $(1, 1, 0, 1, 2, 0, 0, 1)$&$(0,1/2)\oplus2(0,3/2)\oplus(0,5/2)$&$(1, 1, 0, 2, 1, 0, 0, 1)$&$(0,1/2)\oplus(0,3/2)$\\ \hline 
 $(1, 1, 1, 1, 1, 0, 0, 1)$&$2(0,1/2)\oplus(0,3/2)$&$(2, 2, 0, 0, 0, 0, 0, 1)$&$(0,1/2)\oplus(0,3/2)$\\ \hline 
 $(2, 3, 0, 0, 0, 0, 0, 1)$&$(0,1/2)\oplus(0,3/2)\oplus(0,5/2)$&$(2, 2, 0, 0, 1, 0, 0, 1)$&$(0,1/2)\oplus(0,3/2)$\\ \hline 
 $(3, 2, 0, 0, 0, 0, 0, 1)$&$(0,3/2)$&$(0, 3, 0, 0, 0, 0, 0, 2)$&$(0,5/2)$\\ \hline 
 $(0, 4, 0, 0, 0, 0, 0, 2)$&$(0,5/2)\oplus(0,7/2)\oplus(1/2,4)$&$(0, 0, 0, 0, 0, 3, 0, 2)$&$(0,5/2)$\\ \hline 
 $(0, 0, 0, 0, 0, 3, 1, 2)$&$(0,3/2)\oplus(0,5/2)$&$(0, 0, 0, 0, 0, 4, 0, 2)$&$(0,5/2)\oplus(0,7/2)\oplus(1/2,4)$\\ \hline 
 $(0, 0, 0, 0, 1, 0, 0, 2)$&$(0,5/2)$&$(0, 1, 0, 0, 1, 0, 0, 2)$&$(0,3/2)\oplus(0,5/2)$\\ \hline 
 $(0, 2, 0, 0, 1, 0, 0, 2)$&$(0,1/2)\oplus(0,3/2)\oplus(0,5/2)$&$(0, 3, 0, 0, 1, 0, 0, 2)$&$(0,1/2)\oplus2(0,3/2)\oplus2(0,5/2)\oplus(0,7/2)$\\ \hline 
 $(0, 0, 0, 0, 1, 1, 0, 2)$&$(0,3/2)\oplus(0,5/2)$&$(0, 1, 0, 0, 1, 1, 0, 2)$&$(0,1/2)\oplus2(0,3/2)\oplus(0,5/2)$\\ \hline 
 $(0, 2, 0, 0, 1, 1, 0, 2)$&$2(0,1/2)\oplus2(0,3/2)\oplus(0,5/2)$&$(0, 0, 0, 0, 1, 1, 1, 2)$&$(0,3/2)\oplus(0,5/2)$\\ \hline 
 $(0, 1, 0, 0, 1, 1, 1, 2)$&$(0,1/2)\oplus2(0,3/2)\oplus(0,5/2)$&$(0, 0, 0, 0, 1, 2, 0, 2)$&$(0,1/2)\oplus(0,3/2)\oplus(0,5/2)$\\ \hline 
 $(0, 1, 0, 0, 1, 2, 0, 2)$&$2(0,1/2)\oplus2(0,3/2)\oplus(0,5/2)$&$(0, 0, 0, 0, 1, 2, 1, 2)$&$(0,1/2)\oplus2(0,3/2)\oplus(0,5/2)$\\ \hline 
 $(0, 0, 0, 0, 1, 3, 0, 2)$&$(0,1/2)\oplus2(0,3/2)\oplus2(0,5/2)\oplus(0,7/2)$&$(0, 0, 0, 0, 2, 0, 0, 2)$&$(0,5/2)\oplus(0,7/2)\oplus(1/2,4)$\\ \hline 
 $(0, 1, 0, 0, 2, 0, 0, 2)$&$(0,3/2)\oplus3(0,5/2)\oplus2(0,7/2)\oplus(1/2,3)\oplus(1/2,4)$&$(0, 2, 0, 0, 2, 0, 0, 2)$&$(0,1/2)\oplus3(0,3/2)\oplus4(0,5/2)\oplus2(0,7/2)\oplus(1/2,2)\oplus(1/2,3)\oplus(1/2,4)$\\ \hline 
 $(0, 0, 0, 0, 2, 1, 0, 2)$&$(0,3/2)\oplus3(0,5/2)\oplus2(0,7/2)\oplus(1/2,3)\oplus(1/2,4)$&$(0, 1, 0, 0, 2, 1, 0, 2)$&$(0,1/2)\oplus5(0,3/2)\oplus7(0,5/2)\oplus3(0,7/2)\oplus(1/2,2)\oplus2(1/2,3)\oplus(1/2,4)$\\ \hline 
 $(0, 0, 0, 0, 2, 1, 1, 2)$&$(0,3/2)\oplus3(0,5/2)\oplus2(0,7/2)\oplus(1/2,3)\oplus(1/2,4)$&$(0, 0, 0, 0, 2, 2, 0, 2)$&$(0,1/2)\oplus3(0,3/2)\oplus4(0,5/2)\oplus2(0,7/2)\oplus(1/2,2)\oplus(1/2,3)\oplus(1/2,4)$\\ \hline 
 $(0, 0, 0, 0, 3, 0, 0, 2)$&$(0,5/2)\oplus(0,7/2)\oplus2(0,9/2)\oplus(1/2,4)\oplus(1/2,5)\oplus(1,11/2)$&$(0, 1, 0, 0, 3, 0, 0, 2)$&$(0,3/2)\oplus3(0,5/2)\oplus5(0,7/2)\oplus3(0,9/2)\oplus(1/2,3)\oplus3(1/2,4)\oplus2(1/2,5)\oplus(1,9/2)\oplus(1,11/2)$\\ \hline 
 $(0, 0, 0, 0, 3, 1, 0, 2)$&$(0,3/2)\oplus3(0,5/2)\oplus5(0,7/2)\oplus3(0,9/2)\oplus(1/2,3)\oplus3(1/2,4)\oplus2(1/2,5)\oplus(1,9/2)\oplus(1,11/2)$&$(0, 0, 0, 0, 4, 0, 0, 2)$&$(0,5/2)\oplus(0,7/2)\oplus2(0,9/2)\oplus2(0,11/2)\oplus(1/2,4)\oplus(1/2,5)\oplus2(1/2,6)\oplus(1,11/2)\oplus(1,13/2)\oplus(3/2,7)$\\ \hline 
 $(0, 0, 0, 1, 1, 0, 0, 2)$&$(0,3/2)\oplus(0,5/2)$&$(0, 1, 0, 1, 1, 0, 0, 2)$&$(0,1/2)\oplus2(0,3/2)\oplus(0,5/2)$\\ \hline 
 $(0, 2, 0, 1, 1, 0, 0, 2)$&$2(0,1/2)\oplus2(0,3/2)\oplus(0,5/2)$&$(0, 0, 0, 1, 1, 1, 0, 2)$&$(0,1/2)\oplus2(0,3/2)\oplus(0,5/2)$\\ \hline 
 $(0, 1, 0, 1, 1, 1, 0, 2)$&$3(0,1/2)\oplus3(0,3/2)\oplus(0,5/2)$&$(0, 0, 0, 1, 1, 1, 1, 2)$&$(0,1/2)\oplus2(0,3/2)\oplus(0,5/2)$\\ \hline 
 $(0, 0, 0, 1, 1, 2, 0, 2)$&$2(0,1/2)\oplus2(0,3/2)\oplus(0,5/2)$&$(0, 0, 0, 1, 2, 0, 0, 2)$&$(0,3/2)\oplus3(0,5/2)\oplus2(0,7/2)\oplus(1/2,3)\oplus(1/2,4)$\\ \hline 
 $(0, 1, 0, 1, 2, 0, 0, 2)$&$(0,1/2)\oplus5(0,3/2)\oplus7(0,5/2)\oplus3(0,7/2)\oplus(1/2,2)\oplus2(1/2,3)\oplus(1/2,4)$&$(0, 0, 0, 1, 2, 1, 0, 2)$&$(0,1/2)\oplus5(0,3/2)\oplus7(0,5/2)\oplus3(0,7/2)\oplus(1/2,2)\oplus2(1/2,3)\oplus(1/2,4)$\\ \hline 
 $(0, 0, 0, 1, 3, 0, 0, 2)$&$(0,3/2)\oplus3(0,5/2)\oplus5(0,7/2)\oplus3(0,9/2)\oplus(1/2,3)\oplus3(1/2,4)\oplus2(1/2,5)\oplus(1,9/2)\oplus(1,11/2)$&$(0, 0, 0, 2, 1, 0, 0, 2)$&$(0,1/2)\oplus(0,3/2)\oplus(0,5/2)$\\ \hline 
 $(0, 1, 0, 2, 1, 0, 0, 2)$&$2(0,1/2)\oplus2(0,3/2)\oplus(0,5/2)$&$(0, 0, 0, 2, 1, 1, 0, 2)$&$2(0,1/2)\oplus2(0,3/2)\oplus(0,5/2)$\\ \hline 
 $(0, 0, 0, 2, 2, 0, 0, 2)$&$(0,1/2)\oplus3(0,3/2)\oplus4(0,5/2)\oplus2(0,7/2)\oplus(1/2,2)\oplus(1/2,3)\oplus(1/2,4)$&$(0, 0, 0, 3, 0, 0, 0, 2)$&$(0,5/2)$\\ \hline 
 $(0, 0, 0, 3, 1, 0, 0, 2)$&$(0,1/2)\oplus2(0,3/2)\oplus2(0,5/2)\oplus(0,7/2)$&$(0, 0, 0, 4, 0, 0, 0, 2)$&$(0,5/2)\oplus(0,7/2)\oplus(1/2,4)$\\ \hline 
 $(0, 0, 1, 1, 1, 0, 0, 2)$&$(0,3/2)\oplus(0,5/2)$&$(0, 1, 1, 1, 1, 0, 0, 2)$&$(0,1/2)\oplus2(0,3/2)\oplus(0,5/2)$\\ \hline 
 $(0, 0, 1, 1, 1, 1, 0, 2)$&$(0,1/2)\oplus2(0,3/2)\oplus(0,5/2)$&$(0, 0, 1, 1, 2, 0, 0, 2)$&$(0,3/2)\oplus3(0,5/2)\oplus2(0,7/2)\oplus(1/2,3)\oplus(1/2,4)$\\ \hline 
 $(0, 0, 1, 2, 1, 0, 0, 2)$&$(0,1/2)\oplus2(0,3/2)\oplus(0,5/2)$&$(0, 0, 1, 3, 0, 0, 0, 2)$&$(0,3/2)\oplus(0,5/2)$\\ \hline 
 $(1, 3, 0, 0, 0, 0, 0, 2)$&$(0,3/2)\oplus(0,5/2)$&$(1, 1, 0, 0, 1, 0, 0, 2)$&$(0,3/2)\oplus(0,5/2)$\\ \hline 
 $(1, 2, 0, 0, 1, 0, 0, 2)$&$(0,1/2)\oplus2(0,3/2)\oplus(0,5/2)$&$(1, 1, 0, 0, 1, 1, 0, 2)$&$(0,1/2)\oplus2(0,3/2)\oplus(0,5/2)$\\ \hline 
 $(1, 1, 0, 0, 2, 0, 0, 2)$&$(0,3/2)\oplus3(0,5/2)\oplus2(0,7/2)\oplus(1/2,3)\oplus(1/2,4)$&$(1, 1, 0, 1, 1, 0, 0, 2)$&$(0,1/2)\oplus2(0,3/2)\oplus(0,5/2)$\\ \hline 
 $(0, 0, 0, 0, 1, 0, 0, 3)$&$(0,7/2)$&$(0, 1, 0, 0, 1, 0, 0, 3)$&$(0,5/2)\oplus(0,7/2)$\\ \hline 
 $(0, 2, 0, 0, 1, 0, 0, 3)$&$(0,3/2)\oplus(0,5/2)\oplus(0,7/2)$&$(0, 0, 0, 0, 1, 1, 0, 3)$&$(0,5/2)\oplus(0,7/2)$\\ \hline 
 $(0, 1, 0, 0, 1, 1, 0, 3)$&$(0,3/2)\oplus2(0,5/2)\oplus(0,7/2)$&$(0, 0, 0, 0, 1, 1, 1, 3)$&$(0,5/2)\oplus(0,7/2)$\\ \hline 
 $(0, 0, 0, 0, 1, 2, 0, 3)$&$(0,3/2)\oplus(0,5/2)\oplus(0,7/2)$&$(0, 0, 0, 0, 2, 0, 0, 3)$&$(0,5/2)\oplus(0,7/2)\oplus2(0,9/2)\oplus(1/2,4)\oplus(1/2,5)\oplus(1,11/2)$\\ \hline 
 $(0, 1, 0, 0, 2, 0, 0, 3)$&$(0,3/2)\oplus3(0,5/2)\oplus5(0,7/2)\oplus3(0,9/2)\oplus(1/2,3)\oplus3(1/2,4)\oplus2(1/2,5)\oplus(1,9/2)\oplus(1,11/2)$&$(0, 0, 0, 0, 2, 1, 0, 3)$&$(0,3/2)\oplus3(0,5/2)\oplus5(0,7/2)\oplus3(0,9/2)\oplus(1/2,3)\oplus3(1/2,4)\oplus2(1/2,5)\oplus(1,9/2)\oplus(1,11/2)$\\ \hline 
 $(0, 0, 0, 0, 3, 0, 0, 3)$&$(0,3/2)\oplus(0,5/2)\oplus3(0,7/2)\oplus3(0,9/2)\oplus4(0,11/2)\oplus(1/2,3)\oplus2(1/2,4)\oplus3(1/2,5)\oplus3(1/2,6)\oplus(1/2,7)\oplus(1,9/2)\oplus2(1,11/2)\oplus3(1,13/2)\oplus(3/2,6)\oplus(3/2,7)\oplus(2,15/2)$&$(0, 0, 0, 1, 1, 0, 0, 3)$&$(0,5/2)\oplus(0,7/2)$\\ \hline 
 $(0, 1, 0, 1, 1, 0, 0, 3)$&$(0,3/2)\oplus2(0,5/2)\oplus(0,7/2)$&$(0, 0, 0, 1, 1, 1, 0, 3)$&$(0,3/2)\oplus2(0,5/2)\oplus(0,7/2)$\\ \hline 
 $(0, 0, 0, 1, 2, 0, 0, 3)$&$(0,3/2)\oplus3(0,5/2)\oplus5(0,7/2)\oplus3(0,9/2)\oplus(1/2,3)\oplus3(1/2,4)\oplus2(1/2,5)\oplus(1,9/2)\oplus(1,11/2)$&$(0, 0, 0, 2, 1, 0, 0, 3)$&$(0,3/2)\oplus(0,5/2)\oplus(0,7/2)$\\ \hline 
 $(0, 0, 1, 1, 1, 0, 0, 3)$&$(0,5/2)\oplus(0,7/2)$&$(1, 1, 0, 0, 1, 0, 0, 3)$&$(0,5/2)\oplus(0,7/2)$\\ \hline 
 $(0, 0, 0, 0, 1, 0, 0, 4)$&$(0,9/2)$&$(0, 1, 0, 0, 1, 0, 0, 4)$&$(0,7/2)\oplus(0,9/2)$\\ \hline 
 $(0, 0, 0, 0, 1, 1, 0, 4)$&$(0,7/2)\oplus(0,9/2)$&$(0, 0, 0, 0, 2, 0, 0, 4)$&$(0,5/2)\oplus(0,7/2)\oplus2(0,9/2)\oplus2(0,11/2)\oplus(1/2,4)\oplus(1/2,5)\oplus2(1/2,6)\oplus(1,11/2)\oplus(1,13/2)\oplus(3/2,7)$\\ \hline 
 $(0, 0, 0, 1, 1, 0, 0, 4)$&$(0,7/2)\oplus(0,9/2)$&$(0, 0, 0, 0, 1, 0, 0, 5)$&$(0,11/2)$\\ \hline 
  \end{supertabular}}
}  
\footnotesize{

\tablehead{Header of first column & Header of second column \\}
 \tablefirsthead{%
   \hline
   \multicolumn{1}{|c}{ ${\beta}$} &
   \multicolumn{1}{|c||}{$\oplus N^{{\beta}}_{j_l,j_r}(j_l,j_r)$} &
   ${\beta}$ &
   \multicolumn{1}{c|}{$\oplus N^{{\beta}}_{j_l,j_r}(j_l,j_r)$} \\
   \hline}
 \tablehead{%
   \hline
   \multicolumn{1}{|c}{ ${\beta}$} &
   \multicolumn{1}{|c||}{$\oplus N^{{\beta}}_{j_l,j_r}(j_l,j_r)$} &
   ${\beta}$ &
   \multicolumn{1}{c|}{$\oplus N^{{\beta}}_{j_l,j_r}(j_l,j_r)$} \\
   \hline}
 \tabletail{%
   \hline
   \multicolumn{4}{r}{\small\emph{continued on next page}}\\
   }
 \tablelasttail{\hline}
 \bottomcaption{Refined BPS invariants of 6d $E_7$ minimal SCFT.}
\center{
 \begin{supertabular}{|c|p{5.0cm}||c|p{5.0cm}|}\label{tb:E_7-BPS}$(0, 0, 0, 0, 0, 0, 0, 0, 1)$&$(0,1/2)$&$(0, 0, 0, 0, 0, 0, 0, 1, 1)$&$(0,1/2)$\\ \hline 
 $(0, 0, 0, 0, 0, 0, 0, 2, 1)$&$(0,3/2)$&$(0, 0, 0, 0, 0, 0, 0, 3, 1)$&$(0,5/2)$\\ \hline 
 $(0, 0, 0, 0, 0, 0, 0, 4, 1)$&$(0,7/2)$&$(0, 0, 0, 0, 0, 0, 0, 5, 1)$&$(0,9/2)$\\ \hline 
 $(0, 0, 0, 0, 1, 0, 0, 0, 1)$&$(0,1/2)$&$(0, 0, 0, 0, 1, 1, 0, 0, 1)$&$(0,1/2)$\\ \hline 
 $(0, 0, 0, 0, 1, 1, 1, 0, 1)$&$(0,1/2)$&$(0, 0, 0, 0, 2, 0, 0, 0, 1)$&$(0,3/2)$\\ \hline 
 $(0, 0, 0, 0, 2, 1, 0, 0, 1)$&$(0,1/2)\oplus(0,3/2)$&$(0, 0, 0, 0, 2, 1, 1, 0, 1)$&$(0,1/2)\oplus(0,3/2)$\\ \hline 
 $(0, 0, 0, 0, 2, 2, 0, 0, 1)$&$(0,1/2)\oplus(0,3/2)$&$(0, 0, 0, 0, 2, 2, 1, 0, 1)$&$2(0,1/2)\oplus(0,3/2)$\\ \hline 
 $(0, 0, 0, 0, 2, 3, 0, 0, 1)$&$(0,3/2)$&$(0, 0, 0, 0, 3, 0, 0, 0, 1)$&$(0,5/2)$\\ \hline 
 $(0, 0, 0, 0, 3, 1, 0, 0, 1)$&$(0,3/2)\oplus(0,5/2)$&$(0, 0, 0, 0, 3, 1, 1, 0, 1)$&$(0,3/2)\oplus(0,5/2)$\\ \hline 
 $(0, 0, 0, 0, 3, 2, 0, 0, 1)$&$(0,1/2)\oplus(0,3/2)\oplus(0,5/2)$&$(0, 0, 0, 0, 4, 0, 0, 0, 1)$&$(0,7/2)$\\ \hline 
 $(0, 0, 0, 0, 4, 1, 0, 0, 1)$&$(0,5/2)\oplus(0,7/2)$&$(0, 0, 0, 0, 5, 0, 0, 0, 1)$&$(0,9/2)$\\ \hline 
 $(0, 0, 0, 1, 0, 0, 0, 0, 1)$&$(0,3/2)$&$(0, 0, 0, 1, 0, 0, 0, 1, 1)$&$(0,1/2)\oplus(0,3/2)$\\ \hline 
 $(0, 0, 0, 1, 0, 0, 0, 2, 1)$&$(0,1/2)\oplus(0,3/2)$&$(0, 0, 0, 1, 0, 0, 0, 3, 1)$&$(0,3/2)\oplus(0,5/2)$\\ \hline 
 $(0, 0, 0, 1, 0, 0, 0, 4, 1)$&$(0,5/2)\oplus(0,7/2)$&$(0, 0, 0, 1, 1, 0, 0, 0, 1)$&$(0,1/2)\oplus(0,3/2)$\\ \hline 
 $(0, 0, 0, 1, 1, 0, 0, 1, 1)$&$2(0,1/2)\oplus(0,3/2)$&$(0, 0, 0, 1, 1, 0, 0, 2, 1)$&$(0,1/2)\oplus(0,3/2)$\\ \hline 
 $(0, 0, 0, 1, 1, 0, 0, 3, 1)$&$(0,3/2)\oplus(0,5/2)$&$(0, 0, 0, 1, 1, 1, 0, 0, 1)$&$(0,1/2)\oplus(0,3/2)$\\ \hline 
 $(0, 0, 0, 1, 1, 1, 0, 1, 1)$&$2(0,1/2)\oplus(0,3/2)$&$(0, 0, 0, 1, 1, 1, 0, 2, 1)$&$(0,1/2)\oplus(0,3/2)$\\ \hline 
 $(0, 0, 0, 1, 1, 1, 1, 0, 1)$&$(0,1/2)\oplus(0,3/2)$&$(0, 0, 0, 1, 1, 1, 1, 1, 1)$&$2(0,1/2)\oplus(0,3/2)$\\ \hline 
 $(0, 0, 0, 1, 2, 0, 0, 0, 1)$&$(0,1/2)\oplus(0,3/2)$&$(0, 0, 0, 1, 2, 0, 0, 1, 1)$&$(0,1/2)\oplus(0,3/2)$\\ \hline 
 $(0, 0, 0, 1, 2, 1, 0, 0, 1)$&$2(0,1/2)\oplus(0,3/2)$&$(0, 0, 0, 1, 2, 1, 0, 1, 1)$&$2(0,1/2)\oplus(0,3/2)$\\ \hline 
 $(0, 0, 0, 1, 2, 1, 1, 0, 1)$&$2(0,1/2)\oplus(0,3/2)$&$(0, 0, 0, 1, 2, 2, 0, 0, 1)$&$(0,1/2)\oplus(0,3/2)$\\ \hline 
 $(0, 0, 0, 1, 3, 0, 0, 0, 1)$&$(0,3/2)\oplus(0,5/2)$&$(0, 0, 0, 1, 3, 0, 0, 1, 1)$&$(0,3/2)\oplus(0,5/2)$\\ \hline 
 $(0, 0, 0, 1, 3, 1, 0, 0, 1)$&$(0,1/2)\oplus2(0,3/2)\oplus(0,5/2)$&$(0, 0, 0, 1, 4, 0, 0, 0, 1)$&$(0,5/2)\oplus(0,7/2)$\\ \hline 
 $(0, 0, 0, 2, 0, 0, 0, 0, 1)$&$(0,5/2)$&$(0, 0, 0, 2, 0, 0, 0, 1, 1)$&$(0,3/2)\oplus(0,5/2)$\\ \hline 
 $(0, 0, 0, 2, 0, 0, 0, 2, 1)$&$(0,1/2)\oplus(0,3/2)\oplus(0,5/2)$&$(0, 0, 0, 2, 0, 0, 0, 3, 1)$&$(0,1/2)\oplus(0,3/2)\oplus(0,5/2)$\\ \hline 
 $(0, 0, 0, 2, 1, 0, 0, 0, 1)$&$(0,3/2)\oplus(0,5/2)$&$(0, 0, 0, 2, 1, 0, 0, 1, 1)$&$(0,1/2)\oplus2(0,3/2)\oplus(0,5/2)$\\ \hline 
 $(0, 0, 0, 2, 1, 0, 0, 2, 1)$&$2(0,1/2)\oplus2(0,3/2)\oplus(0,5/2)$&$(0, 0, 0, 2, 1, 1, 0, 0, 1)$&$(0,3/2)\oplus(0,5/2)$\\ \hline 
 $(0, 0, 0, 2, 1, 1, 0, 1, 1)$&$(0,1/2)\oplus2(0,3/2)\oplus(0,5/2)$&$(0, 0, 0, 2, 1, 1, 1, 0, 1)$&$(0,3/2)\oplus(0,5/2)$\\ \hline 
 $(0, 0, 0, 2, 2, 0, 0, 0, 1)$&$(0,1/2)\oplus(0,3/2)\oplus(0,5/2)$&$(0, 0, 0, 2, 2, 0, 0, 1, 1)$&$2(0,1/2)\oplus2(0,3/2)\oplus(0,5/2)$\\ \hline 
 $(0, 0, 0, 2, 2, 1, 0, 0, 1)$&$(0,1/2)\oplus2(0,3/2)\oplus(0,5/2)$&$(0, 0, 0, 2, 3, 0, 0, 0, 1)$&$(0,1/2)\oplus(0,3/2)\oplus(0,5/2)$\\ \hline 
 $(0, 0, 0, 3, 0, 0, 0, 0, 1)$&$(0,7/2)$&$(0, 0, 0, 3, 0, 0, 0, 1, 1)$&$(0,5/2)\oplus(0,7/2)$\\ \hline 
 $(0, 0, 0, 3, 0, 0, 0, 2, 1)$&$(0,3/2)\oplus(0,5/2)\oplus(0,7/2)$&$(0, 0, 0, 3, 1, 0, 0, 0, 1)$&$(0,5/2)\oplus(0,7/2)$\\ \hline 
 $(0, 0, 0, 3, 1, 0, 0, 1, 1)$&$(0,3/2)\oplus2(0,5/2)\oplus(0,7/2)$&$(0, 0, 0, 3, 1, 1, 0, 0, 1)$&$(0,5/2)\oplus(0,7/2)$\\ \hline 
 $(0, 0, 0, 3, 2, 0, 0, 0, 1)$&$(0,3/2)\oplus(0,5/2)\oplus(0,7/2)$&$(0, 0, 0, 4, 0, 0, 0, 0, 1)$&$(0,9/2)$\\ \hline 
 $(0, 0, 0, 4, 0, 0, 0, 1, 1)$&$(0,7/2)\oplus(0,9/2)$&$(0, 0, 0, 4, 1, 0, 0, 0, 1)$&$(0,7/2)\oplus(0,9/2)$\\ \hline 
 $(0, 0, 0, 5, 0, 0, 0, 0, 1)$&$(0,11/2)$&$(0, 0, 1, 0, 0, 0, 0, 0, 1)$&$(0,1/2)$\\ \hline 
 $(0, 0, 1, 1, 0, 0, 0, 0, 1)$&$(0,1/2)\oplus(0,3/2)$&$(0, 0, 1, 1, 0, 0, 0, 1, 1)$&$2(0,1/2)\oplus(0,3/2)$\\ \hline 
 $(0, 0, 1, 1, 0, 0, 0, 2, 1)$&$(0,1/2)\oplus(0,3/2)$&$(0, 0, 1, 1, 0, 0, 0, 3, 1)$&$(0,3/2)\oplus(0,5/2)$\\ \hline 
 $(0, 0, 1, 1, 1, 0, 0, 0, 1)$&$2(0,1/2)\oplus(0,3/2)$&$(0, 0, 1, 1, 1, 0, 0, 1, 1)$&$3(0,1/2)\oplus(0,3/2)$\\ \hline 
 $(0, 0, 1, 1, 1, 0, 0, 2, 1)$&$(0,1/2)\oplus(0,3/2)$&$(0, 0, 1, 1, 1, 1, 0, 0, 1)$&$2(0,1/2)\oplus(0,3/2)$\\ \hline 
 $(0, 0, 1, 1, 1, 1, 0, 1, 1)$&$3(0,1/2)\oplus(0,3/2)$&$(0, 0, 1, 1, 1, 1, 1, 0, 1)$&$2(0,1/2)\oplus(0,3/2)$\\ \hline 
 $(0, 0, 1, 1, 2, 0, 0, 0, 1)$&$(0,1/2)\oplus(0,3/2)$&$(0, 0, 1, 1, 2, 0, 0, 1, 1)$&$(0,1/2)\oplus(0,3/2)$\\ \hline 
 $(0, 0, 1, 1, 2, 1, 0, 0, 1)$&$2(0,1/2)\oplus(0,3/2)$&$(0, 0, 1, 1, 3, 0, 0, 0, 1)$&$(0,3/2)\oplus(0,5/2)$\\ \hline 
 $(0, 0, 1, 2, 0, 0, 0, 0, 1)$&$(0,3/2)\oplus(0,5/2)$&$(0, 0, 1, 2, 0, 0, 0, 1, 1)$&$(0,1/2)\oplus2(0,3/2)\oplus(0,5/2)$\\ \hline 
 $(0, 0, 1, 2, 0, 0, 0, 2, 1)$&$2(0,1/2)\oplus2(0,3/2)\oplus(0,5/2)$&$(0, 0, 1, 2, 1, 0, 0, 0, 1)$&$(0,1/2)\oplus2(0,3/2)\oplus(0,5/2)$\\ \hline 
 $(0, 0, 1, 2, 1, 0, 0, 1, 1)$&$4(0,1/2)\oplus4(0,3/2)\oplus(0,5/2)$&$(0, 0, 1, 2, 1, 1, 0, 0, 1)$&$(0,1/2)\oplus2(0,3/2)\oplus(0,5/2)$\\ \hline 
 $(0, 0, 1, 2, 2, 0, 0, 0, 1)$&$2(0,1/2)\oplus2(0,3/2)\oplus(0,5/2)$&$(0, 0, 1, 3, 0, 0, 0, 0, 1)$&$(0,5/2)\oplus(0,7/2)$\\ \hline 
 $(0, 0, 1, 3, 0, 0, 0, 1, 1)$&$(0,3/2)\oplus2(0,5/2)\oplus(0,7/2)$&$(0, 0, 1, 3, 1, 0, 0, 0, 1)$&$(0,3/2)\oplus2(0,5/2)\oplus(0,7/2)$\\ \hline 
 $(0, 0, 1, 4, 0, 0, 0, 0, 1)$&$(0,7/2)\oplus(0,9/2)$&$(0, 0, 2, 0, 0, 0, 0, 0, 1)$&$(0,3/2)$\\ \hline 
 $(0, 0, 2, 1, 0, 0, 0, 0, 1)$&$(0,1/2)\oplus(0,3/2)$&$(0, 0, 2, 1, 0, 0, 0, 1, 1)$&$(0,1/2)\oplus(0,3/2)$\\ \hline 
 $(0, 0, 2, 1, 1, 0, 0, 0, 1)$&$(0,1/2)\oplus(0,3/2)$&$(0, 0, 2, 1, 1, 0, 0, 1, 1)$&$(0,1/2)\oplus(0,3/2)$\\ \hline 
 $(0, 0, 2, 1, 1, 1, 0, 0, 1)$&$(0,1/2)\oplus(0,3/2)$&$(0, 0, 2, 2, 0, 0, 0, 0, 1)$&$(0,1/2)\oplus(0,3/2)\oplus(0,5/2)$\\ \hline 
 $(0, 0, 2, 2, 0, 0, 0, 1, 1)$&$2(0,1/2)\oplus2(0,3/2)\oplus(0,5/2)$&$(0, 0, 2, 2, 1, 0, 0, 0, 1)$&$2(0,1/2)\oplus2(0,3/2)\oplus(0,5/2)$\\ \hline 
 $(0, 0, 2, 3, 0, 0, 0, 0, 1)$&$(0,3/2)\oplus(0,5/2)\oplus(0,7/2)$&$(0, 0, 3, 0, 0, 0, 0, 0, 1)$&$(0,5/2)$\\ \hline 
 $(0, 0, 3, 1, 0, 0, 0, 0, 1)$&$(0,3/2)\oplus(0,5/2)$&$(0, 0, 3, 1, 0, 0, 0, 1, 1)$&$(0,3/2)\oplus(0,5/2)$\\ \hline 
 $(0, 0, 3, 1, 1, 0, 0, 0, 1)$&$(0,3/2)\oplus(0,5/2)$&$(0, 0, 3, 2, 0, 0, 0, 0, 1)$&$(0,1/2)\oplus(0,3/2)\oplus(0,5/2)$\\ \hline 
 $(0, 0, 4, 0, 0, 0, 0, 0, 1)$&$(0,7/2)$&$(0, 0, 4, 1, 0, 0, 0, 0, 1)$&$(0,5/2)\oplus(0,7/2)$\\ \hline 
 $(0, 0, 5, 0, 0, 0, 0, 0, 1)$&$(0,9/2)$&$(0, 1, 1, 0, 0, 0, 0, 0, 1)$&$(0,1/2)$\\ \hline 
 $(0, 1, 1, 1, 0, 0, 0, 0, 1)$&$(0,1/2)\oplus(0,3/2)$&$(0, 1, 1, 1, 0, 0, 0, 1, 1)$&$2(0,1/2)\oplus(0,3/2)$\\ \hline 
 $(0, 1, 1, 1, 0, 0, 0, 2, 1)$&$(0,1/2)\oplus(0,3/2)$&$(0, 1, 1, 1, 1, 0, 0, 0, 1)$&$2(0,1/2)\oplus(0,3/2)$\\ \hline 
 $(0, 1, 1, 1, 1, 0, 0, 1, 1)$&$3(0,1/2)\oplus(0,3/2)$&$(0, 1, 1, 1, 1, 1, 0, 0, 1)$&$2(0,1/2)\oplus(0,3/2)$\\ \hline 
 $(0, 1, 1, 1, 2, 0, 0, 0, 1)$&$(0,1/2)\oplus(0,3/2)$&$(0, 1, 1, 2, 0, 0, 0, 0, 1)$&$(0,3/2)\oplus(0,5/2)$\\ \hline 
 $(0, 1, 1, 2, 0, 0, 0, 1, 1)$&$(0,1/2)\oplus2(0,3/2)\oplus(0,5/2)$&$(0, 1, 1, 2, 1, 0, 0, 0, 1)$&$(0,1/2)\oplus2(0,3/2)\oplus(0,5/2)$\\ \hline 
 $(0, 1, 1, 3, 0, 0, 0, 0, 1)$&$(0,5/2)\oplus(0,7/2)$&$(0, 1, 2, 0, 0, 0, 0, 0, 1)$&$(0,1/2)\oplus(0,3/2)$\\ \hline 
 $(0, 1, 2, 1, 0, 0, 0, 0, 1)$&$2(0,1/2)\oplus(0,3/2)$&$(0, 1, 2, 1, 0, 0, 0, 1, 1)$&$2(0,1/2)\oplus(0,3/2)$\\ \hline 
 $(0, 1, 2, 1, 1, 0, 0, 0, 1)$&$2(0,1/2)\oplus(0,3/2)$&$(0, 1, 2, 2, 0, 0, 0, 0, 1)$&$(0,1/2)\oplus2(0,3/2)\oplus(0,5/2)$\\ \hline 
 $(0, 1, 3, 0, 0, 0, 0, 0, 1)$&$(0,3/2)\oplus(0,5/2)$&$(0, 1, 3, 1, 0, 0, 0, 0, 1)$&$(0,1/2)\oplus2(0,3/2)\oplus(0,5/2)$\\ \hline 
 $(0, 1, 4, 0, 0, 0, 0, 0, 1)$&$(0,5/2)\oplus(0,7/2)$&$(0, 2, 2, 0, 0, 0, 0, 0, 1)$&$(0,1/2)\oplus(0,3/2)$\\ \hline 
 $(0, 2, 2, 1, 0, 0, 0, 0, 1)$&$(0,1/2)\oplus(0,3/2)$&$(0, 2, 3, 0, 0, 0, 0, 0, 1)$&$(0,1/2)\oplus(0,3/2)\oplus(0,5/2)$\\ \hline 
 $(0, 3, 2, 0, 0, 0, 0, 0, 1)$&$(0,3/2)$&$(1, 1, 1, 0, 0, 0, 0, 0, 1)$&$(0,1/2)$\\ \hline 
 $(1, 1, 1, 1, 0, 0, 0, 0, 1)$&$(0,1/2)\oplus(0,3/2)$&$(1, 1, 1, 1, 0, 0, 0, 1, 1)$&$2(0,1/2)\oplus(0,3/2)$\\ \hline 
 $(1, 1, 1, 1, 1, 0, 0, 0, 1)$&$2(0,1/2)\oplus(0,3/2)$&$(1, 1, 1, 2, 0, 0, 0, 0, 1)$&$(0,3/2)\oplus(0,5/2)$\\ \hline 
 $(1, 1, 2, 0, 0, 0, 0, 0, 1)$&$(0,1/2)\oplus(0,3/2)$&$(1, 1, 2, 1, 0, 0, 0, 0, 1)$&$2(0,1/2)\oplus(0,3/2)$\\ \hline 
 $(1, 1, 3, 0, 0, 0, 0, 0, 1)$&$(0,3/2)\oplus(0,5/2)$&$(1, 2, 2, 0, 0, 0, 0, 0, 1)$&$2(0,1/2)\oplus(0,3/2)$\\ \hline 
 $(0, 0, 0, 0, 0, 0, 0, 3, 2)$&$(0,5/2)$&$(0, 0, 0, 0, 0, 0, 0, 4, 2)$&$(0,5/2)\oplus(0,7/2)\oplus(1/2,4)$\\ \hline 
 $(0, 0, 0, 0, 3, 0, 0, 0, 2)$&$(0,5/2)$&$(0, 0, 0, 0, 3, 1, 0, 0, 2)$&$(0,3/2)\oplus(0,5/2)$\\ \hline 
 $(0, 0, 0, 0, 4, 0, 0, 0, 2)$&$(0,5/2)\oplus(0,7/2)\oplus(1/2,4)$&$(0, 0, 0, 1, 0, 0, 0, 0, 2)$&$(0,5/2)$\\ \hline 
 $(0, 0, 0, 1, 0, 0, 0, 1, 2)$&$(0,3/2)\oplus(0,5/2)$&$(0, 0, 0, 1, 0, 0, 0, 2, 2)$&$(0,1/2)\oplus(0,3/2)\oplus(0,5/2)$\\ \hline 
 $(0, 0, 0, 1, 0, 0, 0, 3, 2)$&$(0,1/2)\oplus2(0,3/2)\oplus2(0,5/2)\oplus(0,7/2)$&$(0, 0, 0, 1, 1, 0, 0, 0, 2)$&$(0,3/2)\oplus(0,5/2)$\\ \hline 
 $(0, 0, 0, 1, 1, 0, 0, 1, 2)$&$(0,1/2)\oplus2(0,3/2)\oplus(0,5/2)$&$(0, 0, 0, 1, 1, 0, 0, 2, 2)$&$2(0,1/2)\oplus2(0,3/2)\oplus(0,5/2)$\\ \hline 
 $(0, 0, 0, 1, 1, 1, 0, 0, 2)$&$(0,3/2)\oplus(0,5/2)$&$(0, 0, 0, 1, 1, 1, 0, 1, 2)$&$(0,1/2)\oplus2(0,3/2)\oplus(0,5/2)$\\ \hline 
 $(0, 0, 0, 1, 1, 1, 1, 0, 2)$&$(0,3/2)\oplus(0,5/2)$&$(0, 0, 0, 1, 2, 0, 0, 0, 2)$&$(0,1/2)\oplus(0,3/2)\oplus(0,5/2)$\\ \hline 
 $(0, 0, 0, 1, 2, 0, 0, 1, 2)$&$2(0,1/2)\oplus2(0,3/2)\oplus(0,5/2)$&$(0, 0, 0, 1, 2, 1, 0, 0, 2)$&$(0,1/2)\oplus2(0,3/2)\oplus(0,5/2)$\\ \hline 
 $(0, 0, 0, 1, 3, 0, 0, 0, 2)$&$(0,1/2)\oplus2(0,3/2)\oplus2(0,5/2)\oplus(0,7/2)$&$(0, 0, 0, 2, 0, 0, 0, 0, 2)$&$(0,5/2)\oplus(0,7/2)\oplus(1/2,4)$\\ \hline 
 $(0, 0, 0, 2, 0, 0, 0, 1, 2)$&$(0,3/2)\oplus3(0,5/2)\oplus2(0,7/2)\oplus(1/2,3)\oplus(1/2,4)$&$(0, 0, 0, 2, 0, 0, 0, 2, 2)$&$(0,1/2)\oplus3(0,3/2)\oplus4(0,5/2)\oplus2(0,7/2)\oplus(1/2,2)\oplus(1/2,3)\oplus(1/2,4)$\\ \hline 
 $(0, 0, 0, 2, 1, 0, 0, 0, 2)$&$(0,3/2)\oplus3(0,5/2)\oplus2(0,7/2)\oplus(1/2,3)\oplus(1/2,4)$&$(0, 0, 0, 2, 1, 0, 0, 1, 2)$&$(0,1/2)\oplus5(0,3/2)\oplus7(0,5/2)\oplus3(0,7/2)\oplus(1/2,2)\oplus2(1/2,3)\oplus(1/2,4)$\\ \hline 
 $(0, 0, 0, 2, 1, 1, 0, 0, 2)$&$(0,3/2)\oplus3(0,5/2)\oplus2(0,7/2)\oplus(1/2,3)\oplus(1/2,4)$&$(0, 0, 0, 2, 2, 0, 0, 0, 2)$&$(0,1/2)\oplus3(0,3/2)\oplus4(0,5/2)\oplus2(0,7/2)\oplus(1/2,2)\oplus(1/2,3)\oplus(1/2,4)$\\ \hline 
 $(0, 0, 0, 3, 0, 0, 0, 0, 2)$&$(0,5/2)\oplus(0,7/2)\oplus2(0,9/2)\oplus(1/2,4)\oplus(1/2,5)\oplus(1,11/2)$&$(0, 0, 0, 3, 0, 0, 0, 1, 2)$&$(0,3/2)\oplus3(0,5/2)\oplus5(0,7/2)\oplus3(0,9/2)\oplus(1/2,3)\oplus3(1/2,4)\oplus2(1/2,5)\oplus(1,9/2)\oplus(1,11/2)$\\ \hline 
 $(0, 0, 0, 3, 1, 0, 0, 0, 2)$&$(0,3/2)\oplus3(0,5/2)\oplus5(0,7/2)\oplus3(0,9/2)\oplus(1/2,3)\oplus3(1/2,4)\oplus2(1/2,5)\oplus(1,9/2)\oplus(1,11/2)$&$(0, 0, 0, 4, 0, 0, 0, 0, 2)$&$(0,5/2)\oplus(0,7/2)\oplus2(0,9/2)\oplus2(0,11/2)\oplus(1/2,4)\oplus(1/2,5)\oplus2(1/2,6)\oplus(1,11/2)\oplus(1,13/2)\oplus(3/2,7)$\\ \hline 
 $(0, 0, 1, 1, 0, 0, 0, 0, 2)$&$(0,3/2)\oplus(0,5/2)$&$(0, 0, 1, 1, 0, 0, 0, 1, 2)$&$(0,1/2)\oplus2(0,3/2)\oplus(0,5/2)$\\ \hline 
 $(0, 0, 1, 1, 0, 0, 0, 2, 2)$&$2(0,1/2)\oplus2(0,3/2)\oplus(0,5/2)$&$(0, 0, 1, 1, 1, 0, 0, 0, 2)$&$(0,1/2)\oplus2(0,3/2)\oplus(0,5/2)$\\ \hline 
 $(0, 0, 1, 1, 1, 0, 0, 1, 2)$&$3(0,1/2)\oplus3(0,3/2)\oplus(0,5/2)$&$(0, 0, 1, 1, 1, 1, 0, 0, 2)$&$(0,1/2)\oplus2(0,3/2)\oplus(0,5/2)$\\ \hline 
 $(0, 0, 1, 1, 2, 0, 0, 0, 2)$&$2(0,1/2)\oplus2(0,3/2)\oplus(0,5/2)$&$(0, 0, 1, 2, 0, 0, 0, 0, 2)$&$(0,3/2)\oplus3(0,5/2)\oplus2(0,7/2)\oplus(1/2,3)\oplus(1/2,4)$\\ \hline 
 $(0, 0, 1, 2, 0, 0, 0, 1, 2)$&$(0,1/2)\oplus5(0,3/2)\oplus7(0,5/2)\oplus3(0,7/2)\oplus(1/2,2)\oplus2(1/2,3)\oplus(1/2,4)$&$(0, 0, 1, 2, 1, 0, 0, 0, 2)$&$(0,1/2)\oplus5(0,3/2)\oplus7(0,5/2)\oplus3(0,7/2)\oplus(1/2,2)\oplus2(1/2,3)\oplus(1/2,4)$\\ \hline 
 $(0, 0, 1, 3, 0, 0, 0, 0, 2)$&$(0,3/2)\oplus3(0,5/2)\oplus5(0,7/2)\oplus3(0,9/2)\oplus(1/2,3)\oplus3(1/2,4)\oplus2(1/2,5)\oplus(1,9/2)\oplus(1,11/2)$&$(0, 0, 2, 1, 0, 0, 0, 0, 2)$&$(0,1/2)\oplus(0,3/2)\oplus(0,5/2)$\\ \hline 
 $(0, 0, 2, 1, 0, 0, 0, 1, 2)$&$2(0,1/2)\oplus2(0,3/2)\oplus(0,5/2)$&$(0, 0, 2, 1, 1, 0, 0, 0, 2)$&$2(0,1/2)\oplus2(0,3/2)\oplus(0,5/2)$\\ \hline 
 $(0, 0, 2, 2, 0, 0, 0, 0, 2)$&$(0,1/2)\oplus3(0,3/2)\oplus4(0,5/2)\oplus2(0,7/2)\oplus(1/2,2)\oplus(1/2,3)\oplus(1/2,4)$&$(0, 0, 3, 0, 0, 0, 0, 0, 2)$&$(0,5/2)$\\ \hline 
 $(0, 0, 3, 1, 0, 0, 0, 0, 2)$&$(0,1/2)\oplus2(0,3/2)\oplus2(0,5/2)\oplus(0,7/2)$&$(0, 0, 4, 0, 0, 0, 0, 0, 2)$&$(0,5/2)\oplus(0,7/2)\oplus(1/2,4)$\\ \hline 
 $(0, 1, 1, 1, 0, 0, 0, 0, 2)$&$(0,3/2)\oplus(0,5/2)$&$(0, 1, 1, 1, 0, 0, 0, 1, 2)$&$(0,1/2)\oplus2(0,3/2)\oplus(0,5/2)$\\ \hline 
 $(0, 1, 1, 1, 1, 0, 0, 0, 2)$&$(0,1/2)\oplus2(0,3/2)\oplus(0,5/2)$&$(0, 1, 1, 2, 0, 0, 0, 0, 2)$&$(0,3/2)\oplus3(0,5/2)\oplus2(0,7/2)\oplus(1/2,3)\oplus(1/2,4)$\\ \hline 
 $(0, 1, 2, 1, 0, 0, 0, 0, 2)$&$(0,1/2)\oplus2(0,3/2)\oplus(0,5/2)$&$(0, 1, 3, 0, 0, 0, 0, 0, 2)$&$(0,3/2)\oplus(0,5/2)$\\ \hline 
 $(1, 1, 1, 1, 0, 0, 0, 0, 2)$&$(0,3/2)\oplus(0,5/2)$&$(0, 0, 0, 1, 0, 0, 0, 0, 3)$&$(0,7/2)$\\ \hline 
 $(0, 0, 0, 1, 0, 0, 0, 1, 3)$&$(0,5/2)\oplus(0,7/2)$&$(0, 0, 0, 1, 0, 0, 0, 2, 3)$&$(0,3/2)\oplus(0,5/2)\oplus(0,7/2)$\\ \hline 
 $(0, 0, 0, 1, 1, 0, 0, 0, 3)$&$(0,5/2)\oplus(0,7/2)$&$(0, 0, 0, 1, 1, 0, 0, 1, 3)$&$(0,3/2)\oplus2(0,5/2)\oplus(0,7/2)$\\ \hline 
 $(0, 0, 0, 1, 1, 1, 0, 0, 3)$&$(0,5/2)\oplus(0,7/2)$&$(0, 0, 0, 1, 2, 0, 0, 0, 3)$&$(0,3/2)\oplus(0,5/2)\oplus(0,7/2)$\\ \hline 
 $(0, 0, 0, 2, 0, 0, 0, 0, 3)$&$(0,5/2)\oplus(0,7/2)\oplus2(0,9/2)\oplus(1/2,4)\oplus(1/2,5)\oplus(1,11/2)$&$(0, 0, 0, 2, 0, 0, 0, 1, 3)$&$(0,3/2)\oplus3(0,5/2)\oplus5(0,7/2)\oplus3(0,9/2)\oplus(1/2,3)\oplus3(1/2,4)\oplus2(1/2,5)\oplus(1,9/2)\oplus(1,11/2)$\\ \hline 
 $(0, 0, 0, 2, 1, 0, 0, 0, 3)$&$(0,3/2)\oplus3(0,5/2)\oplus5(0,7/2)\oplus3(0,9/2)\oplus(1/2,3)\oplus3(1/2,4)\oplus2(1/2,5)\oplus(1,9/2)\oplus(1,11/2)$&$(0, 0, 0, 3, 0, 0, 0, 0, 3)$&$(0,3/2)\oplus(0,5/2)\oplus3(0,7/2)\oplus3(0,9/2)\oplus4(0,11/2)\oplus(1/2,3)\oplus2(1/2,4)\oplus3(1/2,5)\oplus3(1/2,6)\oplus(1/2,7)\oplus(1,9/2)\oplus2(1,11/2)\oplus3(1,13/2)\oplus(3/2,6)\oplus(3/2,7)\oplus(2,15/2)$\\ \hline 
 $(0, 0, 1, 1, 0, 0, 0, 0, 3)$&$(0,5/2)\oplus(0,7/2)$&$(0, 0, 1, 1, 0, 0, 0, 1, 3)$&$(0,3/2)\oplus2(0,5/2)\oplus(0,7/2)$\\ \hline 
 $(0, 0, 1, 1, 1, 0, 0, 0, 3)$&$(0,3/2)\oplus2(0,5/2)\oplus(0,7/2)$&$(0, 0, 1, 2, 0, 0, 0, 0, 3)$&$(0,3/2)\oplus3(0,5/2)\oplus5(0,7/2)\oplus3(0,9/2)\oplus(1/2,3)\oplus3(1/2,4)\oplus2(1/2,5)\oplus(1,9/2)\oplus(1,11/2)$\\ \hline 
 $(0, 0, 2, 1, 0, 0, 0, 0, 3)$&$(0,3/2)\oplus(0,5/2)\oplus(0,7/2)$&$(0, 1, 1, 1, 0, 0, 0, 0, 3)$&$(0,5/2)\oplus(0,7/2)$\\ \hline 
 $(0, 0, 0, 1, 0, 0, 0, 0, 4)$&$(0,9/2)$&$(0, 0, 0, 1, 0, 0, 0, 1, 4)$&$(0,7/2)\oplus(0,9/2)$\\ \hline 
 $(0, 0, 0, 1, 1, 0, 0, 0, 4)$&$(0,7/2)\oplus(0,9/2)$&$(0, 0, 0, 2, 0, 0, 0, 0, 4)$&$(0,5/2)\oplus(0,7/2)\oplus2(0,9/2)\oplus2(0,11/2)\oplus(1/2,4)\oplus(1/2,5)\oplus2(1/2,6)\oplus(1,11/2)\oplus(1,13/2)\oplus(3/2,7)$\\ \hline 
 $(0, 0, 1, 1, 0, 0, 0, 0, 4)$&$(0,7/2)\oplus(0,9/2)$&$(0, 0, 0, 1, 0, 0, 0, 0, 5)$&$(0,11/2)$\\ \hline 
  \end{supertabular}}
}  
\footnotesize{

\tablehead{Header of first column & Header of second column \\}
 \tablefirsthead{%
   \hline
   \multicolumn{1}{|c}{ ${\beta}$} &
   \multicolumn{1}{|c||}{$\oplus N^{{\beta}}_{j_l,j_r}(j_l,j_r)$} &
   ${\beta}$ &
   \multicolumn{1}{c|}{$\oplus N^{{\beta}}_{j_l,j_r}(j_l,j_r)$} \\
   \hline}
 \tablehead{%
   \hline
   \multicolumn{1}{|c}{ ${\beta}$} &
   \multicolumn{1}{|c||}{$\oplus N^{{\beta}}_{j_l,j_r}(j_l,j_r)$} &
   ${\beta}$ &
   \multicolumn{1}{c|}{$\oplus N^{{\beta}}_{j_l,j_r}(j_l,j_r)$} \\
   \hline}
 \tabletail{%
   \hline
   \multicolumn{4}{r}{\small\emph{continued on next page}}\\
   }
 \tablelasttail{\hline}
 \bottomcaption{Refined BPS invariants of 6d $E_8$ minimal SCFT.}
\center{
 \begin{supertabular}{|c|p{4.7cm}||c|p{4.7cm}|}\label{tb:E_8-BPS}$(0, 0, 0, 0, 0, 0, 0, 0, 0, 1)$&$(0,1/2)$&$(0, 0, 0, 0, 0, 0, 0, 0, 1, 1)$&$(0,1/2)$\\ \hline 
 $(0, 0, 0, 0, 0, 0, 0, 0, 2, 1)$&$(0,3/2)$&$(0, 0, 0, 0, 0, 0, 0, 0, 3, 1)$&$(0,5/2)$\\ \hline 
 $(0, 0, 0, 0, 0, 0, 0, 0, 4, 1)$&$(0,7/2)$&$(0, 0, 0, 0, 0, 0, 0, 0, 5, 1)$&$(0,9/2)$\\ \hline 
 $(0, 0, 0, 0, 0, 0, 1, 0, 0, 1)$&$(0,1/2)$&$(0, 0, 0, 0, 0, 0, 1, 1, 0, 1)$&$(0,1/2)$\\ \hline 
 $(0, 0, 0, 0, 0, 0, 2, 0, 0, 1)$&$(0,3/2)$&$(0, 0, 0, 0, 0, 0, 2, 1, 0, 1)$&$(0,1/2)\oplus(0,3/2)$\\ \hline 
 $(0, 0, 0, 0, 0, 0, 2, 2, 0, 1)$&$(0,1/2)\oplus(0,3/2)$&$(0, 0, 0, 0, 0, 0, 2, 3, 0, 1)$&$(0,3/2)$\\ \hline 
 $(0, 0, 0, 0, 0, 0, 3, 0, 0, 1)$&$(0,5/2)$&$(0, 0, 0, 0, 0, 0, 3, 1, 0, 1)$&$(0,3/2)\oplus(0,5/2)$\\ \hline 
 $(0, 0, 0, 0, 0, 0, 3, 2, 0, 1)$&$(0,1/2)\oplus(0,3/2)\oplus(0,5/2)$&$(0, 0, 0, 0, 0, 0, 4, 0, 0, 1)$&$(0,7/2)$\\ \hline 
 $(0, 0, 0, 0, 0, 0, 4, 1, 0, 1)$&$(0,5/2)\oplus(0,7/2)$&$(0, 0, 0, 0, 0, 0, 5, 0, 0, 1)$&$(0,9/2)$\\ \hline 
 $(0, 0, 0, 0, 0, 1, 0, 0, 0, 1)$&$(0,3/2)$&$(0, 0, 0, 0, 0, 1, 0, 0, 1, 1)$&$(0,1/2)\oplus(0,3/2)$\\ \hline 
 $(0, 0, 0, 0, 0, 1, 0, 0, 2, 1)$&$(0,1/2)\oplus(0,3/2)$&$(0, 0, 0, 0, 0, 1, 0, 0, 3, 1)$&$(0,3/2)\oplus(0,5/2)$\\ \hline 
 $(0, 0, 0, 0, 0, 1, 0, 0, 4, 1)$&$(0,5/2)\oplus(0,7/2)$&$(0, 0, 0, 0, 0, 1, 1, 0, 0, 1)$&$(0,1/2)\oplus(0,3/2)$\\ \hline 
 $(0, 0, 0, 0, 0, 1, 1, 0, 1, 1)$&$2(0,1/2)\oplus(0,3/2)$&$(0, 0, 0, 0, 0, 1, 1, 0, 2, 1)$&$(0,1/2)\oplus(0,3/2)$\\ \hline 
 $(0, 0, 0, 0, 0, 1, 1, 0, 3, 1)$&$(0,3/2)\oplus(0,5/2)$&$(0, 0, 0, 0, 0, 1, 1, 1, 0, 1)$&$(0,1/2)\oplus(0,3/2)$\\ \hline 
 $(0, 0, 0, 0, 0, 1, 1, 1, 1, 1)$&$2(0,1/2)\oplus(0,3/2)$&$(0, 0, 0, 0, 0, 1, 1, 1, 2, 1)$&$(0,1/2)\oplus(0,3/2)$\\ \hline 
 $(0, 0, 0, 0, 0, 1, 2, 0, 0, 1)$&$(0,1/2)\oplus(0,3/2)$&$(0, 0, 0, 0, 0, 1, 2, 0, 1, 1)$&$(0,1/2)\oplus(0,3/2)$\\ \hline 
 $(0, 0, 0, 0, 0, 1, 2, 1, 0, 1)$&$2(0,1/2)\oplus(0,3/2)$&$(0, 0, 0, 0, 0, 1, 2, 1, 1, 1)$&$2(0,1/2)\oplus(0,3/2)$\\ \hline 
 $(0, 0, 0, 0, 0, 1, 2, 2, 0, 1)$&$(0,1/2)\oplus(0,3/2)$&$(0, 0, 0, 0, 0, 1, 3, 0, 0, 1)$&$(0,3/2)\oplus(0,5/2)$\\ \hline 
 $(0, 0, 0, 0, 0, 1, 3, 0, 1, 1)$&$(0,3/2)\oplus(0,5/2)$&$(0, 0, 0, 0, 0, 1, 3, 1, 0, 1)$&$(0,1/2)\oplus2(0,3/2)\oplus(0,5/2)$\\ \hline 
 $(0, 0, 0, 0, 0, 1, 4, 0, 0, 1)$&$(0,5/2)\oplus(0,7/2)$&$(0, 0, 0, 0, 0, 2, 0, 0, 0, 1)$&$(0,5/2)$\\ \hline 
 $(0, 0, 0, 0, 0, 2, 0, 0, 1, 1)$&$(0,3/2)\oplus(0,5/2)$&$(0, 0, 0, 0, 0, 2, 0, 0, 2, 1)$&$(0,1/2)\oplus(0,3/2)\oplus(0,5/2)$\\ \hline 
 $(0, 0, 0, 0, 0, 2, 0, 0, 3, 1)$&$(0,1/2)\oplus(0,3/2)\oplus(0,5/2)$&$(0, 0, 0, 0, 0, 2, 1, 0, 0, 1)$&$(0,3/2)\oplus(0,5/2)$\\ \hline 
 $(0, 0, 0, 0, 0, 2, 1, 0, 1, 1)$&$(0,1/2)\oplus2(0,3/2)\oplus(0,5/2)$&$(0, 0, 0, 0, 0, 2, 1, 0, 2, 1)$&$2(0,1/2)\oplus2(0,3/2)\oplus(0,5/2)$\\ \hline 
 $(0, 0, 0, 0, 0, 2, 1, 1, 0, 1)$&$(0,3/2)\oplus(0,5/2)$&$(0, 0, 0, 0, 0, 2, 1, 1, 1, 1)$&$(0,1/2)\oplus2(0,3/2)\oplus(0,5/2)$\\ \hline 
 $(0, 0, 0, 0, 0, 2, 2, 0, 0, 1)$&$(0,1/2)\oplus(0,3/2)\oplus(0,5/2)$&$(0, 0, 0, 0, 0, 2, 2, 0, 1, 1)$&$2(0,1/2)\oplus2(0,3/2)\oplus(0,5/2)$\\ \hline 
 $(0, 0, 0, 0, 0, 2, 2, 1, 0, 1)$&$(0,1/2)\oplus2(0,3/2)\oplus(0,5/2)$&$(0, 0, 0, 0, 0, 2, 3, 0, 0, 1)$&$(0,1/2)\oplus(0,3/2)\oplus(0,5/2)$\\ \hline 
 $(0, 0, 0, 0, 0, 3, 0, 0, 0, 1)$&$(0,7/2)$&$(0, 0, 0, 0, 0, 3, 0, 0, 1, 1)$&$(0,5/2)\oplus(0,7/2)$\\ \hline 
 $(0, 0, 0, 0, 0, 3, 0, 0, 2, 1)$&$(0,3/2)\oplus(0,5/2)\oplus(0,7/2)$&$(0, 0, 0, 0, 0, 3, 1, 0, 0, 1)$&$(0,5/2)\oplus(0,7/2)$\\ \hline 
 $(0, 0, 0, 0, 0, 3, 1, 0, 1, 1)$&$(0,3/2)\oplus2(0,5/2)\oplus(0,7/2)$&$(0, 0, 0, 0, 0, 3, 1, 1, 0, 1)$&$(0,5/2)\oplus(0,7/2)$\\ \hline 
 $(0, 0, 0, 0, 0, 3, 2, 0, 0, 1)$&$(0,3/2)\oplus(0,5/2)\oplus(0,7/2)$&$(0, 0, 0, 0, 0, 4, 0, 0, 0, 1)$&$(0,9/2)$\\ \hline 
 $(0, 0, 0, 0, 0, 4, 0, 0, 1, 1)$&$(0,7/2)\oplus(0,9/2)$&$(0, 0, 0, 0, 0, 4, 1, 0, 0, 1)$&$(0,7/2)\oplus(0,9/2)$\\ \hline 
 $(0, 0, 0, 0, 0, 5, 0, 0, 0, 1)$&$(0,11/2)$&$(0, 0, 0, 0, 1, 0, 0, 0, 0, 1)$&$(0,1/2)$\\ \hline 
 $(0, 0, 0, 0, 1, 1, 0, 0, 0, 1)$&$(0,1/2)\oplus(0,3/2)$&$(0, 0, 0, 0, 1, 1, 0, 0, 1, 1)$&$2(0,1/2)\oplus(0,3/2)$\\ \hline 
 $(0, 0, 0, 0, 1, 1, 0, 0, 2, 1)$&$(0,1/2)\oplus(0,3/2)$&$(0, 0, 0, 0, 1, 1, 0, 0, 3, 1)$&$(0,3/2)\oplus(0,5/2)$\\ \hline 
 $(0, 0, 0, 0, 1, 1, 1, 0, 0, 1)$&$2(0,1/2)\oplus(0,3/2)$&$(0, 0, 0, 0, 1, 1, 1, 0, 1, 1)$&$3(0,1/2)\oplus(0,3/2)$\\ \hline 
 $(0, 0, 0, 0, 1, 1, 1, 0, 2, 1)$&$(0,1/2)\oplus(0,3/2)$&$(0, 0, 0, 0, 1, 1, 1, 1, 0, 1)$&$2(0,1/2)\oplus(0,3/2)$\\ \hline 
 $(0, 0, 0, 0, 1, 1, 1, 1, 1, 1)$&$3(0,1/2)\oplus(0,3/2)$&$(0, 0, 0, 0, 1, 1, 2, 0, 0, 1)$&$(0,1/2)\oplus(0,3/2)$\\ \hline 
 $(0, 0, 0, 0, 1, 1, 2, 0, 1, 1)$&$(0,1/2)\oplus(0,3/2)$&$(0, 0, 0, 0, 1, 1, 2, 1, 0, 1)$&$2(0,1/2)\oplus(0,3/2)$\\ \hline 
 $(0, 0, 0, 0, 1, 1, 3, 0, 0, 1)$&$(0,3/2)\oplus(0,5/2)$&$(0, 0, 0, 0, 1, 2, 0, 0, 0, 1)$&$(0,3/2)\oplus(0,5/2)$\\ \hline 
 $(0, 0, 0, 0, 1, 2, 0, 0, 1, 1)$&$(0,1/2)\oplus2(0,3/2)\oplus(0,5/2)$&$(0, 0, 0, 0, 1, 2, 0, 0, 2, 1)$&$2(0,1/2)\oplus2(0,3/2)\oplus(0,5/2)$\\ \hline 
 $(0, 0, 0, 0, 1, 2, 1, 0, 0, 1)$&$(0,1/2)\oplus2(0,3/2)\oplus(0,5/2)$&$(0, 0, 0, 0, 1, 2, 1, 0, 1, 1)$&$4(0,1/2)\oplus4(0,3/2)\oplus(0,5/2)$\\ \hline 
 $(0, 0, 0, 0, 1, 2, 1, 1, 0, 1)$&$(0,1/2)\oplus2(0,3/2)\oplus(0,5/2)$&$(0, 0, 0, 0, 1, 2, 2, 0, 0, 1)$&$2(0,1/2)\oplus2(0,3/2)\oplus(0,5/2)$\\ \hline 
 $(0, 0, 0, 0, 1, 3, 0, 0, 0, 1)$&$(0,5/2)\oplus(0,7/2)$&$(0, 0, 0, 0, 1, 3, 0, 0, 1, 1)$&$(0,3/2)\oplus2(0,5/2)\oplus(0,7/2)$\\ \hline 
 $(0, 0, 0, 0, 1, 3, 1, 0, 0, 1)$&$(0,3/2)\oplus2(0,5/2)\oplus(0,7/2)$&$(0, 0, 0, 0, 1, 4, 0, 0, 0, 1)$&$(0,7/2)\oplus(0,9/2)$\\ \hline 
 $(0, 0, 0, 0, 2, 0, 0, 0, 0, 1)$&$(0,3/2)$&$(0, 0, 0, 0, 2, 1, 0, 0, 0, 1)$&$(0,1/2)\oplus(0,3/2)$\\ \hline 
 $(0, 0, 0, 0, 2, 1, 0, 0, 1, 1)$&$(0,1/2)\oplus(0,3/2)$&$(0, 0, 0, 0, 2, 1, 1, 0, 0, 1)$&$(0,1/2)\oplus(0,3/2)$\\ \hline 
 $(0, 0, 0, 0, 2, 1, 1, 0, 1, 1)$&$(0,1/2)\oplus(0,3/2)$&$(0, 0, 0, 0, 2, 1, 1, 1, 0, 1)$&$(0,1/2)\oplus(0,3/2)$\\ \hline 
 $(0, 0, 0, 0, 2, 2, 0, 0, 0, 1)$&$(0,1/2)\oplus(0,3/2)\oplus(0,5/2)$&$(0, 0, 0, 0, 2, 2, 0, 0, 1, 1)$&$2(0,1/2)\oplus2(0,3/2)\oplus(0,5/2)$\\ \hline 
 $(0, 0, 0, 0, 2, 2, 1, 0, 0, 1)$&$2(0,1/2)\oplus2(0,3/2)\oplus(0,5/2)$&$(0, 0, 0, 0, 2, 3, 0, 0, 0, 1)$&$(0,3/2)\oplus(0,5/2)\oplus(0,7/2)$\\ \hline 
 $(0, 0, 0, 0, 3, 0, 0, 0, 0, 1)$&$(0,5/2)$&$(0, 0, 0, 0, 3, 1, 0, 0, 0, 1)$&$(0,3/2)\oplus(0,5/2)$\\ \hline 
 $(0, 0, 0, 0, 3, 1, 0, 0, 1, 1)$&$(0,3/2)\oplus(0,5/2)$&$(0, 0, 0, 0, 3, 1, 1, 0, 0, 1)$&$(0,3/2)\oplus(0,5/2)$\\ \hline 
 $(0, 0, 0, 0, 3, 2, 0, 0, 0, 1)$&$(0,1/2)\oplus(0,3/2)\oplus(0,5/2)$&$(0, 0, 0, 0, 4, 0, 0, 0, 0, 1)$&$(0,7/2)$\\ \hline 
 $(0, 0, 0, 0, 4, 1, 0, 0, 0, 1)$&$(0,5/2)\oplus(0,7/2)$&$(0, 0, 0, 0, 5, 0, 0, 0, 0, 1)$&$(0,9/2)$\\ \hline 
 $(0, 0, 0, 1, 1, 0, 0, 0, 0, 1)$&$(0,1/2)$&$(0, 0, 0, 1, 1, 1, 0, 0, 0, 1)$&$(0,1/2)\oplus(0,3/2)$\\ \hline 
 $(0, 0, 0, 1, 1, 1, 0, 0, 1, 1)$&$2(0,1/2)\oplus(0,3/2)$&$(0, 0, 0, 1, 1, 1, 0, 0, 2, 1)$&$(0,1/2)\oplus(0,3/2)$\\ \hline 
 $(0, 0, 0, 1, 1, 1, 1, 0, 0, 1)$&$2(0,1/2)\oplus(0,3/2)$&$(0, 0, 0, 1, 1, 1, 1, 0, 1, 1)$&$3(0,1/2)\oplus(0,3/2)$\\ \hline 
 $(0, 0, 0, 1, 1, 1, 1, 1, 0, 1)$&$2(0,1/2)\oplus(0,3/2)$&$(0, 0, 0, 1, 1, 1, 2, 0, 0, 1)$&$(0,1/2)\oplus(0,3/2)$\\ \hline 
 $(0, 0, 0, 1, 1, 2, 0, 0, 0, 1)$&$(0,3/2)\oplus(0,5/2)$&$(0, 0, 0, 1, 1, 2, 0, 0, 1, 1)$&$(0,1/2)\oplus2(0,3/2)\oplus(0,5/2)$\\ \hline 
 $(0, 0, 0, 1, 1, 2, 1, 0, 0, 1)$&$(0,1/2)\oplus2(0,3/2)\oplus(0,5/2)$&$(0, 0, 0, 1, 1, 3, 0, 0, 0, 1)$&$(0,5/2)\oplus(0,7/2)$\\ \hline 
 $(0, 0, 0, 1, 2, 0, 0, 0, 0, 1)$&$(0,1/2)\oplus(0,3/2)$&$(0, 0, 0, 1, 2, 1, 0, 0, 0, 1)$&$2(0,1/2)\oplus(0,3/2)$\\ \hline 
 $(0, 0, 0, 1, 2, 1, 0, 0, 1, 1)$&$2(0,1/2)\oplus(0,3/2)$&$(0, 0, 0, 1, 2, 1, 1, 0, 0, 1)$&$2(0,1/2)\oplus(0,3/2)$\\ \hline 
 $(0, 0, 0, 1, 2, 2, 0, 0, 0, 1)$&$(0,1/2)\oplus2(0,3/2)\oplus(0,5/2)$&$(0, 0, 0, 1, 3, 0, 0, 0, 0, 1)$&$(0,3/2)\oplus(0,5/2)$\\ \hline 
 $(0, 0, 0, 1, 3, 1, 0, 0, 0, 1)$&$(0,1/2)\oplus2(0,3/2)\oplus(0,5/2)$&$(0, 0, 0, 1, 4, 0, 0, 0, 0, 1)$&$(0,5/2)\oplus(0,7/2)$\\ \hline 
 $(0, 0, 0, 2, 2, 0, 0, 0, 0, 1)$&$(0,1/2)\oplus(0,3/2)$&$(0, 0, 0, 2, 2, 1, 0, 0, 0, 1)$&$(0,1/2)\oplus(0,3/2)$\\ \hline 
 $(0, 0, 0, 2, 3, 0, 0, 0, 0, 1)$&$(0,1/2)\oplus(0,3/2)\oplus(0,5/2)$&$(0, 0, 0, 3, 2, 0, 0, 0, 0, 1)$&$(0,3/2)$\\ \hline 
 $(0, 0, 1, 1, 1, 0, 0, 0, 0, 1)$&$(0,1/2)$&$(0, 0, 1, 1, 1, 1, 0, 0, 0, 1)$&$(0,1/2)\oplus(0,3/2)$\\ \hline 
 $(0, 0, 1, 1, 1, 1, 0, 0, 1, 1)$&$2(0,1/2)\oplus(0,3/2)$&$(0, 0, 1, 1, 1, 1, 1, 0, 0, 1)$&$2(0,1/2)\oplus(0,3/2)$\\ \hline 
 $(0, 0, 1, 1, 1, 2, 0, 0, 0, 1)$&$(0,3/2)\oplus(0,5/2)$&$(0, 0, 1, 1, 2, 0, 0, 0, 0, 1)$&$(0,1/2)\oplus(0,3/2)$\\ \hline 
 $(0, 0, 1, 1, 2, 1, 0, 0, 0, 1)$&$2(0,1/2)\oplus(0,3/2)$&$(0, 0, 1, 1, 3, 0, 0, 0, 0, 1)$&$(0,3/2)\oplus(0,5/2)$\\ \hline 
 $(0, 0, 1, 2, 2, 0, 0, 0, 0, 1)$&$2(0,1/2)\oplus(0,3/2)$&$(0, 1, 1, 1, 1, 0, 0, 0, 0, 1)$&$(0,1/2)$\\ \hline 
 $(0, 1, 1, 1, 1, 1, 0, 0, 0, 1)$&$(0,1/2)\oplus(0,3/2)$&$(0, 1, 1, 1, 2, 0, 0, 0, 0, 1)$&$(0,1/2)\oplus(0,3/2)$\\ \hline 
 $(1, 1, 1, 1, 1, 0, 0, 0, 0, 1)$&$(0,1/2)$&$(0, 0, 0, 0, 0, 0, 0, 0, 3, 2)$&$(0,5/2)$\\ \hline 
 $(0, 0, 0, 0, 0, 0, 0, 0, 4, 2)$&$(0,5/2)\oplus(0,7/2)\oplus(1/2,4)$&$(0, 0, 0, 0, 0, 0, 3, 0, 0, 2)$&$(0,5/2)$\\ \hline 
 $(0, 0, 0, 0, 0, 0, 3, 1, 0, 2)$&$(0,3/2)\oplus(0,5/2)$&$(0, 0, 0, 0, 0, 0, 4, 0, 0, 2)$&$(0,5/2)\oplus(0,7/2)\oplus(1/2,4)$\\ \hline 
 $(0, 0, 0, 0, 0, 1, 0, 0, 0, 2)$&$(0,5/2)$&$(0, 0, 0, 0, 0, 1, 0, 0, 1, 2)$&$(0,3/2)\oplus(0,5/2)$\\ \hline 
 $(0, 0, 0, 0, 0, 1, 0, 0, 2, 2)$&$(0,1/2)\oplus(0,3/2)\oplus(0,5/2)$&$(0, 0, 0, 0, 0, 1, 0, 0, 3, 2)$&$(0,1/2)\oplus2(0,3/2)\oplus2(0,5/2)\oplus(0,7/2)$\\ \hline 
 $(0, 0, 0, 0, 0, 1, 1, 0, 0, 2)$&$(0,3/2)\oplus(0,5/2)$&$(0, 0, 0, 0, 0, 1, 1, 0, 1, 2)$&$(0,1/2)\oplus2(0,3/2)\oplus(0,5/2)$\\ \hline 
 $(0, 0, 0, 0, 0, 1, 1, 0, 2, 2)$&$2(0,1/2)\oplus2(0,3/2)\oplus(0,5/2)$&$(0, 0, 0, 0, 0, 1, 1, 1, 0, 2)$&$(0,3/2)\oplus(0,5/2)$\\ \hline 
 $(0, 0, 0, 0, 0, 1, 1, 1, 1, 2)$&$(0,1/2)\oplus2(0,3/2)\oplus(0,5/2)$&$(0, 0, 0, 0, 0, 1, 2, 0, 0, 2)$&$(0,1/2)\oplus(0,3/2)\oplus(0,5/2)$\\ \hline 
 $(0, 0, 0, 0, 0, 1, 2, 0, 1, 2)$&$2(0,1/2)\oplus2(0,3/2)\oplus(0,5/2)$&$(0, 0, 0, 0, 0, 1, 2, 1, 0, 2)$&$(0,1/2)\oplus2(0,3/2)\oplus(0,5/2)$\\ \hline 
 $(0, 0, 0, 0, 0, 1, 3, 0, 0, 2)$&$(0,1/2)\oplus2(0,3/2)\oplus2(0,5/2)\oplus(0,7/2)$&$(0, 0, 0, 0, 0, 2, 0, 0, 0, 2)$&$(0,5/2)\oplus(0,7/2)\oplus(1/2,4)$\\ \hline 
 $(0, 0, 0, 0, 0, 2, 0, 0, 1, 2)$&$(0,3/2)\oplus3(0,5/2)\oplus2(0,7/2)\oplus(1/2,3)\oplus(1/2,4)$&$(0, 0, 0, 0, 0, 2, 0, 0, 2, 2)$&$(0,1/2)\oplus3(0,3/2)\oplus4(0,5/2)\oplus2(0,7/2)\oplus(1/2,2)\oplus(1/2,3)\oplus(1/2,4)$\\ \hline 
 $(0, 0, 0, 0, 0, 2, 1, 0, 0, 2)$&$(0,3/2)\oplus3(0,5/2)\oplus2(0,7/2)\oplus(1/2,3)\oplus(1/2,4)$&$(0, 0, 0, 0, 0, 2, 1, 0, 1, 2)$&$(0,1/2)\oplus5(0,3/2)\oplus7(0,5/2)\oplus3(0,7/2)\oplus(1/2,2)\oplus2(1/2,3)\oplus(1/2,4)$\\ \hline 
 $(0, 0, 0, 0, 0, 2, 1, 1, 0, 2)$&$(0,3/2)\oplus3(0,5/2)\oplus2(0,7/2)\oplus(1/2,3)\oplus(1/2,4)$&$(0, 0, 0, 0, 0, 2, 2, 0, 0, 2)$&$(0,1/2)\oplus3(0,3/2)\oplus4(0,5/2)\oplus2(0,7/2)\oplus(1/2,2)\oplus(1/2,3)\oplus(1/2,4)$\\ \hline 
 $(0, 0, 0, 0, 0, 3, 0, 0, 0, 2)$&$(0,5/2)\oplus(0,7/2)\oplus2(0,9/2)\oplus(1/2,4)\oplus(1/2,5)\oplus(1,11/2)$&$(0, 0, 0, 0, 0, 3, 0, 0, 1, 2)$&$(0,3/2)\oplus3(0,5/2)\oplus5(0,7/2)\oplus3(0,9/2)\oplus(1/2,3)\oplus3(1/2,4)\oplus2(1/2,5)\oplus(1,9/2)\oplus(1,11/2)$\\ \hline 
 $(0, 0, 0, 0, 0, 3, 1, 0, 0, 2)$&$(0,3/2)\oplus3(0,5/2)\oplus5(0,7/2)\oplus3(0,9/2)\oplus(1/2,3)\oplus3(1/2,4)\oplus2(1/2,5)\oplus(1,9/2)\oplus(1,11/2)$&$(0, 0, 0, 0, 0, 4, 0, 0, 0, 2)$&$(0,5/2)\oplus(0,7/2)\oplus2(0,9/2)\oplus2(0,11/2)\oplus(1/2,4)\oplus(1/2,5)\oplus2(1/2,6)\oplus(1,11/2)\oplus(1,13/2)\oplus(3/2,7)$\\ \hline 
 $(0, 0, 0, 0, 1, 1, 0, 0, 0, 2)$&$(0,3/2)\oplus(0,5/2)$&$(0, 0, 0, 0, 1, 1, 0, 0, 1, 2)$&$(0,1/2)\oplus2(0,3/2)\oplus(0,5/2)$\\ \hline 
 $(0, 0, 0, 0, 1, 1, 0, 0, 2, 2)$&$2(0,1/2)\oplus2(0,3/2)\oplus(0,5/2)$&$(0, 0, 0, 0, 1, 1, 1, 0, 0, 2)$&$(0,1/2)\oplus2(0,3/2)\oplus(0,5/2)$\\ \hline 
 $(0, 0, 0, 0, 1, 1, 1, 0, 1, 2)$&$3(0,1/2)\oplus3(0,3/2)\oplus(0,5/2)$&$(0, 0, 0, 0, 1, 1, 1, 1, 0, 2)$&$(0,1/2)\oplus2(0,3/2)\oplus(0,5/2)$\\ \hline 
 $(0, 0, 0, 0, 1, 1, 2, 0, 0, 2)$&$2(0,1/2)\oplus2(0,3/2)\oplus(0,5/2)$&$(0, 0, 0, 0, 1, 2, 0, 0, 0, 2)$&$(0,3/2)\oplus3(0,5/2)\oplus2(0,7/2)\oplus(1/2,3)\oplus(1/2,4)$\\ \hline 
 $(0, 0, 0, 0, 1, 2, 0, 0, 1, 2)$&$(0,1/2)\oplus5(0,3/2)\oplus7(0,5/2)\oplus3(0,7/2)\oplus(1/2,2)\oplus2(1/2,3)\oplus(1/2,4)$&$(0, 0, 0, 0, 1, 2, 1, 0, 0, 2)$&$(0,1/2)\oplus5(0,3/2)\oplus7(0,5/2)\oplus3(0,7/2)\oplus(1/2,2)\oplus2(1/2,3)\oplus(1/2,4)$\\ \hline 
 $(0, 0, 0, 0, 1, 3, 0, 0, 0, 2)$&$(0,3/2)\oplus3(0,5/2)\oplus5(0,7/2)\oplus3(0,9/2)\oplus(1/2,3)\oplus3(1/2,4)\oplus2(1/2,5)\oplus(1,9/2)\oplus(1,11/2)$&$(0, 0, 0, 0, 2, 1, 0, 0, 0, 2)$&$(0,1/2)\oplus(0,3/2)\oplus(0,5/2)$\\ \hline 
 $(0, 0, 0, 0, 2, 1, 0, 0, 1, 2)$&$2(0,1/2)\oplus2(0,3/2)\oplus(0,5/2)$&$(0, 0, 0, 0, 2, 1, 1, 0, 0, 2)$&$2(0,1/2)\oplus2(0,3/2)\oplus(0,5/2)$\\ \hline 
 $(0, 0, 0, 0, 2, 2, 0, 0, 0, 2)$&$(0,1/2)\oplus3(0,3/2)\oplus4(0,5/2)\oplus2(0,7/2)\oplus(1/2,2)\oplus(1/2,3)\oplus(1/2,4)$&$(0, 0, 0, 0, 3, 0, 0, 0, 0, 2)$&$(0,5/2)$\\ \hline 
 $(0, 0, 0, 0, 3, 1, 0, 0, 0, 2)$&$(0,1/2)\oplus2(0,3/2)\oplus2(0,5/2)\oplus(0,7/2)$&$(0, 0, 0, 0, 4, 0, 0, 0, 0, 2)$&$(0,5/2)\oplus(0,7/2)\oplus(1/2,4)$\\ \hline 
 $(0, 0, 0, 1, 1, 1, 0, 0, 0, 2)$&$(0,3/2)\oplus(0,5/2)$&$(0, 0, 0, 1, 1, 1, 0, 0, 1, 2)$&$(0,1/2)\oplus2(0,3/2)\oplus(0,5/2)$\\ \hline 
 $(0, 0, 0, 1, 1, 1, 1, 0, 0, 2)$&$(0,1/2)\oplus2(0,3/2)\oplus(0,5/2)$&$(0, 0, 0, 1, 1, 2, 0, 0, 0, 2)$&$(0,3/2)\oplus3(0,5/2)\oplus2(0,7/2)\oplus(1/2,3)\oplus(1/2,4)$\\ \hline 
 $(0, 0, 0, 1, 2, 1, 0, 0, 0, 2)$&$(0,1/2)\oplus2(0,3/2)\oplus(0,5/2)$&$(0, 0, 0, 1, 3, 0, 0, 0, 0, 2)$&$(0,3/2)\oplus(0,5/2)$\\ \hline 
 $(0, 0, 1, 1, 1, 1, 0, 0, 0, 2)$&$(0,3/2)\oplus(0,5/2)$&$(0, 0, 0, 0, 0, 1, 0, 0, 0, 3)$&$(0,7/2)$\\ \hline 
 $(0, 0, 0, 0, 0, 1, 0, 0, 1, 3)$&$(0,5/2)\oplus(0,7/2)$&$(0, 0, 0, 0, 0, 1, 0, 0, 2, 3)$&$(0,3/2)\oplus(0,5/2)\oplus(0,7/2)$\\ \hline 
 $(0, 0, 0, 0, 0, 1, 1, 0, 0, 3)$&$(0,5/2)\oplus(0,7/2)$&$(0, 0, 0, 0, 0, 1, 1, 0, 1, 3)$&$(0,3/2)\oplus2(0,5/2)\oplus(0,7/2)$\\ \hline 
 $(0, 0, 0, 0, 0, 1, 1, 1, 0, 3)$&$(0,5/2)\oplus(0,7/2)$&$(0, 0, 0, 0, 0, 1, 2, 0, 0, 3)$&$(0,3/2)\oplus(0,5/2)\oplus(0,7/2)$\\ \hline 
 $(0, 0, 0, 0, 0, 2, 0, 0, 0, 3)$&$(0,5/2)\oplus(0,7/2)\oplus2(0,9/2)\oplus(1/2,4)\oplus(1/2,5)\oplus(1,11/2)$&$(0, 0, 0, 0, 0, 2, 0, 0, 1, 3)$&$(0,3/2)\oplus3(0,5/2)\oplus5(0,7/2)\oplus3(0,9/2)\oplus(1/2,3)\oplus3(1/2,4)\oplus2(1/2,5)\oplus(1,9/2)\oplus(1,11/2)$\\ \hline 
 $(0, 0, 0, 0, 0, 2, 1, 0, 0, 3)$&$(0,3/2)\oplus3(0,5/2)\oplus5(0,7/2)\oplus3(0,9/2)\oplus(1/2,3)\oplus3(1/2,4)\oplus2(1/2,5)\oplus(1,9/2)\oplus(1,11/2)$&$(0, 0, 0, 0, 0, 3, 0, 0, 0, 3)$&$(0,3/2)\oplus(0,5/2)\oplus3(0,7/2)\oplus3(0,9/2)\oplus4(0,11/2)\oplus(1/2,3)\oplus2(1/2,4)\oplus3(1/2,5)\oplus3(1/2,6)\oplus(1/2,7)\oplus(1,9/2)\oplus2(1,11/2)\oplus3(1,13/2)\oplus(3/2,6)\oplus(3/2,7)\oplus(2,15/2)$\\ \hline 
 $(0, 0, 0, 0, 1, 1, 0, 0, 0, 3)$&$(0,5/2)\oplus(0,7/2)$&$(0, 0, 0, 0, 1, 1, 0, 0, 1, 3)$&$(0,3/2)\oplus2(0,5/2)\oplus(0,7/2)$\\ \hline 
 $(0, 0, 0, 0, 1, 1, 1, 0, 0, 3)$&$(0,3/2)\oplus2(0,5/2)\oplus(0,7/2)$&$(0, 0, 0, 0, 1, 2, 0, 0, 0, 3)$&$(0,3/2)\oplus3(0,5/2)\oplus5(0,7/2)\oplus3(0,9/2)\oplus(1/2,3)\oplus3(1/2,4)\oplus2(1/2,5)\oplus(1,9/2)\oplus(1,11/2)$\\ \hline 
 $(0, 0, 0, 0, 2, 1, 0, 0, 0, 3)$&$(0,3/2)\oplus(0,5/2)\oplus(0,7/2)$&$(0, 0, 0, 1, 1, 1, 0, 0, 0, 3)$&$(0,5/2)\oplus(0,7/2)$\\ \hline 
 $(0, 0, 0, 0, 0, 1, 0, 0, 0, 4)$&$(0,9/2)$&$(0, 0, 0, 0, 0, 1, 0, 0, 1, 4)$&$(0,7/2)\oplus(0,9/2)$\\ \hline 
 $(0, 0, 0, 0, 0, 1, 1, 0, 0, 4)$&$(0,7/2)\oplus(0,9/2)$&$(0, 0, 0, 0, 0, 2, 0, 0, 0, 4)$&$(0,5/2)\oplus(0,7/2)\oplus2(0,9/2)\oplus2(0,11/2)\oplus(1/2,4)\oplus(1/2,5)\oplus2(1/2,6)\oplus(1,11/2)\oplus(1,13/2)\oplus(3/2,7)$\\ \hline 
 $(0, 0, 0, 0, 1, 1, 0, 0, 0, 4)$&$(0,7/2)\oplus(0,9/2)$&$(0, 0, 0, 0, 0, 1, 0, 0, 0, 5)$&$(0,11/2)$\\ \hline 
  \end{supertabular}}
}

\begin{landscape}
\begin{table}
$F_4$
\vskip 10pt
\begin{center} 
\footnotesize{
\setlength\tabcolsep{1.5pt}
\begin{tabular} {|c|ccccccccccccccccccccccccccccccccccc|} \hline 
$2j_L \backslash 2j_R$  & 0&1&2&3&4&5&6&7&8&9&10&11&12&13&14&15&16&17&18&19&20&21&22&23&24&25&26&27&28&29&30&31&32&33&34 \\  \hline0&&534&&1076&&1649&&2271&&2920&&3495&&3828&&3726&&3081&&2055&&1051&&398&&102&&12&&1&&&&&\\ 1&244&&773&&1412&&2209&&3175&&4247&&5241&&5848&&5705&&4599&&2874&&1326&&431&&88&&8&&&&&&\\ 2&&248&&593&&1117&&1871&&2866&&4050&&5232&&6043&&5997&&4782&&2841&&1185&&332&&56&&3&&&&&\\ 3&39&&146&&341&&688&&1252&&2075&&3140&&4293&&5179&&5278&&4197&&2379&&902&&218&&28&&1&&&&\\ 4&&21&&64&&160&&355&&710&&1285&&2105&&3079&&3919&&4136&&3306&&1802&&617&&125&&12&&&&&\\ 5&1&&6&&22&&64&&160&&356&&712&&1276&&2016&&2728&&2999&&2418&&1266&&390&&66&&4&&&&\\ 6&&&&1&&6&&22&&64&&160&&355&&702&&1210&&1757&&2028&&1660&&839&&229&&30&&1&&&\\ 7&&&&&&&1&&6&&22&&64&&160&&352&&670&&1058&&1295&&1079&&522&&123&&12&&&&\\ 8&&&&&&&&&&1&&6&&22&&64&&159&&338&&588&&772&&660&&308&&62&&4&&&\\ 9&&&&&&&&&&&&&1&&6&&22&&64&&155&&302&&433&&384&&172&&28&&1&&\\ 10&&&&&&&&&&&&&&&&1&&6&&22&&63&&141&&224&&207&&89&&11&&&\\ 11&&&&&&&&&&&&&&&&&&&1&&6&&22&&59&&107&&105&&44&&4&&\\ 12&&&&&&&&&&&&&&&&&&&&&&1&&6&&21&&46&&49&&20&&1&\\ 13&&&&&&&&&&&&&&&&&&&&&&&&&1&&6&&17&&20&&8&&\\ 14&&&&&&&&&&&&&&&&&&&&&&&&&&&&1&&5&&7&&3&\\ 15&&&&&&&&&&&&&&&&&&&&&&&&&&&&&&&1&&2&&1\\ \hline 
\end{tabular} \vskip 3pt  $\beta=(1, 8, 1, 0, 0, 4)$ \vskip 10pt  
} 
\footnotesize{
\setlength\tabcolsep{1.5pt}
\begin{tabular} {|c|cccccccccccccccccccccccccccccc|} \hline 
$2j_L \backslash 2j_R$  & 0&1&2&3&4&5&6&7&8&9&10&11&12&13&14&15&16&17&18&19&20&21&22&23&24&25&26&27&28&29 \\  \hline0&&630&&1287&&2015&&2793&&3492&&3828&&3528&&2572&&1394&&528&&124&&18&&&&&&\\ 1&222&&743&&1456&&2415&&3568&&4687&&5303&&4886&&3425&&1701&&556&&107&&9&&&&&\\ 2&&177&&466&&975&&1783&&2878&&4064&&4835&&4553&&3137&&1455&&414&&62&&4&&&&\\ 3&19&&79&&216&&504&&1034&&1854&&2857&&3625&&3518&&2402&&1048&&258&&31&&1&&&\\ 4&&6&&25&&80&&216&&504&&1016&&1730&&2365&&2390&&1633&&678&&144&&12&&&&\\ 5&&&1&&6&&25&&80&&216&&495&&943&&1406&&1489&&1022&&402&&71&&4&&&\\ 6&&&&&&1&&6&&25&&80&&212&&457&&752&&843&&586&&221&&32&&1&&\\ 7&&&&&&&&&1&&6&&25&&79&&198&&366&&438&&311&&112&&12&&&\\ 8&&&&&&&&&&&&1&&6&&25&&75&&159&&205&&150&&53&&4&&\\ 9&&&&&&&&&&&&&&&1&&6&&24&&61&&86&&66&&23&&1&\\ 10&&&&&&&&&&&&&&&&&&1&&6&&20&&31&&25&&9&&\\ 11&&&&&&&&&&&&&&&&&&&&&1&&5&&9&&8&&3&\\ 12&&&&&&&&&&&&&&&&&&&&&&&&1&&2&&2&&1\\ \hline 
\end{tabular} \vskip 3pt  $\beta=(2, 7, 1, 0, 0, 4)$ \vskip 10pt  
}
\caption{Refined BPS invariants for selected degrees of 6d $F_4$ minimal SCFT.}
\label{BPStable}
 \end{center}
\end{table}
\end{landscape}

\begin{landscape}
\begin{table}
$E_6$
\vskip 10pt
\begin{center}
\tiny{
\setlength\tabcolsep{1.5pt}
\begin{tabular} {|c|ccccccccccccccccccccccccccccccccccc|} \hline
$2j_L \backslash 2j_R$  & 0&1&2&3&4&5&6&7&8&9&10&11&12&13&14&15&16&17&18&19&20&21&22&23&24&25&26&27&28&29&30&31&32&33&34 \\  \hline0&&546&&1101&&1688&&2327&&2991&&3578&&3913&&3807&&3142&&2077&&1063&&400&&99&&15&&&&&&&\\ 1&245&&776&&1420&&2225&&3204&&4293&&5306&&5927&&5788&&4668&&2906&&1335&&435&&88&&8&&&&&&\\ 2&&249&&596&&1126&&1889&&2900&&4104&&5310&&6137&&6095&&4859&&2868&&1194&&336&&53&&4&&&&&\\ 3&39&&146&&341&&689&&1255&&2084&&3159&&4329&&5234&&5347&&4260&&2405&&904&&219&&28&&1&&&&\\ 4&&21&&64&&160&&356&&713&&1294&&2123&&3113&&3969&&4198&&3359&&1817&&618&&128&&11&&&&&\\ 5&1&&6&&22&&64&&160&&356&&713&&1279&&2025&&2746&&3029&&2451&&1280&&390&&66&&4&&&&\\ 6&&&&1&&6&&22&&64&&160&&356&&705&&1219&&1774&&2055&&1686&&845&&227&&31&&1&&&\\ 7&&&&&&&1&&6&&22&&64&&160&&352&&671&&1061&&1303&&1090&&527&&123&&12&&&&\\ 8&&&&&&&&&&1&&6&&22&&64&&159&&339&&591&&780&&670&&311&&61&&4&&&\\ 9&&&&&&&&&&&&&1&&6&&22&&64&&155&&302&&434&&386&&173&&28&&1&&\\ 10&&&&&&&&&&&&&&&&1&&6&&22&&63&&141&&225&&209&&90&&11&&&\\ 11&&&&&&&&&&&&&&&&&&&1&&6&&22&&59&&107&&105&&44&&4&&\\ 12&&&&&&&&&&&&&&&&&&&&&&1&&6&&21&&46&&49&&20&&1&\\ 13&&&&&&&&&&&&&&&&&&&&&&&&&1&&6&&17&&20&&8&&\\ 14&&&&&&&&&&&&&&&&&&&&&&&&&&&&1&&5&&7&&3&\\ 15&&&&&&&&&&&&&&&&&&&&&&&&&&&&&&&1&&2&&1\\ \hline
\end{tabular} \vskip 3pt  \footnotesize{$\beta=(0, 1, 0, 1, 4, 0, 0, 6)$} \vskip 10pt
}
\tiny{
\setlength\tabcolsep{1.5pt}
\begin{tabular} {|c|cccccccccccccccccccccccccccccccccccc|} \hline
$2j_L \backslash 2j_R$  & 0&1&2&3&4&5&6&7&8&9&10&11&12&13&14&15&16&17&18&19&20&21&22&23&24&25&26&27&28&29&30&31&32&33&34&35 \\  \hline0&&992&&1980&&3001&&4100&&5241&&6249&&6824&&6654&&5557&&3777&&2005&&783&&208&&37&&2&&&&&&\\ 1&470&&1483&&2690&&4158&&5899&&7804&&9553&&10592&&10299&&8357&&5332&&2551&&872&&193&&24&&1&&&&&\\ 2&&527&&1227&&2253&&3686&&5532&&7677&&9776&&11154&&10971&&8765&&5308&&2319&&695&&125&&12&&&&&&\\ 3&96&&342&&755&&1458&&2558&&4112&&6067&&8136&&9660&&9729&&7748&&4493&&1791&&470&&69&&4&&&&&\\ 4&&59&&166&&382&&797&&1522&&2653&&4203&&5986&&7449&&7724&&6157&&3430&&1251&&285&&31&&1&&&&\\ 5&5&&22&&65&&168&&387&&807&&1531&&2624&&4003&&5262&&5657&&4543&&2444&&809&&155&&12&&&&&\\ 6&&1&&6&&22&&65&&168&&387&&805&&1505&&2481&&3472&&3896&&3164&&1638&&485&&77&&4&&&&\\ 7&&&&&1&&6&&22&&65&&168&&386&&791&&1420&&2135&&2518&&2074&&1035&&272&&34&&1&&&\\ 8&&&&&&&&1&&6&&22&&65&&168&&382&&754&&1231&&1541&&1297&&625&&143&&13&&&&\\ 9&&&&&&&&&&&1&&6&&22&&65&&167&&368&&661&&889&&768&&356&&69&&4&&&\\ 10&&&&&&&&&&&&&&1&&6&&22&&65&&163&&328&&480&&429&&191&&30&&1&&\\ 11&&&&&&&&&&&&&&&&&1&&6&&22&&64&&148&&241&&225&&97&&12&&&\\ 12&&&&&&&&&&&&&&&&&&&&1&&6&&22&&60&&112&&112&&47&&4&&\\ 13&&&&&&&&&&&&&&&&&&&&&&&1&&6&&21&&46&&49&&20&&1&\\ 14&&&&&&&&&&&&&&&&&&&&&&&&&&1&&6&&17&&20&&8&&\\ 15&&&&&&&&&&&&&&&&&&&&&&&&&&&&&1&&5&&7&&3&\\ 16&&&&&&&&&&&&&&&&&&&&&&&&&&&&&&&&1&&2&&1\\ \hline
\end{tabular} \vskip 3pt  \footnotesize{$\beta=(0, 1, 0, 1, 5, 0, 0, 5)$} \vskip 10pt
}
\caption{Refined BPS invariants for selected degrees of 6d $E_6$ minimal SCFT.}
\label{BPStable}
\end{center}
\end{table}
\end{landscape}

\begin{landscape}
\begin{table}
$E_7$
\vskip 10pt
\begin{center}
\footnotesize{
\setlength\tabcolsep{1.5pt}
\begin{tabular} {|c|cccccccccccccccccccccccccccccc|} \hline
$2j_L \backslash 2j_R$  & 0&1&2&3&4&5&6&7&8&9&10&11&12&13&14&15&16&17&18&19&20&21&22&23&24&25&26&27&28&29 \\  \hline0&&647&&1327&&2079&&2891&&3608&&3957&&3627&&2627&&1417&&531&&131&&14&&1&&&&\\ 1&223&&746&&1466&&2436&&3611&&4754&&5396&&4980&&3489&&1728&&558&&108&&9&&&&&\\ 2&&178&&469&&986&&1806&&2928&&4141&&4945&&4656&&3199&&1476&&412&&66&&3&&&&\\ 3&19&&79&&216&&505&&1037&&1865&&2880&&3670&&3573&&2444&&1066&&258&&31&&1&&&\\ 4&&6&&25&&80&&217&&507&&1027&&1752&&2408&&2438&&1665&&688&&141&&13&&&&\\ 5&&&1&&6&&25&&80&&216&&496&&946&&1416&&1504&&1035&&408&&71&&4&&&\\ 6&&&&&&1&&6&&25&&80&&213&&460&&762&&857&&597&&225&&31&&1&&\\ 7&&&&&&&&&1&&6&&25&&79&&198&&367&&440&&313&&113&&12&&&\\ 8&&&&&&&&&&&&1&&6&&25&&75&&160&&207&&152&&54&&4&&\\ 9&&&&&&&&&&&&&&&1&&6&&24&&61&&86&&66&&23&&1&\\ 10&&&&&&&&&&&&&&&&&&1&&6&&20&&31&&25&&9&&\\ 11&&&&&&&&&&&&&&&&&&&&&1&&5&&9&&8&&3&\\ 12&&&&&&&&&&&&&&&&&&&&&&&&1&&2&&2&&1\\ \hline
\end{tabular} \vskip 3pt  \footnotesize{$\beta=(0, 0, 0, 5, 1, 0, 0, 2, 4)$} \vskip 10pt
}
\tiny{
\setlength\tabcolsep{1.5pt}
\begin{tabular} {|c|cccccccccccccccccccccccccccccccccccc|} \hline
$2j_L \backslash 2j_R$  & 0&1&2&3&4&5&6&7&8&9&10&11&12&13&14&15&16&17&18&19&20&21&22&23&24&25&26&27&28&29&30&31&32&33&34&35 \\  \hline0&&992&&1980&&3001&&4100&&5241&&6249&&6824&&6654&&5557&&3777&&2005&&783&&208&&37&&2&&&&&&\\ 1&470&&1483&&2690&&4158&&5899&&7804&&9553&&10592&&10299&&8357&&5332&&2551&&872&&193&&24&&1&&&&&\\ 2&&527&&1227&&2253&&3686&&5532&&7677&&9776&&11154&&10971&&8765&&5308&&2319&&695&&125&&12&&&&&&\\ 3&96&&342&&755&&1458&&2558&&4112&&6067&&8136&&9660&&9729&&7748&&4493&&1791&&470&&69&&4&&&&&\\ 4&&59&&166&&382&&797&&1522&&2653&&4203&&5986&&7449&&7724&&6157&&3430&&1251&&285&&31&&1&&&&\\ 5&5&&22&&65&&168&&387&&807&&1531&&2624&&4003&&5262&&5657&&4543&&2444&&809&&155&&12&&&&&\\ 6&&1&&6&&22&&65&&168&&387&&805&&1505&&2481&&3472&&3896&&3164&&1638&&485&&77&&4&&&&\\ 7&&&&&1&&6&&22&&65&&168&&386&&791&&1420&&2135&&2518&&2074&&1035&&272&&34&&1&&&\\ 8&&&&&&&&1&&6&&22&&65&&168&&382&&754&&1231&&1541&&1297&&625&&143&&13&&&&\\ 9&&&&&&&&&&&1&&6&&22&&65&&167&&368&&661&&889&&768&&356&&69&&4&&&\\ 10&&&&&&&&&&&&&&1&&6&&22&&65&&163&&328&&480&&429&&191&&30&&1&&\\ 11&&&&&&&&&&&&&&&&&1&&6&&22&&64&&148&&241&&225&&97&&12&&&\\ 12&&&&&&&&&&&&&&&&&&&&1&&6&&22&&60&&112&&112&&47&&4&&\\ 13&&&&&&&&&&&&&&&&&&&&&&&1&&6&&21&&46&&49&&20&&1&\\ 14&&&&&&&&&&&&&&&&&&&&&&&&&&1&&6&&17&&20&&8&&\\ 15&&&&&&&&&&&&&&&&&&&&&&&&&&&&&1&&5&&7&&3&\\ 16&&&&&&&&&&&&&&&&&&&&&&&&&&&&&&&&1&&2&&1\\ \hline
\end{tabular} \vskip 3pt  \footnotesize{$\beta=(0, 0, 1, 5, 1, 0, 0, 0, 5)$} \vskip 10pt
}
\caption{Refined BPS invariants for selected degrees of 6d $E_7$ minimal SCFT.}
\label{BPStable}
\end{center}
\end{table}
\end{landscape}

\begin{landscape}
\begin{table}
$E_8$
\vskip 10pt
\begin{center} 
{
\setlength\tabcolsep{1.5pt}
\begin{tabular} {|c|cccccccccccccccccccc|} \hline 
$2j_L \backslash 2j_R$  & 0&1&2&3&4&5&6&7&8&9&10&11&12&13&14&15&16&17&18&19 \\  \hline0&&3&&7&&13&&21&&27&&28&&17&&4&&1&&\\ 1&&&1&&4&&10&&19&&28&&31&&19&&4&&&\\ 2&&&&&&1&&4&&11&&20&&25&&15&&2&&\\ 3&&&&&&&&&1&&4&&10&&15&&9&&1&\\ 4&&&&&&&&&&&&1&&4&&8&&5&&\\ 5&&&&&&&&&&&&&&&1&&3&&2&\\ 6&&&&&&&&&&&&&&&&&&1&&1\\ \hline 
\end{tabular} \vskip 3pt  $\beta=(1, 1, 1, 1, 1, 3, 0, 0, 0, 4)$ \vskip 10pt  
} 
{
\setlength\tabcolsep{1.5pt}
\begin{tabular} {|c|cccccccccccccccc|} \hline 
$2j_L \backslash 2j_R$  & 0&1&2&3&4&5&6&7&8&9&10&11&12&13&14&15 \\  \hline0&&17&&44&&79&&91&&63&&23&&2&&\\ 1&1&&6&&22&&51&&68&&50&&17&&1&\\ 2&&&&1&&6&&20&&31&&25&&9&&\\ 3&&&&&&&1&&5&&9&&8&&3&\\ 4&&&&&&&&&&1&&2&&2&&1\\ \hline 
\end{tabular} \vskip 3pt  $\beta=(0, 1, 1, 1, 1, 3, 0, 0, 2, 3)$ \vskip 10pt  
}
\caption{Refined BPS invariants for selected degrees of 6d $E_8$ minimal SCFT.}
\label{BPStable}
\end{center}
\end{table}
\end{landscape}



\bibliographystyle{JHEP}
\bibliography{blowupscft}

\end{document}